\documentstyle[vi,psbox,twoside]{mitthesis}

\newcommand{\simlt}{{\stackrel{<}{_\sim}}}
\newcommand{\simgt}{{\stackrel{>}{_\sim}}}
\newcommand{\msun}{${\rm M}_\odot$}
\newcommand{\rsun}{${\rm R}_\odot$}
\newcommand{\mdot}{$\dot{M}$}

\newcommand{\mspyr}{${\rm M}_\odot$ yr$^{-1}$}
\newcommand{\scinot}[2]{${#1}{\times}10^{{#2}}$}
\newcommand{\mscinot}[2]{{{#1}{\times}10^{{#2}}}}
\newcommand{\tento}[1]{$10^{{#1}}$}
\newcommand{\e}[1]{\times 10^{{#1}}}
\newcommand{\lgder}[1]{\frac{{\delta}{#1}}{{#1}}}

\newcommand{\comprod}[2]{{#1}^{\ast}{#2}}
\def\la{\mathrel{\hbox{\rlap{\hbox{\lower4pt\hbox{$\sim$}}}\hbox{$<$}}}}
\def\ga{\mathrel{\hbox{\rlap{\hbox{\lower4pt\hbox{$\sim$}}}\hbox{$>$}}}}

\newcommand{\ave}[1]{\langle{#1}\rangle}
\newcommand{\RXTE}{{\it RXTE\,}}
\newcommand{\EXOSAT}{{\it EXOSAT\,}}
\newcommand{\ASCA}{{\it ASCA\,}}
\newcommand{\ROSAT}{{\it ROSAT\,}}
\newcommand{\Ginga}{{\it Ginga\,}}
\newcommand{\intens}[2]{I\,[#1--#2~keV]}
\newcommand{\softcolor}{\intens{4.8}{6.3}~/ \intens{2.0}{4.8}}
\newcommand{\broadcolor}{\intens{6.3}{13}~/ \intens{2.0}{6.3}}
\newcommand{\hardcolor}{\intens{13}{18}~/ \intens{8.5}{13}}
\newcommand{\hardness}[4]{\intens{#3}{#4}~/ \intens{#1}{#2}}

\begin{document}

% -*-latex-*-
% $Log: cover.tex,v $
% Revision 1.3  93/05/17  17:06:29  starflt
% Added acknowledgements section (suggested by tompalka)
%
%
% Revision 1.2  92/04/22  13:13:13  epeisach
% Fixes for 1991 course 6 requirements
% Phrase "and to grant others the right to do so" has been added to
% permission clause
% Second copy of abstract is not counted as separate pages so numbering works
% out
%
% Revision 1.1  92/04/22  13:08:20  epeisach

\title{
Mass Transfer and Accretion in the Eccentric Neutron-Star Binary Circinus~X-1
}

\author{Robert E. Shirey}
\department{Department of Physics}
% If the thesis is for two degrees simultaneously, list them both separated by
%% \and like this:
% \degree{Doctor of Philosophy \and Master of Science}
\degree{Doctor of Philosophy}
\degreemonth{September}
\degreeyear{1998}
\thesisdate{June 19, 1998}

%% By default, the thesis will be copyrighted to MIT.  If you need to
%% copyright the thesis to yourself, just specify the `vi' documentstyle
%% option.  If for some reason you want to exactly specify the copyright
%% notice text, you can use the \copyrightnoticetext command.
%\copyrightnoticetext{\copyright IBM, 1990.  Do not open till Xmas.}

% If there is more than one supervisor, use the \supervisor command once for
%% each.
\supervisor{Hale V. D. Bradt}{Professor of Physics}

% this is the department committee chairman, not the thesis committee chairman
%\chairman{George Koster}{Chairman, Departmental Committee on Graduate
%Students}
% New for Physics as of 06 Apr 1998:
\chairman{Thomas J. Greytak}{Associate Department Head for Education}

% Make the titlepage based on the above information.  If you need something
% special and can't use the standard form, you can specify the exact text of
% the titlepage yourself.  Put it in a titlepage environment and leave blank
% lines where you want vertical space. The spaces will be adjusted to fill
% the entire page. The dotted lines for the signatures are made with the
% \signature command.
\maketitle

% The abstractpage environment sets up everything on the page except the
% text itself.  The title and other header material are put at the top
% of the page, and the supervisors are listed at the bottom.  A new page
% is begun both before and after.  Of course, an abstract may be more
% than one page itself.  If you need more control over the format of the
% page, you can use the abstract environment, which puts the word
% "Abstract" at the beginning and single spaces its text.

%% You can either \input (*not* \include) your abstract file, or you can put
%% the text of the abstract directly between the \begin{abstractpage} and
%% \end{abstractpage} commands.

% First copy: start a new page, and save the page number.
\newpage
\pagestyle{empty}
\setcounter{savepage}{\thepage}
\begin{abstractpage}
% $Log: abstract.tex,v $
% Revision 1.1  93/05/14  14:56:25  starflt
% Initial revision
% 
% Revision 1.1  90/05/04  10:41:01  lwvanels
% Initial revision
% 
%
%% The text of your abstract and nothing else (other than comments) goes here.
%% It will be single-spaced and the rest of the text that is supposed to go on
%% the abstract page will be generated by the abstractpage environment.  This
%% file should be \input (not \include 'd) from cover.tex.
%%%%%%%%%%%%%%%%%%%%%%%%%%%%%%%%%%%%%%%%%%%%%%%%%%%%%%%%%%%%%%%%%%%%%%%%%%%%

I have carried out a project to study the eccentric neutron-star
binary Circinus~X-1 through an extensive series of observational
studies with the {\em Rossi X-ray Timing Explorer} satellite and
through theoretical computer models I developed to explore mass
transfer and evolution in an eccentric binary.  We also organized two
multi-frequency campaigns to study correlated variability in different
frequency bands.

The X-ray observations showed that the intensity of Cir~X-1 currently
maintains a bright baseline level, with strong flares occurring after
phase zero of each 16.55-day cycle of the source. This behavior is
thought to be due to enhanced mass transfer occurring near periastron
of a highly eccentric binary orbit. Dips below the baseline intensity
level also occur near phase zero.

I modeled the evolution of the energy spectrum during dips with a
variably absorbed bright component plus a fainter unabsorbed
component.  I show that variability not attributable to absorption
dips is related to the spectral/intensity states of the ``Z~source''
class of low-mass X-ray binaries (LMXBs), namely motion along (or
shifts of) the horizontal, normal, and flaring branches of the ``Z''
track in color-color and hardness-intensity diagrams.

I found rapid X-ray variability properties associated with
each spectral/intensity state: On the horizontal branch,
quasi-periodic oscillations (QPOs) in the X-ray intensity shift in
frequency from 1.3 to 35~Hz. On the normal branch, a different QPO
occurs at about 4~Hz. On the flaring branch only strong aperiodic
variability occurs. I modeled the evolution of the energy
spectra associated with each of these branches.

To study mass transfer in an eccentric binary, I developed computer
codes for transfer via Roche-lobe overflow and from a stellar wind.  I
derive theoretical mass accretion profiles and compare them to the
observed profile of the X-ray intensity.

In order to explore the possible evolutionary history of Circinus~X-1,
I developed a binary-evolution computer code for a neutron-star and 
low-mass companion in an eccentric orbit. I use this code in a
population-synthesis study to show that the number of systems in the
Galaxy expected to resemble Cir~X-1 is of order unity, consistent with
its unique status as an LMXB with high eccentricity.

%%%%%%%%%%%%%%%%%%%%%%%%%%%%%%%%%%%%%%%%%%%%%%%%%%%%%%%%%%%%%%%%%%%%%%%%%%%%%

%Radio flux measurements showed marginal evidence for flares at
%about the same time as those in the X-ray band, and infrared
%photometry measurements showed a 1.3 magnitude increase (K~band)
%within a day of the onset of X-ray flaring.  Infrared and optical
%spectroscopy showed Doppler-shifted emission lines that may be related
%to the radial velocity of the system and/or the motion due to the
%binary orbit.

\end{abstractpage}
\newpage

% Second copy: start a new page, and reset the page number.  This way,
% the second copy of the abstract is not counted as separate pages.
% \newpage
% \setcounter{page}{\thesavepage}
% \begin{abstractpage}
% \input{abstract}
% \end{abstractpage}

\newpage

\section*{Acknowledgments}
First, I would like to thank Prof.\ Hale Bradt for giving me the
opportunity to work with him and participate in the {\em Rossi X-ray
Timing Explorer} project. He has demonstrated how to think like a
scientist and produce clear scientific writings. He has always
supported me and encouraged my efforts.

I also want to thank Prof.\ Saul Rappaport for helping with the
several theoretical modeling projects in my thesis.  Although I was
initially reluctant to delve into theoretical calculations, I
eventually found that they were not to be feared and could be quite
interesting.

Ed Morgan got me started on Circinus~X-1 several years ago by
suggesting I work on a proposal to observe Cir~X-1 with \RXTE\ once it
launched. I also learned many analysis techniques from Ed.

Al Levine has provided much-appreciated scientific guidance,
especially in writing papers. His participation has certainly
improved the quality of my work.

In fact, all the members of the \RXTE\ team at M.I.T.\ have
contributed to my education and to this project. Ron Remillard, Wei
Cui, and Deepto Chakrabarty have all given helpful advice and shared
experiences with analyzing and interpreting \RXTE\ data.

The people I've interacted with most at M.I.T. are the other graduate
students, and my office mates in particular. Don Smith came to
M.I.T. at the same time I did, so we've shared all the ups and downs
of the experience. He, Linqing Wen, and Mike Muno have all been great
to work with and to discuss ideas with around the office.  Our former
office mates Charlie Collins and Chris Becker provided a wealth of
knowledge and experience about M.I.T., Boston, and astrophysics.  In
particular my binary evolution code was based on one Chris developed
as part of his thesis.

I would like to thank all the members of our multi-frequency campaigns
for all their efforts to coordinate observations and willingness to
share data: George Nicolson, Ian Glass, Allyn Tennant, Rob Fender,
Kinwah Wu, and Helen Johnston.

My parents have supported my interest in astronomy from the
beginning. They gave me a 6-inch diameter telescope for Christmas when
I was in elementary school (and then let me sell it several years
later when an Atari system seemed more important).

Finally, I want to thank my wife, Anne, for supporting me in so many
ways throughout the whole process. When we were married, exactly five
years ago, I'm sure she didn't know what she was getting into. She has
made many sacrifices to help make this happen.  Now five years and two
children (Ben and Alex) later, we've finally reached this milestone.
I thank her for her patience, support, and love. This thesis is for
Anne.

\begin{flushleft}
\hspace*{4.5in} Robert Shirey \hfill \linebreak
\hspace*{4.5in} Cambridge, MA \hfill \linebreak
\hspace*{4.5in} June 19, 1998 \hfill \linebreak
\end{flushleft}

\pagestyle{headings}
  % -*- Mode:TeX -*-
%% This file simply contains the commands that actually generate the table of
%% contents and lists of figures and tables.  You can omit any or all of
%% these files by simply taking out the appropriate command.  For more
%% information on these files, see appendix C.3.3 of the LaTeX manual. 
\tableofcontents
\newpage
\listoffigures
\newpage
\listoftables

\chapter{Introduction}
\label{ch:intro}

\section{X-ray Astronomy}

% History
X-rays are strongly absorbed by the Earth's atmosphere, making
ground-based observations of astronomical X-ray sources impossible.
Thus, X-ray astronomy did not begin until rockets and high-altitude
balloons became available to carry instruments above a significant
fraction of the atmosphere.  The first celestial X-ray source (other
than the Sun), Scorpius~X-1, was discovered in 1962 using a
rocket-borne detector~\cite{giaconni62}. Additional X-ray sources were
first detected by balloon and rocket experiments during the 1960's.

% X-ray satellites
Many more sources were discovered with the launch of the first
astronomical X-ray satellite, {\em Uhuru}, in 1970. This was followed
by a series of satellites in the 1970's, including {\em Vela~5}, {\em
Copernicus}, {\em Ariel~V}, {\em SAS-3}, {\em OSO-7}, {\em OSO-8},
{\em COS-B}, {\em HEAO-1}, {\em Einstein}, and {\em Hakucho}. In the
1980's {\em Tenma}, \EXOSAT, \Ginga, and {\em Mir-Kvant} were
launched.  Missions in the 1990's include {\em Granat}, {\em ROSAT}
and \ASCA, and most recently, \RXTE\ and {\em BeppoSAX}.  In the next
few years, several major X-ray spectroscopy and imaging missions will
be launched, including {\em AXAF}, {\em XMM}, and {\em Spectrum-X-Gamma}.

\section{X-ray Binaries}

\subsection{Overview}
Most of the bright celestial X-ray sources are thought to consist of a
neutron star or black hole accreting matter from a binary companion
star (see the book {\em X-ray Binaries}~\cite{xrb:book} for a thorough
review of the subject). The gravitational potential energy released
when matter falls onto these highly compact objects produces
high-luminosity radiation (of order \tento{37}--\tento{38}~erg/s) from
a small area ($\sim$10~km radius), resulting in an effective
temperature of
\tento{6}--\tento{7}~K, i.e., primarily X-ray wavelengths
($\lambda\sim$0.1--100~\AA, $E\sim$0.1--100~keV).

% Orbital periods
The binary orbital period can be well established through Doppler
studies of X-ray pulsars (see section~\ref{sec:hmxb} below).  The
period is sometimes detected through modulation of the intensity in
X-ray, optical, or other frequency bands. X-ray binaries have periods
as short as minutes to as long as months.
% Mass function
When the radial component of the orbital velocity of the companion is
also measured, from Doppler-shifted lines in the optical spectrum, a
lower limit to the mass of the compact object (the mass function) can
be derived. A more exact estimate requires knowledge of the mass of
the companion (e.g. measured from Doppler shifts of pulsed X-ray
emission) and the inclination angle of the orbit relative to the line
of sight. Even a lower limit can be very informative, since degeneracy
pressure in a neutron star is not expected to be able to support more
than about 3 times the mass of the sun (\msun). Systems with a mass
function in excess of 3~\msun\ are considered strong candidate
black-holes.

% HMXBs vs LMXBs
X-ray binaries are generally classified into two groups, based on the
spectral type (observed or inferred) of the companion star. High mass
X-ray binaries (HMXBs) have O or B-star companions with masses from
several to about 40~\msun, while low mass
X-ray binaries (LMXBs) have companions later than type A (or even a
white dwarf), with masses usually about 1~\msun\ or less. The two
classes are believed to be produced via different evolutionary
pathways. 

\subsection{HMXBs}
\label{sec:hmxb}
In HMXBs, the OB star can have a substantial radiative-driven stellar wind,
which removes \tento{-10}--\tento{-6}~\msun/yr (1~\msun/yr =
\scinot{6.30}{25} gm/s) with a terminal velocity up to 2000 km/s. The
orbit of the neutron star or black hole can bring it close enough to
the OB star so that it can capture a significant fraction of the
material in the wind and thus power the X-ray source.  Because the
compact object is immersed in the wind material, photo-electric
absorption often attenuates the X-ray intensity. In some HMXBs,
Roche-lobe overflow also contributes to (or dominates) the total mass
transfer (see section~\ref{sec:lmxbs} on LMXBs).
Since OB stars are quite bright in optical/UV light, the optical
counterparts of HMXBs are often easily identified. 

The spindown rate of radio (rotation-powered) pulsars indicates a high
surface magnetic field of ($B\simgt$\tento{12}~G).  Many HMXBs are
X-ray pulsars and are also thought to contain a neutron star with a
high magnetic field.  Such objects channel accreting material along
the magnetic field lines to the magnetic poles. If the spin and
magnetic axes are misaligned, the X-ray beam will rotate (like a
lighthouse beam). If the beam crosses our line of sight, the system
will be observed as an accretion-powered X-ray pulsar (in contrast to
rotation-powered radio pulsars).
The accreted matter produces a torque that can slow or increase the
spin rate (and pulse frequency) of the compact object. In many systems
an accreting white dwarf can be rule out because the rate of change of
the pulse period is too high for an object with a relatively high
moment of inertia (compared to a much smaller neutron star) such as a
white dwarf~\cite{joss84}.  Pulsations are also assumed to require a
surface, allowing black holes to be ruled out in systems that
pulse. Thus, X-ray pulsars are well established as highly-magnetized
accreting neutron stars.

\subsection{LMXBs}
\label{sec:lmxbs}
When the companion star in an X-ray binary lacks a significant stellar
wind, as is the case for most LMXBs and some HMXBs, mass transfer
generally occurs though overflow of the critical equipotential surface
(Roche lobe) of the star. Matter transferred in this manner is driven
through the inner Lagrange point between the stars and has a high
specific angular momentum. The material forms an accretion disk around
the compact object and loses angular momentum though viscosity until
falling onto the surface (or being expelled). 
X-ray heating of the disk and companion dominates the optical light
from the donor star, often making it difficult to determine the
spectral type of the companion.

% Type I bursts
Some LMXBs exhibit flares known as type~I X-ray bursts.  These bursts
typically have rise times from less than a second to $\sim$10~s and
slower decay times, in the range of $\sim$10~s to minutes. Spectral
evolution during bursts suggests a sharp rise in temperature at the
onset of a burst and gradual cooling during the decay. This behavior
is consistent with thermonuclear ignition of accreted matter that has
accumulated on the surface of a neutron star. Because a black hole has
no surface, type~I X-ray bursts provide strong evidence that a system
contains a neutron star.

%\section{Z and Atoll Classes of LMXBs}

Most persistently bright LMXBs are believed to contain low magnetic
field neutron stars ($B \simlt$ \tento{9}~G) since they generally show
X-ray bursts and do not show pulsations (a higher magnetic field would
funnel matter along the field lines onto the poles, where the material
would be burned continuously rather than in bursts, and might produce
pulsations due to the non-uniform emission from the
surface~\cite{joss78}.) 

\begin{figure}
\begin{centering}
\PSbox{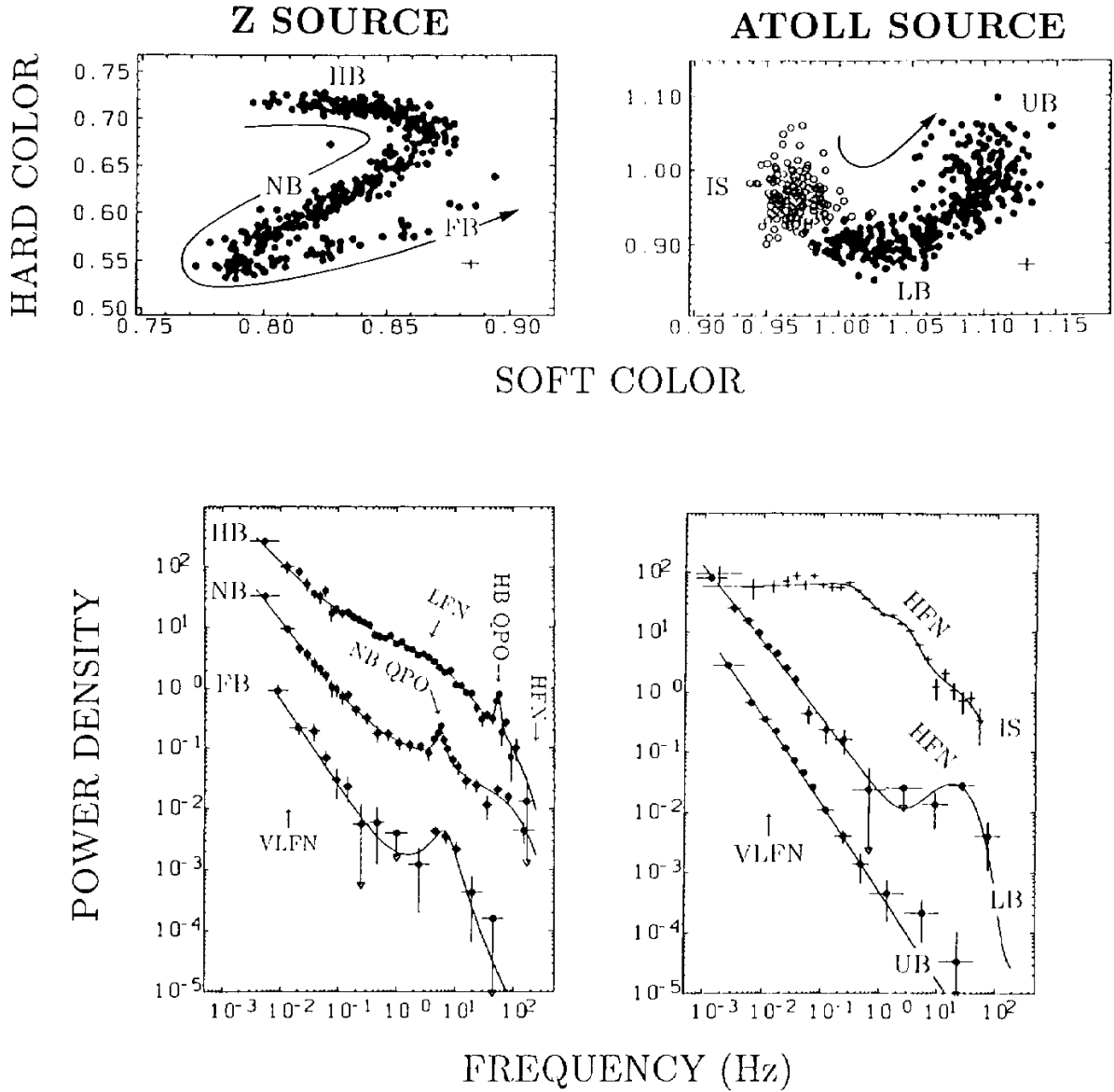 hscale=135 vscale=135 voffset=0 hoffset=-16}{6.2in}{6.2in}
\caption{
X-ray color-color diagrams and power density spectra typical of Z
sources and atoll sources.  The soft color is approximately
\hardness{1}{3}{3}{5} and the hard color
\hardness{5}{6.5}{6.5}{18}. Power spectra are shown for the horizontal, 
normal, and flaring branches (HB/NB/FB) of Z~sources, and the island
state, lower banana, and upper banana (IS/LB/UB) of atoll sources.
(Figure from van~der~Klis 1995~\protect{\cite{klis95}}.)  
}
\label{fig:z_atoll_cc_pds}
\end{centering}
\end{figure}

Neutron-star LMXBs can be classified as either ``Z'' or ``atoll''
sources. These categories are based on the correlated X-ray timing and
spectral properties of these sources~\cite{hk89,klis95} (see
Figure~\ref{fig:z_atoll_cc_pds}). The fast timing properties (on time
scales of seconds or less) of LMXBs and black-hole candidates are
typically studied by computing Fourier power density spectra (PDSs),
which often show several broad-band noise components and sometimes
peaks due to quasi-periodic intensity oscillations (QPOs). The
spectral properties can be examined by forming ratios of count rates
in different energy bands (X-ray hardness ratios or ``colors'').

The six LMXBs known as Z sources are: Sco~X-1, GX~349+2, GX~17+2,
Cyg~X-2, GX~5-1, and GX~340+0. These sources typically trace a ``Z''
pattern in color-color and hardness-intensity diagrams (CDs/HIDs),
where X-ray ``color'' or ``hardness'' refers to the ratio of intensity
in one energy band to that in a lower-energy band, while atoll
sources trace a single arc (see Figure~\ref{fig:z_atoll_cc_pds}).
Z~sources show two types of quasi-periodic oscillations: along the top
``horizontal branch'' (HB) of the Z pattern are 13--60~Hz
horizontal-branch oscillations (HBOs) which increase in frequency
moving to the right on the diagram, and on the diagonal ``normal
branch'' (NB) are $\sim$6~Hz normal-branch oscillations (NBOs). In
some Z sources, the normal-branch oscillations are observed to evolve
into 6--20~Hz QPOs (FBOs) on the lower ``flaring branch'' (FB).  The
broad-band noise components of atoll sources also show correlations
with position in the spectral diagrams.

The most widely held model for these two classes of LMXBs
proposes two physical differences between them: Z~sources have
stronger magnetic fields (1--\scinot{5}{9}~G) and higher mass
accretion rates (near the Eddington limit of $\dot{M}_{Edd} \sim$
\scinot{3}{-8}~\msun/yr, or luminosity $L_{Edd} \sim$ \scinot{1.8}{38}
erg/s) than atoll sources (with magnetic fields $\simlt$\tento{8}~G and
luminosities between \tento{-3}~$L_{Edd}$ and a few
\tento{-1}~$L_{Edd}$)~\cite{klis94}. The suggestion that Z~sources
have stronger magnetic fields than atoll sources was based largely on
the existence of QPOs in Z~sources and their general absence in atoll
sources. In the magnetospheric beat-frequency model for Z-source
horizontal-branch oscillations, the QPOs are the result of clumps of
matter at the inner edge of the accretion disk (at the magnetospheric
radius) which fall in along magnetic field lines to the surface with a
frequency that is the difference between the Keplerian frequency at
the magnetospheric radius and the neutron-star spin
frequency~\cite{alpar85,lamb85}. In this picture, the absence of such
QPOs in atoll sources suggests lower fields so that the disk might
extend almost to the neutron star surface.  Near-Eddington accretion
rates in Z sources are thought to be responsible for the QPOs on the
normal and flaring branches, where a radial accretion inflow occurs
with oscillations in the optical depth~\cite{fortner89}.

Recently kilohertz QPOs ($\sim$300--1200~Hz) have been discovered in a
number of Z and atoll sources with the {\em Rossi X-ray Timing
Explorer} (\RXTE). These QPOs possess many interesting properties (see
van der Klis~\cite{klis98} for a recent review). They often appear in
pairs that shift together in frequency. In some sources, nearly
coherent oscillations have been observed during type~I bursts at a
frequency equal to or double the separation between peaks in kHz QPO
pairs. This suggests that a beat-frequency mechanism is responsible,
e.g., if the higher-frequency QPO peak is related to the Keplerian
frequency at the inner edge of the accretion disk and the oscillations
during bursts reflect the spin frequency (or twice the spin frequency)
of the neutron star, then the lower frequency QPO would occur at the
beat-frequency between the former two frequencies.  One such
beat-frequency is the magnetospheric beat-frequency model, previously
applied to Z-source horizontal branch oscillations (see above).  This
model cannot be correct for both types of QPOs since sometimes HBOs
and kHz QPOs are observed simultaneously; other beat-frequency models
propose a different radius associated with the kHz QPOs, e.g., the
sonic-point model~\cite{miller98}. Important constraints on neutron
star masses, radii, and equations of state can be derived from these
oscillations, since they may reflect neutron-star spin frequencies and
frequencies of orbits very close to the neutron-star surface.

The recent discovery of coherent 2.49-ms pulsations in \RXTE\
observations of the source SAX J1808.4-3658 demonstrates the existence
of millisecond X-ray pulsars~\cite{wijnands98,chakrabarty98}.
Millisecond X-ray pulsars are widely held to be the progenitors of
millisecond radio pulsars.  Angular momentum transfered to the neutron
star by accreting matter is thought to be the mechanism that has
spun-up millisecond radio pulsars, and such sources are expected to
appear as X-ray pulsars during the mass accretion phase.

% Cir X-1 % Cir X-1 % Cir X-1 % Cir X-1 % Cir X-1 % Cir X-1 % Cir X-1 %
\section{Circinus X-1}
\label{sec:intro_cirx1}

\subsection{Intensity Cycle}

The X-ray binary Circinus~X-1 is unique in its complex temporal and
spectral variability. A 16.55~day cycle of flaring is observed in the
X-ray~\cite{kaluzienski76} as well as optical~\cite{moneti92},
IR~\cite{glass78}, and radio bands~\cite{whelan77}. The high degree of
stability of the period of this cycle is evidence that it is the
orbital period. The periodic behavior is believed to be the result of
a highly eccentric binary orbit, in which mass transfer occurs only
near periastron, leading to intermittent obscuration and flaring
near phase zero \cite{murdin80,oosterbroek95,brandt96}.

The X-ray profile and average intensity of the 16.55~day cycle has
varied considerably over time scales of years. The phase zero behavior
has been described as a high-to-low transition, a low-to-high
transition, or complex dips and flares (see e.g.,
\cite{dower82,stewart91,oosterbroek95,shirey96}).
This has lead some authors to suggest that disk precession or apsidal
motion with a time scale of several years plays an important role in
the observed profile~\cite{murdin80,oosterbroek95}.  

\subsection{X-ray Bursts}

Eight X-ray bursts detected during an \EXOSAT\ observation of Cir~X-1
somewhat resembled type I bursts, but had profiles which left their
identification inconclusive~\cite{tennant86a}.  Three additional X-ray
bursts seen during a later \EXOSAT\ observation provided more
convincing evidence for type-I behavior, and thus demonstrated that
Cir~X-1 is a neutron star with a weak magnetic
field~\cite{tennant86b}. This in turn suggests that Cir~X-1 is an
LMXB. Further type~I bursts have not been observed from Cir~X-1 since
the \EXOSAT\ discovery, possibly because the source intensity has been
higher during subsequent observations. (At higher accretion rates, the
material is expected to burn continuously on the surface rather than
accumulate, as is necessary for type I bursts~\cite{joss78}.)

\subsection{Rapid X-ray Variability}

The rapid X-ray variability of Cir~X-1 at times resembles that of both
``atoll'' and ``Z'' low-mass X-ray binaries (LMXBs) as well as
black-hole candidates~\cite{oosterbroek95}. Quasi-periodic
oscillations (QPOs) were reported at 1.4~Hz, 5--20~Hz, and 100--200~Hz
in \EXOSAT\ observations of Cir~X-1 in a bright
state~\cite{tennant87,tennant88}, but other observations at lower
intensity showed no such QPOs~\cite{oosterbroek95}. Based on these
data, it has been suggested that Cir~X-1 is an atoll source that can
uniquely reach the Eddington accretion rate and exhibit normal/flaring
branch QPOs at 5--20~Hz~\cite{oosterbroek95,klis94}. Similar QPOs were
observed in \Ginga\ observations of Cir~X-1~\cite{makino93}. A
significant portion of this current project is devoted to studying the
rapid X-ray variability of Cir~X-1. The current results demonstrate
behavior associated with all three branches of Z-sources (horizontal,
normal, and flaring), including horizontal and normal branch QPOs.  No
evidence for atoll-source behavior is found in the current
observations; this might be related to the currently bright state of
the source.
 
\subsection{Distance}
\label{sec:distance}

Neutral hydrogen clouds orbiting in the Galaxy produce 21-cm
absorption features in the radio spectrum of sources observed through
such clouds. The absorption lines are Doppler shifted by an amount
equal to the radial velocity of clouds in their orbits around the
Galaxy. The maximum velocity observed in these features indicates the
radius of the Galactic orbit tangential to the line of sight, thus
putting a lower limit on the distance to the absorbed source.
H{\small I} absorption features in the spectrum of Cir~X-1 extend to
as much as $-$90 km/s, which puts a lower limit on the distance of
Cir~X-1 at about 80\% of the distance from the Sun to the center of
the Galaxy ($\sim$8--10~kpc, where
1~kpc$\equiv$\scinot{3.086}{21}~cm), or
6.4--8~kpc~\cite{goss77,glass94}.

\subsection{Young Runaway Binary}
Radio images show that Cir~X-1 is embedded in a synchrotron nebula
which trails back toward the nearby ($\sim 1/2^{\,\circ}$, or 70~pc at
8~kpc) supernova remnant (SNR) G~321.9-0.3.  Inside the nebula, there
is a compact source at the position of Cir~X-1, located at (J2000) RA
15~h 20~m 40.99~s, Dec. -57$^\circ$ 09$^{\prime}$
59.91$^{\prime\prime}$. Jet-like radio structures emanate outward from
the compact source for about 30 arcsec before curving back several
arcmin toward the SNR.  This suggests that the system is a
runaway from the explosion that created the neutron star
$\sim$30,000-100,000 years ago~\cite{stewart93}.  The velocity
required for Cir~X-1 to travel $\sim$70~pc in this time is
700--2200~km/s. Although this velocity is quite high for a binary
system, it is not unprecedented. Furthermore, there is spectral
evidence that may also support a high velocity for the system; this is
discussed in Chapter~\ref{ch:multifreq} in conjunction with our
multi-frequency observations, as are the prospects for a measurement
of the angular velocity (proper motion) of Cir~X-1.

The possibility of a young age makes the object highly interesting
from the standpoint of binary-system evolution. For example, mass
transfer in a long-period LMXB ($P\simgt$days) is usually initiated
when the donor star evolves to become a Roche-lobe filling giant
star. If the primary star has only recently collapsed into a neutron
star, this scenario implies that the secondary (probably less massive)
star would be close behind in its evolution; this may be an unlikely
scenario. Based on binary evolution calculations
(Chapter~\ref{ch:bin_evol}), I show in this thesis that the system
need not be unusually young to have a high eccentricity and
substantial accretion rate, and in fact would probably have to be much
older than the supernova remnant if the donor star has low mass.

\subsection{Infrared and Optical Counterpart}
\label{sec:counterpart}

An optical counterpart of Cir~X-1 was resolved from a group of three
stars using images and photometry in the V[0.55$\mu$m], R[0.70$\mu$m],
and I[0.90$\mu$m] optical bands and J[1.25~$\mu$m], H[1.65~$\mu$m],
and K[2.2~$\mu$m] infrared bands. This counterpart is highly reddened
and shows variability with the 16.55-d period~\cite{moneti92}.

Each cycle shows an infrared flare beginning shortly after phase
zero. IR magnitudes at peak can reach K=7.2, H=8.5, and J=9.7, while
the non-flaring component has magnitudes K$\sim$11.5, H$\sim$12.1,
J$\sim$13.3~\cite{glass78,glass94}. The magnitude of the non-flaring
component in optical bands is I$\sim$18.4, R$\sim$19.6, and
V$\sim$21.5~\cite{moneti92}. 

The non-flaring V$-$K color is $\sim$10~mag which would suggest a late
type (low-mass) companion. However, the optical spectrum shows no
atmospheric features that would identify the spectral type of the
companion star, suggesting that disk emission dominates over the light
of the companion~\cite{duncan93}. An early type (high-mass) star would
have to be reddened by $A_{V}\sim11$~mag. 
Based on the minimum X-ray absorption column observed
($N_{H}\sim10^{22}$), the interstellar extinction can be estimated to
be $A_{V}\sim4$~mag, favoring a low-mass star. However, other methods
result in $A_{V}\sim5-11$~mag, so that an O or B-type dwarf or giant
(but not supergiant) cannot be ruled out~\cite{glass94}.  
Past X-ray behavior has suggested that Cir~X-1 is a low mass X-ray
binary, and the results in this thesis strengthen this X-ray evidence
by demonstrating Z-source behavior (Chapters~\ref{ch:feb97paper}
and~\ref{ch:fullZ}).

%The distance modulus for a source 8~kpc away is -14.5 
%(-13.9 for 6~kpc, -15.0 for 10~kpc).

\subsection{Radio Ephemeris}

Based on the onset times of radio flares (defined to be phase zero)
observed at Hartebeeshoek Radio Astronomical Observatory (HartRAO)
between 1978 and 1988, G.~Nicolson calculated the following ephemeris
equation, reported by Stewart et~al.~\cite{stewart91}:
\begin{equation} \label{eq:ephem}
JD_0 = 2443076.87 + (16.5768 - 0.0000353 N)N.
\end{equation}
Although radio flares have been too weak during the past decade to
determine an updated radio ephemeris, equation~\ref{eq:ephem} has
proven to be be an excellent predictor of the current X-ray behavior
(see Chapter~\ref{ch:observs}). It should be noted, however, that
phase zero no longer necessarily corresponds to the onset of radio
flares.  Furthermore, if the cycle is indeed orbital, the exact time
of periastron relative to phase zero is not known. In this thesis, all
references to the phase of Cir~X-1 cycles are based on
equation~\ref{eq:ephem}, unless otherwise noted.

The quadratic term in equation~\ref{eq:ephem} implies that the period
is currently decreasing by about two minutes per year
($\dot{P}=$\scinot{4.26}{-6}=134~s/yr). This would imply a
characteristic time scale ($P/2\dot{P}$) of only $\sim$5000~y for the
period to change by an amount equal to the current period, assuming
the period changed at a constant rate. If the period truly is
changing at such a rapid rate, it would further support the notion
that Cir~X-1 is very young.  However, the radio and X-ray data are not
yet able to confirm that a quadratic term is required. The change of
period in equation~\ref{eq:ephem} amounts to a shift of phase zero of
about 25~minutes per year or about 4~hours over a decade.

For convenience, the zero point of equation~\ref{eq:ephem} can be
shifted by 423 cycles to 1995 December~31.04, immediately after the
December~30 launch of \RXTE, by making the substitution $N'=N-423$:
\begin{equation} \label{eq:ephem423}
JD_0 = 2450082.54 + (16.54694 - 0.0000353 N')N'.
\end{equation}
Thus, the current period will be quoted as 16.55~d throughout this
thesis.

%%%%%%%%%%%%%%%%%%%%%%%%%%%%%%%%%%%%%%%%%%%%%%%%%%%%%%%%%%%%%%%%%%%%%%

\section{Objectives and Overview of this Project}

The M.I.T. Center for Space Research was heavily involved in the
pre-launch planning and instrumentation for the {\em Rossi X-ray
Timing Explorer} satellite (\RXTE), and continues to participate in
the operations and scientific endeavors of that observatory now that
it is in orbit.  The capabilities and scientific instruments of \RXTE\
are described in Chapter~\ref{ch:rxte}. Briefly, the main strengths of
\RXTE\ are: excellent time resolution, large collecting area, broad
spectral coverage, flexible scheduling, and ability to monitor the
X-ray sky to track the intensity of known sources and detect transient
phenomena.

A main objective of this project was to apply the unique capabilities
of \RXTE\ toward some of the outstanding problems of Circinus~X-1 and
to relate the results to X-ray binaries in general.
Since the overall intensity of Cir~X-1 is known to evolve over a period
of years, we first sought to establish the current state of the source
and to study the nature of its cycle profile. Chapter~\ref{ch:observs}
gives a summary of the long-term behavior of Cir~X-1 observed with
\RXTE\, as well as an overview of specific studies we carried out 
through observations of specific cycles.

A major outstanding issue for Cir~X-1 is its unique status among LMXBs
as a possible high-\mdot\ atoll source that can at times show some
types of Z-source behavior.  Thus, several studies were carried out to
explore the characteristics of the rapid variability in Cir~X-1, as
well as how those timing properties relate to other properties such as
orbital phase (Chapter~\ref{ch:mar96paper}) or spectral/intensity
state (Chapters \ref{ch:feb97paper} and~\ref{ch:fullZ}). Chapters
\ref{ch:feb97paper} and~\ref{ch:fullZ} also demonstrate how the energy 
spectrum evolves during various stages of the flaring state.

The variability of Cir~X-1 has been described as both flares and dips.
Dips are important to understand for several reasons: (1) they might
be caused by some part of the mass transfer process, (2) they help
probe the local environment of the system, and (3) they can be
confused with other variability if not properly identified. Detailed
spectral analysis of dips in the intensity of Cir~X-1 is presented in
Chapter~\ref{ch:absdips}.

Since the periodic flaring activity is also observed at other
wavelengths, we organized two multi-frequency campaigns to study
correlated variability. These observations are presented in
Chapter~\ref{ch:multifreq}.

An additional objective of this project was to explore how the
eccentric orbit of Cir~X-1 affects mass transfer and the evolution of the
system. In Chapter~\ref{ch:masstrans}, I show the results of simple
computer simulations of mass transfer due to a stellar wind or
Roche-lobe overflow, for systems with an eccentric orbit.  An
eccentric binary evolution code was also developed
(Chapter~\ref{ch:bin_evol}) to explore possible evolutionary paths for
systems similar to Cir~X-1.

The conclusions based on these studies are discussed in
Chapter~\ref{ch:conclusions}, as are several promising future studies
of Cir~X-1.

\chapter{The Rossi X-ray Timing Explorer}
\label{ch:rxte}

\section{Overview}

\begin{figure}
\begin{centering}
\PSbox{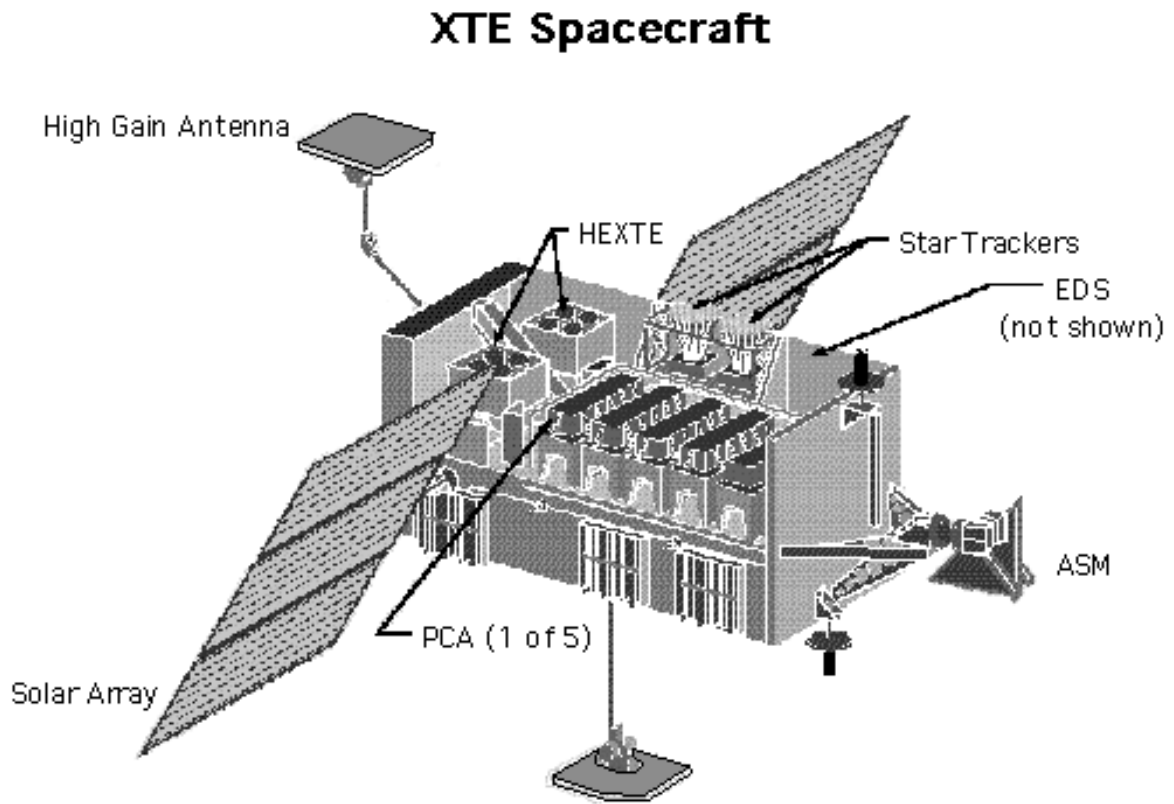 hscale=88 vscale=88 angle=270 
	voffset=450 hoffset=-160}{6.5in}{5.1in}
\caption{{\em Rossi X-ray Timing Explorer} spacecraft.}
\label{fig:spacecraft}
\end{centering}
\end{figure}

The {\em Rossi X-ray Timing Explorer} (\RXTE, see
Figure~\ref{fig:spacecraft}) is a NASA orbiting astrophysical
observatory designed to provide temporal and spectral information
about celestial X-ray sources~\cite{bradt93}.  The satellite was
launched on a Delta~II rocket on 1995 December~30. Its primary targets
are compact objects in our galaxy, such as white dwarfs, neutron
stars, and stellar-mass black holes, as well as active galactic
nuclei, which may contain super-massive black holes. The strengths of
\RXTE\ are its high time resolution, large collecting area, broad
spectral coverage, and ability to detect and respond quickly to
transient phenomena.  These capabilities are achieved through three
instruments: the Proportional Counter Array (PCA), the High-Energy
X-ray Timing Experiment (HEXTE), and the All-Sky Monitor (ASM). In
addition, a specialized on-board computer, the Experiment Data System
(EDS), serves to pre-analyze and compress data from the ASM and PCA to
maximize use of telemetry.  Each instrument is briefly described
below.  For more details see {\em The RXTE Technical
Appendix}~\cite{techapp}.

% % % % % % % % % % % % % % % % % % % % % % % % % % % % % % % % % % % % % % 
\section{Proportional Counter Array}
\label{sec:pca}

The PCA (built by the Goddard Space Flight Center of NASA) consists of
five Proportional Counter Units (PCUs) with a total open area of
$\sim$6250~cm$^2$. This large collecting area allows photons to be
detected at a rate high enough to study variability as short as
milliseconds. The detector electronics record the arrival time of each
photon to an accuracy of 1~$\mu$s, and the EDS is also capable of
processing data with 1~$\mu$s resolution (see section~\ref{sec:eds}).
The PCUs are sensitive to X-ray photons with energies between 2~keV
and 90~keV, and can detect sources down to a flux level of a few
mCrab, where 1.0~Crab $\approx$ 1060~$\mu$Jy at 5.2~keV.  The
intrinsic energy resolution ($\Delta$E/E, FWHM) of the PCA is about
18\% at 6~keV and 10\% at 20~keV\@. The PCUs have no imaging
capability, but collimators restrict the field of view to
$\sim1^{\circ}$. Since the number of bright X-ray sources in the sky
numbers less than 100, the $1^{\circ}$ field of view is small
enough to exclude other bright sources in all but the most crowded
fields (i.e., towards the Galactic center).  

Each PCU detector consists of a chamber of xenon gas (and a small
quantity of methane) with three layers of high-voltage anode wires.
The anodes closest to the detector walls, and a fourth layer on the
bottom form an anti-coincidence detector used to reject events due to
particles which enter through the walls of the detector. X-ray photons
enter through the front window of the chamber and eject a
photo-electron from an atom in the gas. The electron is accelerated
toward an anode wire, creating ion pairs along its path. The
multiplied electron signal is read out by the anode electronics.  A
propane chamber between the main chamber and the detector window is
only sensitive to low-energy photons ($\simlt$3~keV). Anode wires in
the propane layer are also used in the anti-coincidence logic to
screen out electron-induced events.  All data from the PCA is sent to
the EDS for processing before being telemetered to the ground
(see section~\ref{sec:eds}).

The capabilities of the PCA are particularly well-suited for studies
of the rapid variability in X-ray binaries, and we have made use of
these capabilities in our study of Cir~X-1.  Data can be obtained at
high time resolution (fractions of a millisecond) simultaneously in
multiple energy bands to study the variability components (such as the
quasi-periodic oscillations) and their energy-dependence (Chapters
\ref{ch:mar96paper}, \ref{ch:feb97paper}, and~\ref{ch:fullZ}).  All
PCA data are also telemetered as 16-s spectra with 129 channels,
allowing more detailed spectral studies to be made on time scales
shorter than or comparable to most changes in the state of the source
(Chapters \ref{ch:fullZ} and~\ref{ch:absdips}).

% % % % % % % % % % % % % % % % % % % % % % % % % % % % % % % % % % % % % % 
\section{High-Energy X-ray Timing Experiment}
\label{sec:hexte}

The HEXTE detectors (built by the University of California, San Diego)
are sensitive to X-ray photons in the 20--200~keV band, extending the
full capability of \RXTE\ to cover two orders of magnitude in photon
energy (2--200~keV). They are co-aligned with the PCA detectors and
are also collimated to produce a $\sim1^{\circ}$ field-of-view. The
HEXTE consists of two independent clusters of four NaI(Tl)/CsI(Na)
phoswich scintillation detectors. Photons entering such a detector
undergo multiple interactions in the NaI scintillation crystal, and
the resulting signal is transferred through the CsI crystal to a
photo-multiplier tube. The CsI crystal also serves to reject photons
that only partially deposit their energy and events that are induced
by particles. The two detector clusters are alternately rocked on and
off-source to provide continued background monitoring. Each phoswich
detector has a net collecting area of 200~cm$^2$ and an energy
resolution of about 13\% at 60~keV.

HEXTE data is obtained in parallel with all PCA observations. For
Cir~X-1, the HEXTE count rates are relatively low since the X-ray
spectrum of Cir~X-1 is soft, resulting far more counts in the PCA than
HEXTE. Furthermore, the analysis of HEXTE data is complicated by the
rocking detectors and complex calibration issues. Thus, the data we
obtained for Cir~X-1 from HEXTE is not incorporated into this thesis.

% % % % % % % % % % % % % % % % % % % % % % % % % % % % % % % % % % % % % % 
\section{All-Sky Monitor}
\label{sec:asm}

The ASM (built by the Center for Space Research at M.I.T.) consists of
three scanning shadow cameras (SSCs) which regularly monitor all
bright ($\simgt$10~mCrab) celestial X-ray sources (about 10 times per
day), providing long-term light curves of their intensity in three
energy bands between 2~keV and 12~keV~\cite{levine96}.  Each SSC
contains a position-sensitive proportional counter mounted below a
wide-field collimator ($\sim6^{\circ} \times \sim90^{\circ}$ FWHM).
The PSPCs each contain eight carbon-coated quartz fibers which are
used as anode wires in a chamber filled with Xe and CO$_2$.  The anode
electronics read out signals from both ends of each anode and can
determine the position along the wire by comparing the relative
strength of the signals due to the resistance of the anodes.  A mask
covering the outer window of the collimator is coded with rectangular
slit openings which cast X-ray shadows on the detector and allow
positions to be determined (typical error boxes of
$3^{\prime}\times12^{\prime}$).  The three cameras are mounted on a
tripod at one end of the satellite (see Figure~\ref{fig:spacecraft})
and are rotated together, stopping for 90-s dwells every
$6^{\circ}$. A sensitivity of about 30~mCrab (3$\sigma$) is achieved
for each such 90-s exposure. The fields-of-view of two of the cameras
point in the same direction perpendicular to the rotation axis and
cross at a $\sim24^{\circ}$ angle (resulting in positional error boxes
that cross and can be used for further refinement of positions). The
third camera points along the rotation axis. Several of the anodes
lost functionality shortly after launch, but this has not
significantly affected the overall performance.

For Cir~X-1 the ASM has proved invaluable in understanding the general
spectral and intensity profile of the 16.55-d cycle.  The ASM also
plays an important role by placing other observations (e.g., our PCA
observations) in context with the rest of the cycle and by aiding
plans for future observations. With continued successful performance,
the detailed long-term light curve will allow us to track how the
source behavior evolves between its various states (see
section~\ref{sec:intro_cirx1}). So far, the ASM light curve of Cir~X-1
has shown the source to maintain the same high baseline intensity
level for the entire duration of the \RXTE\ mission (see
Chapter~\ref{ch:observs}), with no evidence for evolution on a
few-year time scale as has been suggested based on previous
observations (see section~\ref{sec:intro_cirx1}).

% % % % % % % % % % % % % % % % % % % % % % % % % % % % % % % % % % % % % % 
\section{Experiment Data System}
\label{sec:eds}

The EDS (also built by the CSR at M.I.T.) is a computer containing
eight microprocessor Event Analyzers (EAs), of which six are dedicated
to handling PCA data while the other two are dedicated to the ASM. For
each 90-s dwell, one of the ASM EAs produces position histograms of
the X-ray shadow patterns on each wire of each detector in the three
energy channels mentioned above. The second ASM EA produces
time-series histograms of ``good'' and background count rates, as well
as pulse height histograms useful in monitoring the gain of the
detectors based on an on-board calibration source.

\begin{table}
\begin{center}
\begin{tabular}{cccccc}
\hline
\hline
& \multicolumn{2}{c}{Channel} & \multicolumn{3}{c} {Energy (keV)} \\
Band & Abs & Std2 & Epoch 1 & Epoch 2 & Epoch 3 \\
\hline
A & 0--9    & 0--5              & 0--2.6    & 0--3.0     & 0--3.5 \\
B & 10--13  & 6--9              & 2.6--3.6  & 3.0--4.2   & 3.5--5.0 \\
C & 14--17  & 10--13            & 3.6--4.7  & 4.2--5.5   & 5.0--6.5 \\
D & 18--23  & 14--19            & 4.7--6.3  & 5.5--7.4   & 6.5--8.7 \\
E & 24--35  & 20--31            & 6.3--9.5  & 7.4--11.1  & 8.7--13.1 \\
F & 36--49  & 32--45            & 9.5--13.2 & 11.1--15.6 & 13.1--18.3 \\
G & 50--87  & 46--66            & 13.2--24  & 15.6--28   & 18.3--33 \\
H & 88--249 & 67--127           & 24--71    & 28--84     & 33--99 \\
\hline
\end{tabular}
\end{center}
\caption{
Approximate energy ranges for eight PCA bands (A--H) defined in terms
of absolute (0--255) and Standard2 (0--128) channels. Gain epoch~1
includes all observations before before 1996 March~21, epoch~2 1996
March~21 -- April~15, and epoch~3 after 1996 April~15.  Energy values
shown are derived from the full PCA with all detectors and layers
added.  
}
\label{tab:pca_ch2en}
\end{table}

Two of the PCA EAs process data using standard modes: one that
provides count rates for each layer of the five PCU detectors and
other rates every 1/8~s (Standard1) and another which provides
129-channel spectra every 16~s (Standard2). The other four PCA EAs can
be set to process data in parallel using a wide variety of modes in
order to fit desired information into the telemetry constraints. Most
data modes fall into two categories: event modes which send down the
time and energy of every event and binned modes which send down the
number of photons detected in selectable energy and time bins. Many
data modes process data from within standard sub-bands of the entire
PCA energy range. These sub-bands are defined in terms of the 256
``absolute'' channels of the PCA. Table~\ref{tab:pca_ch2en} shows the
channel and energy ranges for eight standard bands.  The gain of the
PCA detectors was changed twice early in the mission, resulting in a
different channel-to-energy relationship for each of three
epochs. Although the relationship is not strictly linear, a single
channel-to-energy conversion factor is a good approximation for each
epoch: through 1996 March~21, 0.27~keV/chan; 1996 March~21 -- 1996
April~15, 0.32~keV/chan; after 1996 April 15, 0.38~keV/chan. All light
curves, hardness ratios, and Fourier power spectra in this thesis have
been binned into one or several of these standard bands.

The four PCA Event Analyzers not devoted to standard modes are
typically configured, through modes selected by the observer based on
science goals, to provide optimal time resolution and energy
resolution within the available telemetry. For Cir~X-1, we generally
chose two or three ``single-bit'' modes (single-channel binned modes)
which covered absolute channels 0--35 of Table~\ref{tab:pca_ch2en} (up
to 13~keV in epoch~3) using various combinations of energy bands
A--E. These configurations provided a time resolution of 61~$\mu$s or
122~$\mu$s.  Due to lower count rates in the higher channels, the
energy and time of every photon with energy above channel~35 (bands
F--H in Table~\ref{tab:pca_ch2en}) could be recorded with an event
mode having 32 or 64 channels covering absolute energy channels
36--249 and time resolution of 16~$\mu$s or 64~$\mu$s. Sometimes a
binned mode was used to provide 32 low-energy channels at 0.5~s
resolution to allow the possibility of creating spectra on time scales
shorter than the 16~s provided by the Standard2 mode. Thus, all
Cir~X-1 data were obtained with high time resolution in multiple
energy bands for rapid variability studies (Chapters
\ref{ch:mar96paper},
\ref{ch:feb97paper}, and~\ref{ch:fullZ}) and good energy resolution
every 16-s or less for spectral studies (Chapters \ref{ch:fullZ}
and~\ref{ch:absdips}).

\chapter{Overview of RXTE Observations}
\label{ch:observs}

\section{ASM Observations of Circinus X-1}
\label{sec:asmobs}

\subsection{General Features of the ASM Light Curves}

The \RXTE\ All-Sky Monitor (see section~\ref{sec:asm}) has provided
90-s intensity measurements of most bright X-ray sources in three
energy channels (1.5--3~keV, 3--5~keV, and 5-12~keV) about 12 times
daily since early 1996~\cite{levine96}. The intensity for each source
in the ASM field of view, and the diffuse X-ray background, is
obtained from fits to the ASM mask patterns of each 90-s
exposure. Thus background has, in effect, been subtracted from the
source light curves.

\begin{figure}
\begin{centering}
\PSbox{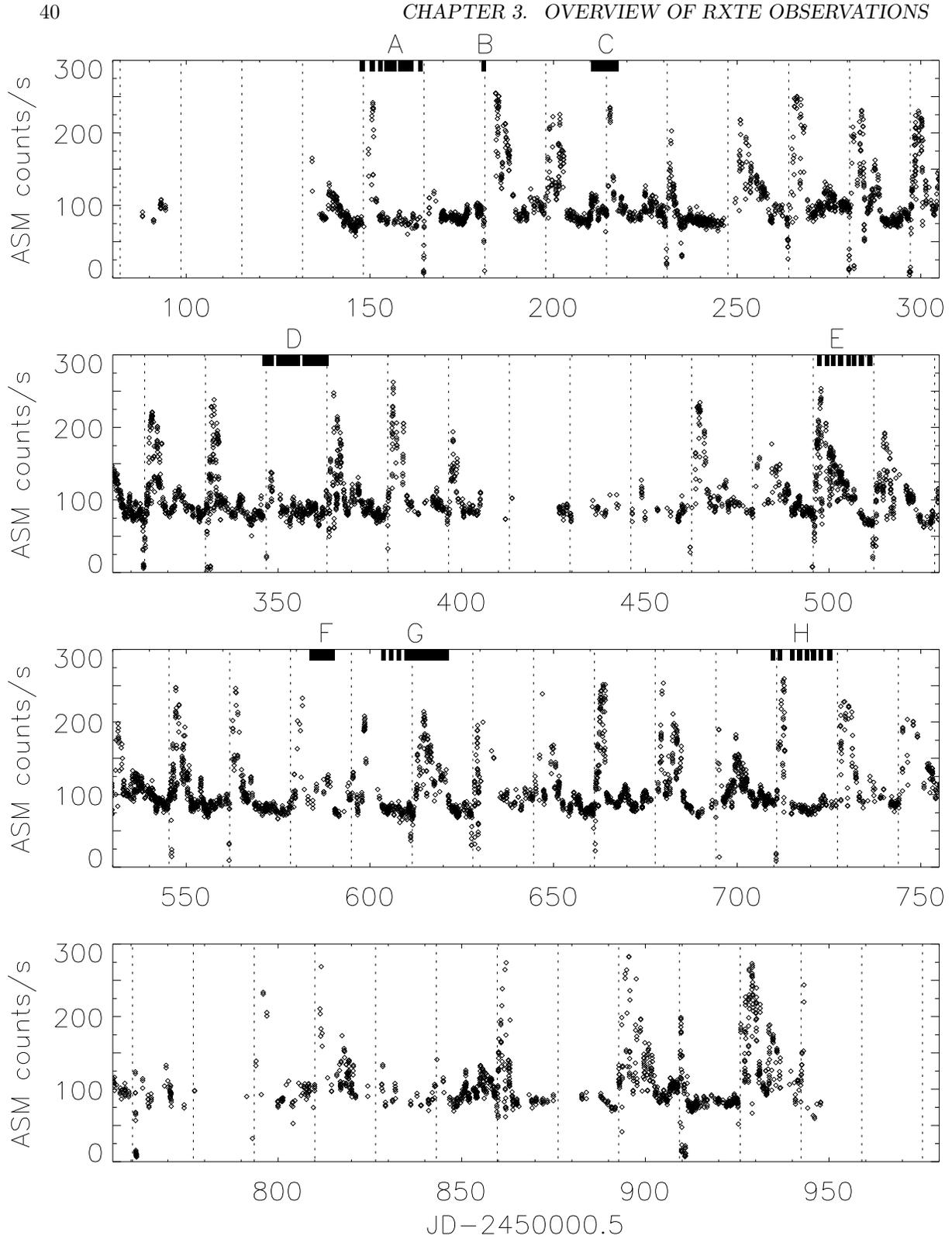 hoffset=-43 voffset=-60}{6.3in}{7.7in}
\caption{ASM light curve (1.5--12~keV; 90-s exposures) covering 
1996 January -- 1998 May.  Phase zero is indicated by vertical dotted
lines. Bars along the top axis mark the times of PCA observations
(labeled A--H; see section~\protect{\ref{sec:pcaobs}}). }
\label{fig:asm_2yr_lc}
\end{centering}
\end{figure}
%(see Table~\protect{\ref{tab:pcaobs}} and Figure~\protect{\ref{fig:pca_by_cycle}}).

The ASM light curve of Cir~X-1 for the entire \RXTE\ mission to date
(through 1998 May) is shown in Figure~\ref{fig:asm_2yr_lc}. (A gap of
$\sim$40~days occurred early in the mission due to instrumental
problems, and a gap of 2--3 weeks has occurred annually, near days 420
and 780, due to sun-angle constraints.)  These observations show
Cir~X-1 in a sustained bright state with a baseline intensity level
very similar to that of the Crab nebula (1.0~Crab $\approx$
75~ASM~counts/s; 1060~$\mu$Jy at 5.2~keV). Each cycle shows strong
flaring wherein the intensity increases by a factor of three or more.
Phase zero of each cycle (based on equation~\ref{eq:ephem}) is
indicated in Figure~\ref{fig:asm_2yr_lc}. The consistent alignment of
brief dips with phase zero and subsequent flaring shortly after phase
zero demonstrate that the decade-old radio ephemeris is very
successful in predicting the current X-ray behavior.

% Description of ASM light curve
The flaring state typically begins during the day following phase
zero, reaching as high as 3.7~Crab (corresponding to a luminosity at
8~kpc that is several times the Eddington limit) and typically lasting
2--5~days.  The profile of the flaring state is actually quite
variable, sometimes showing a secondary flare or an extended main
flare lasting most of the cycle (see Chapter~\ref{ch:feb97paper}).

% Dips
Many cycles observed with the ASM show brief dips below the 1~Crab
level, usually immediately before or shortly after phase zero, but
occasionally occurring later in the cycle as well. Previous satellites
have also observed intensity dips from Cir~X-1 near phase zero. For
example, Brandt et~al.~\cite{brandt96} recently presented evidence for
strong absorption during a phase-zero intensity transition in
\ASCA\ observations of Cir~X-1. More extensive \RXTE\ PCA observations
of dips will be discussed in Chapter~\ref{ch:absdips}.

\subsection{Cycling Hardness Ratios}

\begin{figure}
\begin{centering}
\PSbox{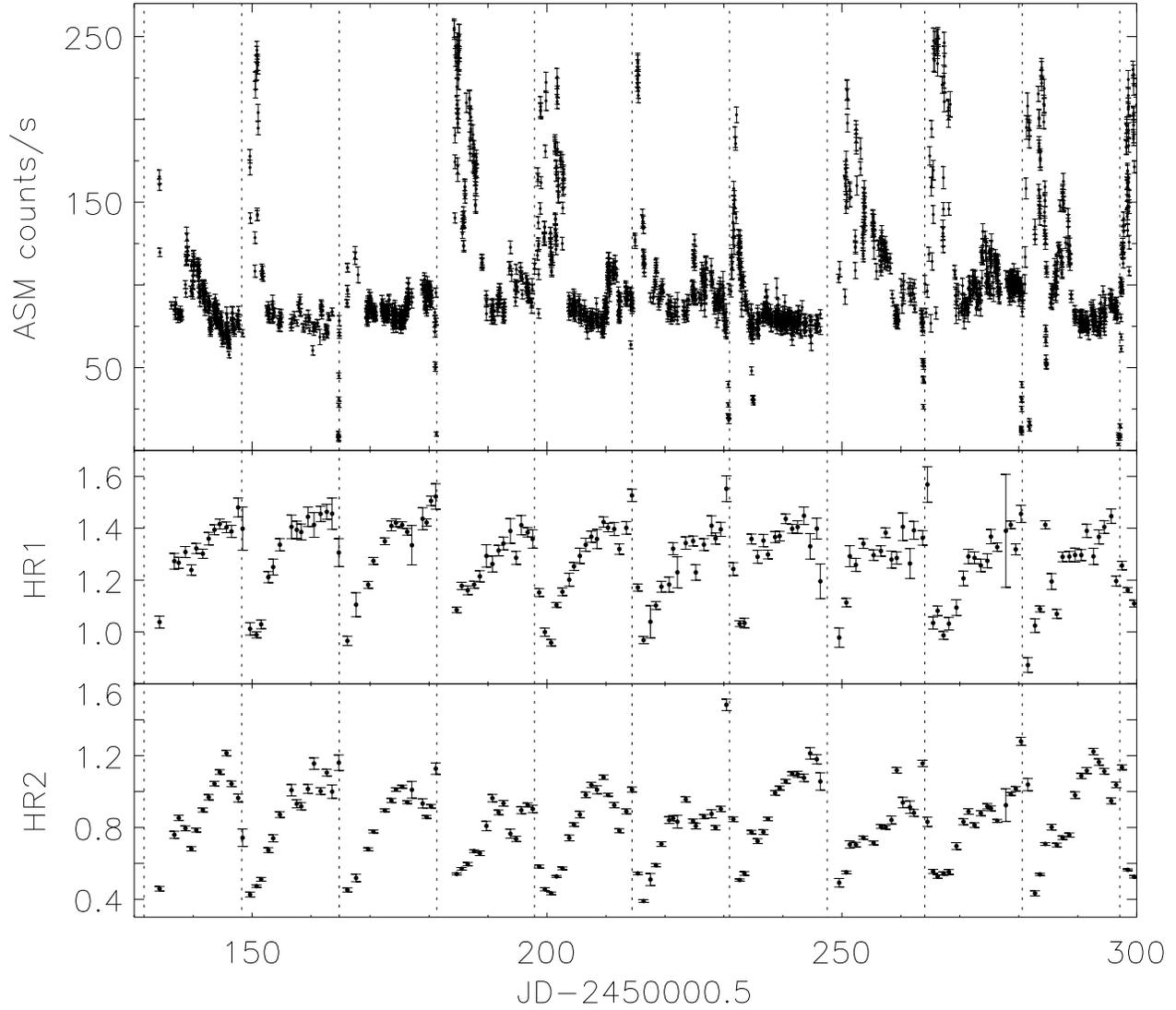
	hscale=100 vscale=100 voffset=-185 hoffset=-40}{6.5in}{6.5in}
\caption{
Light curve and two hardness ratios for the first ten full 16.55-day
cycles of Cir~X-1 observed with the ASM.  Each light-curve point is
the average count rate (\intens{1.5}{12}) from a single 90-s ASM
exposure.  The hardness ratios were obtained from 1-day averages of
the 90-s intensities and are defined as HR1=\hardness{1.5}{3}{3}{5}
and HR2=\hardness{3}{5}{5}{12}. The vertical dotted lines indicate
phase zero.
}
\label{fig:asm_lc_2hr_130_300}
\end{centering}
\end{figure}

% Discussion of ASM HRs and spectral evolution
The light curve and hardness ratios of counts in adjacent energy
channels for the first ten 16.55-d cycles of Cir~X-1 observed with the
ASM are shown in Figure~\ref{fig:asm_lc_2hr_130_300}. These data show
significant spectral evolution during the non-flare phases. The
hardness ratios were found to generally increase (harden) from about
phase 0.2 until phase 0.6-0.85. Beyond this point the hardness ratios
usually flatten, decrease, or dip and rise again before phase
zero. Near phase zero, the ratios vary dramatically (not apparent in
the one-day averages shown) but generally become quite low during the
flaring phases (0 $\simlt \phi \simlt$ 0.2).  The hardening in the two
colors (0.2 $\simlt \phi \simlt$ 0.7) is usually due to both an
increase in the 5--12~keV count rate and a decrease in the 1.5-3~keV
count rate.

% Period
\subsection{16.55-day Period}

The onset of flaring relative to phase zero can vary by of order one
day.  Over the $\sim$50 cycles observed by the ASM, this allows the
period to be determined to within about $1/50~d = 0.02$~d. Since this
uncertainty is similar to the change in period of 0.03~d over 20~yr
implied by equation~\ref{eq:ephem}, it is not yet possible with the
ASM to confirm or reject a period change of that magnitude. However,
to obtain the best-fitting period I used a 64-times
oversampled Fourier power density spectrum from the ASM data in
Figure~\ref{fig:asm_2yr_lc} (and fitting the peak corresponding to the
$\sim$16.55-d period with a Gaussian). This method gave a best-fitting
constant period of 16.555~d. Within the 0.02~d uncertainty, this is
consistent with equation~\ref{eq:ephem}, with or without the quadratic
term. The possibility of using the phase-zero dips as ``markers'' for
the period is explored in the next section (using the average folded
light curve).  Hopefully the ASM will continue to operate for several
more years, potentially allowing the period to be determined with the
accuracy necessary to test the constancy of the period.

\subsection{Folded Cycle Profile}

%-----------------------------------------------
\begin{figure}
\begin{centering}
\PSbox{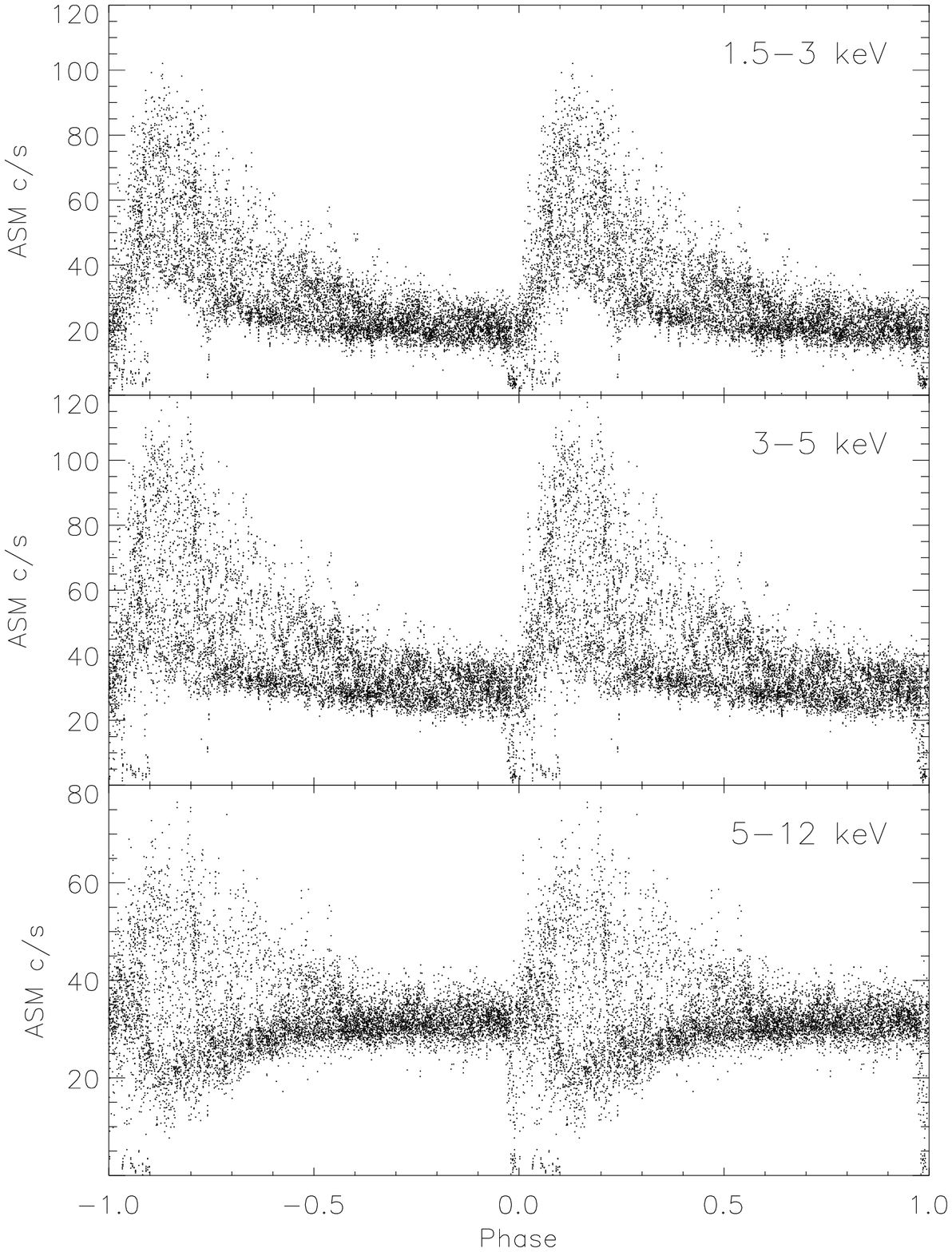 hscale=93 vscale=93 
		hoffset=-36 voffset=-14}{5.86in}{7.8in}
\caption{Folded light curve of Cir~X-1 from the three ASM energy bands.
The data are repeated in order to show two complete cycles.  Each
point is from a 90-s exposure, with a typical error bar of 
1--2~c/s. The folded data spans the entire time period covered by
Figure~\protect{\ref{fig:asm_2yr_lc}}.  }
\label{fig:asm_fold_abc}
\end{centering}
\end{figure}
%-----------------------------------------------

ASM light curves of Cir~X-1 folded at the 16.55-d period (using
equation~\ref{eq:ephem}; Figure~\ref{fig:asm_fold_abc}) show the rapid
rise and more gradual decline of the profile.  The scatter of points
during the ``flaring'' state demonstrates the high degree of
variability at those phases. The 5--12~keV band exhibits flaring above
a baseline that dips below the quiescent level.  This is related to
the spectral softening seen during flaring in
Figure~\ref{fig:asm_lc_2hr_130_300}.  Although dips typically occur
anywhere between phases $-$0.03 and +0.10, they occur most
consistently in a narrow range immediately before phase zero. There is
a clear absence of points at the baseline level in the narrow phase
range 0.985--0.989; however, the total number of points in a phase
range of that width is typically only about 30, so sparse coverage may
play a role in the width of this gap.

\begin{figure}
\PSbox{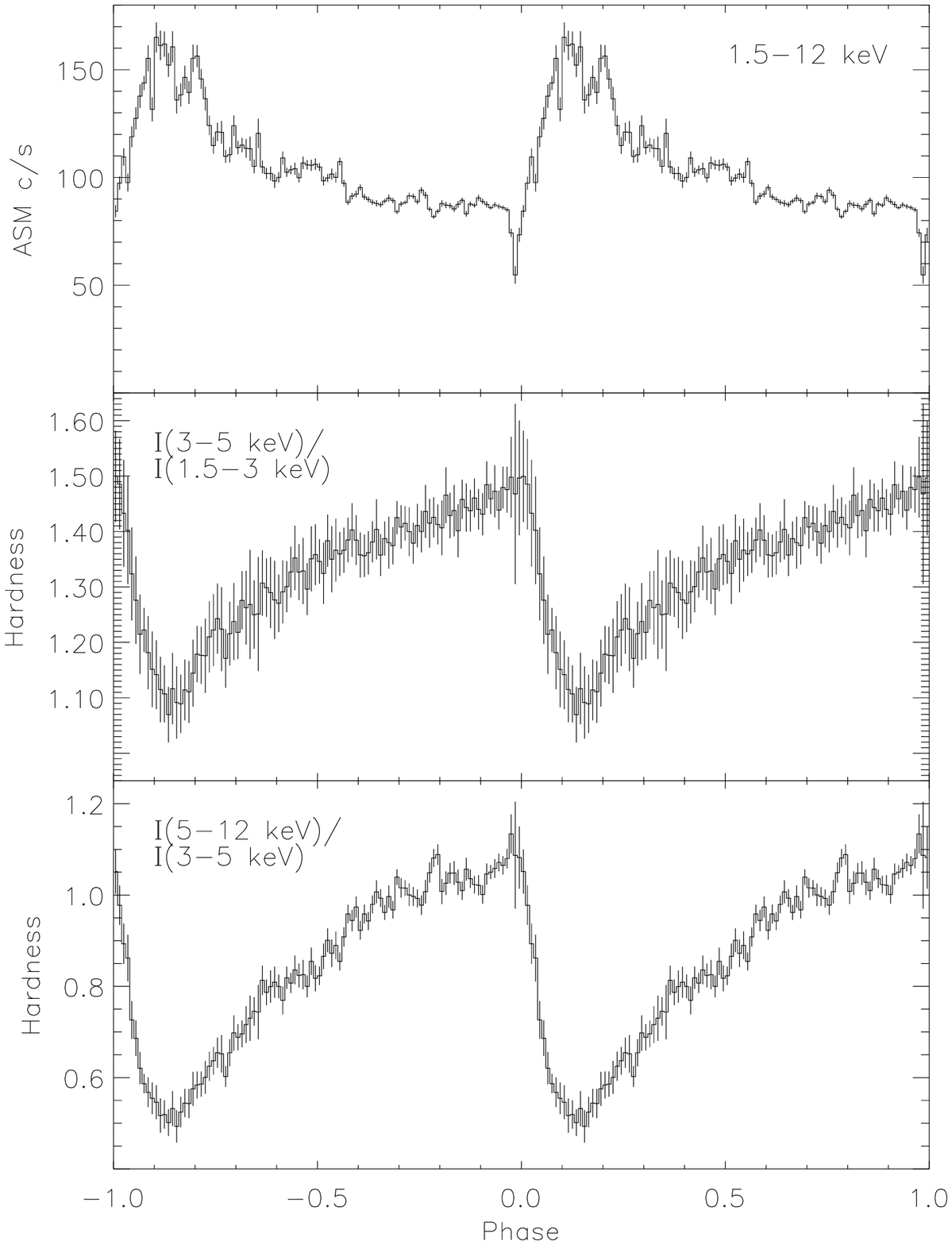 hscale=50 vscale=50 hoffset=-30}{3.3in}{4.25in}
\vspace{-4.25in}\hspace{3.7in}
\PSbox{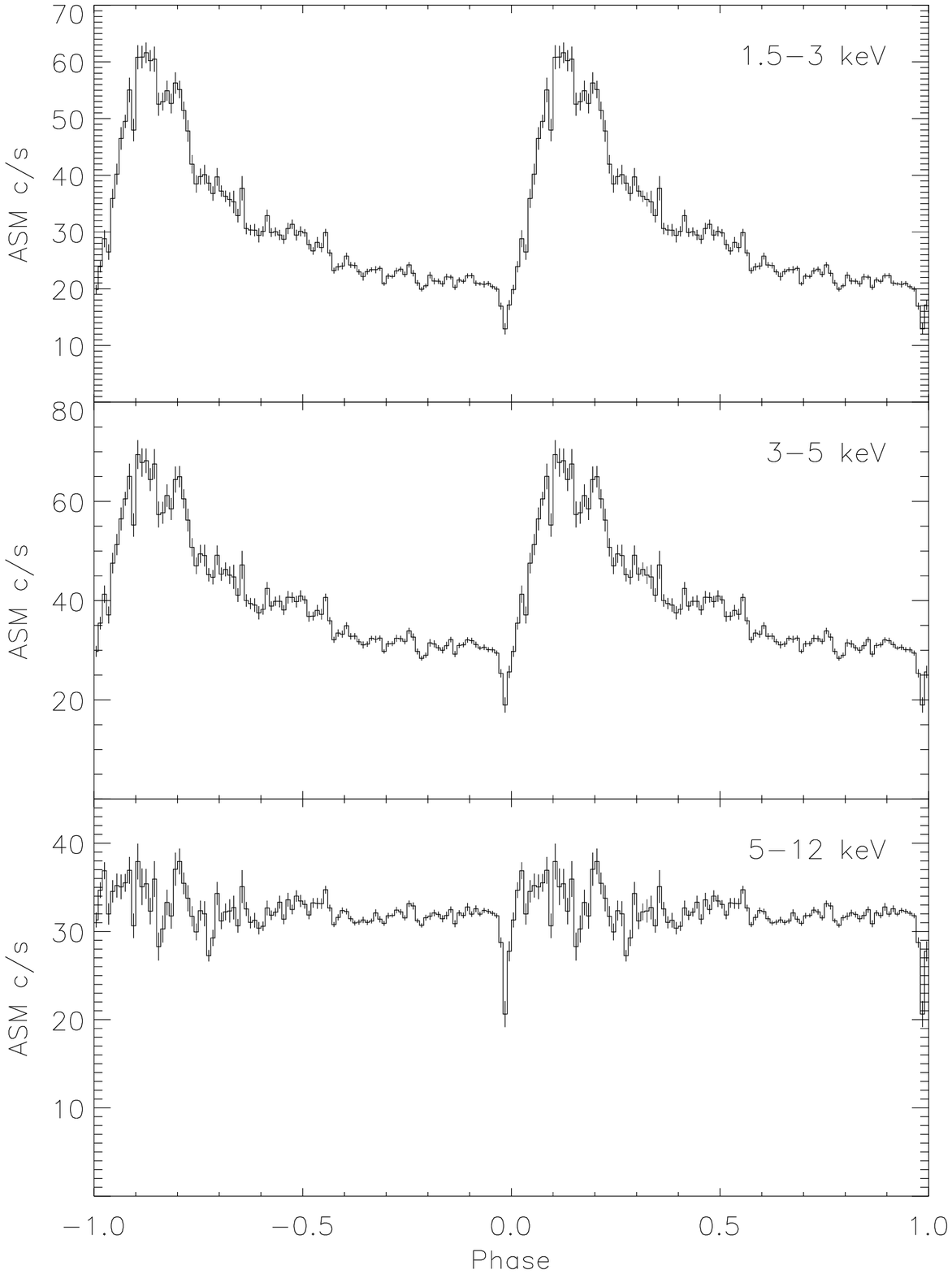 hscale=50 vscale=50 hoffset=205}{3.3in}{4.25in}
\caption{Left: Folded Cir~X-1 light curve from the full ASM energy band
(1.5--12~keV) and two hardness ratios.  Right: Folded light curves
from the three ASM energy bands.  The data are binned into 100 phase
bins per cycle, averaged in each phase bin, and duplicated to show two
complete cycles. Error bars on the light curves indicate the standard
deviation of the mean value derived for each phase bin, and these
error estimates were propagated when computing the hardness
ratios. The folded data spans the entire time period covered by
Figure~\protect{\ref{fig:asm_2yr_lc}}.}
\label{fig:asm_100phbin}
\end{figure}

The data shown in Figure~\ref{fig:asm_fold_abc} for the three ASM
bands were also binned into 100 phase bins and averaged within each
bin.  These average folded profiles, as well as those for the the
total ASM energy band and the two hardness ratios, are shown in
Figure~\ref{fig:asm_100phbin}. The average profile has a very narrow
dip at phase $-$0.012 ($\pm$0.004), followed by a sharp rise over the
first 1.5~d of the cycle and a much more gradual decline.  The
hardness ratios are anti-correlated with the total intensity.  Similar
intensity and hardness profiles were seen in folded \Ginga\ ASM data
from 1987~\cite{tsunemi89}.  The folded profile from the highest
energy ASM band (5--12~keV) does not exhibit the large increase in
flux characteristic of the other two bands and shows only increased
variability during the flaring state, due to the scatter of points
above and below the baseline (see Figure~\ref{fig:asm_fold_abc}).

We attempted to use the narrow dip at phase $-$0.012 in the average
folded curves to constrain the best period. Constant periods of
16.535--16.555~d all showed similar dips, and the dips were deepest
(and most similar to Figure~\ref{fig:asm_100phbin}) using a period of
16.540--16.545. Although the range of periods showing the narrow dip
is lower than the earlier radio periods by about 0.02--0.04~d, we have
no {\em a~priori} knowledge of what shape (if any) this feature should
have, and thus must use caution when interpreting periods determined
by maximizing its depth (or other property).

%-----------------------------------------------
%-----------------------------------------------

%%%%%%%%%%%%%%%%%%%%%%%%%%%%%%%%%%%%%%%%%%%%%%%%%%%%%%%%%%%%%%%%%%%%%%%%%%%%%%

\section{PCA Observations of Circinus X-1}
\label{sec:pcaobs}

\subsection{Overview}

\begin{figure}
\begin{centering}
\PSbox{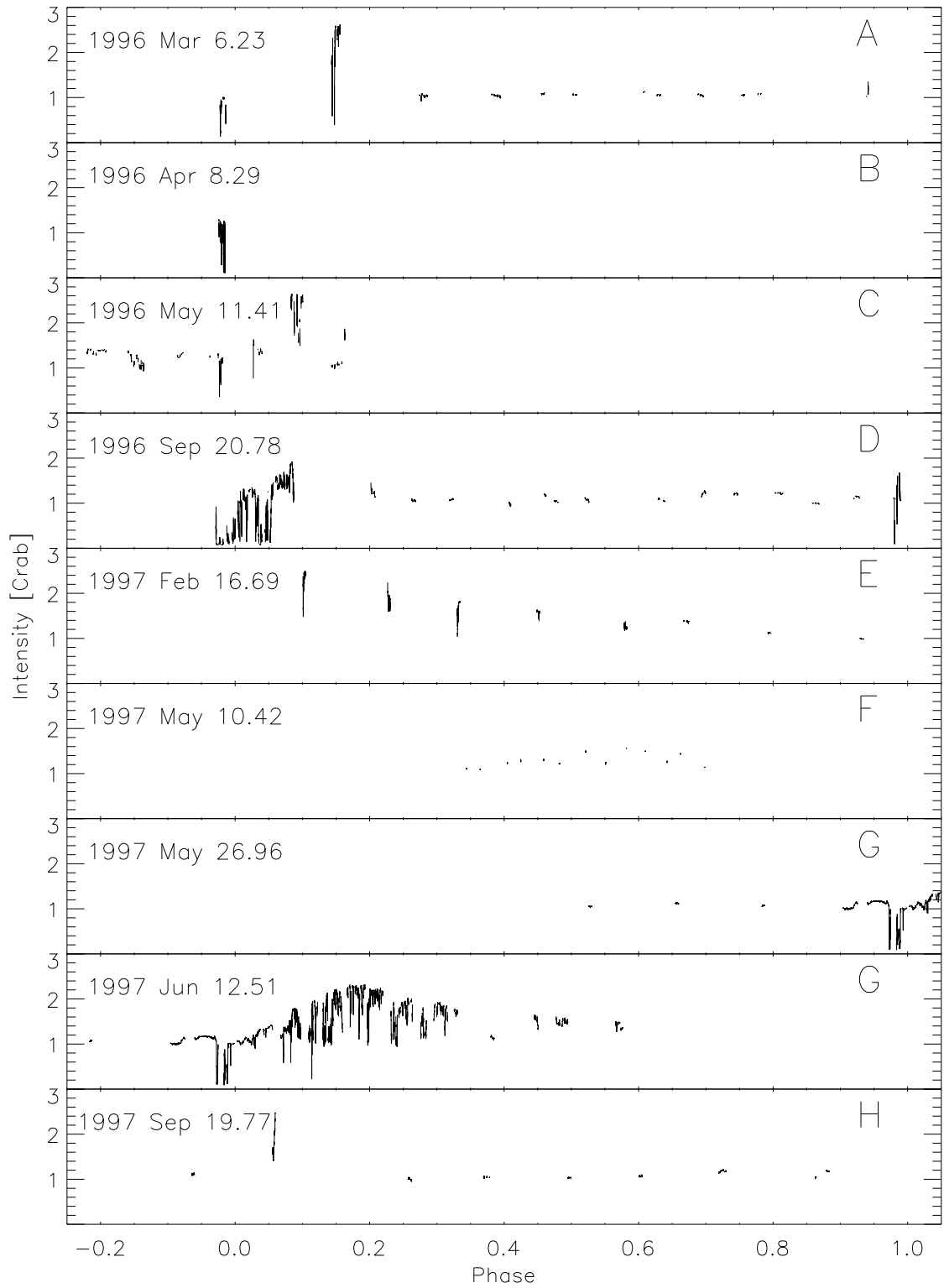 hscale=93 vscale=93 
hoffset=-84 voffset=-78}{5.86in}{7.8in}
\caption{Light curves for each 16.55-d cycle of Cir X-1 observed with the PCA. 
For each cycle, the phase-zero date and letter of the study (see
Table~\protect{\ref{tab:pcaobs}}) are listed. The intensity is for the
full PCA energy range, and a conversion factor of 1~Crab $\approx$
2600~counts\,s$^{-1}$ PCU$^{-1}$ has been used.}
\label{fig:pca_by_cycle}
\end{centering}
\end{figure}

We have proposed and carried out an extensive observation campaign for
Cir~X-1 with the {\em RXTE} PCA, collecting over 800~kiloseconds of
data, as shown in Table~\ref{tab:pcaobs}. The project consisted of
eight separate ``studies'' which used a variety of strategies for
different 16.55-d cycles: We sampled several cycles with brief
observations (typically about 6000~s), distributed at different
phases, to study how the detailed timing and spectral characteristics
evolve with orbital phase and relate to the spectral evolution
indicated by the cycling hardness ratios of the ASM (see above).
High-efficiency observations (i.e., $\sim$60\% coverage) during two
long time segments (2 and 7~days) including phase-zero were used to
study the highly complex dipping and flaring behavior associated with
that portion of the cycle. In order to study correlated behavior at
different wavelengths, we coordinated two multi-frequency campaigns to
provide optical, infrared, and radio observations simultaneous with
our X-ray coverage.  The light curves, from the full PCA energy range,
for each of the studies are shown in Figure~\ref{fig:pca_by_cycle} as
a function of phase for comparison between cycles.  Standard
20-kilosecond light curves of all these observations are collected in
Appendix~\ref{ch:std_pcalc}. The time of each PCA observation is also
indicated on the 2.3-year ASM light curve in Figure~\ref{fig:asm_2yr_lc}.

\begin{table}
\begin{center}
\begin{tabular}{llll}
\hline
       &       & Observing & \\
Study  & Dates & Time      & Description \\
\hline
\hline
A & 1996 Mar. 5 - 21 & 76 ksec & 12 observations over one 16.6-d cycle \\
\hline
B & 1996 Apr. 7 & 10 ksec & 1 observation during phase-zero dips \\
\hline
C &1996 May 7 - 14 & 54 ksec & 7 observations between phase 0.78-0.16, \\
 & & &                        coord.\ with HartRAO (radio) \& SAAO (IR)\\ 
\hline
D & 1996 Sept. 20 - Oct. 8 &190 ksec& 60\% coverage for 2 days (phase 0.97-0.09),\\
 & & &                           13 observations over remainder of 16.6-d cycle \\
\hline
E & 1997 Feb. 18 - Mar 4 & 48 ksec & 8 samples over one 16.6-d cycle \\ 
\hline
F & 1997 May 16 - 21 & 13 ksec & 13 1-ksec observations, twice daily \\ 
\hline
G & 1997 June 4 - 22 & 396 ksec & 56\% coverage for 7 days (phase 0.93-0.33), \\
 & & &                     sampling observations before and after; \\
 & & &                     portions coordinated with AAT (optical), \\
 & & &                     ANU (IR), HartRAO (radio), \& SAAO (IR) \\
\hline
H & 1997 Sept. 18 - Oct. 4 & 50 ksec & 8 samples over one 16.6-d cycle\\
\hline
\hline
 & & 837 ksec & \\
\hline
\end{tabular}
\end{center}
\caption{PCA observations of Cir~X-1 (1996--1997).}
\label{tab:pcaobs}
\end{table}

\subsection{Observation Summaries}

The intensity and activity level of Cir~X-1 had been observed to
change over a period of years (see section~\ref{sec:intro_cirx1}).  We
thus did not know what to expect from Cir~X-1 once \RXTE\ launched.
By the end of 1996 February, it was already clear from the very early
ASM data that Cir~X-1 was currently bright and active. Thus, we
requested that our proposed PCA/HEXTE observations of Cir~X-1 be
activated, and began observations in early 1996 March. The eight
studies (A--H) of our project are described briefly here.

\subsubsection{Study A: 1996 March 5--21}

The first study consisted of twelve observations, each $\sim$6000~s in
duration. They were separated by $\sim$2-day intervals to sample one
complete 16-d cycle (see Figure~\ref{fig:pca_by_cycle}A). The first
observation of study~A occurred during the half day before phase zero
and showed significant dipping behavior. The second observation
occurred shortly after phase zero, and the source was very bright (up
to $\sim$2.6 Crab) and highly variable.  Nine observations carried out
between phases 0.2 and 0.8 showed a quite steady flux, and a final
observation shortly before the next phase zero showed increased
activity again.  These data are presented in
Chapter~\ref{ch:mar96paper}, with a focus on the evolution of the
timing and spectral properties during the non-flaring phases.

\subsubsection{Study B: 1996 April 7}

\begin{figure}
\begin{centering}
\PSbox{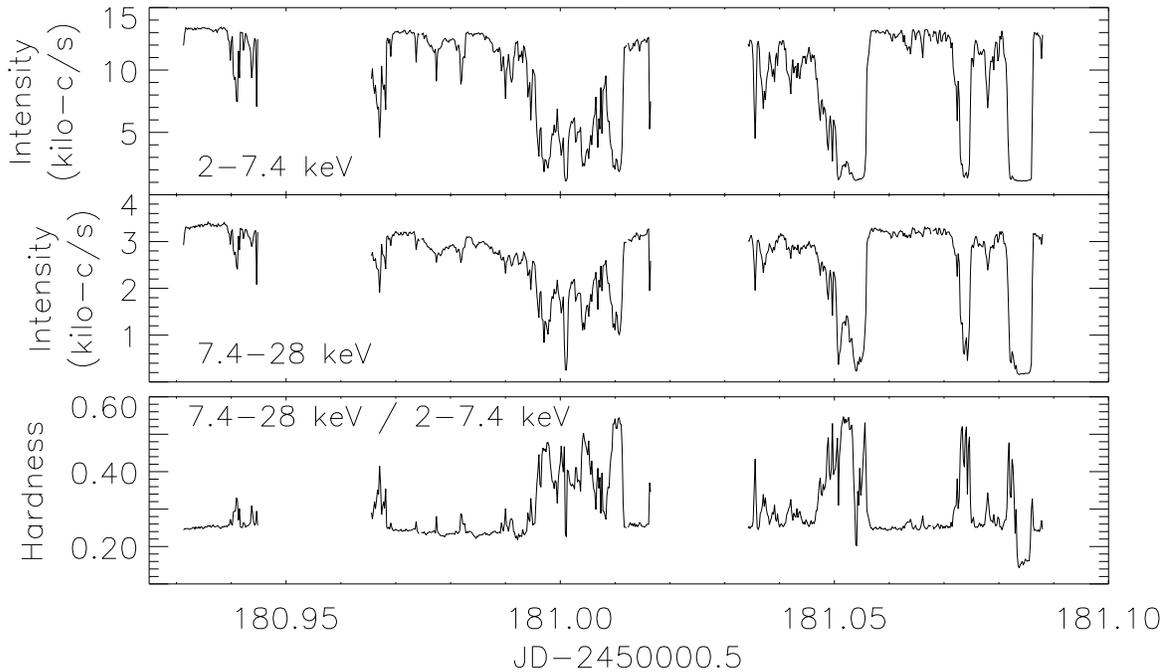 voffset=-56 hscale=90 vscale=90 hoffset=-10}{6.5in}{3.4in}
\caption{
PCA light curves in two energy bands (16-s bins), and the ratio of
intensity in the hard band to that in the soft band, for the
observation on 1996 April 7 (study~B), spanning 15~ks. The lowest
intensity in both bands during dips is well above the background level
($\sim$30~c/s and $\sim$45~c/s for 2--7.4~keV and 7.4--28~keV
respectively, not subtracted).  }
\label{fig:apr96_pcalc_hr}
\end{centering}
\end{figure}

This study consisted of a single 10-ks observation carried out on 1996
April 7, approximately one full cycle after the end of the initial
monitoring campaign (see Figure~\ref{fig:apr96_pcalc_hr}). This
observation also showed significant dipping below the $\sim$1.0~Crab
baseline ($\sim$13~kilocounts/s) during the half day before phase
zero.  The dips are chaotic and accompanied by dramatic spectral
evolution (as indicated by the hardness ratio). Similar dipping
behavior from study~D (see below) was studied in detail and is
presented in Chapter~\ref{ch:absdips}. This observation occurred
during PCA gain epoch~2 (which defines the channel-to-energy
conversion factor, see Table~\ref{tab:pca_ch2en}).  All following
observations of Cir~X-1 occurred during gain epoch~3.

\subsubsection{Study C: 1996 May 7--14}

In 1996 May, the first of two multi-frequency campaigns was carried
out, with coordinated infrared, radio, and \RXTE\ observations.  These
observations spanned several days on either side of phase zero, and
showed an increase in intensity in all three frequency regimes during
the day following phase zero. These multi-frequency observations are
discussed in more detail in Chapter~\ref{ch:multifreq}.

\subsubsection{Study D: 1996 Sept.\ 20 -- Oct.\ 8}
\begin{figure}
\begin{centering}
\PSbox{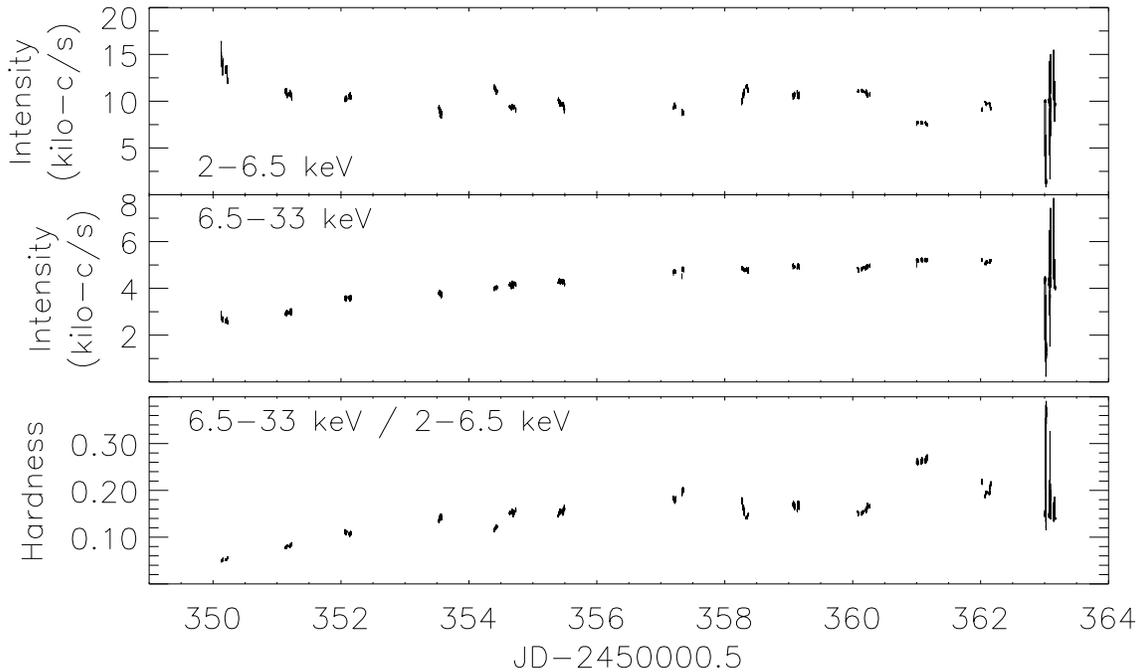 voffset=-56 hscale=90 vscale=90 hoffset=-10}{6.5in}{3.4in}
\caption{
PCA light curves in two energy bands (16-s bins), and the ratio of
intensity in the hard band to that in the soft band, for the
observations on 1996 September~24 -- October~8 (study~D). Due to some
observations with a PCU not operating, all points in this plot were
derived from PCUs 0, 1, \& 2 and adjusted by 5/3 for comparison with
other figures in this section. Phase zeros occurred at days 346.782
and 363.328.}
\label{fig:fall96_pcalc_hr}
\end{centering}
\end{figure}

In 1996 September, a high efficiency observation was carried out,
providing 60\% coverage of the 48-hour period beginning 0.5 days before
phase zero. These data showed extensive dipping both before and after
phase zero and are the focus of Chapter~\ref{ch:absdips}, in which the
spectral evolution of dips is studied in detail. Following this
extensive phase zero observation, 13 sampling observations were made
over the remainder of the 16.55-d cycle. These are shown in
Figure~\ref{fig:fall96_pcalc_hr} and demonstrate a gradual hardening
of the spectrum with orbital phase, as well as more dramatic spectral
evolution during dips in the last observation.

\subsubsection{Study E: 1997 Feb.\ 18 -- Mar.\ 4}
Eight 6-kilosecond observations were carried out during 1997 February
18 -- March 4, sampling a complete cycle at approximately 2-day
intervals (see Figure~\ref{fig:pca_by_cycle}E). The intensity of this
cycle declined from its peak value gradually over $\sim$10~days. These
observations are used in Chapter~\ref{ch:feb97paper} to study the
evolution of timing properties as a function of intensity and spectral
state.

\subsubsection{Study F: 1997 May 16 -- May 21}
In 1997 May, a series of short (1~ks) public observations of Cir~X-1
were carried out between phases 0.35 and 0.7 (the data were originally
part of another observer's program, but were made publicly available
due to a scheduling conflict).  Although these data were not part of
our proposed observations, they fit well with our other sampling
studies and were thus incorporated into the project.  The intensity
level during each observation was relatively steady, but the average
level varied between 1.0 and 1.6~Crab.

\subsubsection{Study G: 1997 June 4--22}
Our most extensive set of observations were carried out during 1997
June. These included a 7-day period of 56\% coverage that included dips
before phase zero and most of a very active flaring state. This
excellent coverage allowed us to observe continuous transitions between
the various spectral/intensity states we found in the earlier 1997
February--March data. These results are presented in Chapter~\ref{ch:fullZ}.

Some of our June PCA observations were coordinated with with radio,
optical, and infrared observations made by collaborators in South
Africa and Australia.  The results from this second multi-frequency
campaign are presented in Chapter~\ref{ch:multifreq} along with the
previous campaign (study~C).

\subsubsection{Study H: 1997 Sept.\ 18 -- Oct.\ 4}
\begin{figure}
\begin{centering}
\PSbox{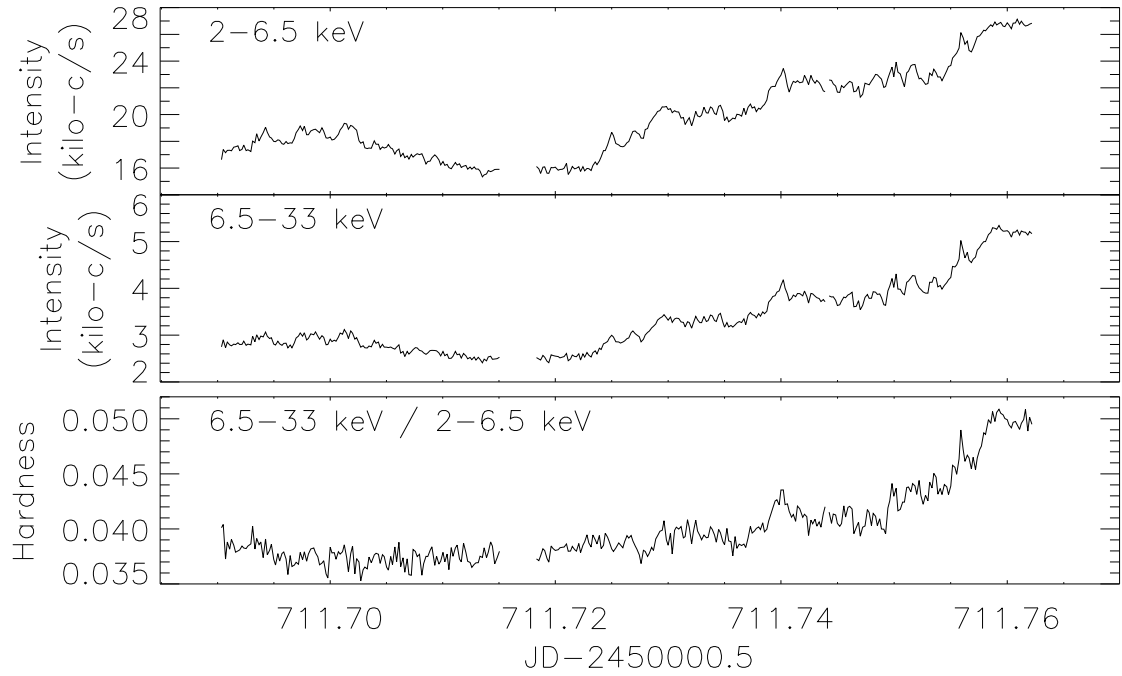 voffset=-56 hscale=90 vscale=90 hoffset=-10}{6.5in}{3.4in}
\caption{
PCA light curves in two energy bands (16-s bins;
13~kilocounts/s=1.0~Crab for the combined 2-33~keV band), and the
ratio of intensity in the hard band to that in the soft band, for the
observation on 1997 September 20 (part of study~H), spanning 7000~s.  }
\label{fig:sept97_pcalc_hr}
\end{centering}
\end{figure}

A final set of eight 6-kilosecond observations were carried out during
1997 September--October to again sample a complete 16-d cycle. These
observations all showed the intensity to be near the 1.0~Crab
baseline, except for the sample at phase 0.06, where the intensity
reached 2.5 Crab. The soft and hard light curves and hardness ratio
for the flaring observation are shown in
Figure~\ref{fig:sept97_pcalc_hr}. There is strong variability on time
scales of hundreds of seconds during the climb to 2.5~Crab. The
hardness ratio is generally anti-correlated with intensity during the
first segment and then correlated with intensity during the remainder
of the observation. This bi-modal behavior can be understood as
relating to different branches in a hardness-intensity diagram.  Both
types of behavior are associated with the spectral/intensity
states that are discussed in Chapters~\ref{ch:feb97paper}
and~\ref{ch:fullZ} in the context of studies~E and~G.

\chapter{X-ray Timing and Spectral Evolution vs.\ Orbital Phase}
\label{ch:mar96paper}

\begin{center}
{\em This chapter has been published (in a slightly modified
form) in the September 20, 1996 issue of\, {\em The Astrophysical Journal}
(Vol. 469)}~\cite{shirey96}.
\end{center}

\section{Overview}

We have carried out a study of Cir X-1 (study~A, see
Figure~\ref{fig:pca_by_cycle}) through detailed sampling over a single
16.55-d intensity cycle with the {\it Rossi X-ray Timing Explorer}
(\RXTE) Proportional Counter Array (PCA). We report here the current
state of the source and focus primarily upon the evolution of the
emission characteristics away from the flaring activity, i.e., at
phases 0.2 $\simlt \phi \simlt$ 0.9 (based on the 1991 radio ephemeris
of Stewart et~al.~\cite{stewart91}).  Heretofore, these phases have
not been systematically sampled with an instrument of large
aperture. In the eccentric binary scenario, the source would be
relatively remote from the secondary during these phases.

During the non-flaring phases, Cir X-1 remained unusually bright
($\sim$1.0~Crab) and relatively steady. The Fourier power density
spectrum of the source was observed to vary with strong correlations
among low-frequency flat-topped power ($\simlt$ 1--10 Hz), a QPO peak
centered at 1.3--12~Hz, and a broad QPO peak centered at $\sim$20~Hz
up to $\sim$100~Hz. As orbital phase increased within the cycle, the
rms amplitude of the flat-topped power generally (but not
monotonically) decreased, while the QPO features generally evolved
toward higher frequency. The PCA spectrum was observed to generally
harden during the non-flaring phases, consistent with results from the
\RXTE\ All-Sky Monitor (see section~\ref{sec:asmobs}).

\section{Observations and Results}

\begin{table}
\begin{center}
\begin{tabular}{ccccccc}
\hline
\hspace{0pt} & Julian Date & Orbital & & & Julian Date & Orbital \\
Obs.\footnotemark[1] &
$-$2450000.5\footnotemark[2] &
Phase\footnotemark[3] &
\hspace{0pt} & Obs. & $-$2450000.5 & Phase \\
\hline
\hline
1 & 147.9 & 0.98 & & 7 & 158.3 & 0.61 \\
2 & 150.7 & 0.15 & & 8 & 158.7 & 0.63 \\
3 & 152.8 & 0.28 & & 9 & 159.7 & 0.69 \\
4 & 154.6 & 0.39 & & 10 & 160.7 & 0.75 \\
5 & 155.8 & 0.46 & & 11 & 161.1 & 0.78 \\
6 & 156.6 & 0.50 & & 12 & 163.8 & 0.94 \\
 \hline
\end{tabular}
\end{center}
\caption{March 1996 PCA observations of Cir~X-1}
\label{tab:mar96_obs}
\end{table}
\footnotetext[1]{Each observation lasted 1--5 hr and produced 1.4--10
 ks of data. Dates and phases listed are the centroids of the
 observations.}
\footnotetext[2]{JD-2450000.5=147 corresponds to 1996 March 5.}
\footnotetext[3]{Orbital phase based on the radio ephemeris of
Stewart et~al.\ 1991~\cite{stewart91}. Due to faint radio flares 
in recent years, this ephemeris is still currently in use as the 
best available (G. Nicolson 1996, private communication).}

\begin{figure}
\begin{centering}
\PSbox{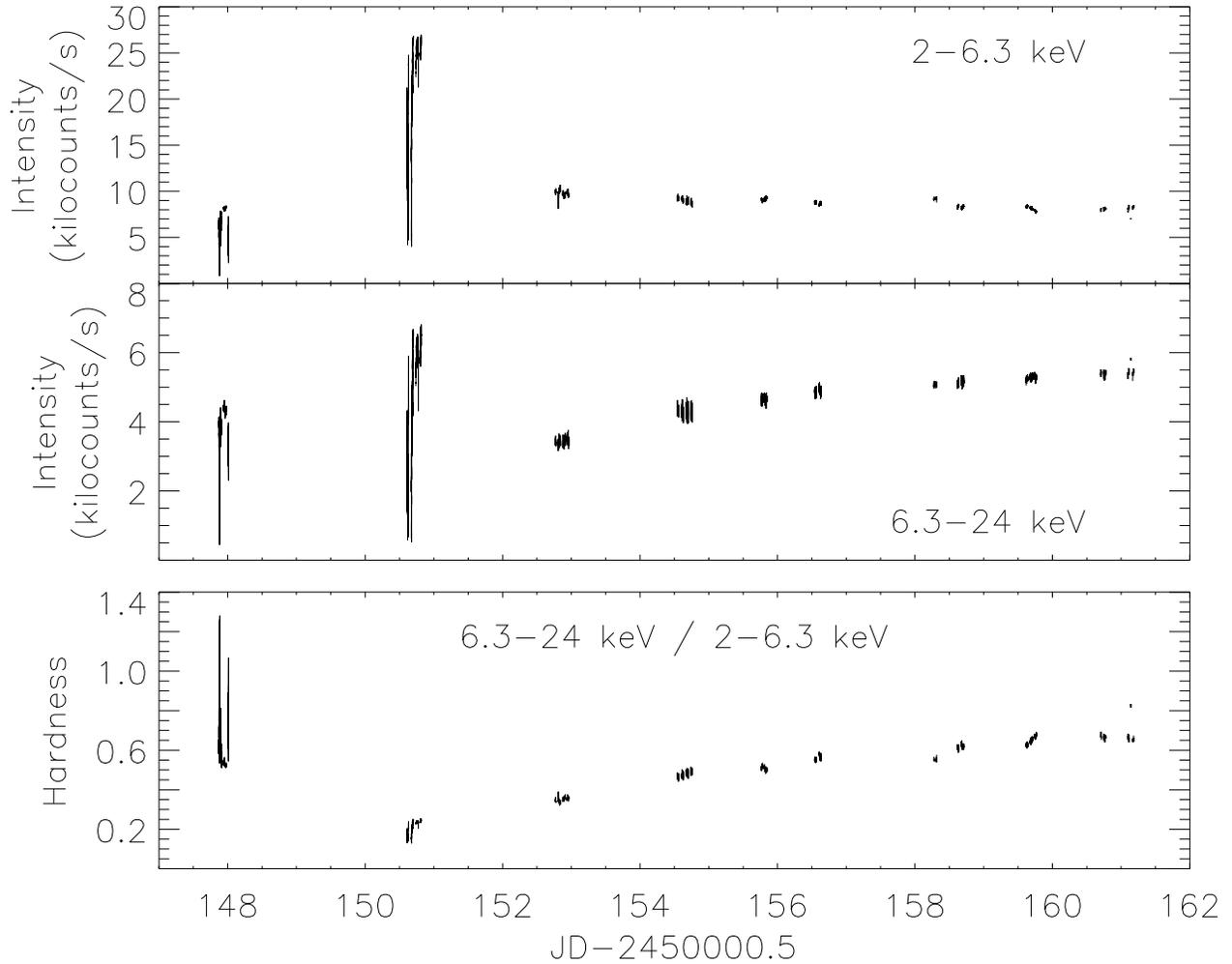 voffset=-250 hoffset=-40}{6.5in}{5.2in}
\caption{
PCA light curves in two energy bands (16-s bins), and the ratio of
intensity in the hard band to that in the soft band, for observations
1--11 of 1996 March. Phase zero corresponds to day~148.227. The
intensity is well above the background level in both bands
($\sim$25~c/s and $\sim$40~c/s for 2--6.3~keV and 6.3--24~keV
respectively, not subtracted) even during the dips. Observation~12 is
omitted since it was carried out with only three PCUs and a different
gain setting.  
}
\label{fig:mar96_2lc_hr}
\end{centering}
\end{figure}

In 1996 March, Cir~X-1 was sampled with the PCA (effective area
$\sim$7000 cm$^2$, see section~\ref{sec:pca}) twelve times during one
16.55-day orbital period. The observation times and orbital phases are
summarized in Table~\ref{tab:mar96_obs}. The low and high-energy light
curves for observations 1--11 are shown in
Figure~\ref{fig:mar96_2lc_hr}, along with the hardness ratio derived
from those intensities. Observation~12 used a different PCA gain
setting (gain epoch~2, see Table~\ref{tab:pca_ch2en} on
page~\pageref{tab:pca_ch2en}) and only 3~PCUs, so it is not included
in Figure~\ref{fig:mar96_2lc_hr}. 
The hardness ratio generally increases during the non-flaring phases as
phase increases up to $\sim$0.7, beyond which the ratios are more
steady. This confirms the evolution observed in the ASM data (see
section~\ref{sec:asmobs}).

\begin{figure}
\begin{centering}
\PSbox{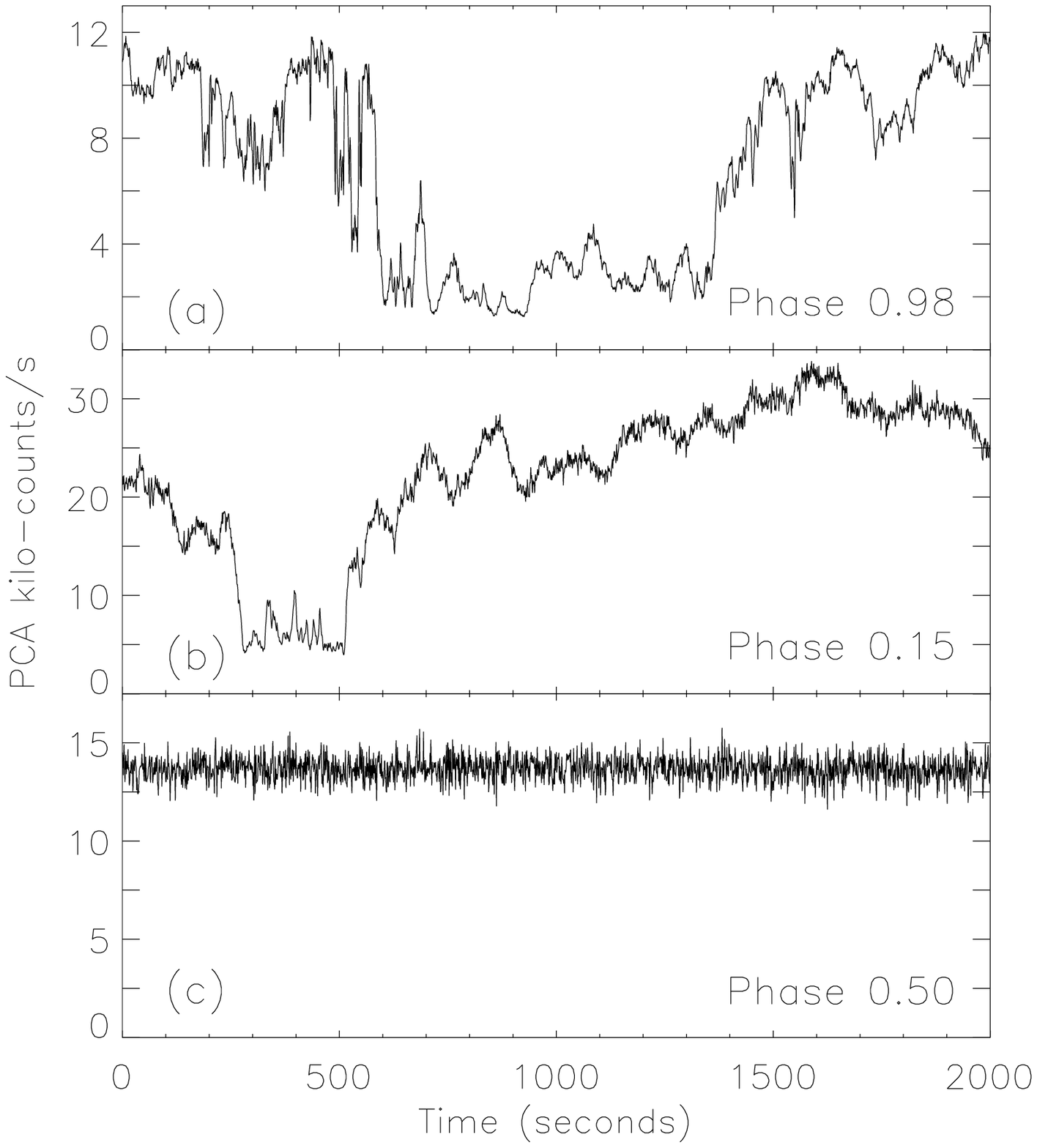 voffset=-100 hoffset=-50}{6.0in}{7in}
\caption{
Sample PCA light curves (2--60~keV) comparing 2000-s
segments showing (a) strong variability and dips near phase zero
(phase 0.98; Obs.~1), (b) a brighter portion of the bright flaring
state three days later (phase 0.15; Obs.~2), and (c) a typical
observation away from phase zero (phase 0.50; Obs.~6). Background
($\sim$100 c/s) is not subtracted and no deadtime corrections have
been made.
}
\label{fig:mar96_pcalc}
\end{centering}
\end{figure}

Three detailed sample light-curve segments are shown in
Figure~\ref{fig:mar96_pcalc}. In the full PCA energy band (2--60~keV),
Cir~X-1 was found to be extremely active, remaining bright ($\simgt$
1.0~Crab) throughout the 16.55-d cycle (1 Crab $\approx$ 13,000
c/s). In the vicinity of phase zero (Obs.~1, 2,
\& 12), dramatic flares and dips occurred
(Figure~\ref{fig:mar96_pcalc}a,b). In Obs.~2 (phase 0.15), the count
rate climbed to more than 2.5~Crab
(Figure~\ref{fig:mar96_pcalc}b). Away from phase zero (0.2 $\simlt
\phi \simlt$ 0.8; Obs.~3--11), the count rate remained fairly steady
(within 15\%) at about 1.0~Crab. Figure 2c shows a sample light curve
for phase 0.5, which is typical of these observations. The source is
relatively steady but exhibits flickering greatly in excess of Poisson
statistics.

Power density spectra (PDS) were produced from PCA data covering
2--8.6 keV in 61-$\mu$s time bins. For each 96-min \RXTE\ orbit, a
continuous observation of 1 to 4 ks was divided into 64-s segments,
and the PDS was calculated for each segment. All PDS from the segments
of a single continuous observation were averaged together, weighted by
the total counts. The resulting average power spectrum from each {\it
RXTE} orbit was then logarithmically rebinned, with
dead-time-corrected Poisson noise subtracted \cite{zhang95} and
normalized to give the fractional rms amplitude squared per~Hz (after
Belloni \& Hasinger~\cite{belloni90b}).

\begin{figure}
\begin{centering}
\PSbox{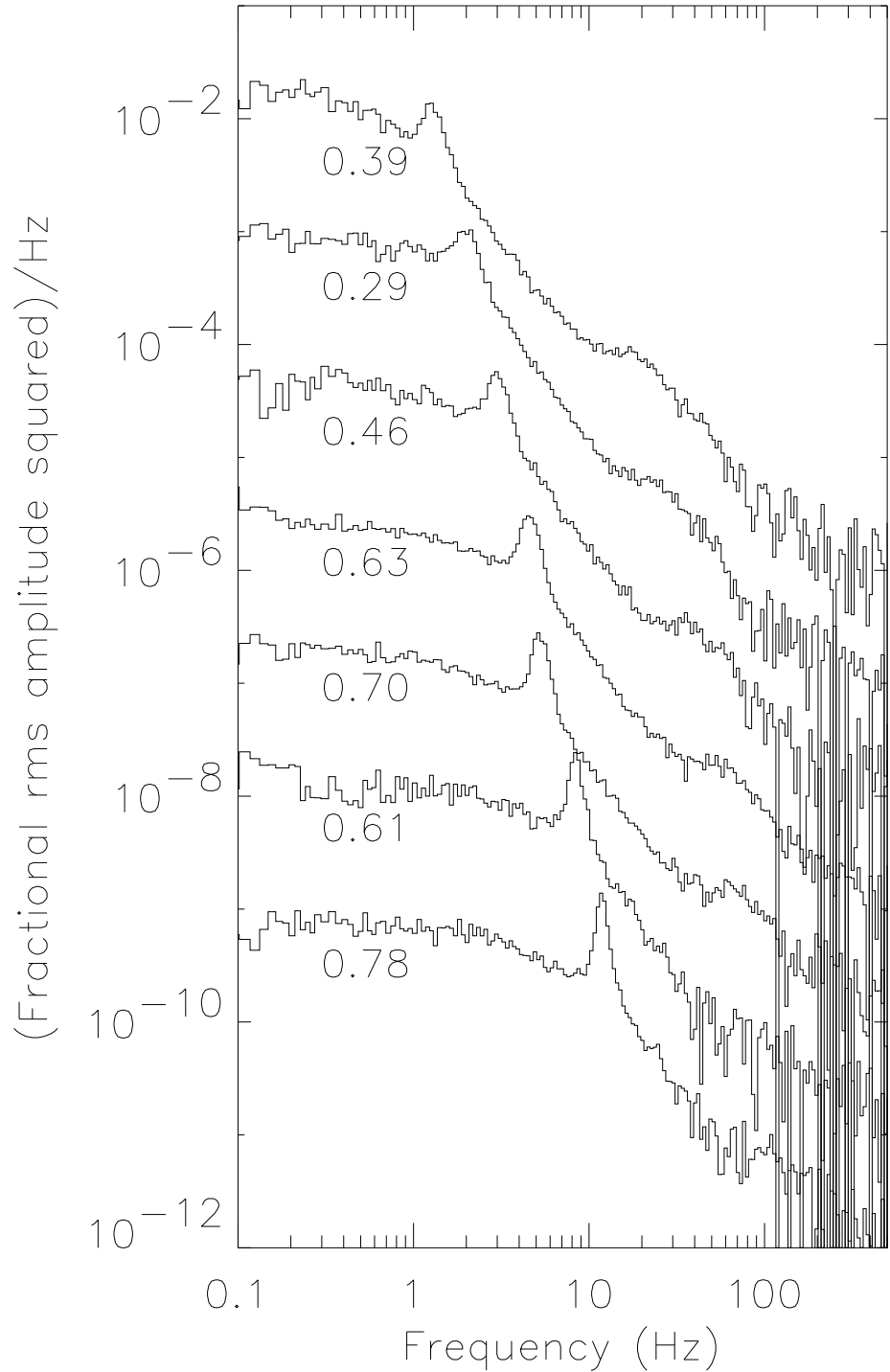 	
	hscale=90 vscale=90 voffset=0 hoffset=0}{4.8in}{7.4in}
\caption{
Power density spectra from seven \RXTE\ orbits away
from phase zero, offset downward at decade intervals. (The ordinate
scale applies to the top curve.) The curves are ordered by the
frequency of the 1.3--12~Hz QPO peak and are labeled with orbital
phase.
}
\label{fig:mar96_pds}
\end{centering}
\end{figure}

The PDS from seven orbits are shown in Figure~\ref{fig:mar96_pds},
offset downward by one-decade intervals. These samples show
significant low-frequency flat-topped power which cuts off above
1-10~Hz and also a QPO peak which sits on the high-frequency edge of
the flat-topped power at 1.3--12~Hz (compared to 1.4~Hz and 5-17~Hz
QPO reported near phase zero by \cite{tennant87,tennant88}. The PDS in
Figure~\ref{fig:mar96_pds} are ordered by the frequency of the QPO
peak. All the PDS during the non-flaring phases exhibit these
features. The flat-topped power level systematically decreases as the
cutoff and QPO frequencies increase (note that each flat-top level is
more than one decade below the previous level in
Figure~\ref{fig:mar96_pds}). Above the cut-off frequencies, the curves
all track together with roughly the same power. A few PDS not
illustrated here have a steep low-frequency component below
0.1-0.4~Hz.

A broad high-frequency QPO is also apparent in
Figure~\ref{fig:mar96_pds} centered at $\sim$20~Hz to $\sim$100~Hz or
more. For observations in which the 1.3--12~Hz QPO frequency is below
$\sim$8~Hz, the high-frequency broad peak is always clearly present;
otherwise, it is sometimes marginally detectable above 100~Hz. Note
that this high-frequency QPO peak moves to higher frequencies with the
low-frequency QPO peak.

\begin{figure}
\begin{centering}
\PSbox{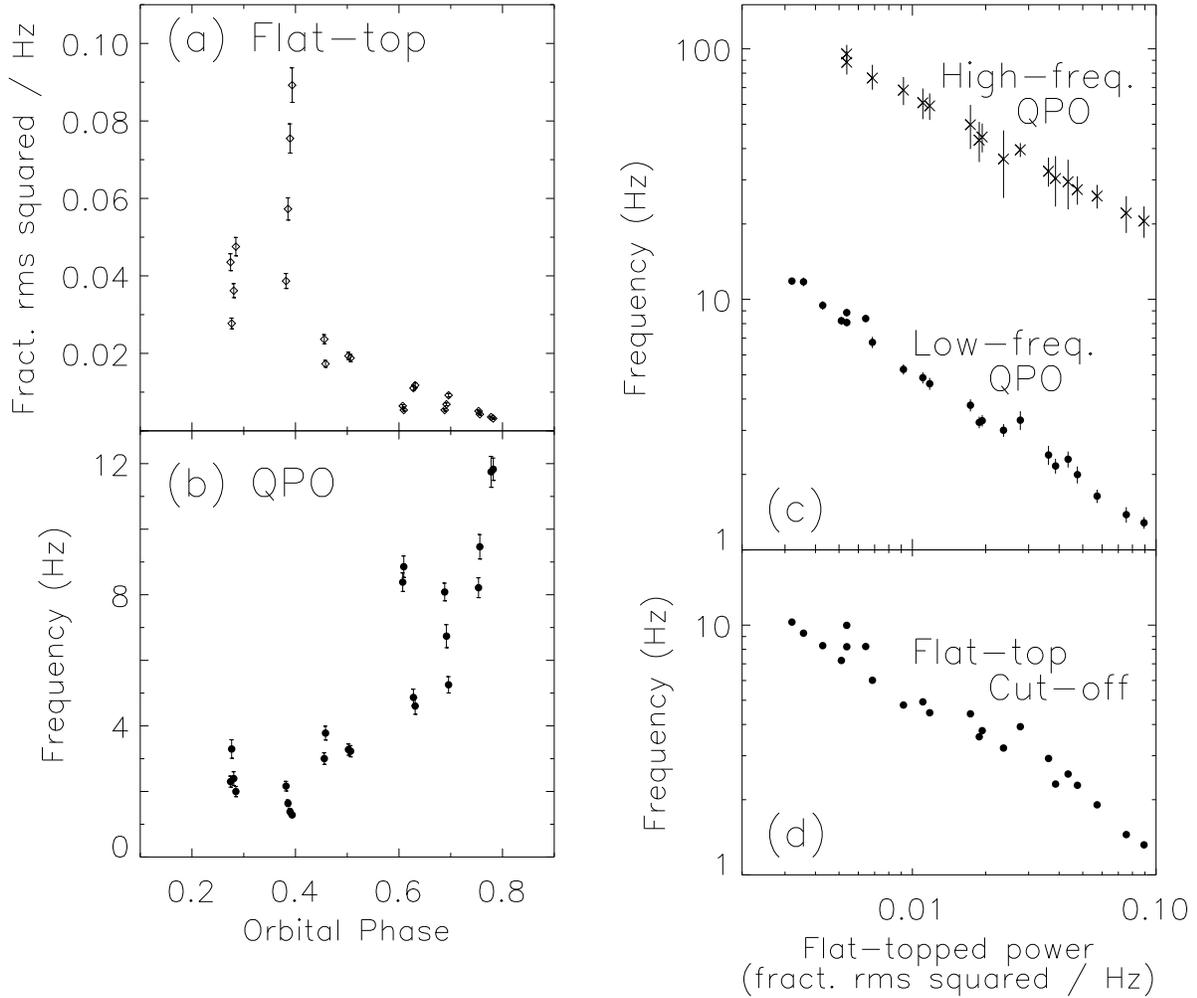
	hscale=90 vscale=90 voffset=-210 hoffset=-30}{6.0in}{5.1in}
\caption{
Flat-top power level (a) and QPO frequency (b) as a
function of orbital phase over a single orbital cycle. (c) Frequency
of the narrow low-frequency QPO and the broad high-frequency QPO vs.
the flat-top power level. (d) Cut-off frequency of the flat-topped
power vs.\ flat-top power level.
}
\label{fig:mar96_pds_trends}
\end{centering}
\end{figure}

All the PDS in the non-flaring phases were analyzed to further study
these trends. The flat-topped component was fit with a zero-centered
Lorentzian profile, the two peaked features were also each fit with a
Lorentzian, and the high-frequency tail was fit with a power
law. Although the results of the fits indicate that this model is
crude (via large values of the associated chi-squared statistic), the
fits were adequate to determine approximate centroid frequencies and
power levels. Errors (rms) for QPO centroid frequencies were estimated
from 90\% confidence intervals of the fits or 20\% of the FWHM of the
peak, whichever was greater. Errors for the flat-top power level were
estimated to be $\sim$5\%. As the orbital phase increased, the power
level of the flat-topped noise generally decreased
(Figure~\ref{fig:mar96_pds_trends}a) while the narrow QPO frequency
increased (Figure~\ref{fig:mar96_pds_trends}b), in accord with the
trends noted above. The trends are not monotonic though; a large
excursion occurs at $\phi$ = 0.4 in both curves. Remarkably, the
correlation between these two quantities (flat-topped level and narrow
QPO frequency) in Figure~\ref{fig:mar96_pds_trends}c is very strong
(and negative). Moreover, the frequency of the 20--100~Hz peak scales
as $\sim$13 times the frequency of the 1.3--12~Hz peak
(Figure~\ref{fig:mar96_pds_trends}c) with a weak tendency for the
frequency ratio to decrease (from $\sim$16 to $\sim$11) with
increasing frequency. The frequency difference between the two peaks
changes by more than a factor of four over this range, from less than
20~Hz to more than 80~Hz. The cutoff frequency of the flat-topped
component (as measured by the width of the best fitting zero-centered
Lorentzian) also moved to higher frequency with the narrow QPO peak
(Figure~\ref{fig:mar96_pds_trends}d). These two frequencies agree
within 20\% for all observations.

\section{Discussion}

The 1.4 and 5-17~Hz QPO reported by Tennant et al.
\cite{tennant87,tennant88} were found during observations near phase
zero. It was suggested that the QPO were bimodal, with the 1.4~Hz QPO
independent of luminosity and the 5--17~Hz component correlated with
the estimated 2--10~keV unabsorbed energy flux
\cite{tennant87,tennant88}. Our observations show a gradual evolution
of the QPO from 1.3 to 12~Hz over the {\it non}-flaring phases. Some
of the phase zero {\it EXOSAT} observations which exhibited 5--17~Hz
QPO also had a QPO peak at 100--200~Hz~\cite{tennant87}. This low/high
frequency QPO pair is likely to be the evolving QPO pair we have
observed, but extended to higher frequencies.

The correlation between the flat-topped power level and the cut-off
frequency (Figure~\ref{fig:mar96_pds_trends}d) is similar to the
effect observed by Belloni and Hasinger \cite{belloni90a} in the the
black hole candidate Cyg X-1 and also observed in the neutron star
4U~1608-522 \cite{yoshida93} by Yoshida et~al. In fact, several
properties of 4U~1608-522 show correlations similar to those we have
observed in Cir~X-1. In two of the three observations of 4U~1608-522
by Yoshida et~al.~\cite{yoshida93}, QPO were seen (at 0.4 and 2.0~Hz
respectively) just above the knee in the power spectrum, and a broader
peaked noise feature was observed at higher frequencies of 1.0 and
5.1~Hz respectively \cite{yoshida93}. We note that the ratio of the
two peaks is the same in both observations, just as the two peaks in
our Cir~X-1 observations move to higher frequency with a nearly
constant frequency ratio (although the ratio is different for the two
sources). In both sources, the spectrum generally (but not always)
hardens as these these features move out to higher frequencies and the
flat-topped power level decreases.

The \RXTE\ observations of Cir~X-1 have (1) confirmed a general
hardening of the spectrum during the non-flaring phases 0.2--0.7, (2)
demonstrated a trend (with moderate scatter) for the power
density features to shift to higher frequency and the amplitude of the
low-frequency power to decrease with increasing orbital phase during
the non-flaring phases, and (3) demonstrated extremely tight
correlations among the QPO features and the flat-topped power. In the
eccentric-orbit model, these characteristics may reflect the evolving
state of the accretion disk while it is being depleted with little or
no replenishment. For example, the leveling of the hardness ratios
after phase 0.6-0.8 may reflect a time constant (of order 10 days) of
the disk structure. 
%The spectral hardening and increasing QPO frequencies suggest that, as
%the depletion occurs, the inner portion of the accretion disk, with
%its higher temperatures and greater Keplerian frequencies, becomes
%relatively more important in the emission process.

\chapter{QPOs Associated with Spectral Branches}
\label{ch:feb97paper}

\begin{center}
{\em This chapter is scheduled to be published (in a slightly modified
form) in the October 10, 1998 issue of\, {\em The Astrophysical Journal}
(Vol. 506)~\cite{shirey98:feb97}.}
\end{center}

%\begin{center}
%\begin{large}
%{\bf Quasi-periodic Oscillations Associated with Spectral Branches in \RXTE\ Observations of Circinus~X-1} \\
%\end{large}
%Robert E. Shirey, Hale V. Bradt, Alan M. Levine, \& Edward H. Morgan
%\copyright 1998. The American Astronomical Society. All rights reserved.
%\end{center}

\section{Overview}

We present {\it Rossi X-ray Timing Explorer}\ (\RXTE) All-Sky Monitor
observations of the X-ray binary Circinus~X-1 which illustrate the
variety of intensity profiles associated with the 16.55~d flaring
cycle of the source. We also present eight observations of Cir~X-1
made with the \RXTE\ Proportional Counter Array over the course of a
cycle wherein the average intensity of the flaring state decreased
gradually over $\sim$12 days (study~E, see
Figure~\ref{fig:pca_by_cycle}). This unusually slow transition allows
us to demonstrate how the time-variability properties of the source
are related to its intensity and its spectral properties (In
Chapter~\ref{ch:mar96paper}, we characterized only the quiescent
phases of a typical cycle).  Fourier power density spectra for these
observations show a narrow quasi-periodic oscillation (QPO) peak which
shifts in frequency between 6.8~Hz and 32~Hz, as well as a broad QPO
peak that remains roughly stationary at $\sim$4~Hz. We identify these
as Z-source horizontal and normal branch oscillations (HBOs/NBOs)
respectively. Color-color and hardness-intensity diagrams (CDs/HIDs)
show curvilinear tracks for each of the observations. The properties
of the QPOs and very low frequency noise allow us to identify segments
of these tracks with Z-source horizontal, normal, and flaring branches
which shift location in the CDs and HIDs over the course of the
16.55~d cycle. These results contradict a previous prediction, based
on the hypothesis that Cir~X-1 is a high-$\dot{M}$ atoll source, that
HBOs should never occur in this source~\cite{oosterbroek95,klis94}.

\section{Observations}

\begin{figure}
\PSbox{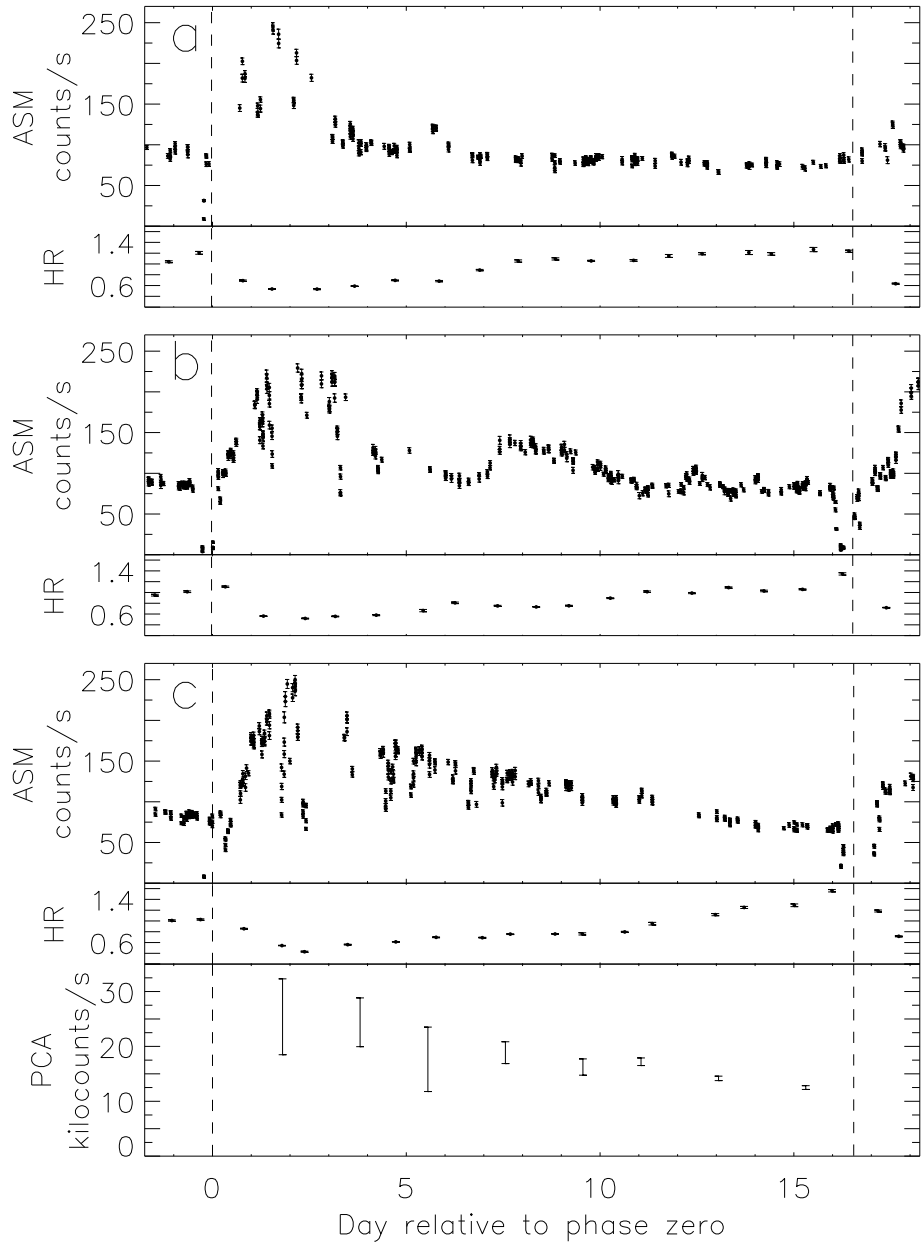 hscale=75 vscale=75 hoffset=36}{4.7in}{6.1in}
\caption{ 
\RXTE\ ASM light curves (1.5--12~keV) for three 16.55~d cycles of
Cir~X-1 showing different flaring profiles. Each intensity point
corresponds to a 90-s exposure by one of the three ASM cameras, and
the hardness ratio (HR), defined as the ratio of counting rates for
5--12~keV to 3--5~keV, is shown in one-day averages. The 3--5~keV to
1.5--3~keV hardness ratio exhibits very similar behavior and is not
shown here. The intensities are for Cir~X-1 after background and other
sources in the field of view have been subtracted. The Crab nebula
yields $\sim$75~c/s.  Vertical dashed lines indicate phase zero based
on the radio ephemeris of Stewart et~al. 1991. Day
zero corresponds to (a) 1997 April~23.87 (b) 1996 August~2.14 and (c)
1997 February~16.69. For cycle (c), the intensity ranges
(\intens{2.0}{18}) seen in the eight \RXTE\ PCA observations (I--VIII
in time order) are also shown.}
\label{fig:feb97_asm}
\end{figure}

% ASM observations
The \RXTE\ ASM has provided 2--12~keV light curves of Cir~X-1 since
1996 February (see Figure~\ref{fig:asm_2yr_lc}). Throughout the ASM
observations, the baseline intensity of Cir~X-1 has remained near
1.0~Crab. However, the profile of each 16.55-d flaring cycle can vary
considerably. The variety of intensity profiles is illustrated in
Figure~\ref{fig:feb97_asm}, which shows ASM light curves and hardness
ratios for three individual cycles.  In many cycles, after 3--5~days
in the flaring state, the intensity is quite steady for the remainder
of the cycle (e.g., Figure~\ref{fig:feb97_asm}a).  In addition to the
main flaring episode, some cycles show a mid-phase flare (not always
at the same phase) to as high as 2~Crab
(Figure~\ref{fig:feb97_asm}b). Occasionally, the flaring state begins
after phase zero and continues for most of the cycle with a gradually
decreasing intensity (Figure~\ref{fig:feb97_asm}c).
Despite the variety of intensity profiles, all cycles observed with
the ASM show the general pattern of spectral hardening mentioned in
section~\ref{sec:asm}. During the half day before phase zero, and
continuing intermittently for up to two days, brief dips occur in many
cycles (perhaps in all cycles, since the ASM coverage is incomplete).
These dips are seen as isolated low points in the ASM light curves of
Figure~\ref{fig:feb97_asm}.

%-----------------------------------------------
%-----------------------------------------------

%----------------------------------------------- 
\begin{table}
\begin{center}
\begin{tabular}{cccc}
\hline
\hline
Obs. & Julian Date\footnotemark[1] & Phase & Mean Intensity \\
 &  &  & (Crab)\footnotemark[2]\\
\hline
I & 2450497.90 & 0.10 & 2.3 \\
II & 2450499.98 & 0.23 & 1.8 \\    
III & 2450501.69 & 0.33 & 1.6 \\   
IV & 2450503.66 & 0.45 & 1.5 \\      
V & 2450505.80 & 0.58 & 1.2 \\      
VI & 2450507.31 & 0.67 & 1.3 \\      
VII & 2450509.35 & 0.79 & 1.1 \\      
VIII & 2450511.62 & 0.93 & 1.0 \\      
\hline
\end{tabular}
\end{center}
\caption{PCA observations of Cir~X-1 during 1997 February~18 -- March~4.}
\label{tab:feb97_obs}
\end{table}
\footnotetext[1]{Midpoint of 2--3~hr observation ($\sim$6~ksec of data per 
		observation).}
\footnotetext[2]{1.0~Crab $\approx$ 13,000~counts/s (2--32~keV, all 5 PCUs)}
%----------------------------------------------- 

% PCA observations
Eight PCA observations ($\sim$6~ksec each) were carried out at roughly
two-day intervals (Table~\ref{tab:feb97_obs}) during 1997 February~18
-- March~4 to sample the 16.55~day cycle shown in
Figure~\ref{fig:feb97_asm}c. The very gradual decline of the
flaring-state intensity in this cycle serendipitously provided an
opportunity to study intensity-related source properties. All five
proportional counter units (PCUs) of the PCA operated normally during
each observation, except during the first few minutes of the first
observation when only three PCUs were on. All intensities for that
period have been adjusted by a factor of 5/3, but these data are not
used in color-color and hardness-intensity diagrams due to gain
differences between detectors.

%-----------------------------------------------
\begin{figure}
\PSbox{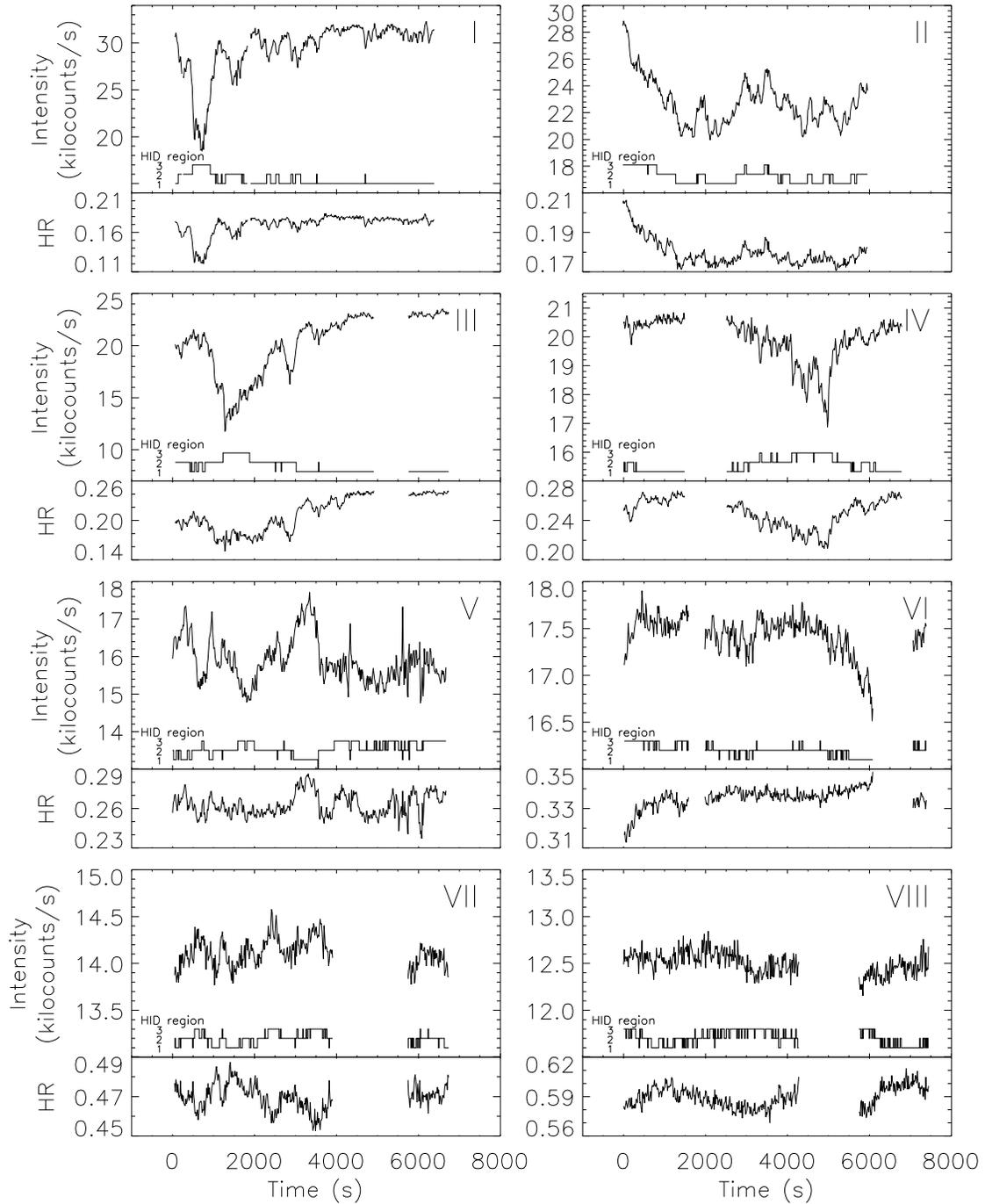 hscale=85 vscale=85}{5.6in}{7.14in}
\caption{ 
PCA light curves (2--18~keV, 5~PCUs) and hardness ratios (HR =
\broadcolor) in 16~s time bins for the eight observations made during
1997 February~18 -- March~4. A count rate of 13~kcts/s $\approx$
1.0~Crab.  The data gaps in Obs.~VI were longer than as shown here;
the second segment of the observation has been shifted left by 4000~s
and the third segment by 5000~s. These data were used to construct the
hardness-intensity diagram in
Figure~\protect{\ref{fig:feb97_cchid_all}}. The association with specific
regions of that diagram is indicated below each light curve.}
\label{fig:feb97_pcalc}
\end{figure}
%-----------------------------------------------

% PCA lightcurves/HRs
Figure~\ref{fig:feb97_pcalc} shows the light curves and hardness
ratios (with 16~s time resolution) for each of the PCA observations
(I--VIII in time order), made as the intensity declined from 2.5~Crab
to 1.0~Crab.  On time-scales of hundreds of seconds, the observations
made at high intensities show strong variability, while observations
at 1.0~Crab ($\sim$13~kilocounts/s) show quite steady count rates. As
expected from the ASM hardness ratios, the PCA hardness ratio
gradually increases from a low value during the early observations
when the source was in the flaring state to a factor four higher as it
reached the quiescent level. The relationship between intensity and
spectral changes is discussed in detail below.

\section{Analysis and Results}

\subsection{Color-color and Hardness-intensity Diagrams}

\begin{figure}
\PSbox{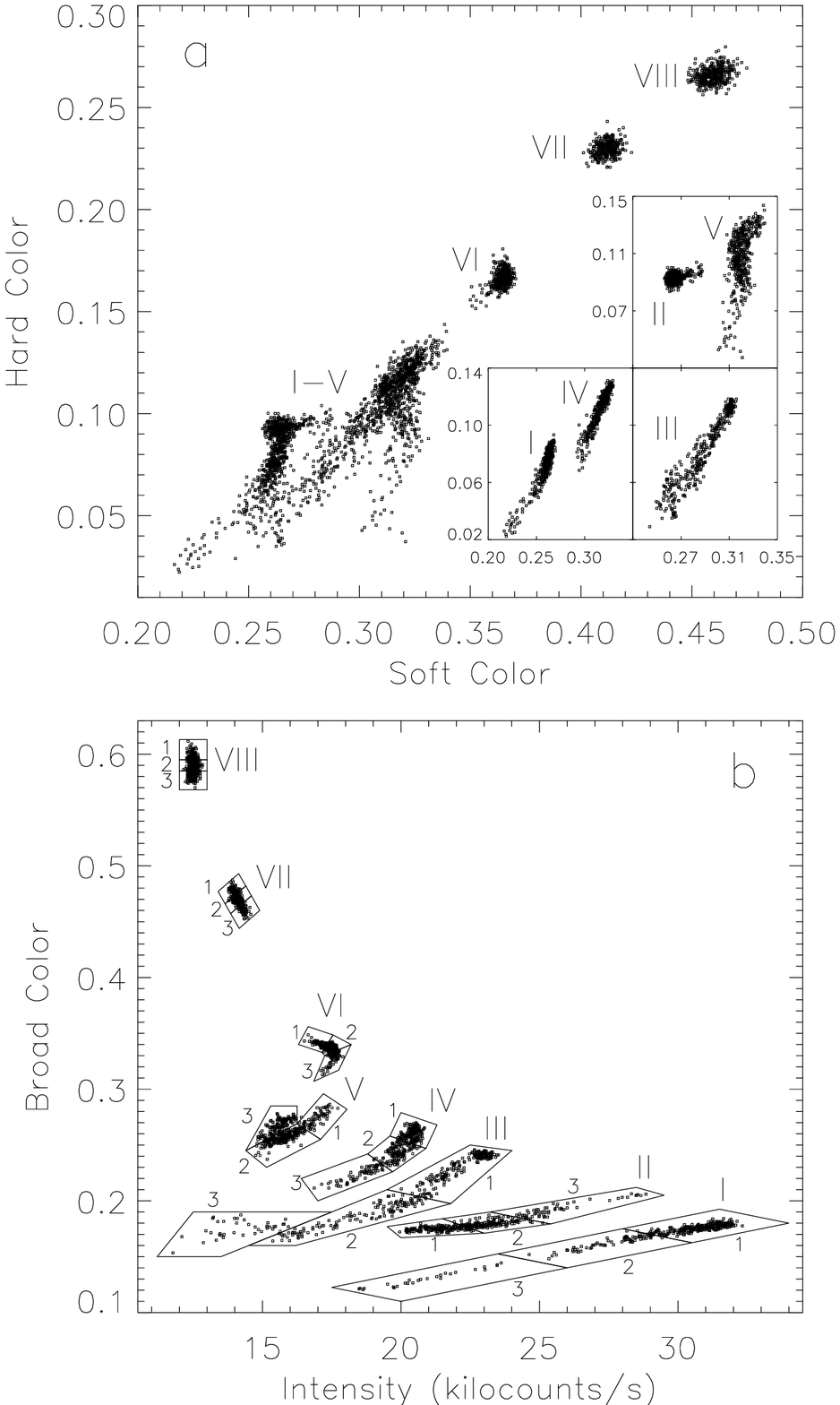 hscale=80 vscale=80 hoffset=50}{4.1in}{7in}
\caption{ 
Color-color diagram (a) and hardness-intensity diagram (b) for all
eight observations (I--VIII). In the CD, soft color is defined as
\softcolor\ and hard color as \hardcolor\@. In the HID, the intensity,
\intens{2.0}{18}, is from all five PCUs and the hardness ratio is a
``broad'' color: \broadcolor\@. Each point corresponds to 16~s of
data. Background has been subtracted, but it does not affect the
intensity or soft color and only slightly affects the hard color. The
three insets in the CD separate overlapping points from observations
I--V. The HID track for each observation has been divided into three
regions (1--3) for timing analysis.}
\label{fig:feb97_cchid_all}
\end{figure}

For the eight PCA observations of 1997 February--March, 16~s intensity
and hardness-ratio measurements were used to construct color-color and
hardness-intensity diagrams (CDs/HIDs, Figure~\ref{fig:feb97_cchid_all}).
The hardness ratios were defined as the ratio of count rates in
selected energy bands: a soft color (\softcolor) and hard color
(\hardcolor) for the CD, and a broad color (\broadcolor) for the HID.
The evolution from flaring to quiescent state produced a large range
of colors and intensities over the entire cycle.  In contrast, each
individual observation yielded a localized cluster or track within the
CDs and HIDs. The spectral branches for each of the observations are
easier to distinguish in the HID than the CD.  The long tracks in the
HID associated with observations~I--V show the color changes
associated with the large intensity variations during the flaring
phases.  The intensity variations are smaller for observations
VI--VIII, but significant color changes do occur during these
observations as well.

%-----------------------------------------------
%-----------------------------------------------

The choice of energy bands used in constructing these diagrams can
affect the appearance of spectral tracks. For observations showing a
single branch, only the length and slope of the branch is affected.
Observations V and VI each show two branches. The orientation of these
branches is discussed in more detail below.

The tracks in the CD and HID are reminiscent of the correlated
spectral/intensity behavior of Z and atoll class LMXBs (see
Figure~\ref{fig:z_atoll_cc_pds}), which also show correlations of
temporal properties with position along tracks or branches in CDs and
HIDs~\cite{hk89}. Thus, we have investigated how the temporal
properties of Cir~X-1 are related to position in the CD or HID. For
this purpose, we divided the HID track for each of the eight
observations into three regions (Figure~\ref{fig:feb97_cchid_all}b).
The choice of numbers for each region was motivated in part by the
timing results discussed below, but the numbers serve mainly as
reference labels rather than as meaningful quantities (such as Z "rank
number").

\subsection{Power Density Spectra}

%-----------------------------------------------
\begin{figure}
\PSbox{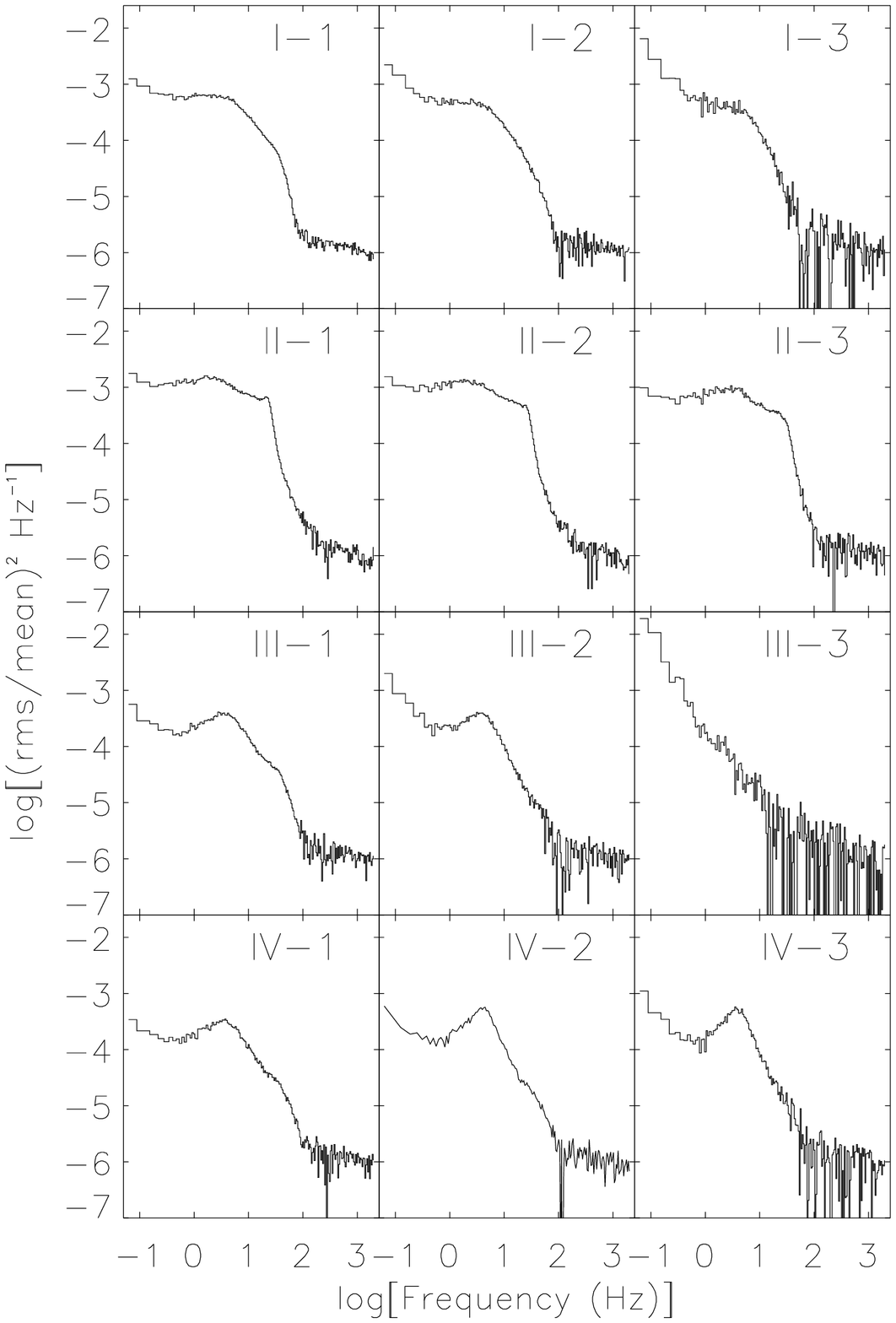 hscale=55 vscale=55 hoffset=-45}{3.25in}{4.6in}
\vspace{-4.6in}\hspace{3.5in}
\PSbox{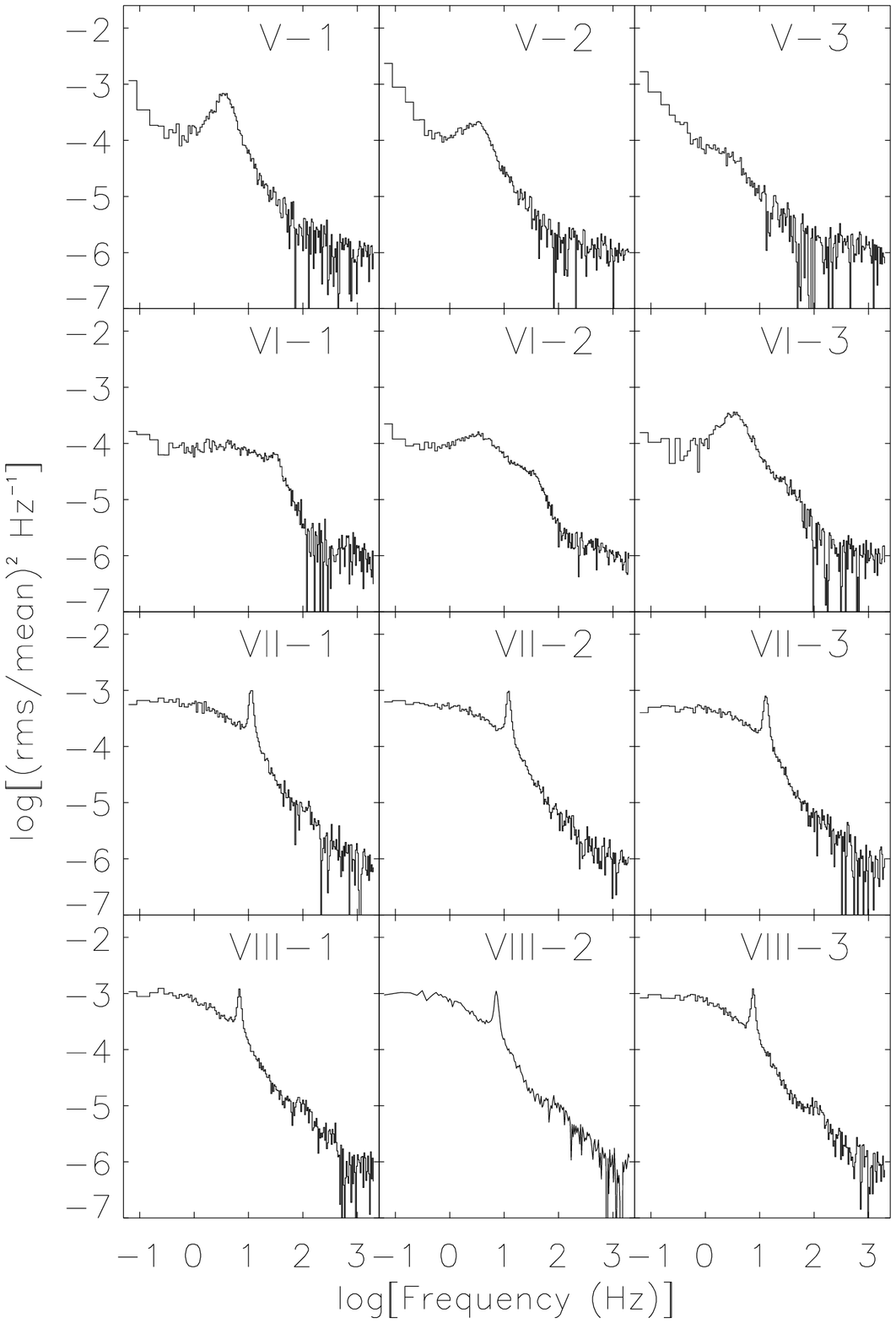 hscale=55 vscale=55 hoffset=192}{3.25in}{4.6in}
\caption{ 
Averaged and rebinned power density spectra (2--32~keV) for each of
the three HID regions for each observation. Poisson noise has been
subtracted from each PDS (see text).}
\label{fig:feb97_pds}
\end{figure}
%-----------------------------------------------

%Procedure
Fourier power density spectra (PDSs) were computed using 16 s segments
with 244~$\mu$s ($2^{-12}$~s) time bins. This was done for both the
full 2--32~keV energy range and for four energy channels:
2.0--4.8~keV, 4.8--13~keV, 13--18~keV, and 18--32~keV. The
Leahy-normalized power spectra~\cite{leahy83} were converted to the
fractional rms normalization by dividing by the background-subtracted
count rate in the selected band. The expected Poisson level, i.e.\ the
level of white noise due to counting statistics, was estimated taking
into account the effects of deadtime~\cite{morgan97,zhang95,zhang96}
and subtracted from each PDS; this method tends to slightly
underestimate the actual Poisson level. For each of the 24 HID regions
defined in Figure~\ref{fig:feb97_cchid_all}, an average PDS was
calculated from the power spectra corresponding to points in that
region. The PDSs were then logarithmically rebinned.

%-----------------------------------------------
\begin{figure}
\PSbox{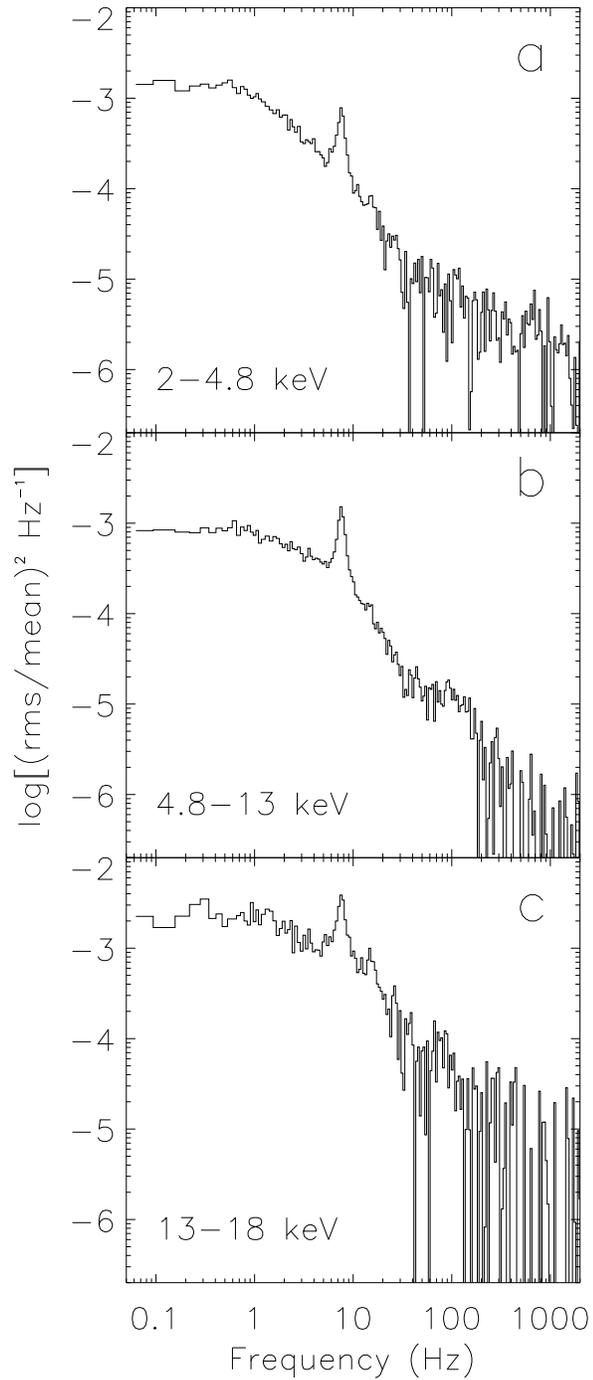 hscale=85 vscale=85 voffset=-10 hoffset=100}{3.2in}{6.8in}
\caption{
Averaged and rebinned power density spectra for HID region VIII-3 in
three energy bands. A harmonic peak of the 7.6~Hz QPO is clearly
visible in the high-energy channel (c).  The broad high-frequency
peak, most clear in (b), occurs near $\sim$100~Hz in this
observation. The low-frequency noise cuts off less sharply as energy
increases.}
\label{fig:feb97_pds3E}
\end{figure}
%-----------------------------------------------

%Basic observations 
The average PDS (2--32 keV) for each HID region is shown in
Figure~\ref{fig:feb97_pds}.  During the extended active state (observations
I-VI), a broad peak is often observed near 4~Hz; this feature is
prominent in PDSs from observations III--VI, weak in observation~II,
and indistinguishable from a flat-topped component in observation~I
(see below). A strong narrow QPO feature is seen at frequencies from
6.8 to 13~Hz in observations VII and VIII.  In some cases, especially
at higher photon energy (see Figure~\ref{fig:feb97_pds3E}), a harmonic peak
is observed at twice the frequency of this QPO\@. A weak narrow QPO
feature is present at frequencies above 20 Hz in regions II-1 and
VI-1.  A sharp ``knee'' is present at similar frequencies in regions
II-2, II-3, III-1, IV-1, VI-2, VI-3, and possibly I-1. Broad
high-frequency noise is sometimes seen, e.g.\ at $\sim$100~Hz in
observation VIII (Figure~\ref{fig:feb97_pds3E}b). There is an underlying red
continuum spectrum of noise in all of the regions of the HID, but the
shape and low frequency slope of the continuum vary over a wide range.

% Identification of components with those seen previously
The narrow QPO peaks and the low frequency noise in the PDSs from the
``quiescent'' 1~Crab observations (VII and VIII) resemble previously
observed PDSs (Chapter~\ref{ch:mar96paper},
\cite{shirey96,oosterbroek95,tennant87}). Those PDSs also contained
narrow QPO features, with centroid frequencies in the range
1.3--20~Hz, and similarly shaped low-frequency noise (e.g.\ see
Figure~\ref{fig:mar96_pds}). The broad high-frequency component that
we detect in the present observations is similar to the 20--100~Hz QPO
seen in earlier PCA observations (Chapter~\ref{ch:mar96paper},
\cite{shirey96}) and to the 100--200~Hz QPO observed with
\EXOSAT~\cite{tennant87}.

% Narrow QPO at high-freq related to more prominent low-freq QPOs
The weak narrow QPO feature above 20--30~Hz in regions II-1 and VI-1
occurs near the knee of the low-frequency noise component. This
similarity to the LFN and prominent QPO at lower frequency in
observations VII and VIII suggests these higher frequency oscillations
are produced by the same physical process as the lower frequency QPOs.
% ``Unpeaked'' narrow QPOs
In observations II and VI, this QPO feature is visible in region 1 as
a small peak that fades in region 2 and becomes only a ``knee'' in
region 3 (see Figure~\ref{fig:feb97_pds}). Thus we assume that this knee is
related to the QPO.  Similar knees are present in regions III-1, IV-1,
and possibly I-1. We include a narrow QPO component in fits of PDSs
which show a knee above 20~Hz, but identify these cases as
``unpeaked'' in the discussion below.

% ``Unpeaked broad'' QPOs
Likewise, although no peak appears in the PDSs from observation I, a
broad noise component has roughly constant power below about 4~Hz and
drops off above that frequency, forming a ``knee'' which might
indicate the presence of the 4~Hz QPO component. The PDS for region
I-1 somewhat resembles those of regions III-1 and IV-1, in that all
show a break in the power spectrum near 4~Hz and a second knee or
change in slope near 30~Hz. We include a broad QPO component in fits
of the PDSs for observation I, but we identify these cases as
``unpeaked''.

% PDS model and fitting
The PDSs were fit with models comprising both broad-band and QPO
components: a power-law for the very low frequency noise (VLFN), an
exponentially cut-off power-law for the broad low-frequency noise, a
Lorentzian for the broad QPO near 4~Hz, Lorentzians for the narrow QPO
and its first harmonic, a broad Lorentzian for the high-frequency
peak, and a second power-law to fit the residual Poisson noise at high
frequency. The model for each PDS consisted of two to five of these
components, depending on which components were necessary for an
acceptable fit. The frequency of the harmonic (when present) of the
narrow QPO was fixed at twice the fundamental frequency.  For the fits
of the PDSs from the four narrower energy channels, the QPO centroid
frequencies were fixed at the values determined from the 2--32~keV
PDSs.  There were generally not enough counts to obtain useful PDS
fits for the 18--32~keV channel.  For use in performing the fits, we
estimated the variance of each power in each binned and averaged PDS
by calculating the sample variance of the powers in the individual
PDSs that were averaged to obtain each point, and dividing the result
by the number of the powers used in computing the sample variance.

% Fitting issues
The centroids of the narrow variable-frequency QPO and the $\sim$4~Hz
QPO were measured accurately whenever a clear peak was
visible. However, in cases where these components are weak or
unpeaked, the centroids were less well-constrained. The centroid of
the broad high-frequency peak and the cut-off frequency of the LFN
were often poorly constrained.

%-----------------------------------------------
\begin{figure}
\PSbox{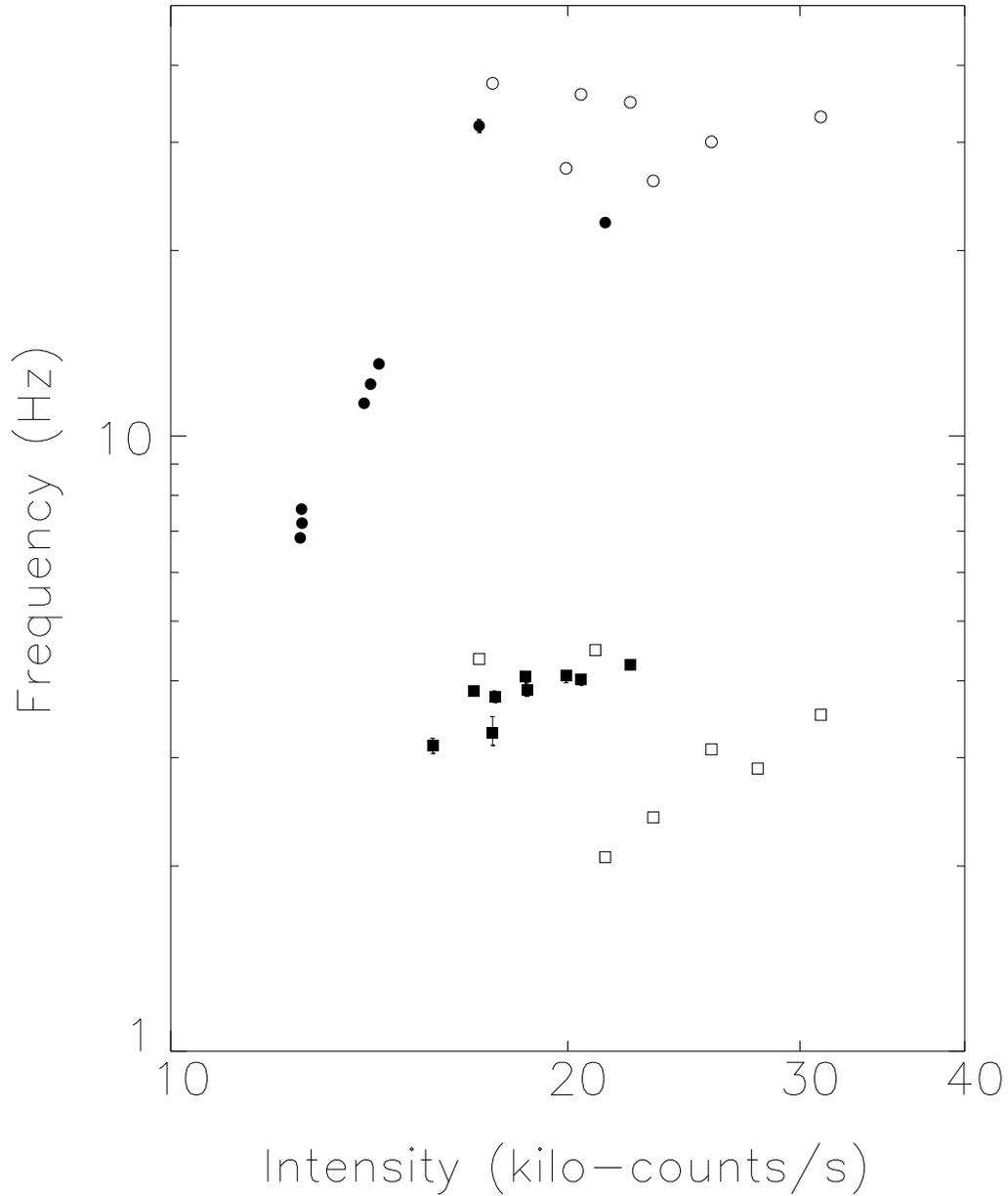 hscale=90 vscale=90 voffset=-80 hoffset=-30}{4.5in}{6in}
\caption{ 
Centroid frequency of the QPOs versus intensity (\intens{2.0}{18})\@.
A filled circle represents the narrow QPO and a filled square
represents the broad $\sim$4~Hz QPO (all points below 5~Hz are the
broad QPO\@). Unfilled circles and squares indicate the approximate
frequency of a knee or very weak peak that may be associated with the
narrow and broad QPO respectively. Error bars on frequency
measurements (filled points only) represent 90\% confidence intervals
for a single parameter ($\Delta\chi^2$=2.7).  In many cases, the error
bar for the QPO frequency is smaller than the plot symbol.}
\label{fig:feb97_freq_vs_I}
\end{figure}
%-----------------------------------------------

% QPO frequencies
Figure~\ref{fig:feb97_freq_vs_I} shows the frequency of the broad and narrow
QPOs versus intensity (2--18~keV). The frequency of the narrow peak is
generally correlated with intensity, starting at 6.8~Hz at 1~Crab and
reaching 32~Hz at 1.3~Crab. At higher intensity, this QPO is sometimes
present above 20~Hz and is often unpeaked (i.e., a knee). In
observations III--VI, the broad QPO is clearly present at 3.3 to
4.3~Hz. This QPO component was included in the fits of the PDSs from
observations~I and II, and the resulting frequencies (2.1--4.5~Hz) are
shown as unfilled squares (indicating a weak peak or a knee) above
20~kcts/s in Figure~\ref{fig:feb97_freq_vs_I}.

% QPO widths
The ratio of the width of the narrow QPO peak to its centroid
frequency ($\Delta\nu/\nu$) is about 0.15 when at 6.8 to 13~Hz. At
higher frequency this QPO becomes broader, with $\Delta\nu/\nu \sim
0.4$. When the broad QPO near 4~Hz is strong, we find that
$\Delta\nu/\nu \sim 1$, and when it is weak $\Delta\nu/\nu \sim 2$ to
$3$.

%-----------------------------------------------
\begin{figure}
\PSbox{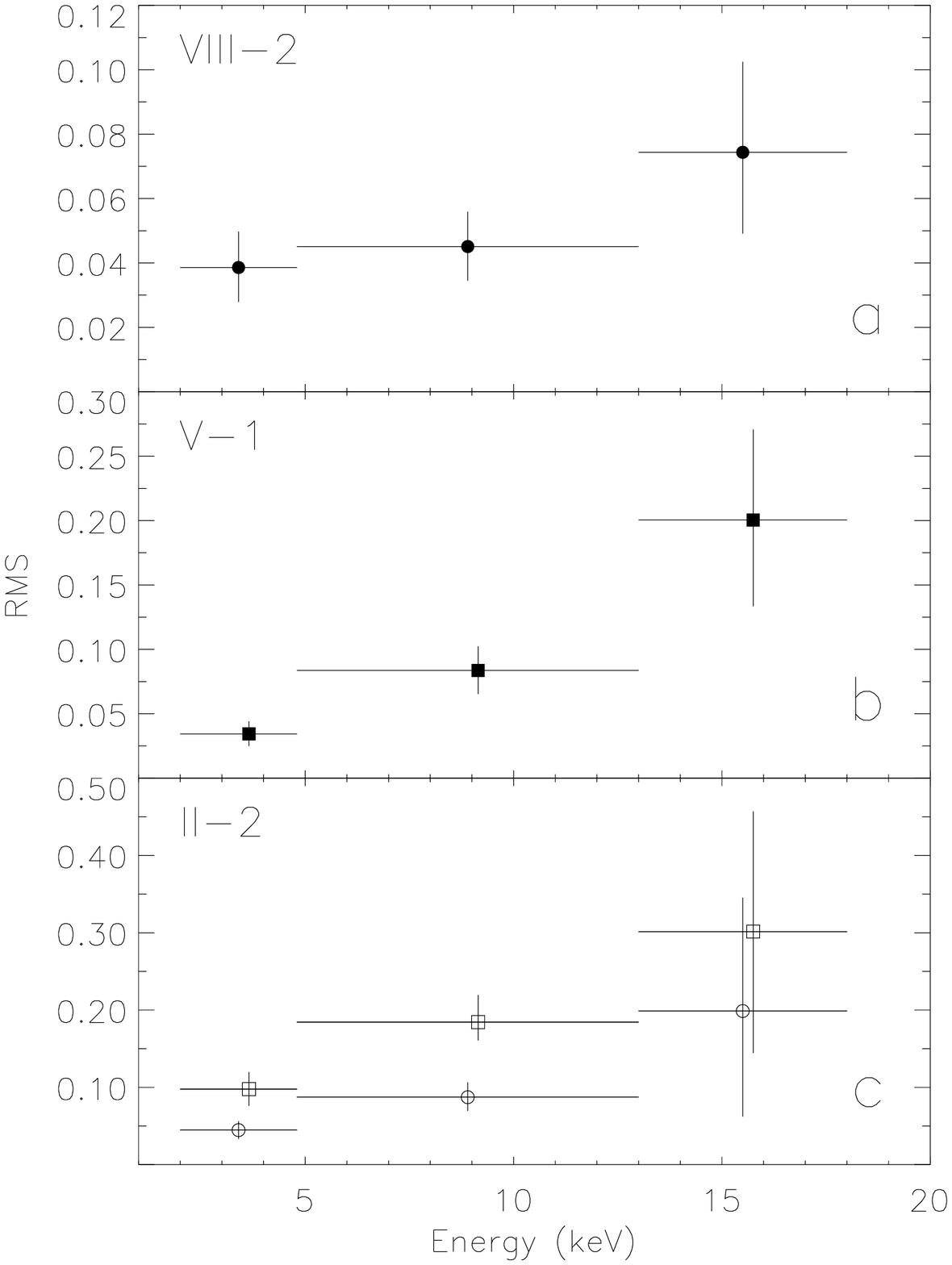 hscale=80 vscale=80 voffset=-10}{4.4in}{6.3in}
\caption{ 
Rms amplitude of QPOs versus photon energy. Panel (a) shows
the rms for the narrow QPO at 7.2~Hz (solid dot) and (b) for an example
of the broad 4~Hz QPO (solid box). In panel (c), unfilled circles and
boxes indicate the rms amplitude of a component forming a knee or very
weak peak that may be associated with the narrow and broad QPOs
respectively. The broad QPO points have been offset slightly to the
right in energy for clarity. Errors on QPO amplitudes represent 90\%
confidence intervals.}
\label{fig:feb97_rms_vs_E}
\end{figure}
%-----------------------------------------------

% QPO amplitudes
Figure~\ref{fig:feb97_rms_vs_E} illustrates the dependence of the rms
amplitude of the QPOs upon photon energy. Typical values for the rms
amplitude of the 6.8--13.1~Hz QPO at 2--4.8~keV, 4.8--13~keV, and
13--18~keV are 4\%, 5\%, and 8\% respectively
(Figure~\ref{fig:feb97_rms_vs_E}a), indicating a weak trend of increasing
rms amplitude at higher photon energy.  The amplitude of the broad QPO
increases significantly at higher photon energy. For clearly peaked
4~Hz QPOs, the rms amplitude is typically about 3\%, 8\%, and 18\% in
these three energy bands (Figure~\ref{fig:feb97_rms_vs_E}b). The rms values
vary considerably when these components are weak or unpeaked but their
amplitudes still generally increase with energy
(Figure~\ref{fig:feb97_rms_vs_E}c).

\subsection{Temporal Behavior versus Position on Spectral Branches}

%-----------------------------------------------
\begin{figure}
\PSbox{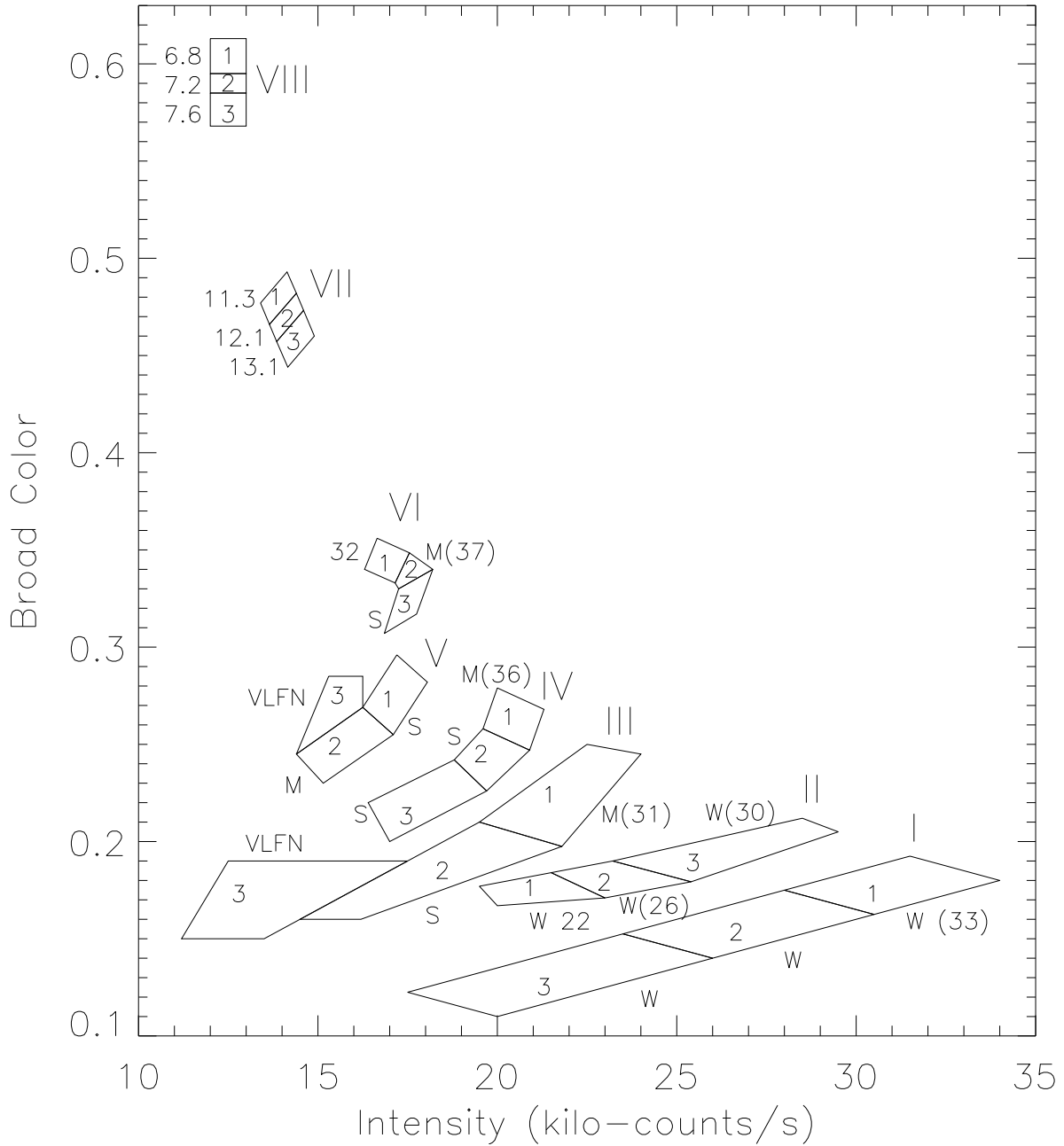 voffset=-120 hoffset=-46}{6.2in}{6.8in}
\caption{ 
Hardness-intensity diagram showing QPO properties for the
regions from Figure~\protect{\ref{fig:feb97_cchid_all}}b. The frequency of
the 6.8 to 32~Hz QPO is labeled (in Hz) beside each region where it is
present. Parenthesized frequencies indicate that this component was
unpeaked, i.e., a knee. Letters indicate the strength of the broad
4~Hz QPO: S---strong, M---medium, and W---weak or unpeaked.}
\label{fig:feb97_hidqpo}
\end{figure}
%-----------------------------------------------

The outlines of the HID regions of Figure~\ref{fig:feb97_cchid_all}b are
reproduced in Figure~\ref{fig:feb97_hidqpo} with labels summarizing the
observed QPO properties.
% HBO: 
As the hardness ratio decreases and the intensity increases along the
HID tracks for observations VIII, VII, and VI, the frequency of the
narrow QPO feature increases from 6.8~Hz to 32~Hz. The feature is
rather weak and knee-like in observation VI, but it appears to have a
width consistent with the width of the prominent QPO peak in
observations VII and VIII. A similar weak and somewhat knee-like
feature is also present in observation II, where it increases in
frequency from 22~Hz to 30~Hz as the intensity increases. The PDSs
from observations I, III, and IV all show a knee above 30~Hz at the
high-intensity, hard end of their HID tracks; these knees may be
related to the narrow QPO features seen in the other observations.

% NBO:
The broad 4~Hz QPO is not present in the ``quiescent'' observations
(VII and VIII). This QPO is strongest in portions of the
intermediate-intensity observations (III--VI) and is weakly present in
the soft, high-intensity observations (II and possibly I).

% VLFN:
Very low frequency noise dominates the power spectrum of regions V-3
and III-3. Both of these regions appear to begin upturned branches at
the low-intensity, soft end of branches showing the more pronounced
4~Hz QPOs.

\section{Discussion}

% Propose Z source 
The combined temporal and spectral-branch properties of the
observations presented here suggest Z-like behavior. We identify the
6.8--32~Hz QPOs as horizontal-branch oscillations (HBOs), the 4~Hz QPO
as normal-branch oscillations (NBOs), and the strong VLFN as
flaring-branch behavior (see discussion below).  These identifications
of characteristic time-variability patterns then help to identify the
tracks in the HID as horizontal, normal, and flaring branches
(HB/NB/FB), where each 6 ks observation of Cir~X-1 appears to have
captured a snapshot of portions of one or two of the branches.  The
spectral branches appear to shift around as the flaring gradually
subsides, rather than forming a stable Z pattern. It is likely that
the shapes of the spectral branches become distorted somewhat during
these large shifts.  We now describe the inferred properties of each
of the spectral branches in more detail.

\subsection{Horizontal Branch}
% Timing evidence
HID regions VIII, VII, and VI-1 show a narrow QPO peak or knee at
6.8--7.6~Hz, 11.3--13.1~Hz, and 32~Hz respectively. This frequency
range overlaps the 13--60~Hz range of typical horizontal branch
QPOs~\cite{klis95}. The associated low-frequency noise and harmonic
peak are also typical of horizontal branch power spectra. The broad
high frequency peak in Cir~X-1 may be related to the high frequency
noise component often observed on the horizontal branch.

% Spectral-branch behavior
The HID track for observation VI shows the narrow QPO at 32~Hz on a
roughly horizontal segment (region VI-1) and a knee at 37~Hz on the
right end of this segment (region VI-2). The apex of region VI-2
brings a transition to the 4~Hz QPO, which is dominant on the downward
branch of this track (region VI-3). This is very similar to the HB/NB
transition in Z sources. 

% Upturned left extension of HB, and March 96 comparison
When Cir~X-1 is in ``quiescence'' in observations VII and VIII, the
``horizontal branch'' turns upward and becomes vertical in the HID.
For comparison, \RXTE\ PCA observations of Cir~X-1 from 1996 March
10--19 which show a narrow QPO peak at 1.3--12~Hz~\cite{shirey96}
(Chapter~\ref{ch:mar96paper}) are almost entirely confined to the
12.3--14.7 kcts/s (2--21 keV) intensity range. The HID tracks for
those observations lie along a nearly vertical line, and probably
represent sections of the ``horizontal'' branch.

%Observation II
Observation II may also be on part of the HB, since a weak narrow QPO
appears to evolve into a knee and increase in frequency from 22 to
30~Hz as the intensity increases. However, the broad QPO is also
weakly visible in PDSs for this observation. The fact that
observations II, VI, VII, and VIII all show little variation of the
hard color used in Figure~\ref{fig:feb97_cchid_all}a suggests that
observation~II may be associated with the other HB observations.

\subsection{Normal Branch}

% Timing and spectral evidence
The 4~Hz QPO is observed when the source intensity rises above the
``quiescent'' 1-Crab level ($\sim$13~kcts/s). It is roughly stationary
in frequency (3.3--4.3~Hz when clearly peaked) and broader than the
HBO. The feature is easily seen in observations III--VI; at these
times the location in the HID moves along diagonal tracks. The
$\sim$4~Hz frequency and motion along diagonal tracks in the HID is
consistent with the 4--7~Hz NBOs observed at nearly constant frequency
on the NB of typical Z sources~\cite{hk89}. We therefore identify
the broad 4~Hz QPO as a normal branch oscillation, and the diagonal
tracks for observations III--VI as shifted normal branches.

% Evolution due to shifts
The broad QPO component may be also present in the highest intensity
observations, as a weak feature in observation~II and in the form of a
break near the 4~Hz QPO frequency in observation~I.  We also note that
at the top of the normal branch (regions I-1, III-1, IV-1, VI-2) a
knee above 30~Hz is present in addition to the NBO component.

% March 96 comparison
A similar broad 4~Hz QPO is present in observations from 1996 March
5--6 (Obs.~1 in Table~\ref{tab:mar96_obs}) made immediately before
phase zero of the cycle showing the 1.3--12~Hz narrow QPO.

\subsection{Flaring Branch}
% Timing behavior
Beyond the left apex of the normal branch a short upturned branch is
observed in HID region~V-3 and possibly III-3. The PDS for these
regions are dominated by very low frequency noise, which is typical
for flaring branches, and no QPO peaks are obviously apparent. We note
that in the well-established Z sources neither NBOs nor HBOs are
present on the flaring branch, except for Sco~X-1 and GX~17+2, in
which the NBO evolves into a 6--20~Hz QPO (see \cite{klis95} and
references therein).

%-----------------------------------------------
\begin{figure}
\PSbox{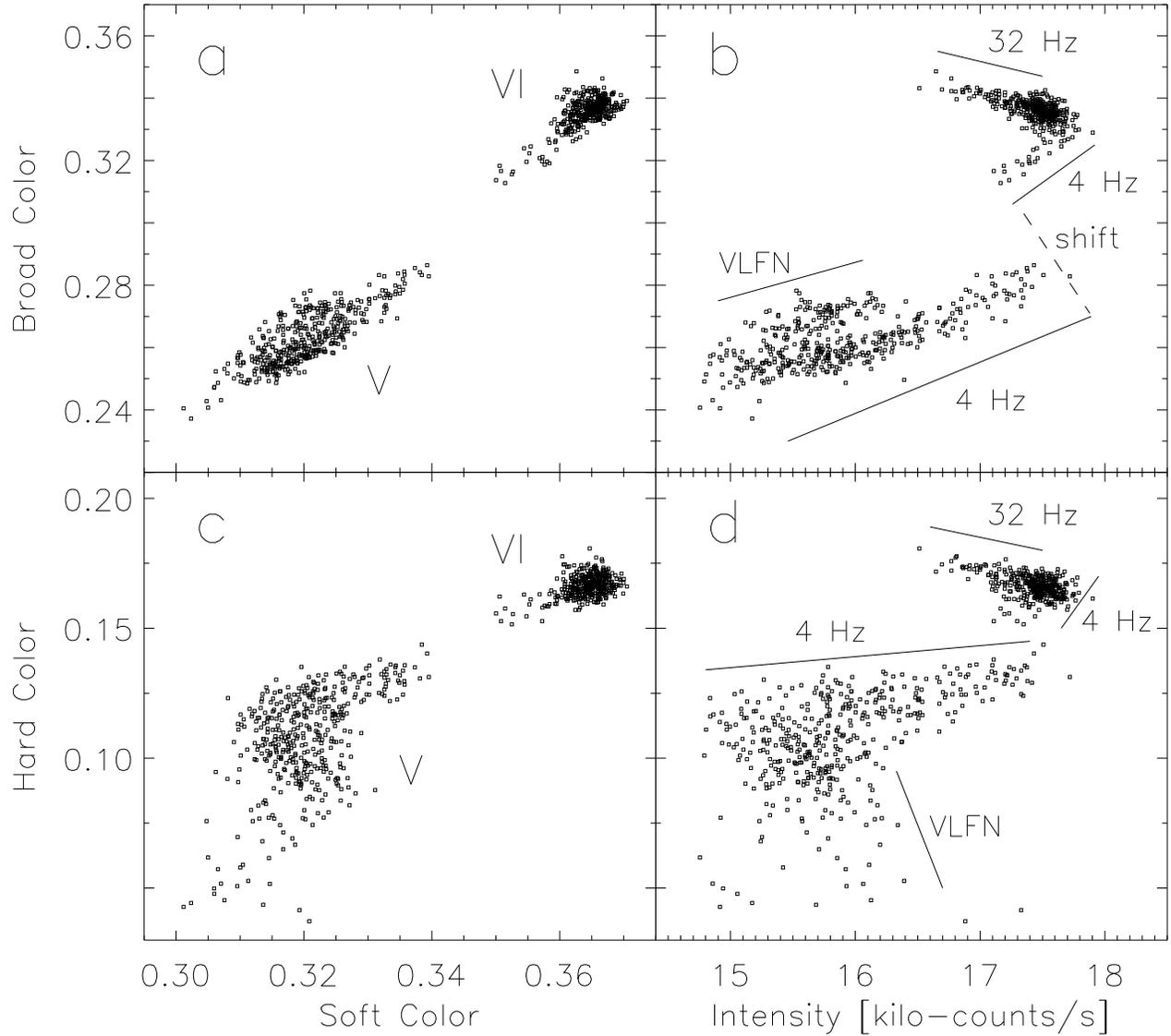 voffset=-200 hoffset=-36}{6.5in}{5.9in}
\caption{ 
Broad-color and hard-color CDs and HIDs for observations~V and VI.  In
all four diagrams, Obs.~V is in the lower left and Obs.~VI in the
upper right. The CD and HID tracks for Obs.~V both turn upward at the
left end in the broad-color diagrams (a,b) but turn downward in the
hard-color diagrams (c,d). In the HIDs, presence of the 32~Hz HBO,
4~Hz NBO, and VLFN is indicated along the branches. An apparent shift
of the normal branch between observations~V and VI is labeled in (b).
The intensity, \intens{2.0}{18}, is from all five PCUs. The soft color
is defined as \softcolor, the broad color as \broadcolor, and the hard
color as \hardcolor\@. Each point corresponds to 16~s of
background-subtracted data.}
\label{fig:feb97_cchid_56medhard}
\end{figure}
%-----------------------------------------------

% Spectral-branch behavior
The left end of the spectral track for observation~V bends upward in
the HID shown in Figure~\ref{fig:feb97_cchid_all}b, but bends downward in
the CD in Figure~\ref{fig:feb97_cchid_all}a. This behavior is demonstrated
more clearly in Figure~\ref{fig:feb97_cchid_56medhard}, which shows CDs and
HIDs for observations V and VI. When a broad color (\broadcolor) is
used as the ordinate of the diagrams
(Figure~\ref{fig:feb97_cchid_56medhard}a,b), the track for observation~V
turns upward on the left end. When a harder color (\hardcolor) is used
as the ordinate (Figure~\ref{fig:feb97_cchid_56medhard}c,d), this branch
turns downward. The CD and particularly the HID version based on the
harder color show the most clear similarity to canonical Z diagrams,
with the temporal behavior of observations V and VI being generally
consistent with horizontal, normal, and flaring branches. The
broad-color HID (Figure~\ref{fig:feb97_cchid_56medhard}b) shows evidence for
a shift of the normal branch that does not show up in the other three
diagrams of that figure.

\subsection{Relation to Other Sources}
\label{sec:feb97_other_sources}

% Shifts of `Z' in other sources
Our observations reveal spectral branches which shift in the CD and
HID as Cir~X-1 evolves from a soft, high-intensity state to a hard,
lower-intensity state. The ASM light curves and hardness ratios
(Figure~\ref{fig:feb97_asm}) show that this evolution occurs
periodically with the 16.55~day cycle, thus suggesting that the CD/HID
shifts may also be periodic. Shifts of the ``Z'' pattern in CDs and
HIDs have been observed in the so-called Cyg-like Z sources:
Cyg~X-2~\cite{kuulkers96:cygx-2}), GX~5-1~\cite{kuulkers94:gx5-1}, and
GX~340+0~\cite{kuulkers96:gx340+0}. However, the shifts do not occur
periodically in those sources, nor do they have the magnitude of the
shifts observed in Cir~X-1.

% Unusual FBs in Z sources
The flaring branch of Cir X-1 turns upward when a soft or broad color
is used on the vertical axis. When a harder color is used, this branch
turns downward but then bends to the left. In the Cyg-like Z sources,
the flaring branch sometimes turns upward or starts toward higher
intensity and then loops back to lower intensity
\cite{kuulkers96:cygx-2,kuulkers94:gx5-1,kuulkers96:gx340+0,penninx91:gx340+0}).
In some cases, these sources are observed to ``dip'' while on the
flaring branch~\cite{kuulkers94:gx5-1,penninx91:gx340+0,wijnands97}),
with tracks which turn down and then to the left, similar to that of
Cir~X-1 in Figure~\ref{fig:feb97_cchid_56medhard}c.

% Upturned HB
The left end of the horizontal branch in Cir~X-1 turns upward and
becomes vertical at low intensity (Figure~\ref{fig:feb97_hidqpo}). On
this section of the branch, HBO frequencies are low: 6.8--13~Hz in
observations VII and VIII and 1.3--12~Hz in the earlier 1996 March
observations. A similar effect was reported in
GX~5-1~\cite{lewin92:gx5-1,kuulkers94:gx5-1}, in which the HB turns
upward at the low-intensity end while HBOs are observed at relatively
low frequency (13--17~Hz). Lewin et~al.~\cite{lewin92:gx5-1} suggested
that other Z sources might show such an upward turn of the HB if their
intensities and QPO frequencies became sufficiently low.  Recent
\RXTE\ observations of Cyg~X-2~\cite{smale98} show a long
vertical extension of the horizontal branch in an HID.

% Our interpretation of the EXOSAT observations
The 5--20~Hz narrow QPO was detected with \EXOSAT\ at an intensity
similar to the quiescent level observed by \RXTE. We note that
absorption dips are responsible for much of the structure seen in the
CD shown for that observation; however, the HIDs show that the narrow
QPO occurred on an upturned left end of a horizontally oriented track
as in our data (see Figures~2--4, 8, \& 10 in \cite{oosterbroek95}).
At higher intensity during the same observation, the narrow QPO was
not present, and we note that some of the high-intensity PDSs show
hints of a broad peak near 4~Hz. We thus conclude that the behavior
observed by \EXOSAT\ during that observation is related to the Z-like
behavior we observe with \RXTE.

Most of the other \EXOSAT\ observations took place when Cir~X-1 was
significantly lower in intensity than the ``quiescent'' level of the
current observations. The CDs and HIDs for these \EXOSAT\ observations
did not show tracks which could clearly be identified as Z or atoll.
Their power spectra were generally dominated by VLFN, typical of atoll
sources in the banana state, and sometimes also showed a broad red
noise component resembling atoll high-frequency
noise~\cite{oosterbroek95}. However, these power-spectral shapes are
not unique to atoll sources: power spectra for black hole candidates
in the high state are dominated by VLFN, as are those of the current
observations on the low-intensity end of the normal branch and on the
flaring branch (i.e., regions III-3 and V-3).

% Unique Status Among Z and Atoll Sources
Cir~X-1 was expected to never show HBOs since atoll-like behavior was
taken as evidence that the magnetic field is not strong enough to
allow the magnetospheric beat frequency mechanism (MBFM) to
operate~\cite{klis94,oosterbroek95}. (However, it is also possible
that the HBOs are not produced by the MBFM\@.) The results presented
here demonstrate both HBOs and NBOs in Cir~X-1 and show no evidence
for atoll behavior. Since the atoll-like behavior observed with
\EXOSAT\ occurred at lower intensity than in the present observations,
it is possible that they do represent a different state of the
source. If Cir~X-1 actually can show atoll behavior as well as the
Z-like behavior shown here, then we would have new clues to the
differences between the two types of sources. Such observations would
challenge the hypothesis that differences in both $\dot{M}$ and
magnetic field distinguish these two classes.

\section{Summary}
Our results from an analysis of \RXTE\ observations of Cir~X-1 reveal
behavior similar to that of Z sources, and, in particular, allow us to
identify temporal and spectral signatures of the horizontal, normal,
and flaring branches.  The spectral variability of Cir~X-1 is seen to
correspond to tracks in a HID which are similar in direction to the
typical direction in the HID of Z sources in general, but the
locations of the tracks corresponding to each branch move from
observation to observation in a systematic manner. 

To be specific, in the current observations of Cir~X-1 the horizontal
branch is characterized by the presence of relatively narrow
6.8--32~Hz QPO features in the PDS. The track in the HID of the
horizontal branch is horizontal at the high intensity end and becomes
vertical at the low intensity end, where the source is ``quiescent'',
i.e., has an intensity near 1~Crab and is characterized by a
relatively low degree of variability on time scales longer than 1~s.
The normal branch is characterized by broad 4~Hz QPOs, and by motion
in the HID which generally falls along tracks which run diagonally
from hard high-intensity locations to soft low-intensity locations.
There are also time intervals when the PDS is dominated by very low
frequency noise. We identify these intervals as excursions onto the
flaring branch.

The large amplitude intensity variations associated with the
active/flaring state of Cir~X-1 can be divided into three categories:
(1) motion across the horizontal portion of the horizontal branch and
along the normal and flaring branches, (2) shifts of the spectral
branches, and (3) absorption dips.  While our \RXTE\ observations have
allowed us to recognize and distinguish these different types of
variability, there is still much to be understood about the physical
mechanisms responsible.

\chapter{Correlated Timing and Spectral Behavior}
\label{ch:fullZ}

\section{Overview}

The eight observations presented in Chapter~\ref{ch:feb97paper}
(study~E) were characterized by well-separated spectral tracks in
color-color and hardness-intensity diagrams. Based on timing
characteristics, the spectral branches were identified with Z-source
horizontal, normal, and flaring branches (HB/NB/FB).  However, due to
the short duration of the observations and the large shifts of the
tracks between them, the complete spectral-branch pattern for could
only be inferred from fragmented pieces.

Those results indicated that the transition between the quiescent and
flaring states of Cir~X-1 is at least partially related to motion
around the ``Z'' track. In 1997 June, we carried out an extensive
\RXTE\ campaign to study in detail the transition to active state, 
including 7~days of 56\% coverage beginning before the phase zero dips
(study~G, see Figure~\ref{fig:pca_by_cycle}). The excellent coverage
provided by those observations allowed us to find portions of the
transition which clearly demonstrate the full spectral track of
Cir~X-1. The evolution of the timing characteristics and energy
spectrum were studied as a function of position along the track. We
found a continuous evolution of the power density spectrum between the
different states defined by the spectral branches.  We also found that
each branch of the spectral track is associated with a specific type
of evolution of the energy spectrum (e.g., pivoting about 7~keV in one
case and increasing at low energy while remaining constant at high
energy in another case). We explored various physical models for the
energy spectrum and parameterized the evolution of the spectrum in
terms of a two-component model consisting of a multi-temperature
``disk blackbody'' and a higher-temperature isothermal blackbody.

\section{1997 June PCA Observations}

\begin{figure}
\begin{centering}
\PSbox{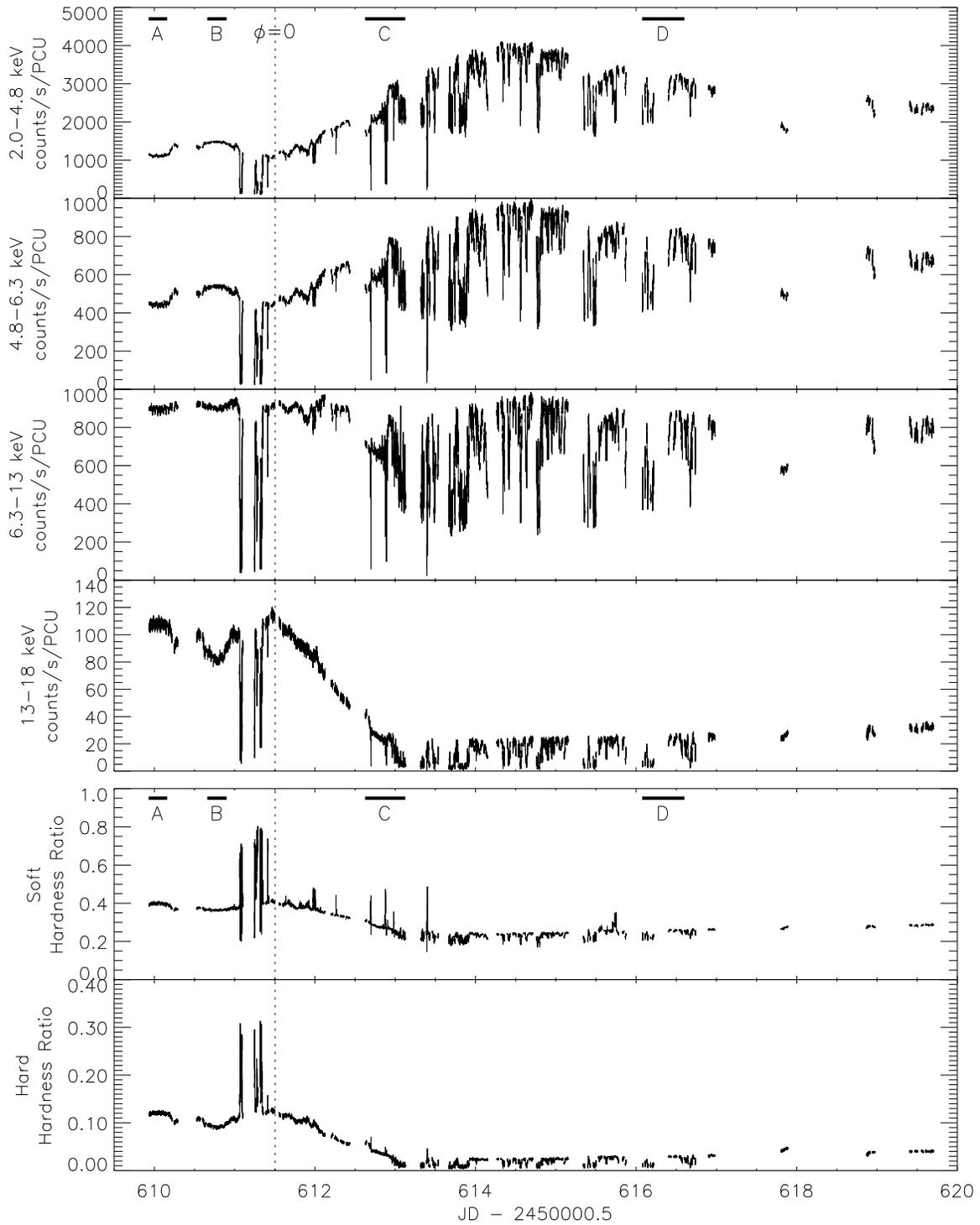 hscale=88 vscale=88 voffset=-14}{6.1in}{7.3in}
\caption{
Light curves in four energy bands and two hardness ratios for PCA
observations of Cir~X-1 from 1997 June 10--20, covering a 10-day
period around phase zero ($\phi=0$). The intensities at the beginning
of these observations (day~610) are typical ``quiescent'' levels.
Each point represents 16~s of background-subtracted data from PCUs 0,
1, and 2.  Ratios of the intensities in the four bands produce soft
(\hardness{2.0}{4.8}{4.8}{6.3}) and hard (\hardness{6.3}{13}{13}{18})
hardness ratios. Segments labeled A, B, C \& D were used for spectral
studies.  
}
\label{fig:june97_10d}
\end{centering}
\end{figure}

The PCA light curves and hardness ratios for the 1997 June
observations are shown in Figure~\ref{fig:june97_10d}. These data show
only moderate variability before phase zero until entering a phase of
significant dipping during the half day before phase zero. The
hardness ratios show that significant spectral evolution (both
hardening and softening) occurs during these dips. Similar dip
behavior from another set of PCA observations was studied in detail
and is presented in Chapter~\ref{ch:absdips}.

By phase zero (day~611.5), the main dipping episode ends, and the
climb to the flaring state begins. While the intensity increases by
more than a factor of three in the lowest energy band (2--4.8~keV),
the intensity between 6.3~keV and 13~keV does not climb at all, and
above 13~keV the intensity actually decreases by a factor of about 10
over the first 1.5~days following phase zero. This anti-correlation of
the low and high-energy intensity during the transition results in
decreased hardness ratios after phase zero, as is observed by the ASM
(section~\ref{sec:asm}). After a relatively smooth transition toward
high total intensity during the first day following phase zero, the
intensity becomes highly variable (i.e., the ``active'' or ``flaring''
state) for the remainder of the observations (about a week).

\section{Complete Spectral Track}

\begin{figure}
\begin{centering}
\PSbox{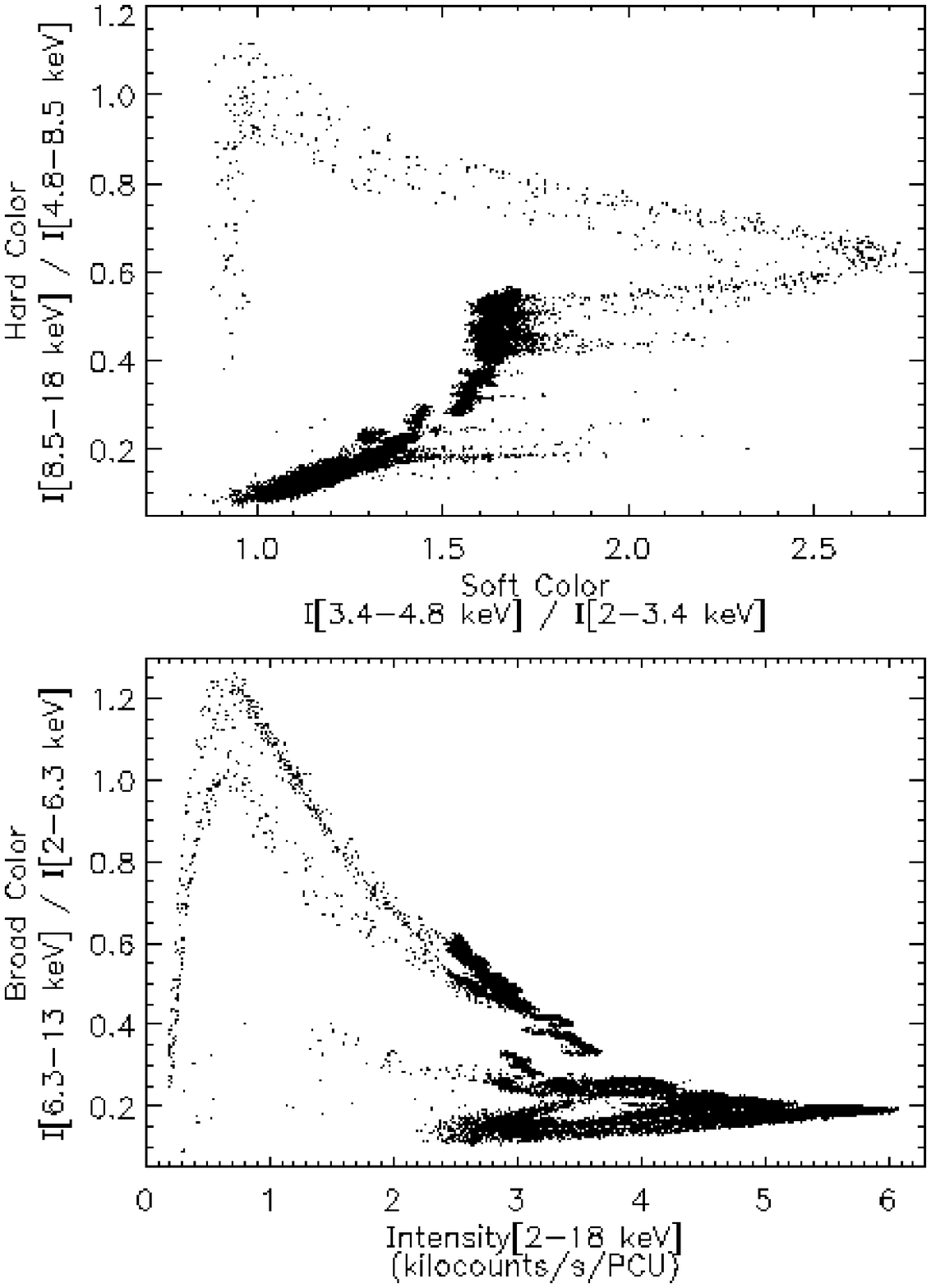 
	hoffset=-84 voffset=-100}{5.9in}{7.8in}
%	hoffset=-16 voffset=-60}{5.9in}{7.8in}
\caption{
Color-color and hardness-intensity diagrams from PCA observations
during 1997 June 10--21 (the entire period covered by
Figure~\protect{\ref{fig:june97_10d}}). Each point represents 16 s of
background-subtracted data from PCUs 0, 1, and 2.
}
\label{fig:june97_cchid_all}
\end{centering}
\end{figure}

% identify dip tracks in CD & HID
The color-color and hardness-intensity diagrams (CDs/HIDs) for all
data in Figure~\ref{fig:june97_10d} are shown in
Figure~\ref{fig:june97_cchid_all}. These data cover a significant
portion (10~d) of an entire 16.55~d cycle. The dips seen in
Figure~\ref{fig:june97_10d} appear as prominent light tracks with two
sharp bends in the CD (initially toward the right of the main
arc-shaped locus) and one sharp bend in the HID
($I<2.3$~kilocounts/s/PCU).  I show in Chapter~\ref{ch:absdips} that
tracks with these shapes are indicative of highly variable absorption
of a bright spectral component and only moderate, fixed absorption of
a fainter component.  Having identified absorption dip signatures, I
will now focus on non-dip spectral behavior (presumably more directly
related to the X-ray source) for the remainder of this chapter.

% overall locus of CD/HID points
Most of the data fall along a single arc-shaped locus in the CD and a
more complicated curved structure in the HID. This behavior is quite
similar to that shown for the eight brief sampling observations from
1997 February--March presented in Chapter~\ref{ch:feb97paper} (see the
CD and HID in Figure~\ref{fig:feb97_cchid_all}).  However, the
high-efficiency coverage of the current observations has filled in
many of the gaps that occurred between the tracks of the earlier
observations.

\begin{figure}
\begin{centering}
\PSbox{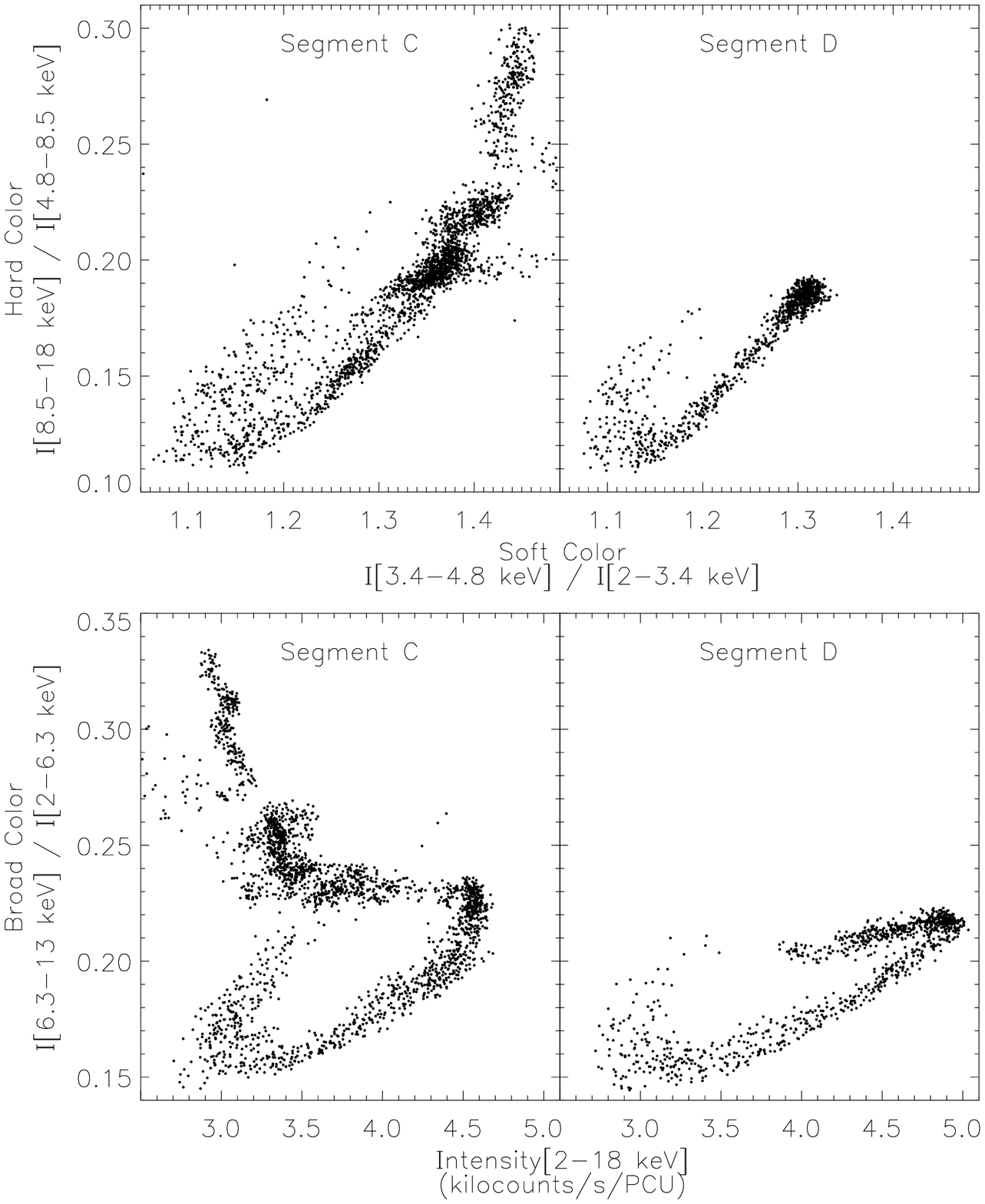 
hscale=97 vscale=97 hoffset=-25 voffset=-46}{6.5in}{7.7in}
\caption{
Color-color and hardness-intensity diagrams from time segments ``C''
(left panels) and ``D'' (right panels) of
Figure~\protect{\ref{fig:june97_10d}} (1997 June 13.625--14.125 and
1997 June 17.075--17.600 respectively).  Each point represents 16 s of
background-subtracted data from PCUs 0, 1, and 2.  }
\label{fig:june97_cchid_13_17}
\end{centering}
\end{figure}

% detail structure in CDs & HIDs
In order to examine the detailed structure with the overall locus of
CD/HID points, the data were divided into shorter time segments (of
order hours) and plotted separately in CDs and HIDs. In general, each
segment produced a fragmented track or tracks, similar to the 1997
February--March observations. In Figure~\ref{fig:june97_cchid_13_17},
the CDs and HIDs are shown for two time segments (labeled ``C'' and
``D'' in Figure~\ref{fig:june97_10d}), during which the source traced
out a significant portion of several connected branches.  These time
segments were each about 12~hours in duration and separated by several
days. Tracks of other time segments generally each resembled some
portion of the entire pattern shown in
Figure~\ref{fig:june97_cchid_13_17}, but often with a shifted position
in the diagrams.

The location of the tracks for segments C and~D in the HID for the
total 10~d observation set is such that the ``hole'' produced by the
looped branches is visible between 3 and 4~kilocounts/s/PCU in
Figure~\ref{fig:june97_cchid_all}.
The data from the segment~C included some absorption dips, resulting
in tracks moving off the right side of the CD and the left side of the
HID (and far beyond the limits of the plot in both cases).  
The HID patterns are consistent with the shape of the full ``Z'' track
as inferred by the fragmented tracks in Chapter~\ref{ch:feb97paper}.
For segment~C, there is a large upturned left extension of the
horizontal branch, the horizontal portion of the HB, the diagonal
normal branch, and a flaring branch that turns above rather than below
the NB. Similar tracks are also present in the HID for segment~D,
except there is only a small hint an upward turn at the left end of
the HB. The plots for the two segments have the same limits and show a
significant shift of the HB and upper NB between the two time
segments. The HID for segment~D is remarkably similar to that derived
from \RXTE\ PCA observations of the Z~source
Cyg~X-2~\cite{smale98}. The Cyg~X-2 HID shows a very prominent
vertical extension of the HB.  

The branches in Figure~\ref{fig:june97_cchid_13_17} are less well
separated in the color-color diagrams, as noted in
Chapter~\ref{ch:feb97paper}. However, the flaring branch clearly turns
above the normal branch in the current diagrams, and the upturned left
extension of HID horizontal branch of segment~C appears to also have
an increased slope in the CD. The upturned flaring branch is very
similar to the flaring branch observed in the color-color diagram for
the Z~source GX~349+2 in recent \RXTE\ PCA observations~\cite{zhang98}.

\begin{figure}
\begin{centering}
\PSbox{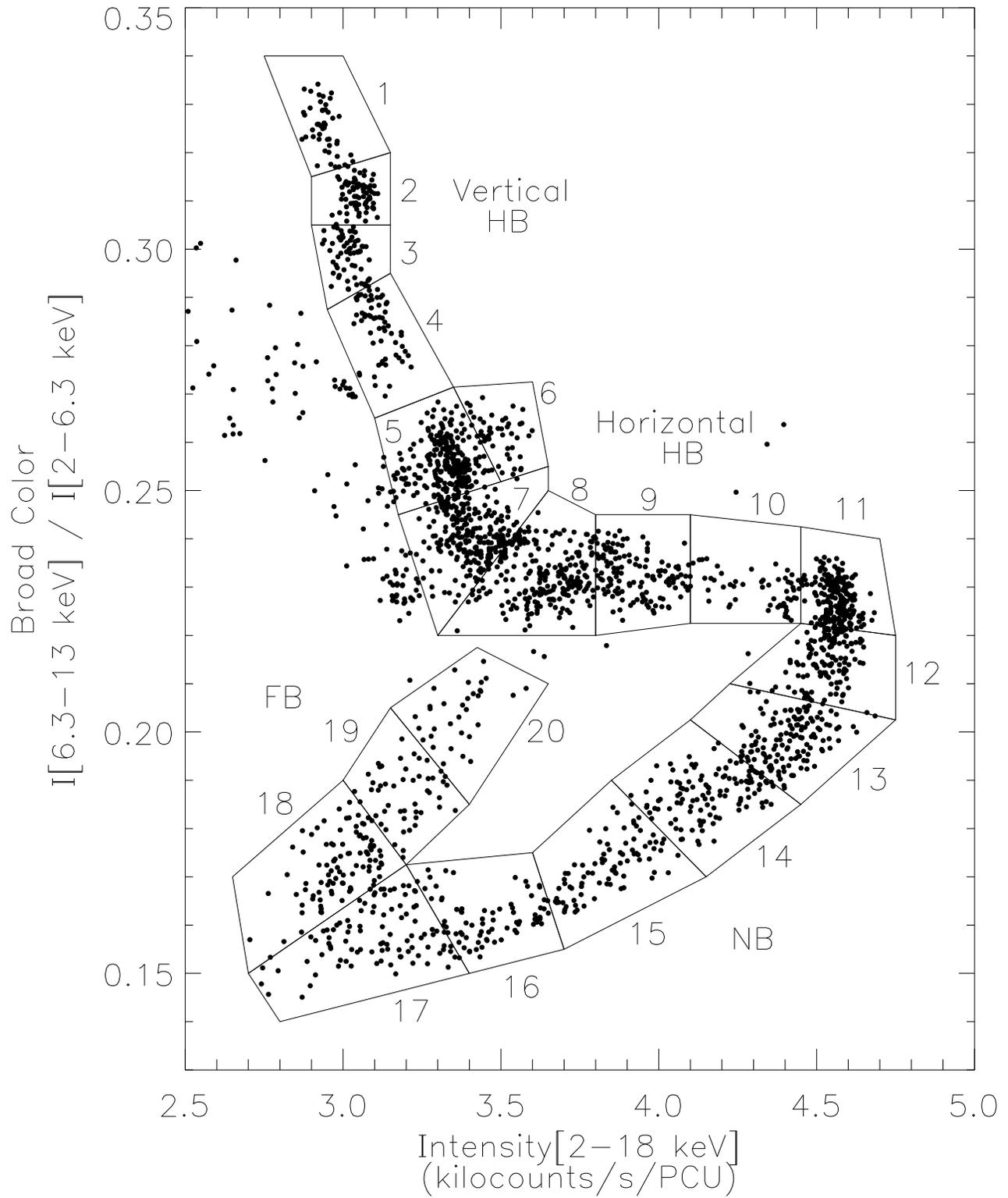 hoffset=-32 voffset=-50}{6.5in}{7.6in}
\caption{Hardness-intensity diagram from time segment~C  
(1997 June 13.625--14.125; day 12.625--13.125). The HID track has been
divided into 20 regions from which power density spectra and energy
spectra were constructed.  }
\label{fig:june97_hid20reg}
\end{centering}
\end{figure}
Twenty regions along the HID track of time segment~C, shown in the
lower left panel of Figure~\ref{fig:june97_cchid_13_17}, were selected
for further timing and spectral analysis. These regions are shown in
Figure~\ref{fig:june97_hid20reg} and labeled with increasing numbers
from the left HB, through the NB, to the FB. We will refer to this
label as the ``rank number'' after similar work done by other authors
on the standard Z sources; however, the numbers at the various apexes
have no special significance. It should also be noted that region~6
does not fit in with the monotonic increase of rank number around the
spectral track, and may be an indication of a shifted horizontal
portion of the HB.

\begin{figure}
\begin{centering}
\PSbox{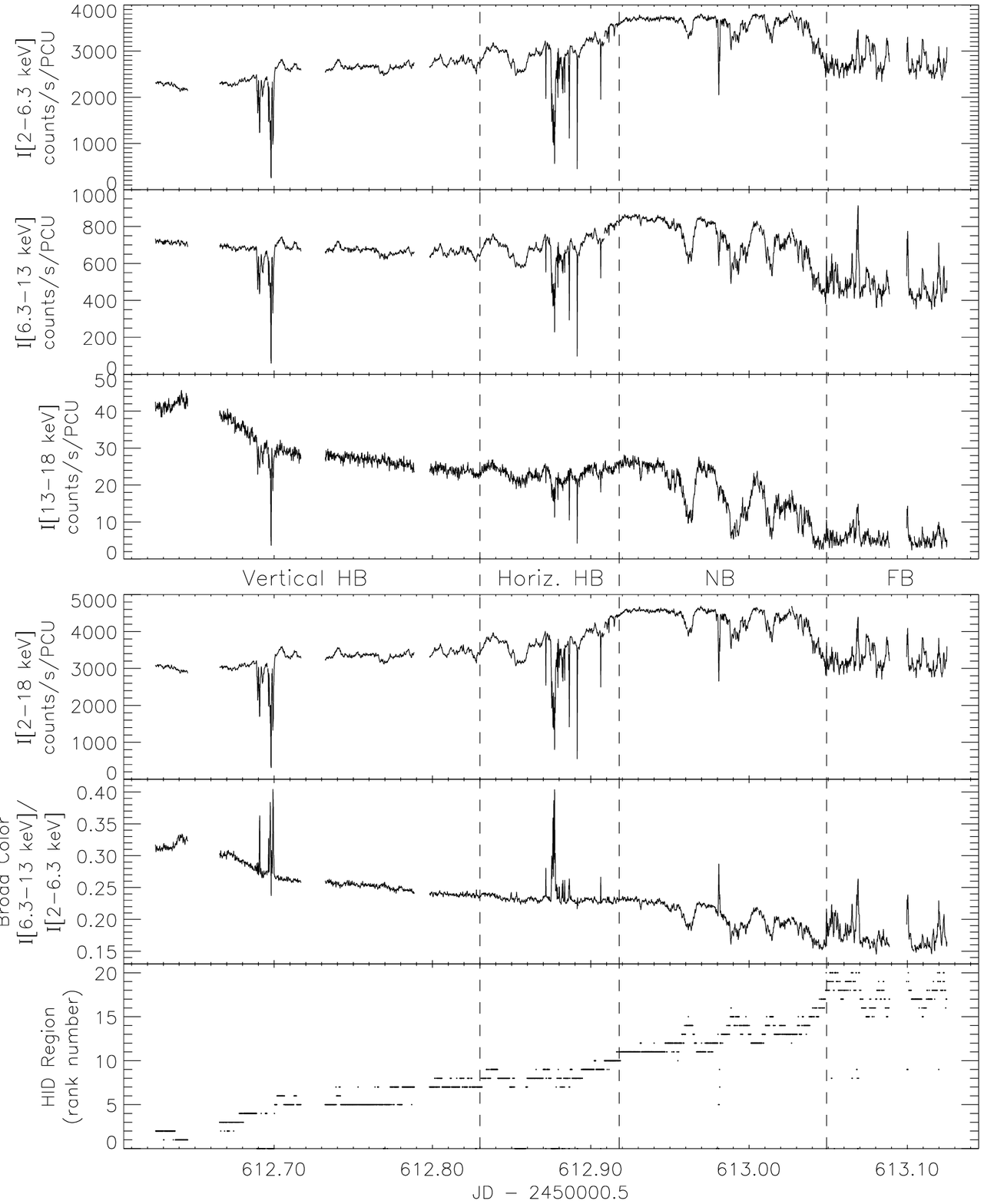 
	hscale=94 vscale=94 hoffset=-14 voffset=-26}{6.1in}{7.45in}
\caption{
Top: light curves in 3 energy channels and bottom: total 2--18~keV
light curve, broad color, and HID regions for time segment~C (1997
June 13.625-14.125; see Figure~\protect{\ref{fig:june97_10d}}).  Based
on HID region numbers, the predominant spectral branch is identified
for each portion of the data. Absorption dips (omitted from HID
regions) are clearly identified by decreased intensity coupled with
upward broad-color spikes.}
\label{fig:june97_lc_hr_reg}
\end{centering}
\end{figure}

% description of light curves for the HID regions
The light curves, hardness ratio, and HID regions versus time for the
data in Figure~\ref{fig:june97_hid20reg} (time segment~C) are shown in
Figure~\ref{fig:june97_lc_hr_reg}. During this half-day segment, the
source generally moves toward higher rank number around the spectral
track as the observation progresses. Thus, the data have been divided
into four sub-segments which predominantly correspond to each portion
of the HID track: the horizontal and vertical portions of the HB, the
NB, and the FB.  Brief absorption dips occur in all but the flaring
branch during this particular data set; these are easily identified by
sharp intensity dips coupled with a spiked increase in broad color.

The different branches are characterized by the following
characteristics, excluding the behavior associated with the dips: The
upturned left portion of the HB evolves relatively smoothly, with a
slight increase in 2--6.3~keV intensity and a decrease of almost a
factor of two in the 13--18~keV band. The horizontal portion of the HB
shows a substantial increase in soft intensity and on average shows 
relatively steady hard intensity. The normal branch shows increased
variability; it is bright in the soft bands but shows a decrease in
the hard band. The NB/FB transition occurs at lower intensity in all
bands compared to most of the NB. The flaring branch itself is then
produced by high-variability ``mini-flares'' or bursts above the NB/FB
apex level.

% exception points due to dips
Although the HID regions were defined in
Figure~\ref{fig:june97_hid20reg} such that obvious absorption tracks
were avoided, one brief dip, on day 612.98, occurred from region 12 on
the normal branch and placed a few points artificially across regions
9, 7, and 5. These points are easily identified in
Figure~\ref{fig:june97_lc_hr_reg} and are thus not included in
subsequent timing and spectral analysis.

% exception points due to flaring 
Likewise, the highest mini-flares on the flaring branch actually
extend beyond region~20 and cross regions 8 and 9. In fact a few such
points can even be seen above region~10 in
Figure~\ref{fig:june97_hid20reg}. The FB points that fell into HB
regions can also be clearly identified as points with rank numbers of
8 or 9 in the FB portion of Figure~\ref{fig:june97_lc_hr_reg}. These
are not included in subsequent timing and spectral analysis.

\section{Evolution of the Power Density Spectrum}

\begin{figure}
\PSbox{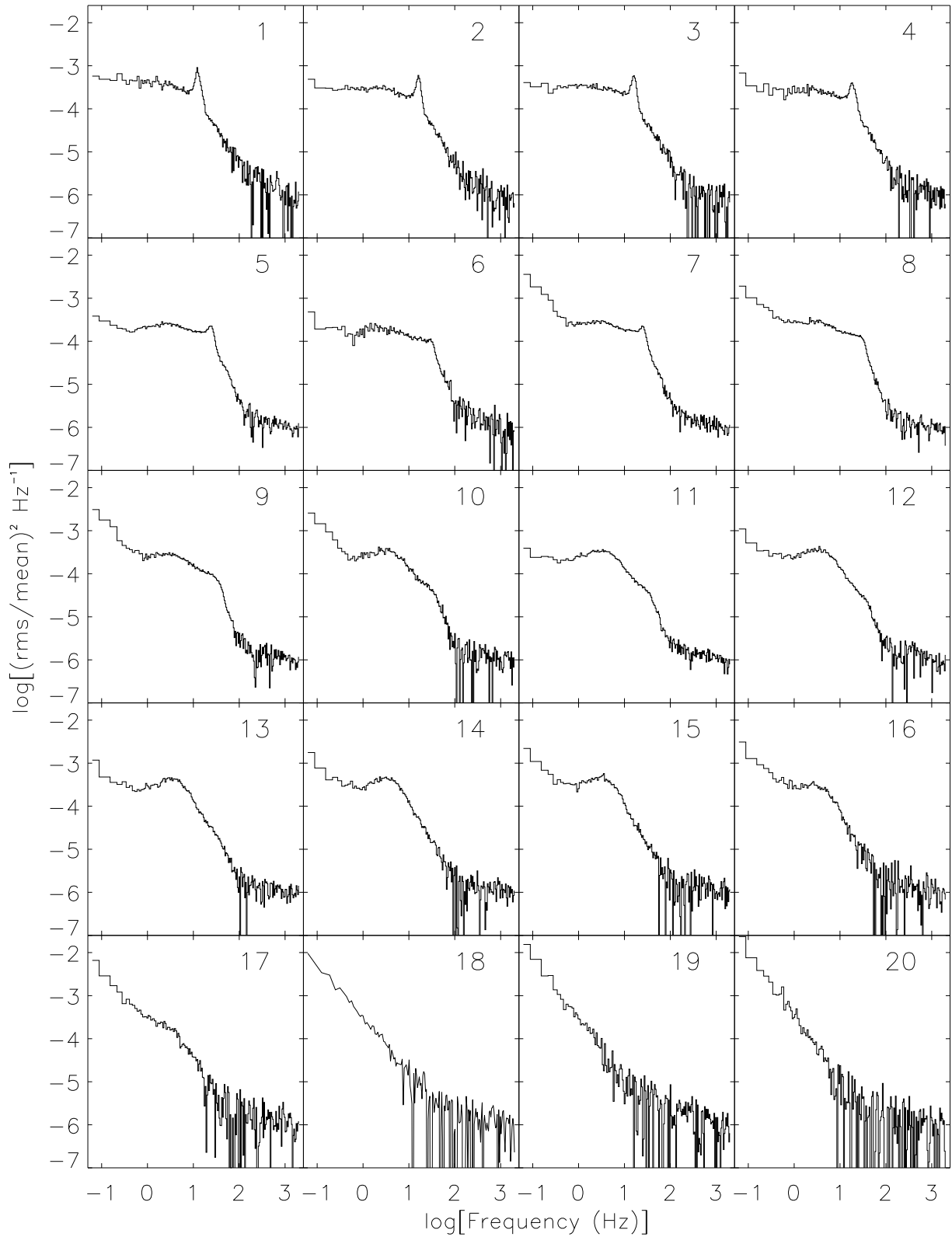 hoffset=-36 voffset=-52}{6.3in}{7.85in}
\caption{ 
Averaged and rebinned power density spectra (2--32~keV) for each of
the 20 regions along the HID track in
Figure~\protect{\ref{fig:june97_hid20reg}}. The estimated Poisson
noise level has been subtracted from each PDS.}
\label{fig:june97_20pds}
\end{figure}

%Procedure
Fourier power density spectra (PDSs) were computed for each 16~s of
time segment~C (1997 June 13.625--14.125).  Each transform used
$2^{16}$ 244-$\mu$s ($2^{-12}$~s) time bins and covered the full
2--32~keV energy range.
%This was done for both the full 2--32~keV energy range and for four 
%energy channels: 2.0--4.8~keV, 4.8--13~keV, 13--18~keV, and 18--32~keV.
The expected Poisson level, i.e.\ the level of white noise due to
counting statistics, was estimated taking into account the effects of
deadtime~\cite{morgan97,zhang95,zhang96} and subtracted from each PDS;
this method tends to slightly underestimate the actual Poisson
level. For each of the 20 HID regions defined in
Figure~\ref{fig:june97_hid20reg}, an average PDS was calculated from
the power spectra corresponding to points in that region. The PDSs
were then logarithmically rebinned and are shown in
Figure~\ref{fig:june97_20pds}.

% Basic observations
The general features of the power spectra are similar to those
observed in previous PCA observations (see Figures \ref{fig:mar96_pds} and
\ref{fig:feb97_pds}), although here a continuous evolution is observed between 
each timing state in a single observation. The narrow QPO is observed
to evolve from 12--25 Hz moving down the upturned left extension of
the HB (regions 1--5 and 7; region 6 may be a shifted version of
8). Across the horizontal portion of the HB (8--11), the narrow QPO
remains close to 30~Hz and fades into a knee, while the broad QPO
gradually rises up near 4~Hz. The broad QPO is present all along the
normal branch (regions 8--16) but is most prominently peaked in the
middle of the branch. On the flaring branch (regions 18--20), no QPOs
are present and the power spectrum shows only strong very low frequency
noise.

\section{Evolution of the Energy Spectrum}
\label{sec:june97_spec_qual}

%-----------------------------------------------
\begin{figure}
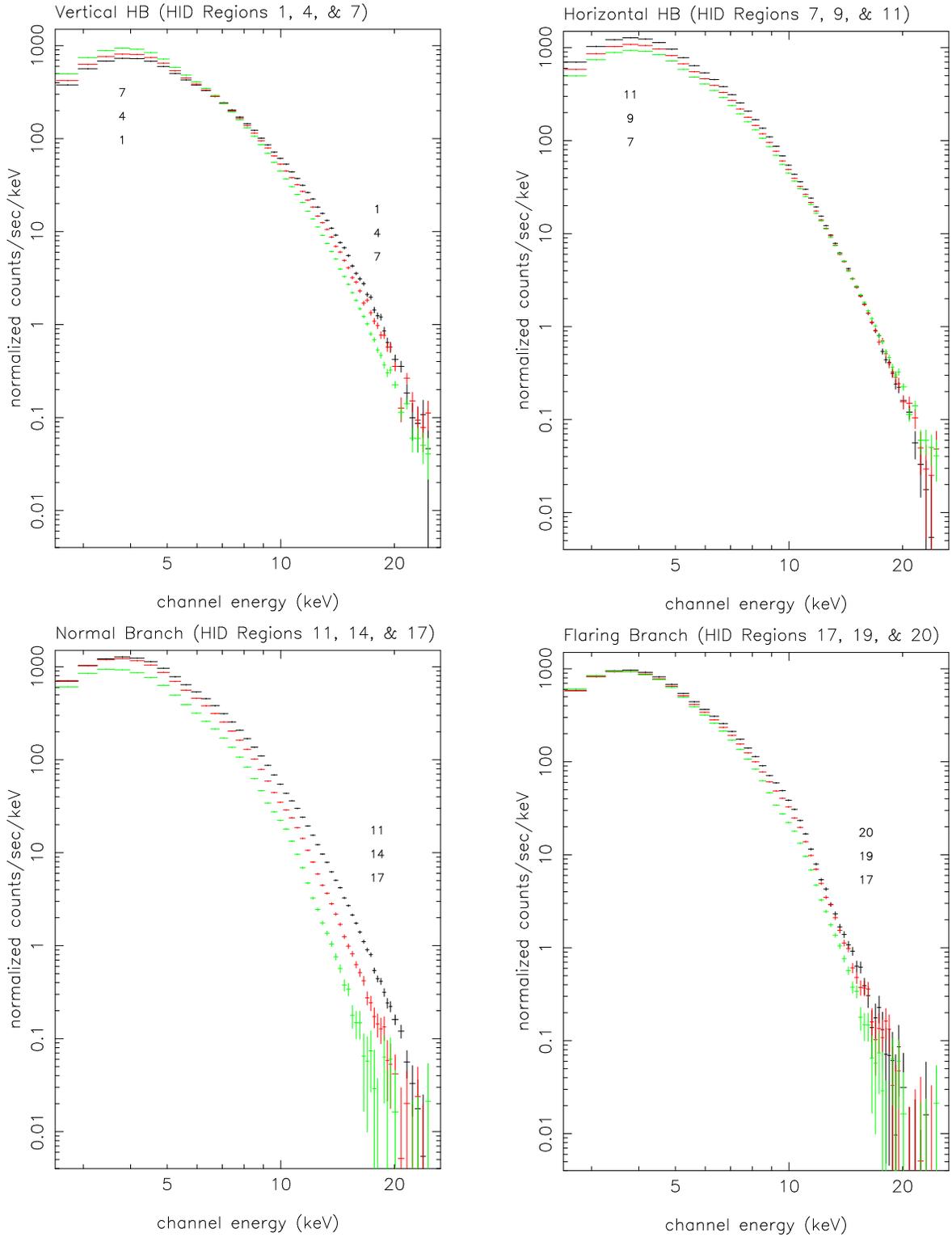

\vspace{-0.2in}
\PSbox{figures/june97_spec_vHB.eps hscale=40 vscale=40 hoffset=-15}{3.0in}{4.0in}
\vspace{-4.0in}\hspace{3.2in}
\PSbox{figures/june97_spec_hHB.eps hscale=40 vscale=40 hoffset=222}{3.0in}{4.0in}
\PSbox{figures/june97_spec_NB.eps hscale=40 vscale=40 hoffset=-15}{3.0in}{4.0in}
\vspace{-4.0in}\hspace{3.2in}
\PSbox{figures/june97_spec_FB.eps hscale=40 vscale=40 hoffset=222}{3.0in}{4.0in}
\vspace{-0.2in}
\caption{ 
Cir~X-1 energy spectra (2.5--25~keV; PCU~0 only) for several regions
on each segment of the HID track in
Figure~\protect{\ref{fig:june97_hid20reg}}.  The region numbers within
the figures are ordered vertically to match the relative intensities
at the low and/or high-energy ends of the spectra. 
}
\label{fig:june97_spec_branches}
\end{figure}
%-----------------------------------------------

% Procedure
The Standard2 data mode of the PCA produces 129-channel energy spectra
every 16 s.  A parallel background file was constructed using the
``pcabackest'' program provided with the FTOOLS analysis package
(version 4.0) and three background model files provided by the PCA
instrument team at NASA/GSFC (pca\_bkgd\_q6\_e03v01.mdl,
pca\_bkgd\_xray\_e03v02.mdl, pca\_bkgd\_activ\_e03v03.mdl). Average
energy spectra (and background spectra) were constructed for each of
the 20 HID regions, separately for each of the five PCUs. A 1\%
systematic error estimate was added to each channel of the spectra to
account for calibration uncertainties.

% Qualitative Evolution of Spectra Along Branches
Representative energy spectra for each segment of the HID track are
shown in Figure~\ref{fig:june97_spec_branches}. They demonstrate how
the spectrum evolves along each branch. One spectrum in each panel is
included in the next (since it is derived from the intersection of
branches), allowing the comparison to be boot-strapped around the
entire HID track. The region numbers in the figures are ordered
vertically to aid the comparison.

% vertical HB
The spectrum is hardest in region~1, at the top of the vertical
extension of the horizontal branch.  Motion down the branch
(softening, regions~1--7) is due to pivoting of the spectrum about
$\sim$7~keV, i.e., increasing intensity below $\sim$7~keV and
decreasing intensity above that energy.
% horizontal HB
Motion to the right across the horizontal portion of the HB
(regions~8--11) is due to continued increasing low-energy intensity,
but with a nearly constant spectrum above 12~keV.  The numerator of
the hardness ratio (\hardness{2}{6.3}{6.3}{13}) does not extend very
far into this band of nearly constant flux and furthermore is
dominated by photons with energies at the low end of the 6.3--13~keV
band where the intensity increases by a factor more similar to that of
the denominator. So, although the 2.5--25~keV spectrum softens
significantly (e.g., comparing the intensity at 5~keV to 18~keV), the
hardness ratio in Figure~\ref{fig:june97_spec_branches} decreases only
slightly from region~7 to region~11.

% NB
While moving down the normal branch (regions 11--17), the spectrum
decreases across the entire 2.5--25~keV band, but most strongly at
high energy (thus further softening). Towards the bottom of the NB,
the spectrum begins to show an abrupt steepening in slope above
10~keV.
% FB
Motion up the FB (regions 18--20) is produced by increasing intensity
above $\sim$4~keV and a relatively constant spectrum below that energy
(thus hardening).  The change in slope above 10~keV evolves into a
much more prominent step feature, with structure more complicated than
merely a change in slope (see below).

\section{Selection of Spectral Models}
\label{sec:june97_spec_models}

In order to explore possible models for use in fitting the spectra
from the HID regions, two high-quality spectra were constructed from
long steady segments (17--19~ks) at the beginning of the 1997 June
observations (time segments A and B in Figure~\ref{fig:june97_10d},
from days 609.93--610.16 and 610.66--610.90 respectively).  Based on
timing properties measured throughout these data sets, time segment~A
falls on the vertical portion of the HB (strong narrow QPO at
8.4--11.5~Hz) and time segment~B falls near the HB/NB apex (weak
narrow QPO above 30~Hz and/or the broad 4~Hz QPO).  Variability in
both of these segments was limited to less than 10\% in all energy
bands between 2.5--18~keV. We thus constructed a single (averaged)
spectrum for each segment, and we will refer to them as spectrum~A and
spectrum~B.  Errors in these spectra are dominated by the 1\%
systematics at all energies up to $\sim$20~keV.

% Fit procedure
Based on a study of the PCA response matrices using Crab nebula data,
R.~Remillard has recommended limiting fits to 2.5--25 keV and using
only PCUs 0, 1, and 4 (pvt.\ comm., currently available at {\tt
http://lheawww.gsfc.nasa.gov/users/keith/ronr.txt}). Spectra from each
of these detectors are fit separately. Fit parameters reported are the
average values for PCUs 0, 1, and when possible 4 (see below), and
errors are conservatively estimated from the entire range allowed by
90\% confidence intervals from each of the detectors. PCU~4
consistently gives lower normalizations for fitted spectral
components, so normalizations and flux values from that detector are
not included when computing the average values and their errors.
Furthermore, spectrum~B was not constructed for PCU~4 since that
detector was turned off during part of time segment~B.

% Single-component fits
Several single-component models were fit to spectra~A and B (all
models included an additional component for interstellar
absorption). Blackbody and power-law models fit very poorly in both
cases, as did a multi-temperature ``disk blackbody'' spectrum (summed
emission from various radii of an accretion
disk~\cite{mitsuda84,makishima86}; model ``diskbb'' in XSPEC), with
reduced $\chi^2$ ($\chi^2_r$) values of 22--545.  A thermal
bremsstrahlung model (emission due to acceleration of electrons by
protons or ions in an optically thin plasma cloud) provided a better
fit to spectrum~B ($\chi^2_r=4.0$), but fit spectrum~A poorly
($\chi^2_r=34$).
A relatively good fit was achieved for both spectra with a modified
bremsstrahlung model (see Table~\ref{tab:june97_fitsAB}), which
includes the self-Comptonization of bremsstrahlung photons to higher
energy due to interactions with electrons in an optically thick plasma
cloud~\cite{compls} (model ``compLS'' in XSPEC).

A number of two-component models were also fit to these two spectra.
A model using a disk blackbody and power law did not fit well
($\chi^2_r$=3--5), mainly because the spectrum does not flatten to a
single slope at high energy.
A hard (high-temperature, $\simgt$2~keV) blackbody is often used to fit
the spectra of LMXBs thought to contain a neutron star, where emission
near the surface might produce a high-temperature blackbody with small
effective area.  
A power law with high-energy exponential cutoff plus a hard blackbody
fit well (see Table~\ref{tab:june97_fitsAB}) but parameters for
the cutoff power law were poorly constrained since the cutoff energy
($E_{cut}\approx1.7~keV$) was below the PCA bandpass.  

%\newpage
\begin{table}
\scriptsize
%\small
\renewcommand{\arraystretch}{1.5}
\begin{centering}
\begin{tabular}{ccccccccc}
\hline
 & $N_H/10^{22}$~\footnotemark[1] & & & & & & Flux/10$^{-8}$~\footnotemark[3] & \\ 
& (cm$^{-2}$) &  \multicolumn{5}{c}{Model Components\footnotemark[2]} 
& erg\,cm$^{-2}$\,s$^{-1}$ &
 $\chi^2_r$~\footnotemark[4] \\
\hline
\hline

 & & & \multicolumn{3}{c}{Self-Comptonized Bremsstrahlung} & & & \\
\cline{4-6}
 & & &  $T$   & Optical &      & & & \\
 & & & (keV)  & depth   & norm & & & \\
\cline{3-7}
A & $2.80^{+0.22}_{-0.23}$ & 
  & $2.78^{+0.04}_{-0.05}$ & $11.05^{+0.47}_{-0.36}$ & $4.54^{+0.24}_{-0.23}$ & 
  & $2.73^{+0.02}_{-0.02}$ & 1.16--1.55 \\
B & $3.94^{+0.18}_{-0.21}$ & 
  & $2.89^{+0.12}_{-0.11}$ & $7.20^{+0.71}_{-0.69}$ & $9.67^{+0.37}_{-0.43}$ & 
  & $3.04^{+0.02}_{-0.02}$ & 1.51--2.02 \\
\hline

 & & \multicolumn{3}{c}{Cutoff Power Law~\footnotemark[5]} & \multicolumn{2}{c}{Blackbody~\footnotemark[6]} & & \\
\cline{3-5} \cline{6-7} 
 & & Photon  & $E_{cut}$ &      &  $T$  & $R$  & & \\
 & & index   &   (keV)   & norm & (keV) & (km) & & \\
\cline{3-7}
A &  $0.27^{+0.94}_{-0.27}$ 
  & $-0.65^{+0.72}_{-0.30}$ & $1.73^{+0.58}_{-0.24}$ & $2.24^{+1.53}_{-0.47}$ 
  & $2.44^{+0.10}_{-0.05}$ & $4.40^{+0.34}_{-0.87}$ 
  & $2.73^{+0.02}_{-0.02}$ & 0.79--1.04 \\
B & $1.95^{+0.74}_{-0.86}$ 
  & $-0.34^{+0.46}_{-0.60}$ & $1.67^{+0.31}_{-0.27}$ & $5.97^{+2.70}_{-2.29}$ 
  & $2.29^{+0.09}_{-0.07}$ & $4.61^{+0.67}_{-0.73}$ 
  & $3.04^{+0.02}_{-0.02}$ & 0.79--1.35 \\
\hline

 & & \multicolumn{2}{c}{Blackbody~\footnotemark[6]} & & \multicolumn{2}{c}{Blackbody~\footnotemark[6]} & \\
\cline{3-4} \cline{6-7} 
 & & $T$ & $R$ & & $T$ & $R$ & & \\ & & (keV) & (km)
 & & (keV) & (km) & & \\
\cline{3-7}
A & $0.00^{+0.02}_{-0.00}$ 
  & $1.16^{+0.02}_{-0.02}$ & $24.13^{+0.38}_{-0.37}$ & 
  & $2.33^{+0.02}_{-0.03}$ & $5.38^{+0.16}_{-0.15}$ 
  & $2.72^{+0.02}_{-0.02}$ & 2.51--2.82 \\
B & $0.01^{+0.21}_{-0.01}$ 
  & $1.12^{+0.01}_{-0.02}$ & $30.46^{+0.92}_{-0.40}$ & 
  & $2.20^{+0.03}_{-0.03}$ & $5.59^{+0.27}_{-0.25}$ 
  & $3.04^{+0.02}_{-0.02}$ & 1.42--1.78 \\
\hline

 & & \multicolumn{2}{c}{Disk Blackbody~\footnotemark[7]} & & \multicolumn{2}{c}{Blackbody~\footnotemark[6]} & \\
\cline{3-4} \cline{6-7} 
 & & $T_{in}$ & $R_{in}\cos^{1/2}\theta$ & &  $T$  & $R$  & & \\
 & &  (keV)   &  (km)                    & & (keV) & (km) & & \\
\cline{3-7}
A & $1.44^{+0.26}_{-0.27}$ 
  & $1.81^{+0.08}_{-0.06}$ & $9.39^{+0.64}_{-0.64}$ & 
  & $2.47^{+0.04}_{-0.04}$ & $4.10^{+0.25}_{-0.25}$ 
  & $2.73^{+0.02}_{-0.02}$ & 0.79--1.06 \\
B & $2.49^{+0.23}_{-0.23}$ 
  & $1.54^{+0.04}_{-0.04}$ & $15.63^{+0.93}_{-0.84}$ & 
  & $2.28^{+0.05}_{-0.05}$ & $4.66^{+0.37}_{-0.37}$  
  & $3.04^{+0.02}_{-0.02}$ & 0.75--1.31 \\
\hline
\end{tabular}
\end{centering}
\renewcommand{\arraystretch}{1}
\caption{Fit parameters for spectra~A \& B for four models.}
\label{tab:june97_fitsAB}
\end{table}

\footnotetext[1]{Absorption column density (Hydrogen atoms per cm$^2$).}
\footnotetext[2]{
Errors quoted are 90\% confidence limits for a single parameter
($\Delta\chi^2=2.7$).}
\footnotetext[3]{Total 2.5--25~keV flux.}
\footnotetext[4]{
Reduced $\chi^2$ = $\chi^2/dof$, where dof = degrees of freedom = the
number of spectral bins (52--54/spectrum) minus the number of fit
parameters (4--6).}
\footnotetext[5]{High-energy exponential cut-off.}
\footnotetext[6]{Blackbody radius assumes a distance of 8~kpc.}
\footnotetext[7]{
For the disk blackbody, the inner radius of the accretion disk (times
$\cos^{1/2}\theta$, where $\theta$ is the disk inclination angle and
$\theta=0$ is parallel to the line of sight) is given for a distance of
8~kpc.}
%\newpage

Two blackbodies ($\sim$1.1~keV and $\sim$2.2~keV) fit moderately well
(see Table~\ref{tab:june97_fitsAB}), but required negligible
interstellar absorption. The low absorption is inconsistent with
previous measurements from \ASCA\ and \ROSAT\ (both sensitive below
2~keV where the absorption is most easily constrained) which estimated
the interstellar column density to be $N_H=$(1.8--2.4)$\times
10^{22}$~cm$^{-2}$ \cite{brandt96,predehl95}.

A soft disk blackbody, with temperatures at the inner edge of the disk
of 1.5--1.8~keV, plus the hard blackbody fit both spectra quite well
(see Table~\ref{tab:june97_fitsAB}). This gave absorption column
densities roughly consistent with the \ASCA\ and
\ROSAT\ values.  Other two-component models could also
produce similar fits to those describe above. Thus, the spectrum
cannot be uniquely deconvolved into separate components. However,
for purposes of fitting the HID regions, the model composed of a disk
blackbody and blackbody was used. This model is physically motivated
(a component from an accretion disk and a harder component from near
the surface of a neutron star) and provides good fits with
interstellar column densities roughly consistent with previously
measured values.

\begin{figure}
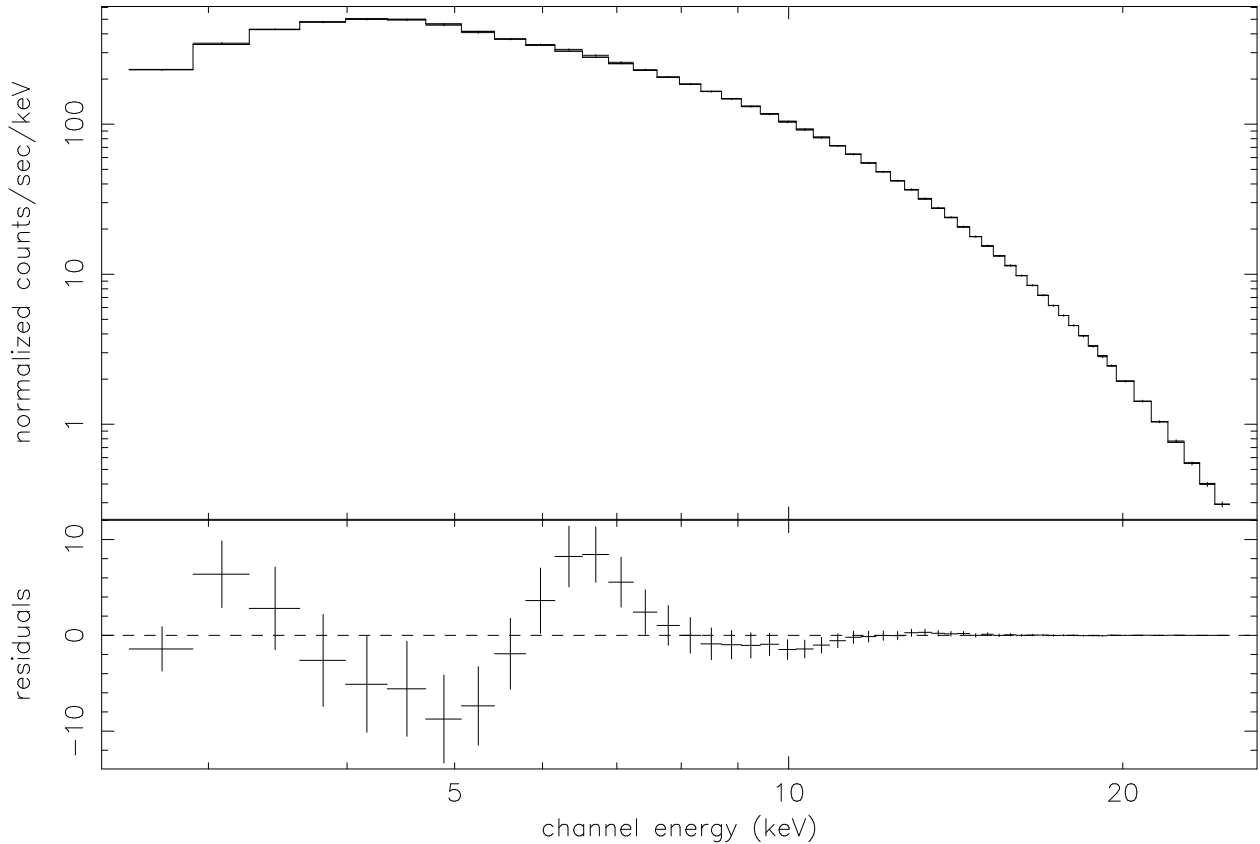

\begin{centering}
\PSbox{figures/june97_specA_diskbb_bb.eps hscale=72 vscale=72 angle=270 voffset=400 hoffset=-52}{6.5in}{4.40in}
\caption{
Spectrum~A and model (histogram) consisting of a disk
blackbody and blackbody (see
Table~\protect{\ref{tab:june97_fitsAB}}). The residuals show a peak at
6--7~keV that may be due to an iron emission line. }
\label{fig:june97_specA_diskbb_bb}
\end{centering}
\end{figure}

The spectral fit for spectrum~A for a disk blackbody plus blackbody is
shown in Figure~\ref{fig:june97_specA_diskbb_bb}. This plot
illustrates the good fit of the model over the entire 2.5--25~keV
band. Peaked residuals at 6--7~keV suggest the presence of an emission
line, probably iron K$\alpha$. Very similar residuals appear in most
of the fits discussed above for both spectra A and B. Addition of a
Gaussian line to the models does in fact improve the fits in almost
all cases; however, the best-fitting line often has an extremely large
Gaussian width ($\sigma>1$~keV). The energy resolution of the PCA is
about 1~keV FWHM at 6~keV; thus it is difficult to place reliable
constraints on the parameters (such as centroid and width) of a narrow
component such as an emission line.  We have not included an emission
line in the fits reported in Table~\ref{tab:june97_fitsAB}. The
presence of an emission line near 6.4~keV is discussed in more detail
in Chapter~\ref{ch:absdips} in conjunction with spectra of absorption
dips, which show the line more prominently.

\section{Fits to Spectra from 20 HID Regions}

\begin{figure}
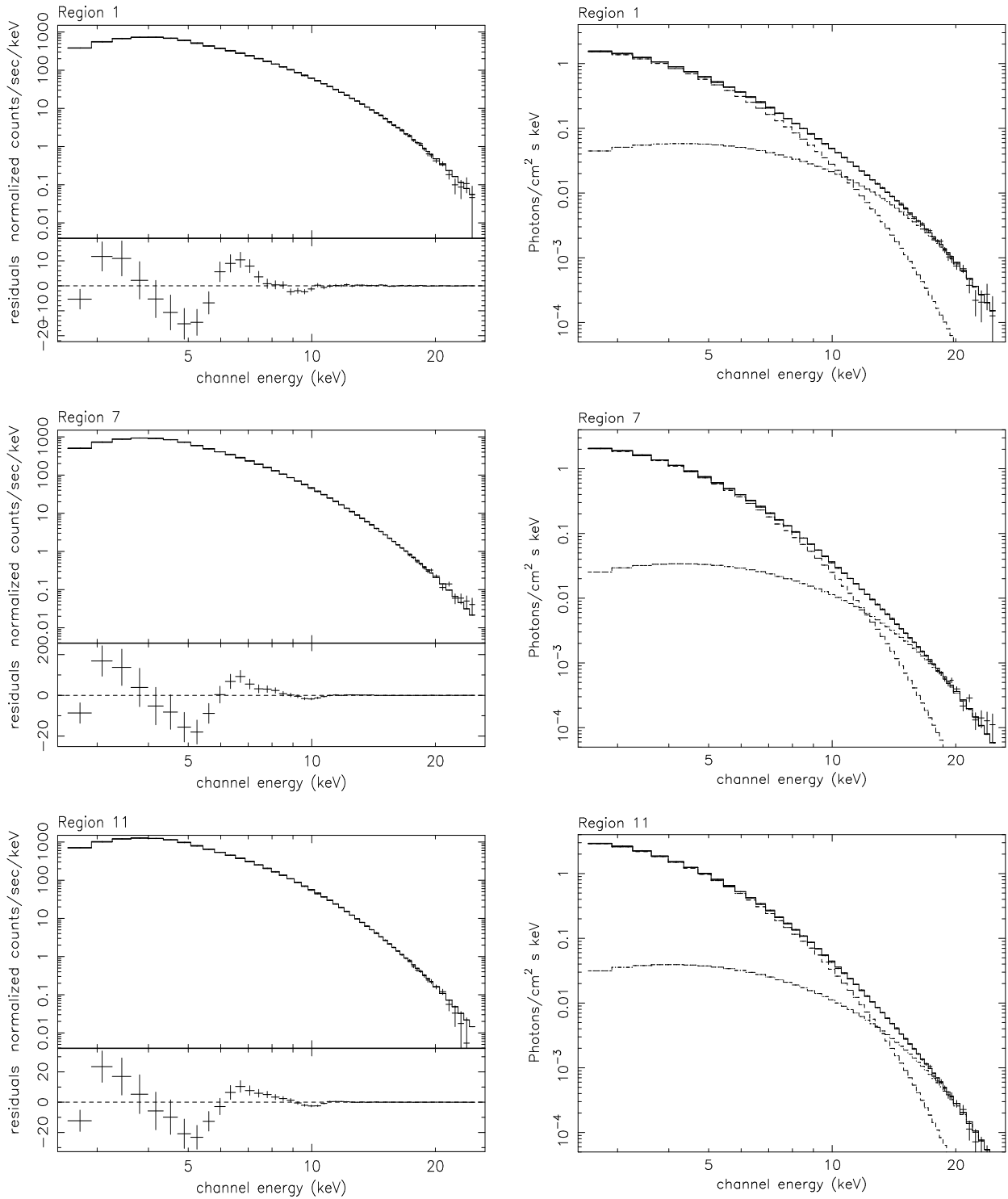

\vspace{0.2in}
\PSbox{figures/june97_spec_reg1.eps hscale=33 vscale=37 voffset=230 hoffset=-25 angle=270}{3.2in}{2.4in}
\vspace{-2.2in}\hspace{3.25in}
\PSbox{figures/june97_specmod_reg1.eps hscale=33 vscale=33  voffset=230 hoffset=218 angle=270}{3.2in}{2.4in}
\PSbox{figures/june97_spec_reg7.eps hscale=33 vscale=37  voffset=230 hoffset=-25 angle=270}{3.2in}{2.4in}
\vspace{-2.2in}\hspace{3.25in}
\PSbox{figures/june97_specmod_reg7.eps hscale=33 vscale=33  voffset=230 hoffset=218 angle=270}{3.2in}{2.4in}
\PSbox{figures/june97_spec_reg11.eps hscale=33 vscale=37  voffset=230 hoffset=-25 angle=270}{3.2in}{2.4in}
\vspace{-2.2in}\hspace{3.25in}
\PSbox{figures/june97_specmod_reg11.eps hscale=33 vscale=33  voffset=230 hoffset=218 angle=270}{3.2in}{2.4in}
\vspace{-0.3in}
\caption{ 
Fitted energy spectra (2.5--25~keV; PCU~0 only) and model components
(disk blackbody and blackbody) for HID regions 1, 7, and 11 of
Figure~\protect{\ref{fig:june97_hid20reg}}.  The disk blackbody
dominates at low energy and the blackbody dominates at high
energy. Fit parameters are listed in
Table~\protect{\ref{tab:june97_fitsHID}}.  
}
\label{fig:june97_spec_hidfits1}
\end{figure}

\begin{figure}
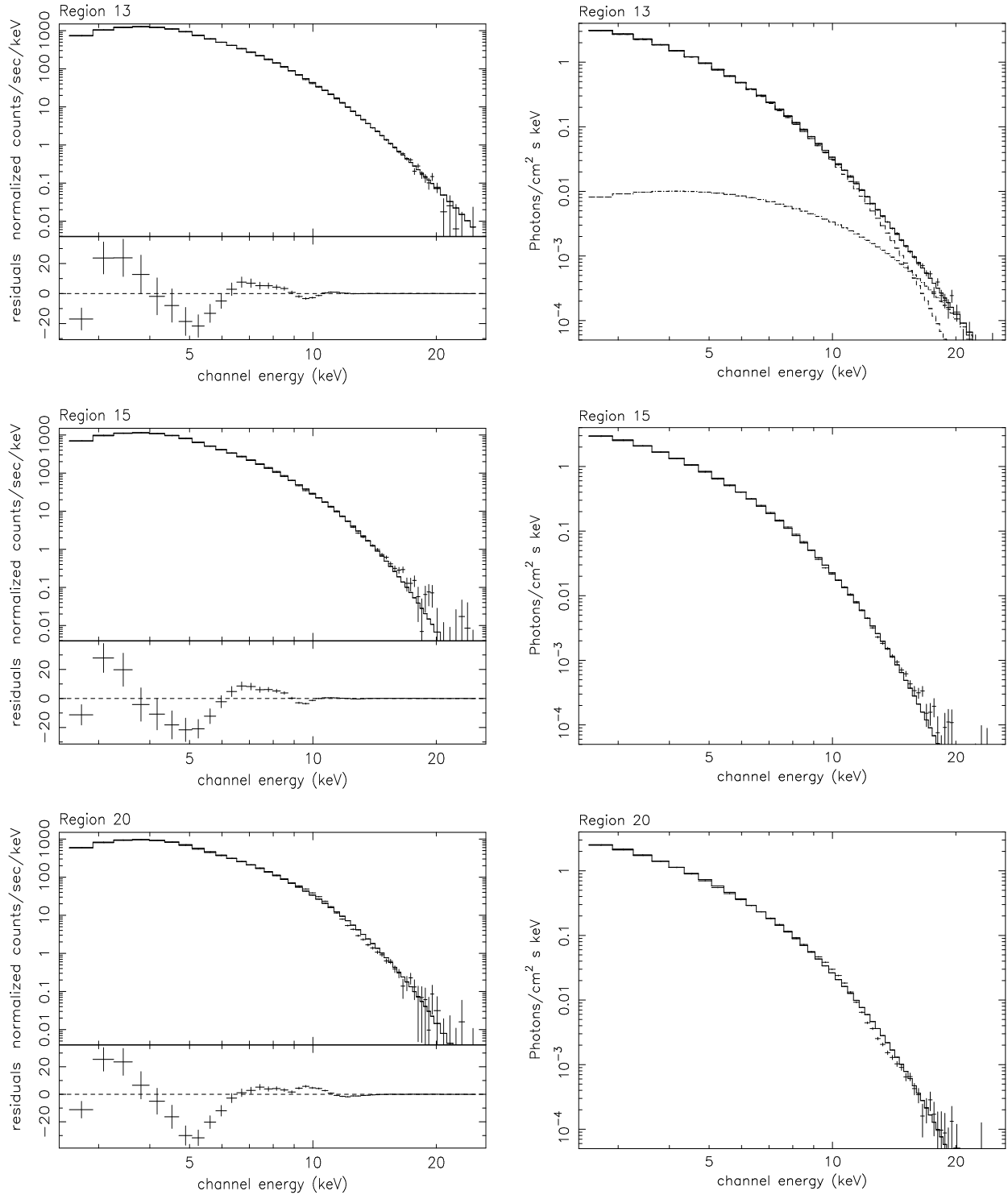

\vspace{0.2in}
\PSbox{figures/june97_spec_reg13.eps hscale=33 vscale=37 voffset=230 hoffset=-25 angle=270}{3.2in}{2.4in}
\vspace{-2.2in}\hspace{3.25in}
\PSbox{figures/june97_specmod_reg13.eps hscale=33 vscale=33  voffset=230 hoffset=218 angle=270}{3.2in}{2.4in}
\PSbox{figures/june97_spec_reg15.eps hscale=33 vscale=37  voffset=230 hoffset=-25 angle=270}{3.2in}{2.4in}
\vspace{-2.2in}\hspace{3.25in}
\PSbox{figures/june97_specmod_reg15.eps hscale=33 vscale=33  voffset=230 hoffset=218 angle=270}{3.2in}{2.4in}
\PSbox{figures/june97_spec_reg20.eps hscale=33 vscale=37  voffset=230 hoffset=-25 angle=270}{3.2in}{2.4in}
\vspace{-2.2in}\hspace{3.25in}
\PSbox{figures/june97_specmod_reg20.eps hscale=33 vscale=33  voffset=230 hoffset=218 angle=270}{3.2in}{2.4in}
\vspace{-0.45in}
\caption{ 
Fitted energy spectra (2.5--25~keV; PCU~0 only) and model components
(disk blackbody and blackbody) for HID regions 13, 15, and 20 of
Figure~\protect{\ref{fig:june97_hid20reg}}.  The blackbody only
contributes to the high-energy end of the spectrum in region 13, and
has faded away entirely in regions 15 and 20.  Fit parameters are
listed in Table~\protect{\ref{tab:june97_fitsHID}}.  }
\label{fig:june97_spec_hidfits2}
\end{figure}

A disk blackbody plus blackbody was fit to the spectra from each of
the 20 HID regions. The resulting fit parameters are listed in
Table~\ref{tab:june97_fitsHID}, and several representative fits are
shown in Figures~\ref{fig:june97_spec_hidfits1}
and~\ref{fig:june97_spec_hidfits2}. We estimated the distance to
Cir~X-1 to be 8~kpc (see section~\ref{sec:intro_cirx1}) in converting
blackbody and disk blackbody normalizations to radii.

% HB
The spectra along the horizontal branch (regions 1--11) were all fit
relatively well, with residuals similar to those for spectra~A and~B
above, suggesting a possible emission line from iron. These spectra
all show column densities of 1.8--2.3$\times10^{22}$~cm$^{-2}$,
consistent with the \ASCA\ and \ROSAT\ values discussed above.  The
temperatures of both components ($\sim$1.3~keV for the inner disk and
$\sim$2.0~keV for the hard blackbody) are quite stable on the HB, with
only slight evidence for cooling down the vertical portion of the
branch (i.e., higher temperatures in regions 1--4). The pivoting
spectrum on the vertical portion of the HB may be related to these
temperature changes.
The inner radius of the disk blackbody component (times a factor of
order unity involving the inclination angle of the disk) increases
from 19 to 33~km, while the radius of hard blackbody remains between 3
and 4~km. Thus, it is mainly changes in the (soft) disk blackbody
radius that produce the HB track. These size scales are consistent
with the hypothesis that these components arise from emission close to
a neutron star, which has a radius of order 10~km.
From region~1 to 11, the total 2.5--25~keV flux increases
monotonically (with the exception of region 6) from
2.89--4.35$\times10^{-8}$~erg~cm$^{-2}$~s$^{-1}$ (corresponding to
1.2--1.8 times the Eddington luminosity limit for a 1.4~\msun\ neutron
star at 8~kpc).

% NB
Moving down the normal branch, the quality of the fits decrease, as
indicated by the increasing $\chi^2$ values in
Table~\ref{tab:june97_fitsHID}.  The absorption column density
gradually decreases by a factor of two, but may be related to the
decreasing fit quality.  
The inner radius and temperature of the disk blackbody change only
slightly on the normal branch. In contrast, the hard blackbody begins
to fade on the upper portion of the normal branch (regions 12--14), as
indicated by a decreasing radius for the emission area. In fact, by
the middle of the normal branch, the hard blackbody has faded entirely
and fits have lower $\chi^2$ values without it. Thus, the hard
blackbody is omitted from the fits for regions 15--20. The 2.5--6~keV
residuals continue to appear similar to those on the HB, but the peaked
$\sim$6.5~keV residual on the left HB becomes broader and more
complicated for higher rank numbers.

\begin{table}
\small
\renewcommand{\arraystretch}{1.2}
\begin{centering}
\begin{tabular}{cccccccc}
\hline
HID~\footnotemark[1]  & $N_H/10^{22}$~\footnotemark[2] & $T_{in}$ & $R_{in}\cos^{1/2}\theta$~\footnotemark[3] & $T$   & $R$~\footnotemark[4]  & Flux/10$^{-8}$~\footnotemark[5] & \\ 
region & (cm$^{-2}$)   & (keV)    & (km)                   & (keV) & (km) & erg\,cm$^{-2}$\,s$^{-1}$ & $\chi^2_r$~\footnotemark[6] \\
\hline
\hline
     1 & $1.83^{+0.28}_{-0.32}$ & $1.45^{+0.05}_{-0.05}$ & $19.29^{+1.31}_{-1.22}$ & $2.16^{+0.09}_{-0.07}$ & $3.65^{+0.50}_{-0.52}$ & $2.89^{+0.03}_{-0.02}$ & 1.41--2.14 \\
     2 & $1.97^{+0.30}_{-0.28}$ & $1.41^{+0.04}_{-0.05}$ & $20.87^{+1.46}_{-1.29}$ & $2.15^{+0.09}_{-0.10}$ & $3.71^{+0.68}_{-0.64}$ & $2.99^{+0.02}_{-0.02}$ & 1.40--1.70 \\
     3 & $2.04^{+0.27}_{-0.35}$ & $1.38^{+0.04}_{-0.04}$ & $22.24^{+1.30}_{-1.16}$ & $2.08^{+0.09}_{-0.08}$ & $4.00^{+0.56}_{-0.52}$ & $2.96^{+0.02}_{-0.02}$ & 1.32--1.81 \\
     4 & $2.18^{+0.37}_{-0.32}$ & $1.36^{+0.05}_{-0.06}$ & $23.46^{+2.27}_{-1.85}$ & $2.07^{+0.18}_{-0.17}$ & $3.89^{+1.27}_{-1.18}$ & $3.02^{+0.03}_{-0.02}$ & 2.21--3.47 \\
     5 & $2.29^{+0.33}_{-0.27}$ & $1.31^{+0.03}_{-0.05}$ & $27.22^{+1.97}_{-1.74}$ & $2.02^{+0.11}_{-0.14}$ & $3.78^{+0.97}_{-0.94}$ & $3.21^{+0.02}_{-0.02}$ & 2.51--3.24 \\
     6 & $2.35^{+0.30}_{-0.24}$ & $1.34^{+0.03}_{-0.04}$ & $26.21^{+1.33}_{-1.19}$ & $2.02^{+0.11}_{-0.13}$ & $3.31^{+0.58}_{-0.53}$ & $3.37^{+0.02}_{-0.02}$ & 1.41--1.55 \\
     7 & $2.31^{+0.24}_{-0.25}$ & $1.30^{+0.02}_{-0.03}$ & $28.67^{+1.20}_{-1.05}$ & $2.03^{+0.09}_{-0.11}$ & $3.30^{+0.65}_{-0.57}$ & $3.27^{+0.02}_{-0.02}$ & 2.22--2.76 \\
     8 & $2.27^{+0.29}_{-0.25}$ & $1.29^{+0.02}_{-0.03}$ & $30.06^{+1.70}_{-1.48}$ & $2.01^{+0.12}_{-0.14}$ & $3.34^{+0.91}_{-0.83}$ & $3.46^{+0.03}_{-0.02}$ & 2.41--2.70 \\
     9 & $2.18^{+0.21}_{-0.24}$ & $1.30^{+0.03}_{-0.03}$ & $30.68^{+1.15}_{-1.07}$ & $2.01^{+0.14}_{-0.14}$ & $3.36^{+0.85}_{-0.69}$ & $3.75^{+0.02}_{-0.02}$ & 2.29--2.44 \\
    10 & $2.01^{+0.23}_{-0.24}$ & $1.31^{+0.02}_{-0.03}$ & $31.35^{+1.31}_{-1.10}$ & $2.01^{+0.12}_{-0.11}$ & $3.02^{+0.86}_{-0.67}$ & $4.15^{+0.02}_{-0.03}$ & 1.82--2.56 \\
    11 & $1.98^{+0.19}_{-0.22}$ & $1.30^{+0.02}_{-0.02}$ & $32.90^{+1.24}_{-1.17}$ & $1.93^{+0.08}_{-0.08}$ & $3.52^{+0.67}_{-0.54}$ & $4.39^{+0.03}_{-0.03}$ & 2.55--3.05 \\
    12 & $1.78^{+0.22}_{-0.26}$ & $1.29^{+0.02}_{-0.02}$ & $33.41^{+1.51}_{-1.34}$ & $1.94^{+0.17}_{-0.12}$ & $3.09^{+0.82}_{-0.72}$ & $4.35^{+0.03}_{-0.03}$ & 2.87--3.06 \\
    13 & $1.63^{+0.21}_{-0.25}$ & $1.28^{+0.02}_{-0.02}$ & $34.02^{+1.09}_{-0.92}$ & $2.12^{+0.22}_{-0.19}$ & $1.51^{+0.70}_{-0.47}$ & $4.21^{+0.03}_{-0.03}$ & 3.74--4.66 \\
    14 & $1.53^{+0.18}_{-0.17}$ & $1.26^{+0.01}_{-0.02}$ & $33.94^{+1.28}_{-1.06}$ & $2.34^{+0.53}_{-0.40}$ & $0.84^{+0.77}_{-0.46}$ & $3.98^{+0.03}_{-0.03}$ & 4.19--5.11 \\
    15 & $1.48^{+0.15}_{-0.16}$ & $1.22^{+0.00}_{-0.00}$ & $35.30^{+0.51}_{-0.49}$ & & & $3.68^{+0.02}_{-0.02}$ & 5.15--6.34 \\
    16 & $1.43^{+0.19}_{-0.17}$ & $1.19^{+0.00}_{-0.00}$ & $35.88^{+0.50}_{-0.48}$ & & & $3.32^{+0.02}_{-0.02}$ & 5.91--7.52 \\
    17 & $0.94^{+0.17}_{-0.15}$ & $1.20^{+0.00}_{-0.00}$ & $32.69^{+0.23}_{-0.53}$ & & & $3.00^{+0.02}_{-0.01}$ & 5.87--7.16 \\
    18 & $0.56^{+0.16}_{-0.18}$ & $1.26^{+0.01}_{-0.01}$ & $28.22^{+0.49}_{-0.48}$ & & & $2.94^{+0.02}_{-0.02}$ & 9.59--11.39 \\
    19 & $0.64^{+0.17}_{-0.16}$ & $1.29^{+0.01}_{-0.01}$ & $27.02^{+0.35}_{-0.36}$ & & & $3.08^{+0.02}_{-0.02}$ & 9.36--10.45 \\
    20 & $0.81^{+0.15}_{-0.19}$ & $1.33^{+0.01}_{-0.01}$ & $26.25^{+0.38}_{-0.37}$ & & & $3.24^{+0.02}_{-0.02}$ & 14.67--17.01 \\
\hline
\end{tabular}
\end{centering}
\renewcommand{\arraystretch}{1}
\caption{Fit parameters for HID regions 1--20 using a model consisting of a disk blackbody and a blackbody.}
\label{tab:june97_fitsHID}
\end{table}
\footnotetext[1]{
Errors quoted are 90\% confidence limits for a single parameter
($\Delta\chi^2=2.7$).}
\footnotetext[2]{Absorption column density (Hydrogen atoms per cm$^2$).}
\footnotetext[3]{
Inner radius of the accretion disk (times $\cos^{1/2}\theta$, where
$\theta$ is the disk inclination angle and $\theta=0$ is parallel
to the line of sight) assuming a distance of 8~kpc.}
\footnotetext[4]{Blackbody radius assuming a distance of 8~kpc.}
\footnotetext[5]{Total 2.5--25~keV flux.}
\footnotetext[6]{
Reduced $\chi^2$ = $\chi^2/dof$, where dof = degrees of freedom = the
number of spectral bins (52--54/spectrum) minus the number of fit
parameters (4--6).}

%-----------------------------------------------
%-----------------------------------------------
%-----------------------------------------------

\begin{figure}
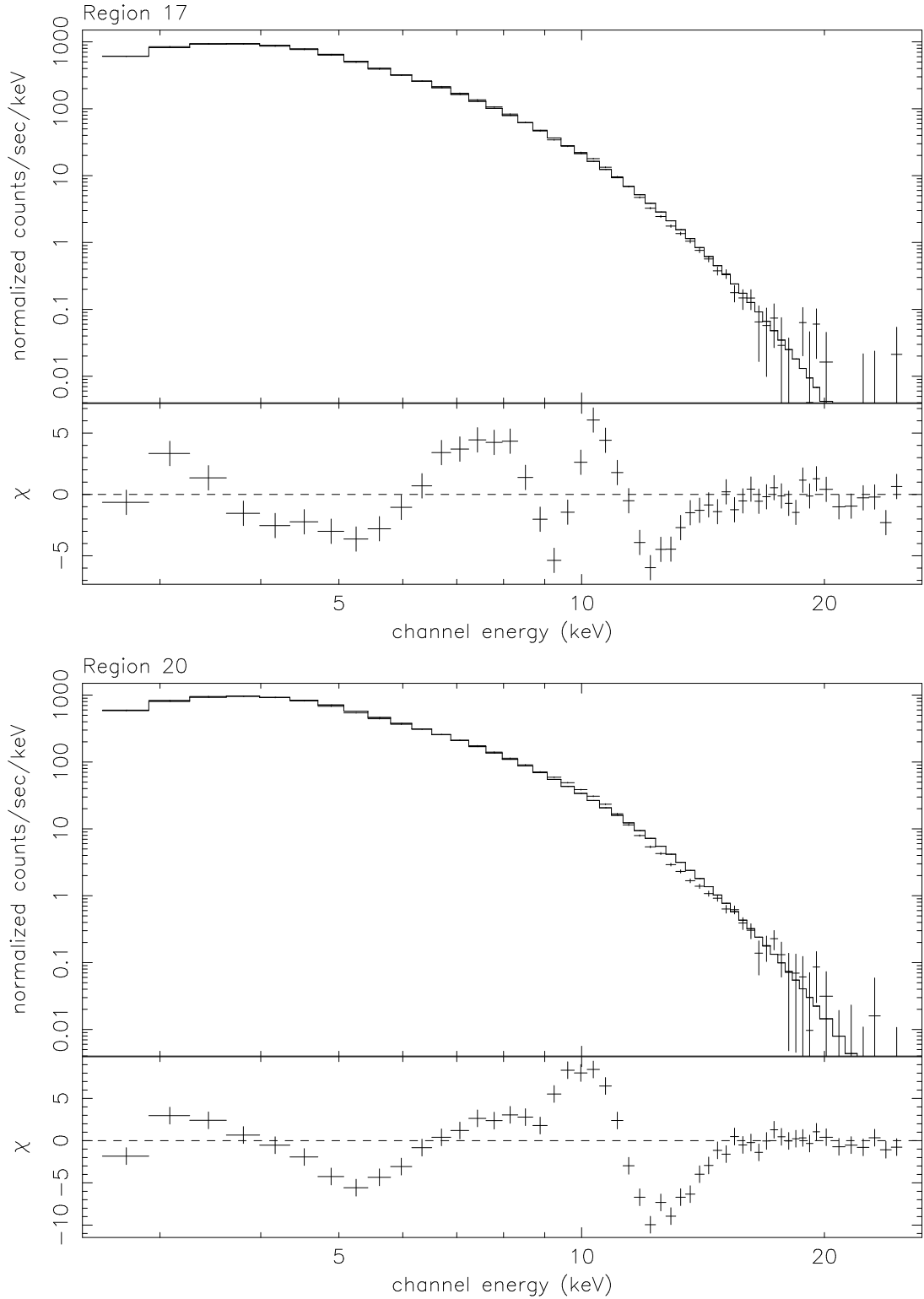

\begin{centering}
\PSbox{figures/june97_specdel_reg17.eps hscale=62 vscale=62 angle=270 voffset=350 hoffset=-25}{5.6in}{3.9in}
\PSbox{figures/june97_specdel_reg20.eps hscale=62 vscale=62 angle=270 voffset=340 hoffset=-25}{5.6in}{3.9in}
\caption{
Spectra from HID regions 17 and 20 (crosses) and fitted models
(histograms) consisting of a disk blackbody and blackbody. The ratio
of residuals to error bars shows a narrow edge or line-like feature
above 10~keV. 
}
\label{fig:june97_specdel}
\end{centering}
\end{figure}

% FB
On the flaring branch (regions 18--20), the fit quality decreases
further, accompanied by very low values for the absorption column
density. A significant contribution to the high $\chi^2$ values on the
flaring branch is due to the feature above 10~keV that was mentioned
above in section~\ref{sec:june97_spec_qual}.  This feature actually
begins to develop on the lower NB, as early as region~14 and is
illustrated by two spectra shown Figure~\ref{fig:june97_specdel}. The
residuals shown for each bin have been divided by the size of the
error bar for that bin.  The feature resembles a narrow line or
absorption edge but its location above 10~keV rules out even
hydrogen-like iron as a possible source.

A number of other spectral models were fit to the HID-region spectra,
and all failed to satisfactorally fit the lower portion of the track
(rank number 14 and greater). In addition no single or
double-component model was found that could reproduce the unusual
feature above 10~keV. For example, a broken power law (model with two
different slopes above and below a cutoff energy) fails to account for
the more step-like nature of the feature.

\section{Summary}

Although color-color and hardness-intensity diagrams for Cir~X-1 are
complicated by absorption dips and shifting spectral tracks, the
source clearly traces out a complete Z-source track during short time
segments (hours). Power spectra taken from regions of the HID track
show continuous evolution from the narrow QPO (increasing in frequency
from 12--30~Hz) on the horizontal branch, to the broad 4~Hz QPO on the
normal branch, to only very low frequency noise on the flaring
branch. Flaring-branch light curves show high levels of variability
due to ``mini-flares''.

The energy spectrum on the horizontal branch is fit well with a
two-component model consisting of a soft disk blackbody and a harder
blackbody (presumably from closer to the surface). In this model,
motion along the HB is mainly associated with an increasing inner
radius of the disk (increasing disk blackbody normalization). This
would imply that, as the luminosity increases across the HB, the inner
edge of the disk is pushed further away from the surface. It is not
clear how this is related to the increasing QPO frequency, which would
typically be expected to require a {\em decreasing} radius if the QPOs
were related to Keplerian motion at the inner edge of the disk.
Energy spectra on the normal branch indicate that the hard blackbody
fades away, leaving only the disk blackbody. On the lower NB, a
feature in the spectrum develops above 10~keV. This feature becomes
more prominent on the flaring branch. Although, there is still much to
be understood about the physical mechanisms responsible for the
spectral and timing evolution of Cir~X-1 around the HID diagram, these
results have demonstrated the specific timing and spectral properties
associated with each branch.

\chapter{Absorption Dips}
\label{ch:absdips}

\section{Overview}
\RXTE\ All-Sky Monitor light curves of Cir~X-1 show that dips occur 
near phase zero of the 16.55-day cycle of the source
(Chapter~\ref{ch:observs}). \RXTE\ Proportional Counter Array
observations carried out between 1996 September~20--22 provided 60\%
observing efficiency for 48~hours around phase zero (study~D,
see Figure~\ref{fig:pca_by_cycle}). These observations showed
significant dipping activity during much of those two days. We have
studied the dramatic spectral evolution associated with the dips, and
show that the spectrum throughout the dips is well-fit by variable and
at times heavy absorption of a bright component, plus a faint
component that is only weakly absorbed. We also show that an iron
emission line near 6.5~keV present during dips maintains a constant
absolute flux level outside the dips as well, indicating that the line
is associated with the faint component. We suggest that these results
are consistent with a model in which the bright component is
reprocessed by a scattering medium, only slightly modifying the
spectral shape, but adding an iron line due to fluorescence. These
results are also generally consistent with a spectral study based on
\ASCA\ observations of a low-to-high transition in
Cir~X-1~\cite{brandt96}, where the spectrum was fit with a
partial-covering model.

\section{Dips in RXTE Light Curves}

\begin{figure}
\begin{centering}
\PSbox{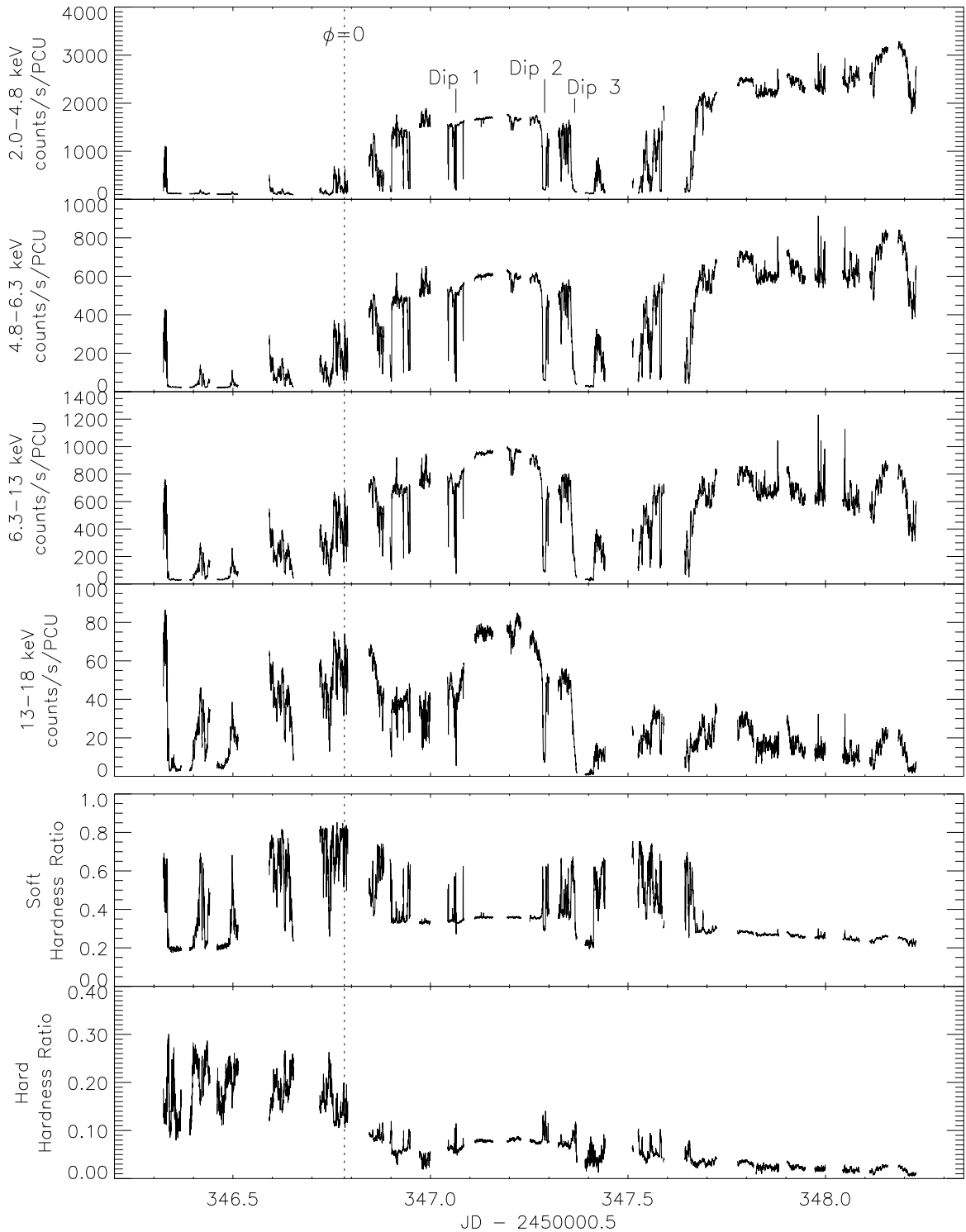 hscale=90 vscale=90 voffset=-18 hoffset=-7}{6.1in}{7.4in}
\caption{
Light curves in four energy channels and two hardness ratios for PCA
observations of Cir~X-1 from 1996 September~20--22, covering a two-day
period around phase zero ($\phi=0$). Three dips have been identified
for further study. Each point represents 16~s of background-subtracted
data from PCUs 0, 1, and 2. Ratios of the intensities produce soft
(\hardness{2.0}{4.8}{4.8}{6.3}) and hard (\hardness{6.3}{13}{13}{18})
hardness ratios. The intensity levels of the segment at
day~$\sim$347.2 (after Dip~1) are close to the level in each band
during the quiescent phases of the orbit.  }
\label{fig:fall96_48hr}
\end{centering}
\end{figure}

Light curves from the ASM and PCA show that the intensity of Cir~X-1
drops below the 1~Crab baseline near phase zero of most (possibly all)
cycles. These low-intensity episodes, which we refer to as ``dips'',
are brief compared to the 16.55-d cycle, lasting as short as seconds
(e.g., Figure~\ref{fig:june97_lc_hr_reg}) and as long as half a day
(e.g., Figure~\ref{fig:fall96_48hr}). They often show abrupt
transitions (rise or fall times much smaller than the total duration),
and are characterized by specific spectral evolution which we describe
in detail below. During the strongest dips observed in PCA
observations, the intensity dropped to levels more than a factor of 10
below the baseline level (2--90~keV).  Dips in the ASM light curves
(2--12~keV) reached similar levels (i.e., as low as $\sim$80~mCrab).

The dips most often occurred during the half day before and after
phase zero, but also sometimes continued through day~1.5 in ASM data
and day~2.5 in PCA data. On a couple of occasions, dips also occurred at
later phases in the ASM light curve, e.g., near day~4 of the cycle.
Dips occurred in all six cycles of Cir~X-1 for which we obtained PCA
observations during the half day before phase zero (see
Figure~\ref{fig:pca_by_cycle}). All-Sky Monitor light curves of
Cir~X-1 show intensity points below the 1~Crab baseline in about half
of the cycles observed (see section~\ref{sec:asm}). Because ASM
coverage is incomplete (about ten to twelve 90-s exposures per day),
it is possible that dips might actually occur in all cycles.

The PCA observations (study~D) made during 1996 September~20--22
(Figure~\ref{fig:fall96_48hr}) covered a 48-hour period around phase
zero (with $\sim$60\% observing efficiency) and showed significant
dipping for about half of the observed time. These observations
provided the opportunity to study dipping behavior in detail, and are
thus the focus of this chapter. Dips in other cycles have also been
examined and show similar behavior to that presented below.

% Brief overview of 48-hr observation
The observations shown in Figure~\ref{fig:fall96_48hr} began about a
half day before phase zero. The intensity of Cir~X-1 was very low (an
extended dip) during the first half day of the observations, and then
returned to the much higher ``quiescent'' level shortly after phase
zero.  During the half day immediately following phase zero, the
intensity underwent shorter intermittent dips.  The strongest dips all
reached similar minimum levels of about 200--350 counts/s/PCU
(2--18~keV) after background ($\sim$10 c/s/PCU) was subtracted.
A second episode of significant extended dipping occurred from about
0.5 to 0.9~days after phase zero, followed by flaring-state activity
for the remaining half day of the observation.  The flaring state is
characterized by (1) a gradual increase of the average intensity in
the low-energy PCA bands (which then dominate the total PCA count
rate) accompanied by a significant drop in the high-energy bands and
(2) individual flares on shorter timescales (less than a few hours)
which show increased intensity in all PCA energy bands. The spectral
and timing evolution during the flaring state of Cir~X-1 was discussed
in Chapters~\ref{ch:feb97paper} and
\ref{ch:fullZ}.

\section{Evolution of Hardness Ratios During Dips}

\begin{figure}
\begin{centering}
\PSbox{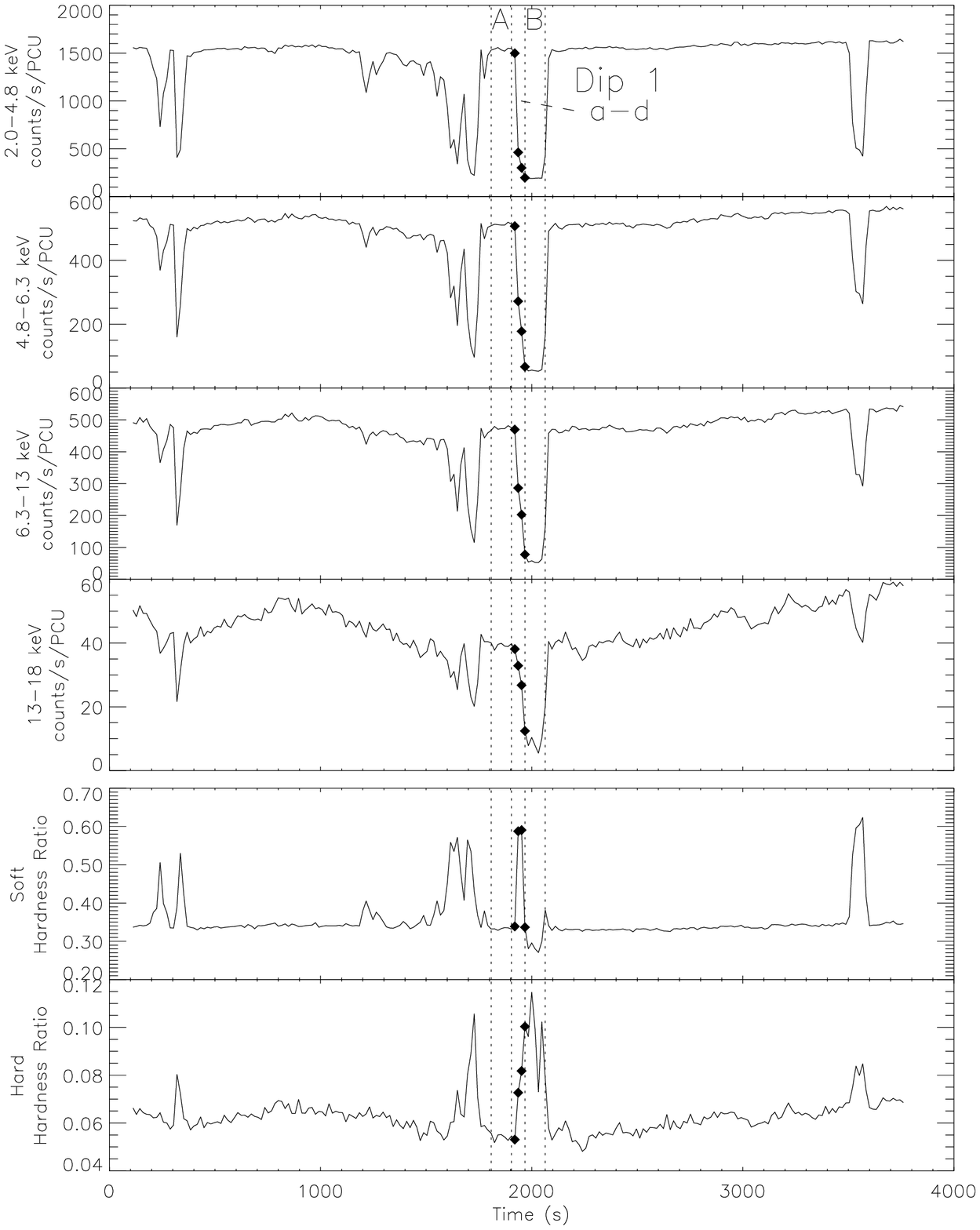 hscale=90 vscale=90 hoffset=-10}{6.1in}{7.7in}
\caption{
Light curves in four energy channels and two hardness ratios spanning
4000~s including Dip~1. Time zero corresponds to day~347.041 in
Figure~\protect{\ref{fig:fall96_48hr}}. The intensity and hardness
ratio points are the same as in
Figure~\protect{\ref{fig:fall96_48hr}}.  Energy spectra were extracted
from two 96-s time segments (A and B), indicated by dotted vertical lines,
and four 16-s segments (a--d) indicated by diamonds.  }
\label{fig:fall96_dip1lc}
\end{centering}
\end{figure}

\begin{figure}
\begin{centering}
\PSbox{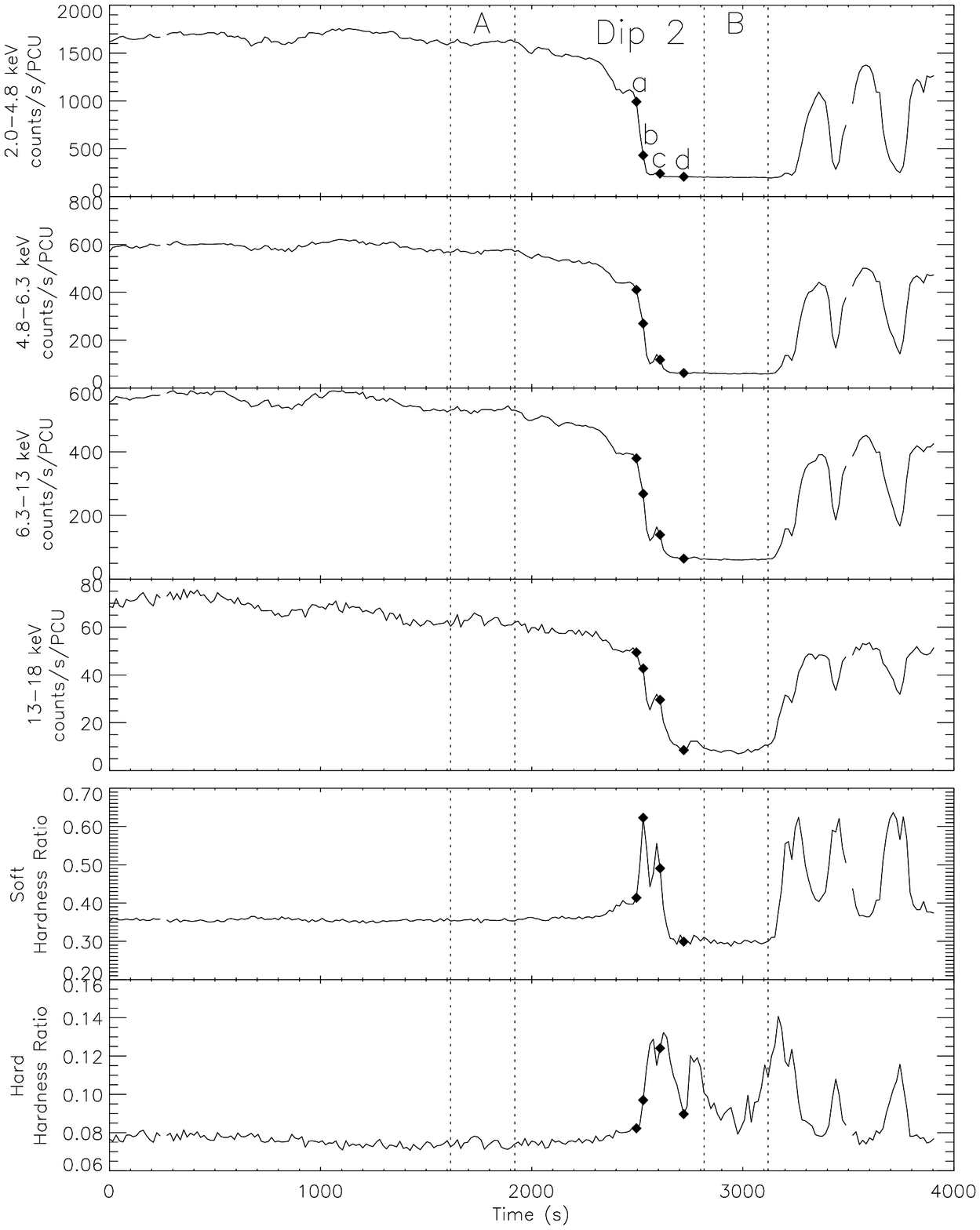 hscale=90 vscale=90 hoffset=-10}{6.1in}{7.7in}
\caption{
Light curves in four energy channels and two hardness ratios spanning
4000~s including Dip~2. Time zero corresponds to day~347.254 in
Figure~\protect{\ref{fig:fall96_48hr}}. The intensity and hardness
ratio points are the same as in
Figure~\protect{\ref{fig:fall96_48hr}}.  Energy spectra were extracted
from two 304-s time segments (A and B), indicated by dotted vertical
lines, and four 16-s segments (a--d), indicated by diamonds.  }
\label{fig:fall96_dip2lc}
\end{centering}
\end{figure}

\begin{figure}
\begin{centering}
\PSbox{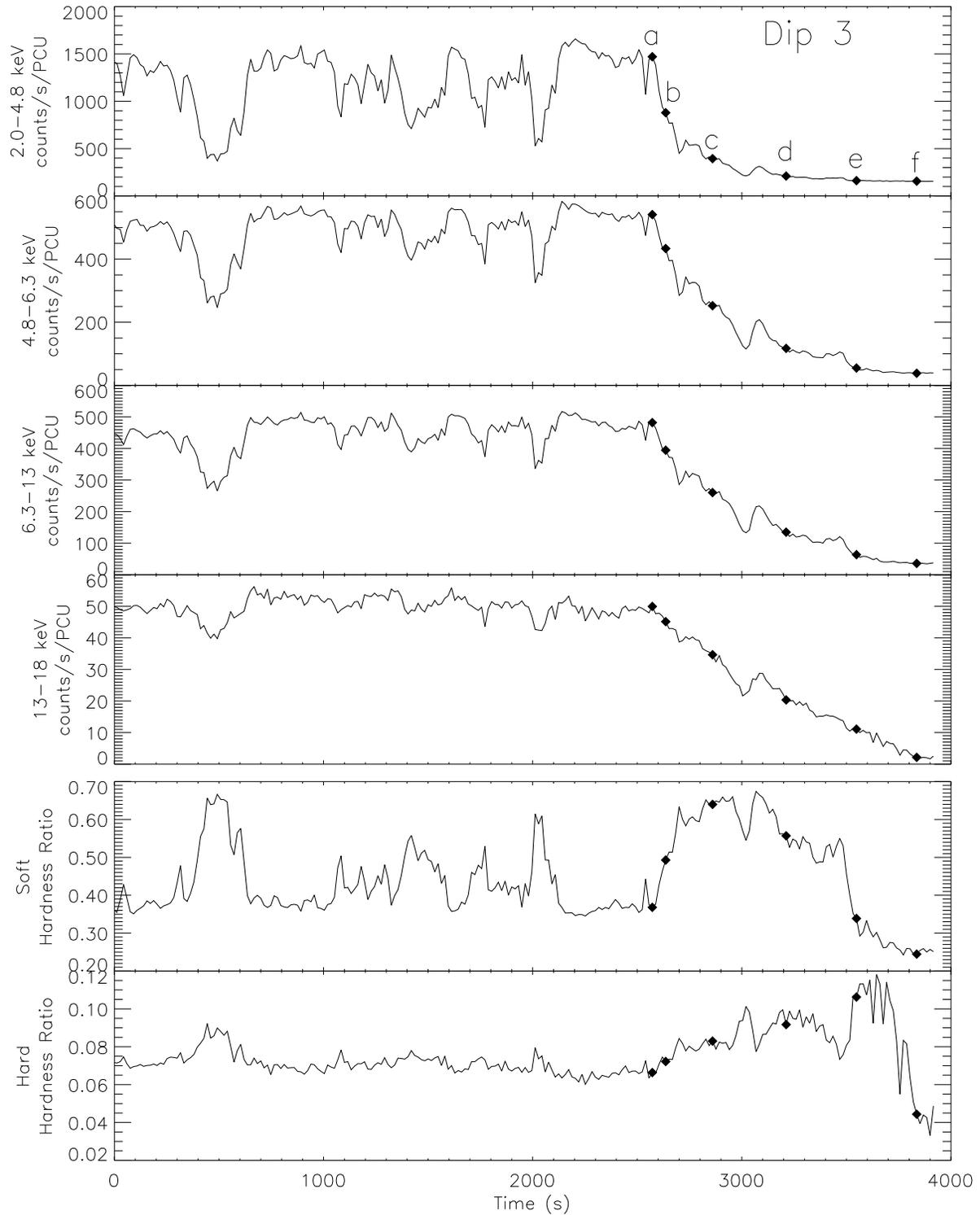 hscale=90 vscale=90 hoffset=-10}{6.1in}{7.7in}
\caption{
Light curves in four energy channels and two hardness ratios spanning
4000~s including Dip~3. Time zero corresponds to day~347.325 in
Figure~\protect{\ref{fig:fall96_48hr}}. The intensity and hardness
ratio points are the same as in
Figure~\protect{\ref{fig:fall96_48hr}}.  Energy spectra were extracted
from six 16-s time segments (a--f), indicated by diamonds.
}
\label{fig:fall96_dip3lc}
\end{centering}
\end{figure}

Three dips in Figure~\ref{fig:fall96_48hr} have been identified for
further study (Dips 1--3). A 4000-s segment of
Figure~\ref{fig:fall96_48hr} is shown for each of these dips in
Figures \ref{fig:fall96_dip1lc},~\ref{fig:fall96_dip2lc},
and~\ref{fig:fall96_dip3lc}.  The light curves in multiple PCA energy
channels show that the transitions into and out of dips occur more
rapidly at low energy that in higher energy bands (best demonstrated
by the gradual ingress of Dip~3 in Figure~\ref{fig:fall96_dip3lc}),
thus hardness ratios initially increase (harden) as the denominator
falls more quickly than the numerator. Low energy count rates are also
the first to reach a fixed bottom level. This causes hardness ratios
to decrease (soften) since the denominator approaches a constant while
the numerator continues to decrease.

\begin{figure}
\begin{centering}
\PSbox{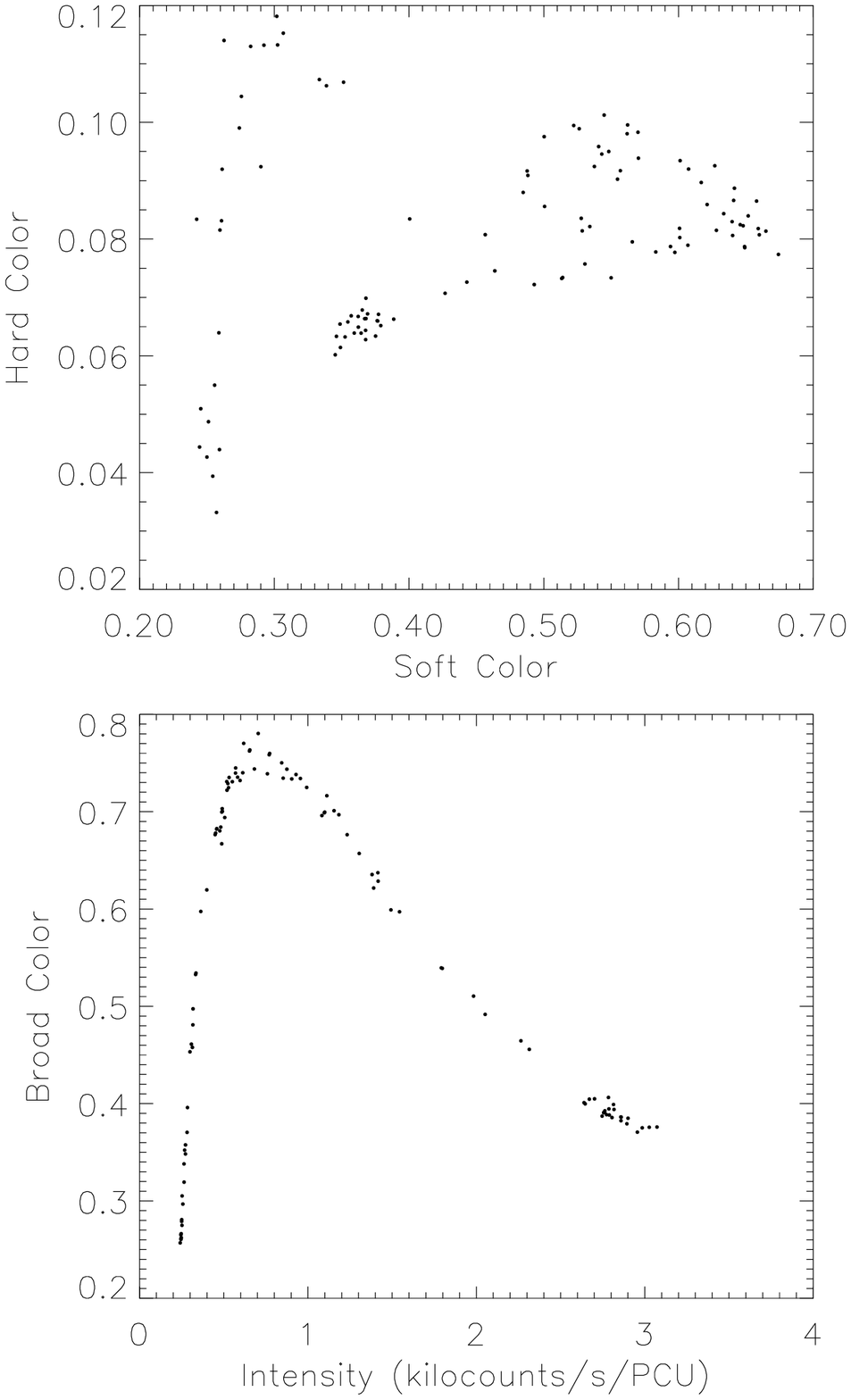 hscale=90 vscale=90 voffset=-16 hoffset=-35}{4.5in}{7.5in}
\caption{
Color-color and hardness-intensity diagrams for the last 1716~s of
data in Figure~\protect{\ref{fig:fall96_dip3lc}}, during which the
intensity gradually transitioned from the non-dip baseline to the
bottom of a dip. Intensity is \intens{2.0}{18}, and the hardness
ratios are defined as soft color:
\hardness{2.0}{4.8}{4.8}{6.3}, hard color: \hardness{6.3}{13}{13}{18},
and broad color:
\hardness{2.0}{6.3}{6.3}{18}.  Each point represents 16~s of
background-subtracted data from PCUs 0, 1, and 2.
}
\label{fig:fall96_cc_hid_85911900_2ks}
\end{centering}
\end{figure}

\begin{figure}
\begin{centering}
\PSbox{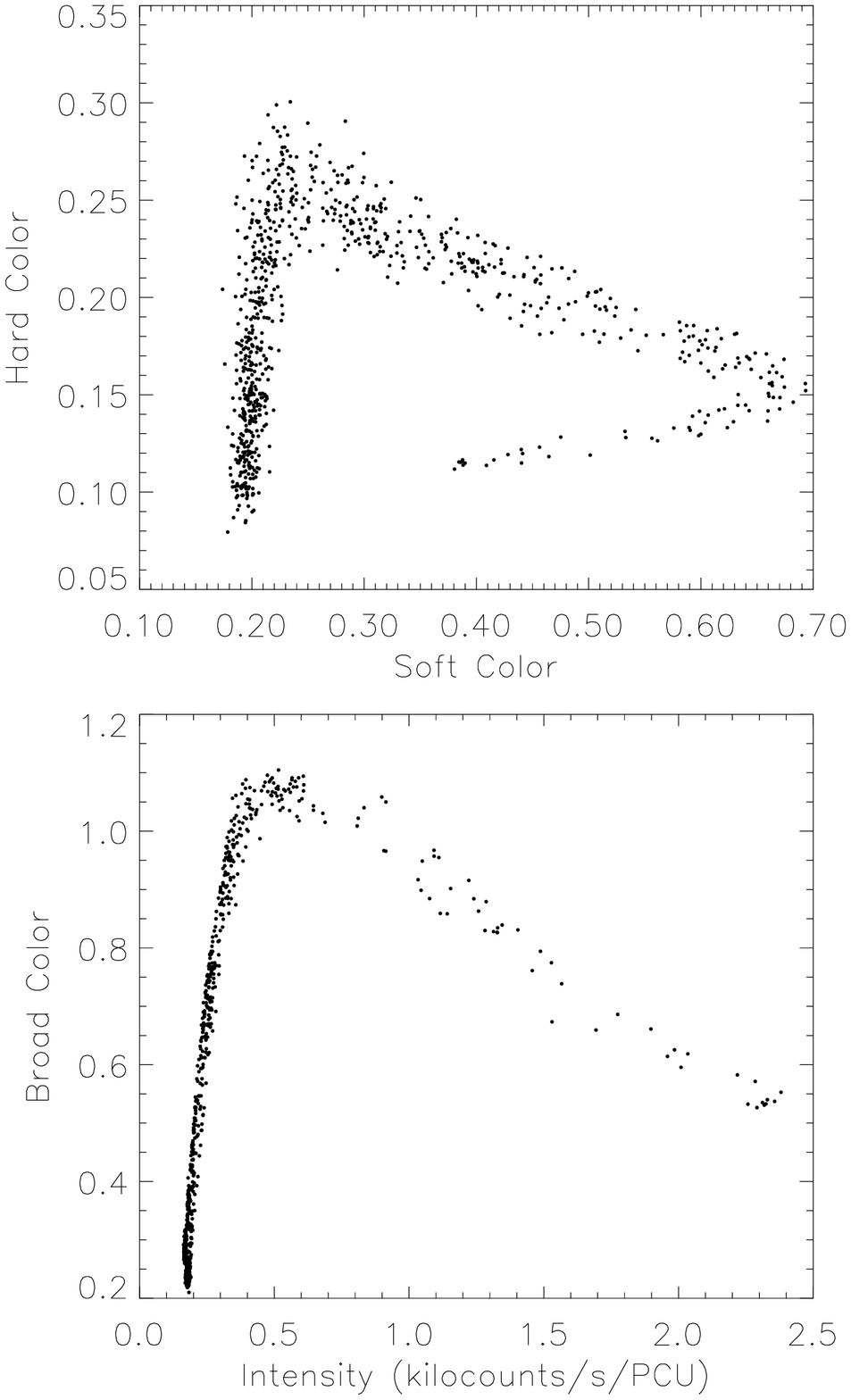 hscale=90 vscale=90 voffset=-16 hoffset=-35}{4.5in}{7.5in}
\caption{
Color-color and hardness-intensity diagrams for the first three segments in
Figure~\protect{\ref{fig:fall96_48hr}} (day 346.31--346.52), during 
which Cir~X-1 was in an extended low/dip state. Intensity is
\intens{2.0}{18}, and the hardness ratios are defined as
soft color: \hardness{2.0}{4.8}{4.8}{6.3}, hard color:
\hardness{6.3}{13}{13}{18}, and broad color: 
\hardness{2.0}{6.3}{6.3}{18}.  Each point represents 16~s of
background-subtracted data from PCUs 0, 1, and 2.
}
\label{fig:fall96_cc_hid_85823000_17ks}
\end{centering}
\end{figure}

The intensity and hardness ratios from Dip~3 (the last 1716~s of data
in Figure~\ref{fig:fall96_dip3lc}) have been used to produce
color-color and hardness-intensity diagrams (CDs/HIDs), shown in
Figure~\ref{fig:fall96_cc_hid_85911900_2ks}. The spectral evolution
during the dip produces tracks with dramatic bends in both the CD and
the HID.  A similar CD and HID has also been produced for the first
three segments of Figure~\ref{fig:fall96_48hr} (day 346.31--346.52),
during which the intensity was mostly in an extended low/dip state
(Figure~\ref{fig:fall96_cc_hid_85823000_17ks}). 

In the HIDs of Figures~\ref{fig:fall96_cc_hid_85911900_2ks}
and~\ref{fig:fall96_cc_hid_85823000_17ks}, dips produce motion to the
left (toward lower intensity) and initially upward (harder), but turn
downward (softer) when the count rate in the denominator reaches its
bottom level. In the CDs (most prominently in
Figure~\ref{fig:fall96_cc_hid_85823000_17ks}), since two hardness
ratios are employed (hard and soft color), motion is initially to the
right and upward (harder in both colors but mostly in the lower
channels).  When the lowest channel reaches the bottom level, the
track turns to the left but continues upward. When intermediate
channels stop decreasing and only the highest channels still drop, the
track finally turns downward. The intensity of the second lowest
energy band rapidly drops by a factor of two (at time 3500~s in
Figure~\ref{fig:fall96_dip3lc}) just before settling to its lowest
level; this resulted in a gap in the the middle CD branch in
Figure~\ref{fig:fall96_cc_hid_85911900_2ks}.

Evolution of the position in CDs and HIDs apart from dips is
interpreted in Chapters~\ref{ch:feb97paper} and~\ref{ch:fullZ} as
motion around a ``Z'' track as well as shifts of the entire ``Z''
pattern. The exact position of a dip track in the diagrams depends on
its starting point, i.e., the baseline intensity and spectrum.
Although starting from different positions, all dips we have observed
in \RXTE\ light curves of Cir~X-1 produce the same general shape in
CDs and HIDs (but not always complete tracks if dips are weak).

\section{Evolution of Energy Spectra During Dips}

\subsection{Spectral Model}

Cold (neutral) matter most effectively absorbs photons at lower
energies, so obscuration by cold matter produces dips that are more
gradual at high energy than at low energy, as we observe in
Cir~X-1. However the fact that strong dips in Cir~X-1 appear to
``bottom-out'' well above background suggests the presence of a faint
component unaffected by the dips. Thus, in modeling spectra of dips in
Cir~X-1, at least two components are necessary: (1) a bright component
modified by variable and at times very heavy absorption and (2) a
faint component seen through a constant and relatively low column
density.  For simplicity, we assume that the spectral shape of the two
components are similar, as might be the case if the faint component is
due to scattering (e.g., by a corona, stellar wind, etc.)\ of the main
bright component back into the line of sight from other initial
emission directions. (Although the terms ``direct'' and ``scattered''
components would be more physical, we will continue to use the terms
``bright'' and ``faint'' since they do not tie the discussion to a
particular assumed model.)

From a study of the energy spectrum during non-dip observations
(Chapter~\ref{ch:fullZ}), we found that the 2.5--25~keV spectrum of
Cir~X-1 was generally fit well with a model consisting of a
multi-temperature ``disk blackbody'' plus a hard blackbody. Thus, we
adopt this model for the bright component during dips and constrain
the faint component to use the same model parameters, but with the
flux multiplied by a constant factor $f$ which is less than unity.

The absorption column density for the faint component is assumed to
remain constant throughout a dip, but the column density for the
bright component is allowed to vary to produce the dips.  The light
curves during strong dips show a significant reduction in intensity
even at high energy (13--18~keV), indicating very high column
densities ($ N_H > 10^{24}$ cm$^{-2}$). Photo-electric absorption is
the dominant process responsible for dips at low energy, but the
photo-electric cross-section decreases with energy. Thus, above about
10~keV, Thomson scattering (with an approximately energy-independent
cross section) becomes important. Both photo-electric and Thomson
cross-sections are included in the absorption calculation for the
bright component.  The complete model used can then be expressed as:
\begin{equation} 
\label{eq:dip_model}
F = \left[\exp(-\sigma_{ph}{N_H}^{(1)})\exp(-\sigma_{Th}{N_H}^{(1)}) + 
	\exp(-\sigma_{ph}{N_H}^{(2)}) f \right] M,
\end{equation}
where $F$ is the observed flux, $\sigma_{ph}$ and $\sigma_{Th}$ are
the photo-electric and Thomson cross-sections, ${N_H}^{(1)}$ and
${N_H}^{(2)}$ are the effective hydrogen column densities of the
bright and faint components respectively, $f$ is the ratio of the flux
of the faint component to that of bright component, and $M$ is the
disk blackbody plus blackbody model. 

%%%%%%%%%%%%%%%%%%%%%%%%%%%%%%%%%%%%%%%%%%%%%%%%%%%%%%%%%
\subsection{Spectral Fitting Strategy}

Because dip spectra are composed of multiple components (variably
absorbed and faint manifestations of the disk blackbody and blackbody)
we make use of joint fits of spectra from inside and outside dips to
simultaneously constrain the parameters of the various components.
The combined $\chi^2$ value obtained when the model is fit to multiple
spectra is minimized by varying the model parameters, but allowing
only the absorption column density of the bright component to vary
between spectra. Since the disk blackbody and blackbody parameters
typically evolve on timescales of hours (see Chapter~\ref{ch:fullZ}),
it is important to compare spectra during a dip to spectra immediately
outside the dip.  

In order to perform joint spectral fits, we extracted PCA energy
spectra from time segments before and during the three dips shown in
Figures \ref{fig:fall96_dip1lc}, \ref{fig:fall96_dip2lc},
and~\ref{fig:fall96_dip3lc}.  The same procedure for constructing and
fitting PCA spectra presented in sections \ref{sec:june97_spec_qual}
and~\ref{sec:june97_spec_models} is used here as well (comparing
results from PCUs 0, 1, and 4).

%Strictly speaking, the faint component should always be included in
%spectral fits, even outside dips, since that emission region is
%probably visible at all times. However, the spectral shape of the
%bright and faint components are similar, so the faint component cannot
%be distinguished from part of the bright component. In fact, it is
%only during dips that the faint component is revealed.

%%%%%%%%%%%%%%%%%%%%%%%%%%%%%%%%%%%%%%%%%%%%%%%%%%%%%%%%%
\subsection{Energy Spectra Inside and Outside Dips}

\begin{figure}
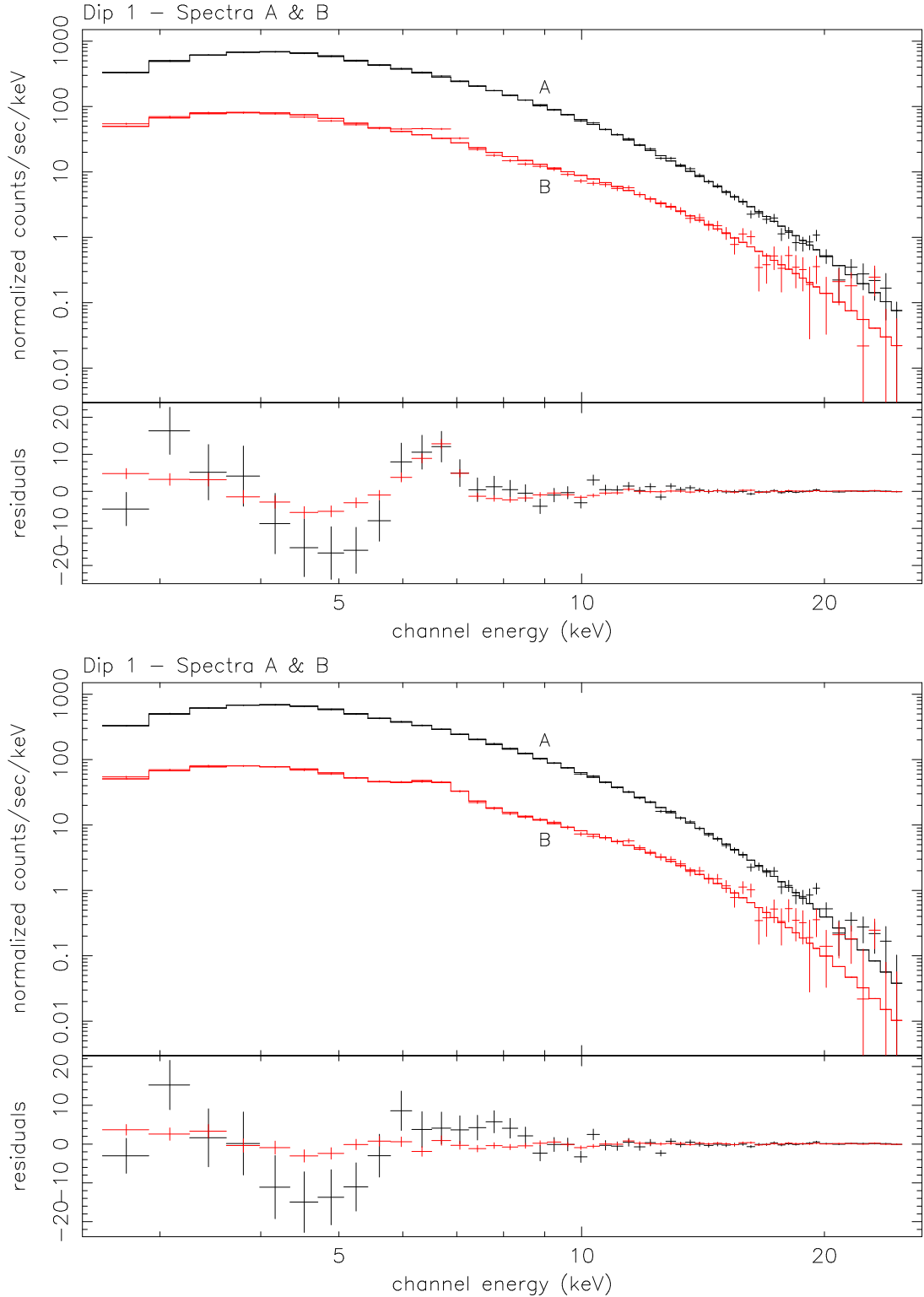

\begin{centering}
\PSbox{figures/dip1_two96s_nogauss.eps hscale=62 vscale=62 angle=270 voffset=350 hoffset=-25}{5.6in}{3.9in}
\PSbox{figures/dip1_two96s_gauss.eps hscale=62 vscale=62 angle=270 voffset=340 hoffset=-25}{5.6in}{3.9in}
\caption{ 
Top: Spectral fits for 96-s segments prior to Dip~1 (spectrum A) and
during Dip~1 (spectrum B). Bottom: same spectra and model (see text),
but with a Gaussian emission-line component included to fit the peaked
residuals near 6.5~keV.  The data shown are from PCU~0 only.}
\label{fig:dip1_two96s}
\end{centering}
\end{figure}

\begin{figure}
\begin{centering}
\PSbox{figures/dip2_two304s_nogauss.eps hscale=62 vscale=62 angle=270 voffset=350 hoffset=-25}{5.6in}{3.9in}
\PSbox{figures/dip2_two304s_gauss.eps hscale=62 vscale=62 angle=270 voffset=340 hoffset=-25}{5.6in}{3.9in}
\caption{ 
Top: Spectral fits for 304-s segments prior to Dip~2 (spectrum A) and
during Dip~2 (spectrum B). Bottom: same spectra and model (see text),
but with a Gaussian emission-line component included to fit the peaked
residuals near 6.5~keV.  The data shown are from PCU~0 only.}
\label{fig:dip2_two304s}
\end{centering}
\end{figure}

We compiled spectra from a 96-s time segments immediately prior to
Dip~1 and from another 96-s segment during the lowest portion of
Dip~1 (segments labeled ``A'' and ``B'' in
Figure~\ref{fig:fall96_dip1lc}). Likewise, we selected a 304-s segment
from before Dip~2, and other segment of the same duration from the
lowest portion of that dip (segments ``A'' and ``B'' in
Figure~\ref{fig:fall96_dip2lc}). These spectra are shown in Figures
\ref{fig:dip1_two96s} and~\ref{fig:dip2_two304s} (top panels), 
along with fitted curves using the model described above
(equation~\ref{eq:dip_model}).  The best-fitting parameters for the
bright and faint component of each fit are listed in the first two
lines of Table~\ref{tab:dip_continua}.  The column density
(${N_H}^{(2)}$) of the faint component is given in the table. The
column density (${N_H}^{(1)}$) of the bright component differs for
spectra A and~B of each fit and is thus given separately in
Table~\ref{tab:dip_absorpAB}.

The disk blackbody and blackbody parameters for both dips are similar
to those obtained for non-dip spectra in Chapter~\ref{ch:fullZ}
(temperatures of 1.5~keV and 2.2~keV respectively). The flux of the
faint component is 10--11\% that of the bright component ($f$). The
column density of the faint component is very low, and consistent with
zero. This low value is inconsistent with estimates of the
interstellar column density measured with \ASCA\ and \ROSAT\
($N_H = $1.8--2.4$\times 10^{22}$~cm$^{-2}$)~\cite{brandt96,predehl95}.
This may indicate that the spectral shape of the faint component is
not exactly the same as that of the bright component. If scattering is
responsible for the faint component, then this process might distort
the spectrum slightly toward lower energy, thereby giving the
appearance of reduced interstellar absorption.  The use of different
spectral models for the bright and faint components becomes
unproductive due to the large number of parameters involved, so we
allow the column density of the faint component to account for small
difference in spectral shape.

\vspace*{0.2in}

\begin{table}
\small
\renewcommand{\arraystretch}{1.2}
\begin{center}
\begin{tabular}{ccccccccc}
\hline
& & \multicolumn{2}{c}{Disk Blackbody~\footnotemark[3]} &
\multicolumn{2}{c}{Blackbody~\footnotemark[4]} &
\multicolumn{2}{c}{Faint Component} & \\
Joint & Iron &
$T_{in}$ & $R_{in}\cos^{1/2}\theta$ & 
$T$   & $R$  & 
 & $N_H^{ (2) }/10^{22}$~\footnotemark[6] & \\
fit\footnotemark[1] & Line~\footnotemark[2] &
(keV) & (km) &
(keV) & (km) & 
$f$~\footnotemark[5] & (cm$^{-2}$) & 
$\chi^2_r$~\footnotemark[7] \\
\hline
\hline
1A--B & no &
	$1.51^{+0.10}_{-0.17}$ & $16.94^{+3.96}_{-2.20}$ &
	$2.20^{+0.62}_{-0.42}$ & $3.77^{+3.33}_{-2.55}$ &
	$0.107^{+0.005}_{-0.006}$ &
	$0.00^{+0.07}_{-0.00}$ & 
	3.89--4.07 \\
2A--B & no &
	$1.51^{+0.04}_{-0.06}$ & $17.54^{+1.25}_{-0.87}$ & 
	$2.32^{+0.16}_{-0.15}$ & $3.47^{+0.86}_{-0.76}$ & 
	$0.103^{+0.003}_{-0.003}$ &
	$0.03^{+0.28}_{-0.03}$ & 
	7.58--7.84 \\
\hline
1A--B & yes & 
	$1.23^{+0.22}_{-0.15}$ & $25.32^{+8.56}_{-7.03}$ & 
	$1.79^{+0.33}_{-0.16}$ & $7.87^{+3.34}_{-4.14}$ & 
	$0.096^{+0.007}_{-0.006}$ &
	$0.00^{+0.10}_{-0.00}$ & 
	1.44--1.62 \\
2A--B & yes &
	$1.35^{+0.06}_{-0.05}$ & $22.45^{+1.83}_{-1.69}$ & 
	$2.10^{+0.12}_{-0.09}$ & $5.23^{+0.92}_{-0.93}$ & 
	$0.092^{+0.002}_{-0.002}$ &
	$0.00^{+0.10}_{-0.00}$ & 
	1.87--2.15 \\
\hline
1a--d & yes &
	$1.18^{+0.19}_{-0.16}$ & $27.04^{+9.39}_{-6.33}$ & 
	$1.72^{+0.22}_{-0.12}$ & $8.22^{+2.69}_{-3.03}$ & 
	$0.103^{+0.009}_{-0.011}$ &
	$0.00^{+0.44}_{-0.00}$ &
	0.99--1.20 \\
2a--d & yes & 
	$1.26^{+0.14}_{-0.14}$ & $25.12^{+7.66}_{-5.20}$ & 
	$2.08^{+0.25}_{-0.15}$ & $5.46^{+1.08}_{-1.08}$ & 
	$0.117^{+0.013}_{-0.014}$ &
	$0.15^{+0.70}_{-0.15}$ & 
	0.80--1.12 \\
3a--f & yes & 
	$1.07^{+0.08}_{-0.09}$ & $42.66^{+8.34}_{-6.60}$ & 
	$1.74^{+0.09}_{-0.09}$ & $9.23^{+1.41}_{-1.32}$ & 
	$0.054^{+0.005}_{-0.005}$ &
	$0.00^{+0.14}_{-0.00}$ & 
	1.27--1.34 \\
\hline
\end{tabular}
\renewcommand{\arraystretch}{1}
\caption{
Top group: joint fit parameters of the bright and faint components of
spectra from outside (A) and inside (B) dips 1 and~2.
Middle group: the same, but with an added Gaussian line.
Third group: joint fit parameters for the four to six 16-s spectra for
dips 1, 2, and~3.
The absorption parameters ($N_H^{(1)}$) for the bright component are
given in Tables~\protect{\ref{tab:dip_absorpAB}} (A--B)
\&~\protect{\ref{tab:dip_absorp16s}} (a-f) and the iron line parameters are
given in Table~\protect{\ref{tab:dip_gauss}}.  }
\label{tab:dip_continua}
\end{center}
\end{table}
\footnotetext[1]{
In joint fits (using the model given in
equation~\protect{\ref{eq:dip_model}}), only the column density
($N_H^{(1)}$) of the bright component is allowed to vary between
spectra (see Tables~\protect{\ref{tab:dip_absorpAB}}
and~\protect{\ref{tab:dip_absorp16s}}). Errors quoted are 90\%
confidence limits for a single parameter ($\Delta\chi^2=2.7$).}
\footnotetext[2]{
A Gaussian emission line ($\sim$6.5~keV) was included in the final
five fits (see Table~\protect{\ref{tab:dip_gauss}}). }
\footnotetext[3]{
For the disk blackbody, the inner radius of the accretion disk (times
$\cos^{1/2}\theta$, where $\theta$ is the disk inclination angle and
$\theta=0$ is parallel to the line of sight) is given for a distance of
8~kpc.}
\footnotetext[4]{The blackbody radius is given assuming a distance of 8~kpc.}
\footnotetext[5]{
The bright and faint components are assumed to have the same disk
blackbody and blackbody parameters, but the faint component is scaled
down by a factor $f$.}
\footnotetext[6]{
Absorption column density on the faint component 
(Hydrogen atoms per cm$^2$).}
\footnotetext[7]{
Reduced $\chi^2$ = $\chi^2/dof$, where dof = degrees of freedom = the
number of spectral bins ($\sim$54/spectrum) minus the number of fit
parameters.}

\vspace*{0.2in}

% nH table for 1A--B & 2A--B
\begin{table}
\renewcommand{\arraystretch}{1.2}
\begin{center}
\begin{tabular}{cccc}
\hline
         & $N_H^{(1)}/10^{22}$ &        & $N_H^{(1)}/10^{22}$\\
Spectrum & (cm$^{-2}$) & Spectrum & (cm$^{-2}$)  \\
\hline
\hline
\multicolumn{4}{c}{No Gaussian} \\
1A & $2.83^{+0.57}_{-0.43}$ 	& 2A & $3.50^{+0.29}_{-0.34}$ 	  \\
1B & $176^{+12}_{-13}$ & 2B & $306^{+21}_{-19}$ \\
\cline{1-4}
\multicolumn{4}{c}{With Gaussian} \\
1A & $3.81^{+0.71}_{-0.81}$ 	& 2A & $4.30^{+0.26}_{-0.26}$ \\
1B & $184^{+15}_{-14}$ & 2B & $283^{+16}_{-14}$ \\
\hline
\end{tabular}
\renewcommand{\arraystretch}{1}
\caption{
Effective hydrogen column density of the bright component outside (A)
and inside (B) Dips 1 and~2 (from fits with and without a
$\sim$6.5~keV Gaussian emission line).}
\label{tab:dip_absorpAB}
\end{center}
\end{table}

% this is here to force table footnotes onto the right page
%\newpage

% this is here to force text off the page of tables
\vspace*{0.07in}

The column density of the bright component outside dips (spectra 1A
and 2A; see Table~\ref{tab:dip_absorpAB}) is about \scinot{3}{22}
cm$^{-2}$, just slightly above the interstellar value. The column
density during the low segments of dips (spectra 1B and 2B) is
extremely high: \scinot{184}{22}~cm$^{-2}$ for spectrum~1B and
\scinot{283}{22}~cm$^{-2}$ for spectrum~2B\@. This high column density 
value is required to produce the observed reduction in flux at 20~keV
and is sufficient to render the absorbed component totally negligible
at lower energies. Thus, the flux in spectra 1B and~2B below 5~keV is
due entirely to the faint component.

\subsection{Iron Emission Line}

The fits for both dips show similar residuals for the spectra from
outside and inside the dip (Figures~\ref{fig:dip1_two96s}
and~\ref{fig:dip2_two304s}). These residuals are also similar to those
obtained in fits of spectra from the horizontal branch or upper normal
branch in the study in Chapter~\ref{ch:fullZ}. In particular, the fits
of these spectra all show peaked residuals near 6.5~keV. In the two
spectra from the bottom of dips (spectra 1B and~2B), the feature is
prominently visible as a ``bump'' in the spectra themselves (see
Figures~\ref{fig:dip1_two96s} and~\ref{fig:dip2_two304s}). The
parameters of this feature are best constrained when only the faint
component is present.  The line component is probably from iron
K$\alpha$ emission, which occurs at 6.4~keV for neutral iron and
higher energy for ionized iron.  The fact that these residuals in the
dip spectra are very similar in absolute flux level to the residuals
immediately outside the dips strongly suggests that the line feature
is associated with the faint component. For example, iron fluorescence
might occur during the scattering process.

Based on the similar peaked residuals inside and outside the dips, a
Gaussian emission line component was added near 6.5~keV (see the
bottom panels of Figures~\ref{fig:dip1_two96s}
and~\ref{fig:dip2_two304s}).  The addition of the line does reduce the
peaked residuals near 6.5~keV, but does not improve the residuals (of
similar strength but indicating data {\em below} the model) between
4--5~keV.  The results of the joint fits, now including the Gaussian
line, are listed in the second group in Tables~\ref{tab:dip_continua}
and \ref{tab:dip_absorpAB}, showing that the other model parameters are
generally consistent with the previous fits within the 90\% confidence
limits. Table~\ref{tab:dip_continua} also shows that a significant
improvement in the reduced $\chi^2$ value was achieved in both
cases. We will continue to include a Gaussian line in all subsequent fits.

% Gaussian Table
\begin{table}
\renewcommand{\arraystretch}{1.2}
\begin{center}
\begin{tabular}{cccc}
\hline
Joint & $E$   & $\sigma$ & normalization\\
Fit   & (keV) & (keV)    & (photons~cm$^{-2}$~s$^{-1}$)\\
\hline
\hline
1A--B & $6.59^{+0.12}_{-0.13}$ & $0.18^{+0.28}_{-0.18}$ & $0.018^{+0.007}_{-0.004}$ \\
2A--B & $6.58^{+0.06}_{-0.07}$ & $0.29^{+0.10}_{-0.13}$ & $0.016^{+0.002}_{-0.002}$ \\
\hline
1a--d & $6.59^{+0.16}_{-0.15}$ & $0.13^{+0.29}_{-0.13}$ & $0.015^{+0.002}_{-0.002}$\\
2a--d & $6.45^{+0.16}_{-0.20}$ & $0.47^{+0.29}_{-0.32}$ & $0.025^{+0.012}_{-0.008}$\\
3a--f & $6.46^{+0.23}_{-0.27}$ & $0.39^{+0.32}_{-0.39}$ & $0.012^{+0.004}_{-0.003}$\\
\hline
\end{tabular}
\renewcommand{\arraystretch}{1}
\caption{
Gaussian emission-line parameters for the joint fits of spectra
outside and inside dips (A--B) and during entry into a dip (a--d,
a--f).  In each case, the same Gaussian line parameters were used for
each spectrum in a joint fit.  }
\label{tab:dip_gauss}
\end{center}
\end{table}

The parameters for the Gaussian line are shown in
Table~\ref{tab:dip_gauss}. Although the best-fitting centroid energy
of the line is close to 6.6~keV, the Gaussian width is large (probably
due to the relatively coarse energy resolution of the PCA: $\sim$1~keV
FWHM at 6~keV), so that we cannot confirm whether the line is from
neutral or partially ionized iron or whether it is actually composed
of multiple unresolved narrow lines.

\subsection{Spectral Evolution During Dip Transitions}

\begin{figure}
%\begin{centering}
\PSbox{figures/dip1_four16s_gauss.eps hscale=95 vscale=95 voffset=-64 hoffset=-56}{6.46in}{7.8in}
\caption{ 
Spectral fits for four 16-s segments (a--d) during the decline into
Dip~1. Only the column density of the bright component varies between
the four jointly-fit curves. The data shown are from PCU~0 only.}
\label{fig:dip1_four16s}
%\end{centering}
\end{figure}

\begin{figure}
%\begin{centering}
\PSbox{figures/dip2_four16s_gauss.eps hscale=95 vscale=95 voffset=-64 hoffset=-56}{6.46in}{7.8in}
\caption{ 
Spectral fits for four 16-s segments (a--d) during the decline into
Dip~2.  Only the column density of the bright component varies between
the four jointly-fit curves. The data shown are from PCU~0 only.}
\label{fig:dip2_four16s}
%\end{centering}
\end{figure}

\begin{figure}
%\begin{centering}
\PSbox{figures/dip3_six16s_gauss.eps hscale=95 vscale=95 voffset=-64 hoffset=-56}{6.46in}{7.8in}
\caption{ 
Spectral fits for six 16-s segments (a--d) during the decline into
Dip~3.  Only the column density of the bright component varies between
the six jointly-fit curves. The data shown are from PCU~0 only.}
\label{fig:dip3_six16s}
%\end{centering}
\end{figure}

Through its large collecting area, the PCA provides good-quality
spectra of bright sources such as Cir~X-1 every 16~s. This enables us
to study the detailed evolution of the spectrum during the transitions
between the high and low states outside and inside dips.  We selected
four to six 16-s segments from the ingress of Dips 1, 2 and 3
(indicated as diamonds on Figures \ref{fig:fall96_dip1lc},
\ref{fig:fall96_dip2lc}, and~\ref{fig:fall96_dip3lc}).
The 16-s spectra from each of these dips are shown in Figures
\ref{fig:dip1_four16s}, \ref{fig:dip2_four16s},
and~\ref{fig:dip3_six16s}, along with the model curves from joint fits
of the 4--6 spectra of each dip. 

As expected from the light curves of the dips (Figures 
\ref{fig:fall96_dip1lc}, \ref{fig:fall96_dip2lc}, 
and~\ref{fig:fall96_dip3lc}), the intensity in each case initially
decreases at low energy, but then reaches a fixed level while the
intensity at higher energy continues to decrease. Intermediate spectra
in each figure show a ``step'' near 7.1~keV, which cannot be accounted
for solely by the emission line, but is naturally fit by the
iron K~absorption edge of the absorption component and indicates a
column density of $N_H >$ few$\times10^{23}$.  As discussed above, the
intensity at the bottom of the dip is significantly reduced (relative
to outside the dip) even at 20~keV\@ due to Thomson scattering,

The fit parameters for the bright and faint components in each joint
fit of the 16-s spectra (1a--d, 2a--d, and 3a--f) are shown in
Table~\ref{tab:dip_continua}.  The flux in the faint component
relative to the bright component ranges from 5\% and 12\% in these
fits, and the column density for the faint component is again
consistent with zero.
The Gaussian line parameters for these fits are listed in
Table~\ref{tab:dip_gauss}, and indicate a line centered at
6.4--6.6~keV. The best-fitting normalization (absolute flux) of the
line differs by a factor of two among the three dips but is consistent
with no change within the errors.
For each set of spectra, the variable absorption column density of the
bright component is shown in Table~\ref{tab:dip_absorp16s}. In each
case the column density increases from a relatively low value ($N_H =$
4--9$\times10^{22}$) at the start of a dip to very high values ($N_H >
10^{24}$) at the lowest part of the dips.

These fits show that the model developed above, where only the
absorption column density on the bright component varies between all
spectra of a given dip, is quite successful in reproducing the
evolution of the spectrum throughout the dips.

% nH table for 1a-d, 2a-d,3a-f
\begin{table}
\renewcommand{\arraystretch}{1.2}
\begin{center}
\begin{tabular}{cccccc}
\hline
      & $N_H^{(1)}/10^{22}$ &  & $N_H^{(1)}/10^{22}$ &  & $N_H^{(1)}/10^{22}$ \\
Spectrum & (cm$^{-2}$) & Spectrum & (cm$^{-2}$) & Spectrum & (cm$^{-2}$)   \\
\hline
\hline
1a & $4.16^{+1.10}_{-0.75}$ & 2a & $8.84^{+1.37}_{-1.09}$ & 3a & $8.31^{+1.09}_{-0.96}$ \\
1b & $25.8^{+1.6}_{-1.3}$ & 2b & $25.8^{+1.4}_{-1.5}$ & 3b & $15.7^{+1.0}_{-1.1}$ \\
1c & $43.6^{+1.7}_{-2.1}$ & 2c & $64.0^{+2.4}_{-2.5}$ & 3c & $33.2^{+1.4}_{-1.5}$\\
1d & $129^{+8}_{-8}$ & 2d &$242^{+83}_{-52}$  & 3d & $66.7^{+1.6}_{-1.5}$\\
   &  &    &  & 3e & $127^{+8}_{-7}$ \\
   &  &    &  & 3f & $>\mscinot{2}{4}$ \\
\hline
\end{tabular}
\renewcommand{\arraystretch}{1}
\caption{
Effective hydrogen column density responsible for the variable
absorption in the bright component, due to photo-electric absorption
and Thompson scattering, during 16-s spectra from Dips 1, 2, and~3.}
\label{tab:dip_absorp16s}
\end{center}
\end{table}

\section{Discussion}

% Similar behavior in other sources
The spectral evolution during dips in Cir~X-1 produces curved tracks
in CDs and HIDs. By identifying these tracks we can exclude absorption
dip behavior from analysis intended to study Z-source spectral tracks,
as we have done in Chapter~\ref{ch:fullZ}.

Similar absorption-dip behavior was recently observed in \RXTE\
observations of the black-hole candidates GRO~J1655-40 and 4U~1630-47
\cite{kuulkers98,tomsick98}. Dips in these sources
show evidence for a faint unabsorbed component and produced curved
tracks in CDs and HIDs similar to those we observe in Cir~X-1.

% ASCA Cir~X-1
Spectral analysis from an \ASCA\ observation of Cir~X-1 during an
intensity transition also showed evidence for an unabsorbed
component~\cite{brandt96}. A partial-covering model was used to fit
the energy spectrum (with two blackbodies), but it was noted that the
model could be interpreted as a direct component plus a scattered
component. The \ASCA\ spectra also showed an iron K~edge due to
absorption with column densities near
\tento{24} cm$^{-2}$. Thus our results are generally consistent 
with the \ASCA\ results, but we are able to demonstrate how the
continuum spectrum evolves throughout dips.  

An iron line at 6.4~keV was seen in the low-state \ASCA\ spectrum, but
no line was detected at a similar flux level in the high-state
spectrum~\cite{brandt96}.  Our analysis shows that an iron emission
line appears to be associated with the faint component, and is present
outside dips at the same flux level.  Since scattering by an extended
cloud of material is probably responsible for the faint component, it
is likely that the iron K$\alpha$ emission is also produced as a
result of the scattering process.

The fact that absorption dips in Cir~X-1 mainly occur within a day of
phase zero and are then followed by significant flaring, suggests that
the dips are associated with the mass transfer process. Thus,
absorption might be due to the mass transfer stream itself or due to a
bulge on the disk produced by the addition of matter. The intermittent
nature of the PCA light curves near phase zero
(Figure~\ref{fig:fall96_48hr}) indicate that we are observing the fine
structure (temporal or geometrical) of the obscuring material.

\chapter{Multi-frequency Observations}
\label{ch:multifreq}

\section{Overview}

The 16.55-d X-ray period of Cir~X-1 is also seen in the radio,
infrared, and optical bands (see Chapter~\ref{ch:intro}). We organized
two multi-frequency campaigns, in 1996 May and 1997 June, to study
correlated variability in different frequency bands. Such observations
have the potential to constrain emission mechanisms and provide
information about the mass accretion rate. For example, in Z~source
LMXBs the optical flux is believed to result from higher energy
photons being reprocessed in the accretion disk. Thus optical
intensity is taken as a measure of mass accretion rate, and has been
used to argue that mass accretion rate increases along Z source
spectral tracks from the horizontal branch, down the normal branch,
and onto the flaring branch~\cite{hasinger90,augusteijn92}.

\section{May 1996 Campaign}

The first campaign (see Table~\ref{tab:multifreq96}) occurred during
1996 May 7--14, covering a week including phase zero on May 11.413 UT
(JD 2450214.913). Radio observations were made from South Africa with
the 28-m dish at Hartebeeshoek Radio Astronomical Observatory
(HartRAO) by George Nicolson. Infrared observations were made with the
1.9-m telescope at the South African Astronomical Observatory (SAAO)
by Ian Glass. In addition to the ongoing All-Sky Monitor observations,
seven \RXTE\ PCA (and HEXTE) observations were made with a combined
exposure time of 54~kiloseconds.

\begin{table}
\begin{center}
\begin{tabular}{cccc}
\hline
\hline
Frequency &              & Type of               &      \\ 
Band      & Observatory  & Observation           & Dates \\
\hline
Radio     & HartRAO      & 8.5~GHz (3 samples/d) & ongoing\\
          &              & 8.5~GHz \& 5.0 GHz    & May 10--15\\
IR        & SAAO (1.9~m) & K-band photometry     & May (7),8--12,(13)\\
X-ray     & RXTE         & ASM                   & ongoing\\
X-ray     & RXTE         & PCA                   & May 7--14\\
\hline
\end{tabular}
\end{center}
\caption{Multi-frequency Observations of Cir~X-1 from 1996 May.
Dates in parentheses were cancelled due to poor weather conditions.}
\label{tab:multifreq96}
\end{table}

\begin{figure}
\begin{centering}
\PSbox{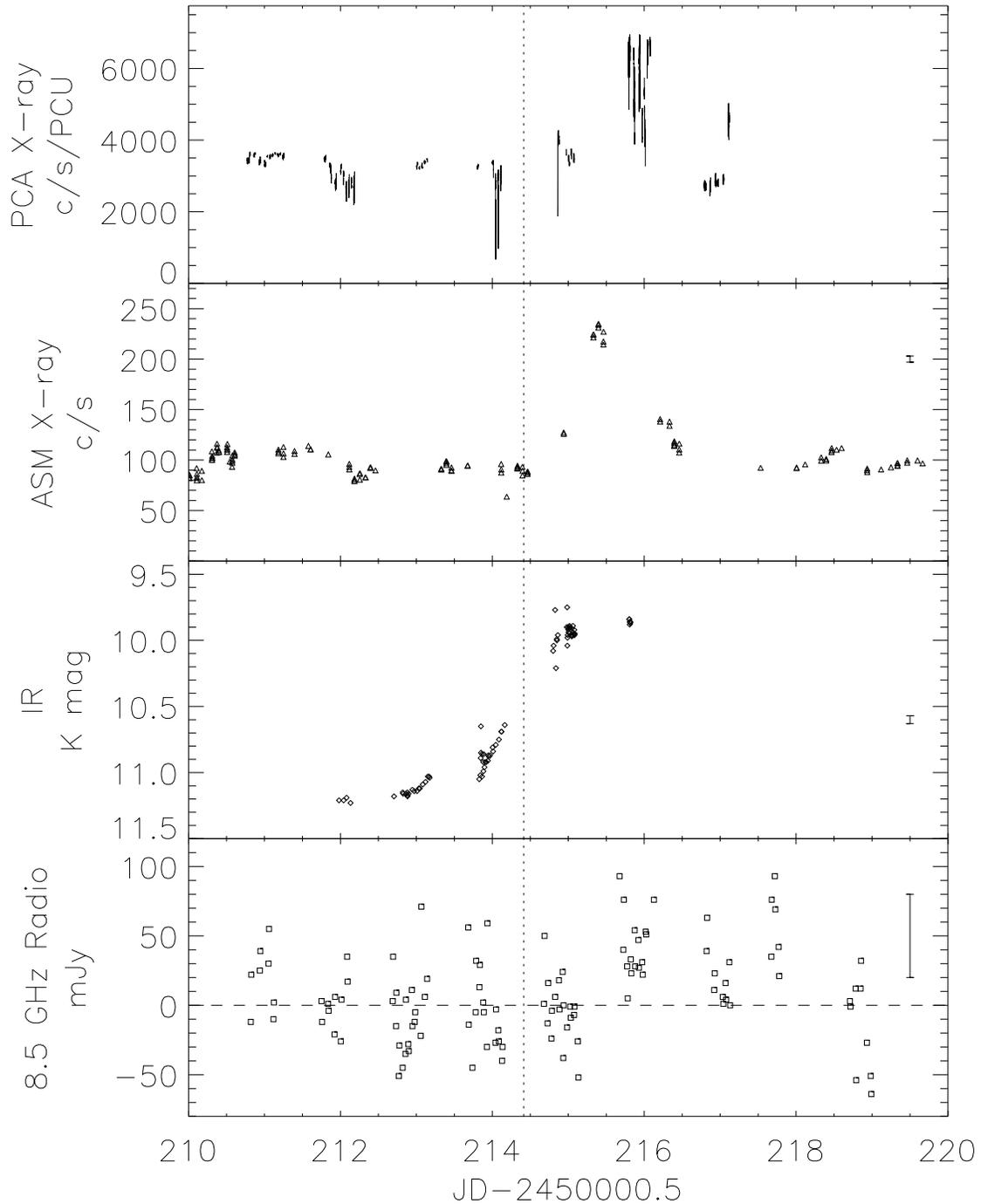 
	hscale=86 vscale=86 hoffset=-7 voffset=-20}{5.64in}{7.0in}
\caption{X-ray, IR, and radio light curves of Cir~X-1 from 1996 May 7--17.
Phase zero is indicated by the vertical dotted line at day 214.413
(1997 May 11.413). PCA intensity is from the 2--32~keV band (16-s
bins) and ASM intensity is from 2--12~keV (90-s exposures). Errors on
the PCA data are negligible. Typical error bars are shown for the ASM,
IR, and radio data: errors on the ASM data are typically 2--5~c/s, on
the IR data are typically about 0.03 magnitudes (worse on days 214 \&
215), and on the radio data are about 30~mJy. The average quiescent
level of 29~mJy has been subtracted from the radio data. (Data
provided by I.~Glass (IR), G.~Nicolson (radio), and the
\RXTE/ASM team.)  }
\label{fig:multifreq96_lc}
\end{centering}
\end{figure}

\subsection{Radio (HartRAO)}
\label{sec:hartrao96}

Radio intensity measurements of Cir~X-1 at 8.5~GHz (3.5~cm) are
routinely carried out as often as several times daily at HartRAO. In
recent years the periodic flares have been weak, i.e. no flares above
100~mJy, compared to 200~mJy -- 2~Jy between 1976--1986 (1 Jansky = 1 Jy =
${\rm 10^{-26}~W\,m^{-2}\,Hz^{-1}}$).  In fact, radio flares since 1986 have
not been strong enough to update the radio ephemeris.

In conjunction with our multi-frequency campaign, HartRAO observations
were made 12--20 times per day at both 8.5~GHz and 5.0~GHz from 1996
May 10--15 and 5--7 times per day at 8.5~GHz for several days prior to
and after that period. Each measurement takes about 8~minutes. The
observations are made as drift scans, i.e., the telescope is stationary
during a scan, which begins pointing ahead of the source. Because
Cir~X-1 lies in the Galactic plane, there is strong confusion from
other sources such as H\,{\small II} regions and the supernova remnant
with which Cir~X-1 is associated. A quiescent average scan is
subtracted from the data and then fit the telescope beam pattern at
the position of Cir~X-1. This method gives a residual flux relative to
the quiescent background rather than the absolute flux, but is
sufficient to monitor the flaring behavior.

The 8.5~GHz data from JD-2450000.5 = 210--219 are shown in
Figure~\ref{fig:multifreq96_lc}. The error on a typical point is about
30~mJy (1 $\sigma$). The average flux from observations between phase
0.2 to 1.0 is 29~mJy, and has been subtracted from the data. There is
no evidence for variability during that phase interval. Between days
215.5--218.0 (phase $\sim$0.05--0.2) the flux is higher.  After
subtracting the 29~mJy quiescent level from the data, then 66\% of the
data points lie between 2--5 sigma above the noise of 19~mJy.  This is
consistent with weak flaring observed in other cycles from earlier in
1996, but flaring may be occurring earlier in the cycle than the
corresponding period in 1995, when the flaring was between phases
0.2--0.3.

\subsection{Infrared Photometry (SAAO)}
\label{sec:saao96}

The infrared observations at SAAO provided K~band (2.2~$\mu$m)
photometry measurements 4--32 times per night for five nights from 1996
May 8--12 (see Figure~\ref{fig:multifreq96_lc}).  An aperture of
9~arcsec diameter was used. The K magnitude was 11.2 on May~8--9 (day
211--212) and gradually brightened by about 1~magnitude during
May~10--12 (day 213--215) to 10.15 (K = 0 mag corresponds to a flux
density of ${\rm 6.3\e{-23}~W\,m^{-2}\,Hz^{-1}}$, and other magnitudes
are related to flux through $m_2 - m_1 = -2.5
\log[F_2/F_1]$). The quiescent and flaring magnitudes are similar to
those obtained by Glass during 1980--1993, but are somewhat lower than
the high values (K=9.5 to 7.7) observed in the late
1970's~\cite{glass94}. Observing conditions during the first two and
last nights were good and the errors are 0.03~mag or better. The data
suggest a possible precursor event during day~214; however the
observing conditions were not as good during the third and fourth
nights.  When the conditions were bad, Cir~X-1 was measured relative
to star~J from the image in Whelan et~al.~\cite{whelan77}, which was
used as a standard taken to be K=9.82.

\subsection{Correlated Multi-frequency Variability}
\label{sec:correlated96}

\begin{figure}
\begin{centering}
\PSbox{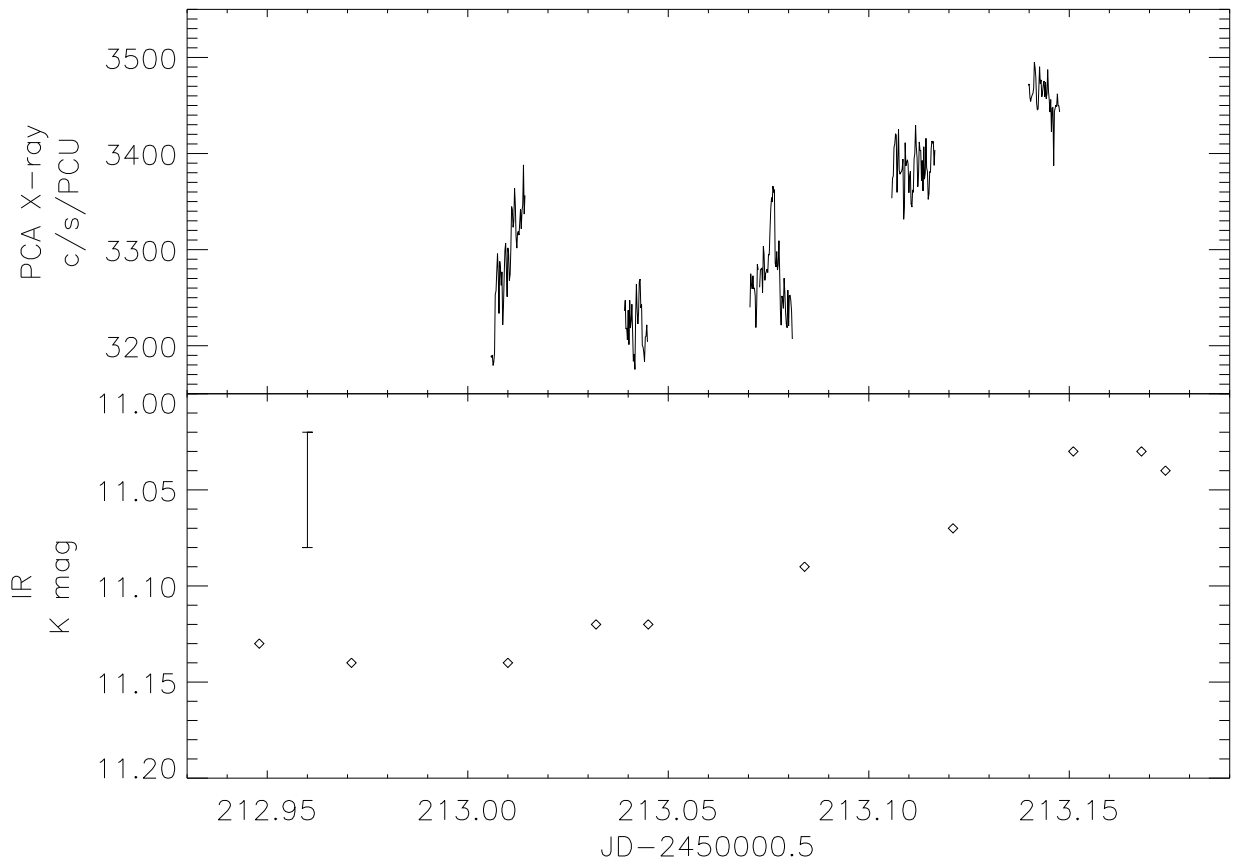 hscale=70 vscale=70 angle=90 voffset=0 hoffset=500}{6.5in}{4.5in}
\caption{X-ray and IR light curves of Cir~X-1 from day 212.93--213.19.
X-ray intensity is from the 2--32~keV PCA band (16-s bins). A typical
IR error bar of $+/-$0.03 magnitudes is shown.  (IR data courtesy of
I.~Glass.)  }
\label{fig:ir_pca_tjd213}
\end{centering}
\end{figure}

The X-ray and radio light curves in Figure~\ref{fig:multifreq96_lc}
reach a peak value during the second day after phase zero, while the
IR reaches a peak value only 0.5~d after phase zero. The PCA and IR
data both show evidence for increasing variability before phase zero.
Furthermore, both show a gradual rise in intensity during the period
from day~213.0 to 213.15, shown in more detail in
Figure~\ref{fig:ir_pca_tjd213}. Fourier power spectra from each of the
five PCA segments in Figure~\ref{fig:ir_pca_tjd213} show a narrow
quasi-periodic oscillation peak at 26.0~Hz, 24.2~Hz, 25.4~Hz,
31.54~Hz, and 37.61~Hz (in time order of the five segments). This
timing behavior indicates that increasing X-ray intensity corresponds
to motion to the right or down the horizontal branch in a
hardness-intensity diagram (see Chapters~\ref{ch:feb97paper} \&
\ref{ch:fullZ} and \cite{shirey98:feb97}). 
Thus, this segment shows IR flux increasing from the horizontal branch
toward the normal branch, similar to the optical trend in Z
sources~\cite{hasinger90,augusteijn92}.

\section{June 1997 Campaign}

The second campaign (see Table~\ref{tab:multifreq97}) occurred during
1997 June 4--22, covering a half cycle before and after phase zero on
June 12.506 UT (JD 2450612.006). Radio observations were made from
South Africa at Hartebeeshoek Radio Astronomical Observatory (HartRAO)
by George Nicolson. Infrared photometry measurements were made with
the 1.9-m telescope at the South African Astronomical Observatory
(SAAO) by Ian Glass. IR spectroscopy at the the ANU 2.3~m telescope in
Siding Springs, Australia and optical spectroscopy at the
Anglo-Australian Telescope (AAT, 3.9~m) were carried out by Rob
Fender, Kinwah Wu, and Helen Johnston. In addition to the ongoing
All-Sky Monitor observations, extensive \RXTE\ PCA (and HEXTE)
observations were made, with 56\% coverage over 7~days around phase
zero, and shorter observations before and after that week.  The X-ray
observations for this campaign are discussed in
Chapter~\ref{ch:fullZ}.

\begin{table}
\begin{center}
\begin{tabular}{cccc}
\hline
\hline
Frequency & & Type of & \\ Band & Observatory & Observation & Dates \\
\hline
Radio     & HartRAO      & 8.5~GHz \& 5.0~GHz  & ongoing\\
IR        & SAAO (1.9~m) & K-band photometry      & June (10--14),15--16\\
          & ANU (2.3~m)  & K-band spectroscopy    & June (12),20\\
Optical   & AAT (3.9~m)  & spectroscopy           & June 4,(12)\\
X-ray     & RXTE         & ASM		          & ongoing\\
          &              & PCA (samples)          & June 4,6,8,18,19,20\\
          &              & PCA ($\sim$continuous) & June 11--17\\
\hline
\end{tabular}
\end{center}
\caption{Multi-frequency Observations of Cir~X-1 from 1997 June.
Dates in parentheses were cancelled due to poor weather conditions.
}
\label{tab:multifreq97}
\end{table}

\begin{figure}
\begin{centering}
\PSbox{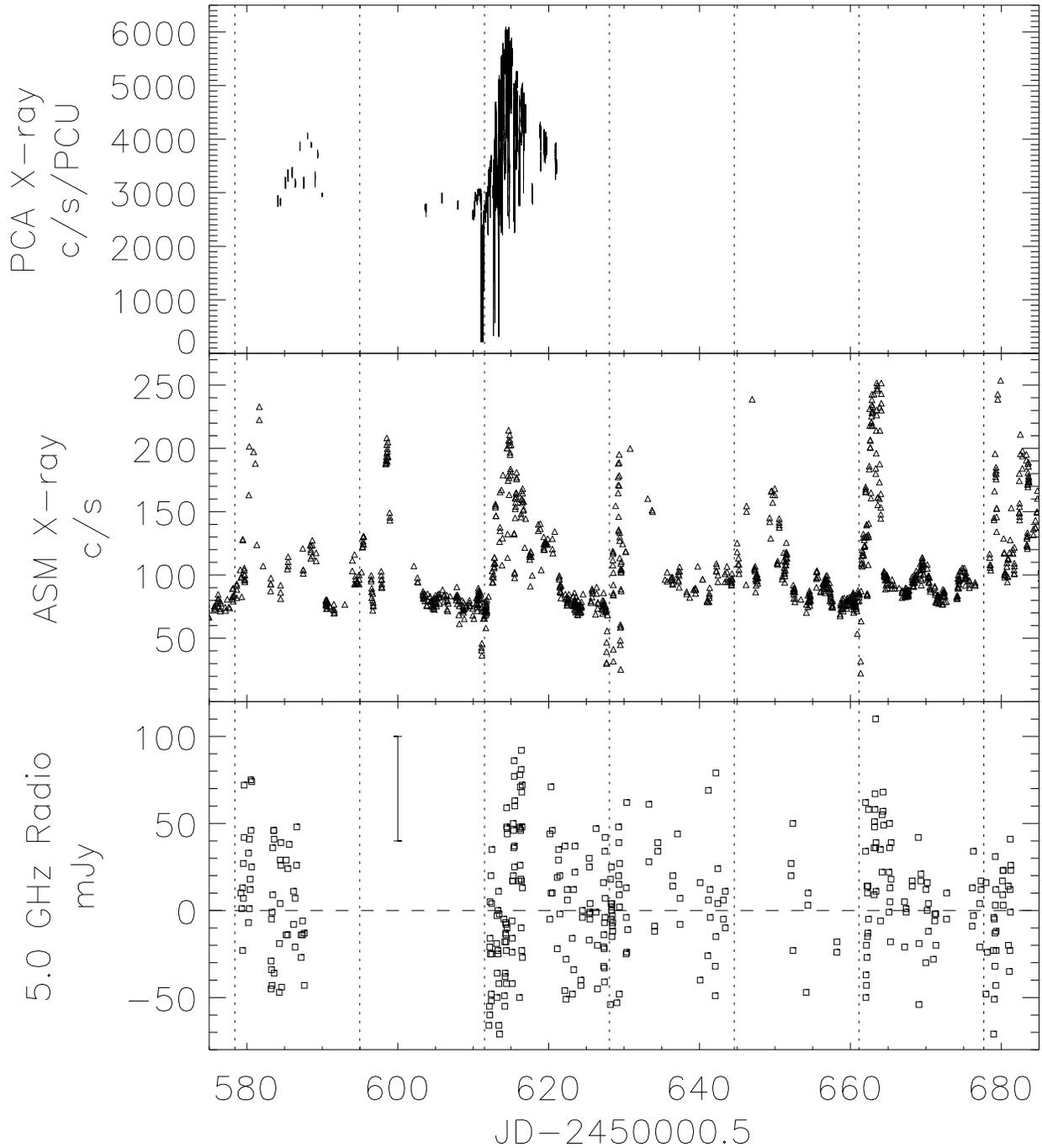 
	hscale=100 vscale=100 hoffset=-30 voffset=-110}{6.45in}{6.85in}
\caption{
X-ray (PCA and ASM) and radio (5.0~GHz) light curves of Cir~X-1 from
1997 May 7 -- August 25.  Phase zero is indicated by the vertical
dotted lines. PCA intensity is from the 2--32~keV band (16-s bins) and
ASM intensity is from 2--12~keV (90-s exposures). Errors on the PCA
data are negligible and errors on the ASM data are typically 2--5~c/s.
For each 16.55-d cycle, the average radio flux for the phase interval
0.3 to 0.9 of that cycle was subtracted from the data; the bias
removed was typically 50--60 mJy and a typical error bar is
shown. (The radio data was provided by G.~Nicolson, and ASM data by
the \RXTE/ASM team. The PCA data are from studies~F and G; see
Table~\protect{\ref{tab:pcaobs}}.)  }
\label{fig:multifreq97_lc}
\end{centering}
\end{figure}

\subsection{Radio (HartRAO)}
\label{sec:hartrao97}

In conjunction with our multi-frequency campaign, coverage during the
ongoing radio (8.5~GHz and 5.0~GHz) monitoring at HartRAO was
increased. The 5.0~GHz radio data for 1997 May -- August are shown in
Figure~\ref{fig:multifreq97_lc}, along with the ASM light curve and
the PCA data from the 1997 June observations (study~G) and the brief
observations from 1997 May (study~F)\@. As in the previous year, the
radio intensity was faint with only minimal variability (a constant DC
bias level has again been subtracted). However, in at least two cycles
(starting at days 611 and 661), including the cycle that was the main
focus of this multi-frequency campaign, there does appear to be weak
evidence for flaring at about the same time as the X-ray flaring
observed with the PCA and ASM.

\subsection{IR Photometry (SAAO)}
\label{sec:saao97}
Infrared photometry observations (see section~\ref{sec:saao96}) were
scheduled at SAAO for seven nights, from 1997 June 10--16
(JD-2450000.5 = 609--615). Due to poor weather conditions, only
limited data were obtained, shown in Table~\ref{tab:ir_june97}.  On
June~14, the JHK magnitudes are similar to typical peak values
observed since 1980. On June~15, the K magnitude is about 0.5~mag
lower, but still well above the typical quiescent value of
K$\approx$11.5.

\begin{table}
\begin{center}
\begin{tabular}{rcccc}
\hline
\hline
JD         &   J   &  H    & K    & L\\
\hline
2450614.35 & 11.97 & 10.68 & 9.88 & 8.98\\
       .42 &       &       & 9.88 & \\
       .48 &       &       & 9.85 & \\
2450615.41 &       &       & 10.34 & \\
\hline
\end{tabular}
\end{center}
\caption{Infrared JHKL photometry measurements from 1997 June 14 \& 15.
Errors are +/$-$~0.03, except on the L magnitude the error is +/$-$~0.05.}
\label{tab:ir_june97}
\end{table}

\subsection{IR Spectroscopy (ANU)}
\label{sec:anu97}

\begin{figure}
\begin{centering}
\PSbox{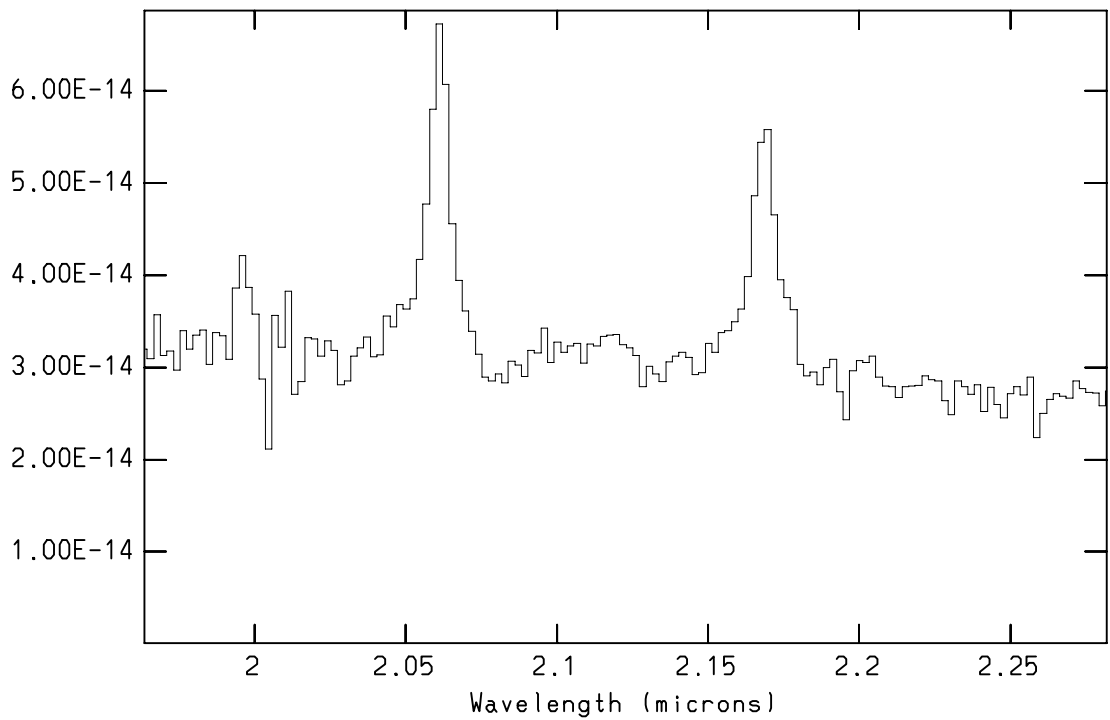 voffset=-255 hoffset=-75}{5.8in}{3.7in}
\caption{
Infrared spectrum of Cir~X-1 from ANU on 1997 June~20 (about
phase~0.5), showing He\,I (2.06 $\mu$m) and H Brackett $\gamma$ (2.165
$\mu$m) emission lines. This spectrum was compiled from the sum of 16
120-s exposures. (Figure courtesy of H. Johnston.)  
}
\label{fig:anu_ir_spec}
\end{centering}
\end{figure}

Infrared K-band spectroscopy was carried out at the ANU 2.3~m
telescope at Siding Spring, Australia. Two nights were scheduled, near
phase~0.0 on June~12 and near phase~0.5 on June~20. Observations on
the first night were cancelled due to bad weather, but conditions were
more favorable on June~20 and good quality K-band spectra of Cir~X-1
were obtained (see Figure~\ref{fig:anu_ir_spec}). The spectrum shows H
Brackett $\gamma$ and He{\small I} emission lines. Both lines are
red-shifted by about +430 km/s, and there is a hint of asymmetry on the
blue (shorter wavelength) wing.

\subsection{Optical Spectroscopy (AAT)}
\label{sec:aat97}

\begin{figure}
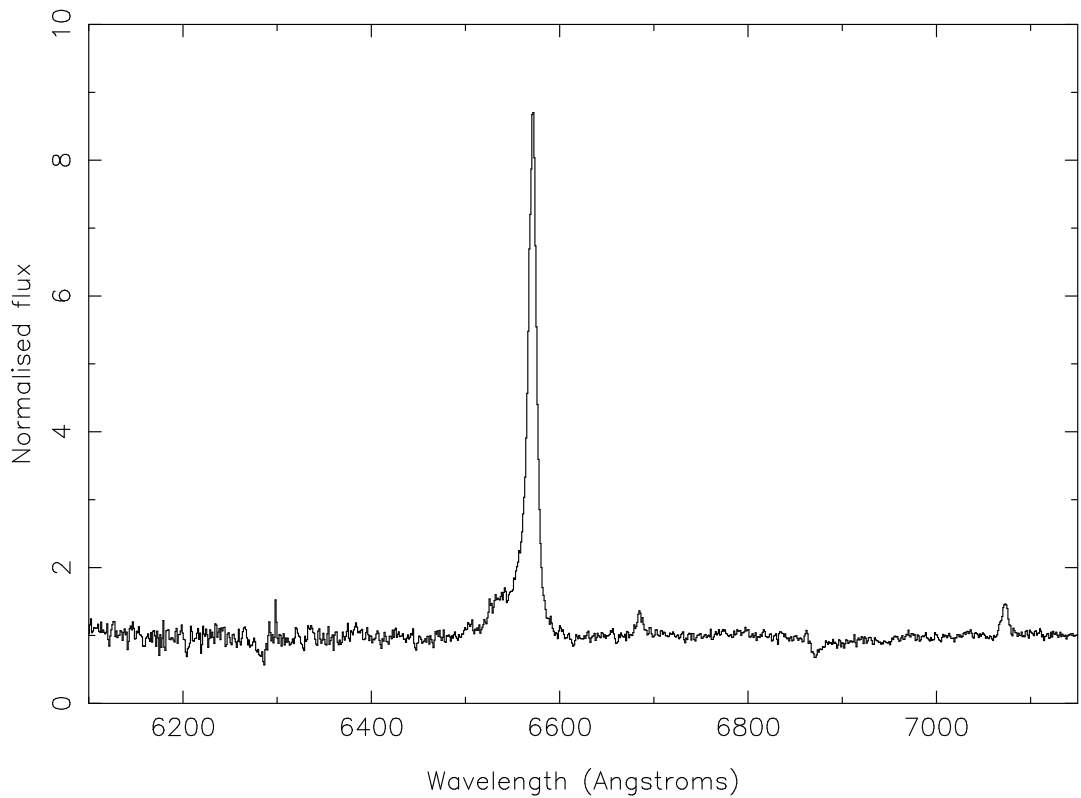

\begin{centering}
%\PSbox{figures/aat_optical_spec.eps hscale=72 vscale=72 angle=270 voffset=410 hoffset=-52}{6.5in}{4.9in}
\PSbox{figures/aat_optical_spec.eps hscale=60 vscale=60 angle=270 voffset=340 hoffset=-25}{5.4in}{4.0in}
\caption{
Optical spectrum of Cir~X-1 from AAT on 1997 June~4 (about phase~0.5),
showing a prominent H$\alpha$ emission line, as well as weaker
He\,I(6678~\AA) and He\,I(7065~\AA) lines. (Figure
courtesy of H. Johnston.)  }
\label{fig:aat_optical_spec}
\end{centering}
\end{figure}

Optical spectroscopy at the H$\alpha$ region (6100--7200~\AA) was
carried out with the AAT 3.9~m telescope. Observations were scheduled
on two nights, near phase~0.5 on June~4 and near phase~0.0 on June~12.
Like the ANU observations on June~12, the AAT observations on that
night were cancelled due to bad weather. The observations on June~4
were successful, and six good-quality spectra were obtained. The
average spectrum, normalized by the continuum, is shown in
Figure~\ref{fig:aat_optical_spec}. This spectrum shows a prominent
H$\alpha$ emission line, as well as weaker He\,{\small I}(6678~\AA)
and He\,{\small I}(7065~\AA) lines. The H$\alpha$ line is asymmetric,
as suggested by previous
observations~\cite{duncan93,mignani97,nicolson80,whelan77}, and can be
fit well by two Gaussians lines: a broad component blue-shifted from
the rest wavelength by -300~km/s and a narrow component red-shifted
from the rest wavelength by +375~km/s.  The two narrow He\,{\small I}
lines are red-shifted by an amount similar to the narrow H$\alpha$
line, and thus all three show a similar red-shift to the lines in the
IR spectrum in section~\ref{sec:anu97}.  All three narrow lines have
widths of about 9.5~\AA, or $\sim$400~km/s, while the width of the
broad H$\alpha$ component is 46~\AA, or 2100~km/s.

\subsection{Comparison of the AAT Optical Spectrum to HST Results}

The optical spectrum obtained with AAT during our 1997 June campaign
is comparable to the spectrum of Cir~X-1 obtained with the Hubble
Space Telescope (HST) Faint Object Spectrograph (FOS) in 1995 June by
Mignani, Caraveo, \& Bignami~\cite{mignani97}. The HST spectrum also
showed an asymmetric H$\alpha$ emission line, and was likewise fit by
a broad and a narrow Gaussian line. The broad line is centered on the
H$\alpha$ rest wavelength and has a width of 11.4~\AA. The narrow
component is red-shifted by 380~km/s and has a width of 3.32~\AA.
Based on the HST results, the authors suggested that the broad line
originates in the accretion disk, with temperatures of order
$\sim$20000~K responsible for its large width. They further suggest
that the narrow line arises from a ``hot spot'' where the accretion
stream from the companion impacts the disk.  Thus the narrow line,
observed at phase zero with HST, was expected to be observed at
varying blue/red-shifts at other orbital phases.

The interpretation suggested for the HST spectrum is not compatible
with the additional results obtained with AAT\@, which show the narrow
line at almost the same wavelength as HST, but at
phase~0.5. Furthermore, it is the {\em broad} component which has
shifted, from the rest wavelength to $-$300~km/s blue-shifted.  Clearly,
further observations at various orbital phases are needed.  If the
narrow H$\alpha$ component, and the other optical and IR lines, prove
to be fixed at about +400~km/s, then that velocity might reflect the
radial velocity of the system, providing further support for the
hypothesis that Cir~X-1 is a high-velocity runaway from
SNR~321.9-0.3. Measuring the velocity of the shifting broad component
might help map the binary orbit, about which very little is securely
known.

\section{Conclusions}

The X-ray light curves of Cir~X-1 show significant variability on all
time scales down to seconds (and even shorter time scales including
the QPOs). Unfortunately, most observations at other wavelengths
typically provide measurements on timescales of hours, allowing only
coarse correlations to be studied. Yet even those measurement can be
quite valuable. For instance, our multi-frequency campaign from 1996
has shown that according to the ephemeris equation currently in use
(equation~\ref{eq:ephem}), the X-ray, IR, and radio bands all show
enhanced emission within a day of phase zero (all peaking after phase
zero). The 1997 radio data also show weak evidence for flares after
phase zero, coincident with those in the X-ray band.  The correlated
increase in IR and X-ray intensity over several hours shown in
Figure~\ref{fig:ir_pca_tjd213} demonstrates the need for observations
with improved time-resolution.

Clearly a return of the radio intensity to its previous bright levels
would be very useful in improving the orbital ephemeris equation
(i.e., testing the change in period implied by the current quadratic
term.) Bright radio flares would also be able to be studied in much
greater detail than the current marginal detections, allowing
short-term correlation with other wavelengths to be explored.  Delays
of specific events between wavebands would provide information on how
photons are being reprocessed from one band into another.  We continue
to maintain a Target of Opportunity proposal with \RXTE\ to provide
X-ray coverage, should bright radio flares return.

As mentioned above, one promising follow-up project to these campaigns
is to study the shifting optical/IR emission lines. If such work can
provide information about the binary orbit (through Doppler mapping),
then the X-ray data can be re-examined with more consideration of the
geometry of the system. This work may begin as early as this summer.

A promising radio-interferometry imaging project is the attempt to
detect proper motion of Cir~X-1 away from the nearby supernova remnant
(SNR), as is expected in the runaway binary scenario (see
Chapter~\ref{ch:intro}). Based on the 700--2000~km/s velocity implied
by its distance from the SNR, and assuming a distance of 8~kpc,
suggests that proper motion might be of order 1$^{\prime\prime}$ in
20~yr~\cite{stewart93}, which could be detected in only a few years of
precise measurements.  Confirmation of association with the SNR, and
thus of a young age, would be highly significant to our understanding
of the evolution of the system. Preliminary work has recently been
started on this project (by R.~Fender et~al.), with Australia
Telescope Compact Array (ATCA) images taken to establish the location
of an appropriate point source to use as a reference for proper motion
measurements. Such radio imaging may also provide further information
on the jet-like structures seen inside the synchrotron nebula
surrounding Cir~X-1.

\chapter{Mass Transfer in an Eccentric Binary}
\label{ch:masstrans}

\section{Overview}

Mass transfer in X-ray binaries generally occurs through Roche-lobe
overflow or via a stellar wind. Significant stellar winds only occur
in high-mass donor (O or B) stars. Although the X-ray properties of
Cir~X-1 most resemble those of a low-mass X-ray binary, the true
nature of the donor star and the mechanism for mass transfer remain
uncertain (see section~\ref{sec:counterpart}). Furthermore, the high
eccentricity inferred from the 16.55-d intensity cycle of Cir~X-1 will
have a significant effect on how mass transfer occurs. In this
chapter, I explore the effect of eccentricity on mass transfer through
both Roche-lobe overflow and a stellar wind.

\section{Mass Transfer via Roche-Lobe Overflow}
\label{sec:roche}

Mass transfer in most low-mass X-ray binaries is believed to occur
through Roche-lobe overflow, wherein one star fills its critical
equipotential surface and matter from the stellar surface streams
through the inner Lagrange (L1) point (a saddle point in the net
gravitational and centrifugal potential between the stars) toward the
other star. As matter flows in, the Coriolis force causes it to orbit
the compact object, forming an accretion disk wherein matter loses
angular momentum through viscosity and spirals inward.  In an
eccentric orbit, the situation is much more complicated, with such
mass transfer limited to the time near periastron passages.  To help
understand the behavior of mass transfer in an eccentric binary, I
developed a computer code to follow the trajectory of a test particle
started from the surface of a donor star and moving in an eccentric
binary system. The particle was followed until it reaches the
neutron-star accretion disk, returns to the donor, or leaves the
system.

\subsection{Equipotential Surfaces in an Eccentric Binary Orbit}
\label{sec:eccRochePoten}
The standard Roche-lobe framework describes the potential of a binary
pair of stars in circular orbit, co-rotating at the orbital
frequency. In an eccentric orbit, not only does the separation between
the stars vary, but so does the instantaneous angular frequency of the
orbit, making co-rotation impossible.  A generalized Roche potential
for centrally condensed stars in an eccentric binary orbit has been
derived by Avni \cite{avni76}.  A Cartesian coordinate system was
used, with the origin at the center of mass of star~1, of mass
$M_1$, which rotates with an angular velocity ($\vec{\omega}_{rot}$)
along the Z~axis. The reference frame rotates about the center of mass
of star~1 such that star~2 (of mass $M_2$), always lies on the $+$X
axis.  The potential at a point ($X$,$Y$) is:
\begin{equation} \label{eq:eccRochePoten}
\psi = - \frac{G M_1}{d_1} - \frac{G M_2}{d_2} + \frac{G M_2}{D^2} X 
	- \frac{1}{2} \omega_{rot}^2(X^2 + Y^2) 
\end{equation}
%\begin{equation} \label{eq:eccRochePoten2}
%\psi = - \frac{G M_1}{D} \left[\frac{1}{r_1} + \frac{Q}{r_2} - Q x + 
%	\frac{1}{2} \left(\frac{\omega_{rot}}{\omega_K}\right)^{2}
%	\left(\frac{D}{a}\right)^{3}(1 + Q)(x^2 + y^2) \right],
%\end{equation}
where $d_1$ and $d_2$ are the distances from ($X$,$Y$) to the center
of mass of each star and D is the instantaneous separation of the
stars.  To determine the location of the instantaneous inner Lagrange
point, this potential is maximized along the line connecting the
centers of the two stars. Although this potential is derived for the
frame rotating with the orbit (and thus with an angular frequency
continuously changing to match the instantaneous Keplerian frequency)
the location of the L1 point is specified by a distance from the center of
mass of star~1 along the line between the stars, and is thus
well-defined in an inertial frame. Likewise, the quantities in
equation~\ref{eq:eccRochePoten} necessary to calculate the location of
the L1 point are distances and masses, which can be measured in
inertial space, and the angular rotation frequency of star~1, which is
also measured in the inertial frame.  Thus, this equation can be used
to locate the instantaneous L1 point in an inertial frame, which will
prove convenient in the calculation below.

\subsection{Keplerian Orbits}
\label{sec:keplerorb}
Keplerian orbits can be parameterized by a quantity know as the
eccentric anomaly, $u$, which evolves in time according to Kepler's
equation:
\begin{equation} \label{eq:ecc_anom}
\frac{2 \pi t}{P} = u  - e \sin u , 
\end{equation}
where $t$ is the time elapsed since periastron, $P$ is the orbital
period, and $e$ is the eccentricity. The left-hand side of
equation~\ref{eq:ecc_anom} is also know as the mean anomaly.
The separation vector of the two stars is then given by:
\begin{equation} \label{eq:xyorbit}
\begin{array}{c} 
x(u) = a (\cos u - e) \\
y(u) = a \sqrt{1 - e^2} \sin u,
\end{array}
\end{equation}
where $a$ is the semimajor axis of the orbit (half of the maximum
separation distance of the stars), and periastron lies along the $+x$
axis.  For each time step, equation~\ref{eq:ecc_anom} is solved for
$u$ via Newton's method, and the positions of the two stars are
calculated from equations~\ref{eq:xyorbit}. The velocities of the
stars can be calculated by substituting $du/dt$, from the time
derivative of equation~\ref{eq:ecc_anom}, and $u$ into the time
derivatives of equations~\ref{eq:xyorbit}.

\subsection{Trajectory of a Particle in the Binary Potential}

The trajectory of a particle of negligible mass $m$ moving in the
potential of a binary system can be calculated by integrating the
instantaneous velocity and acceleration of the particle in the
inertial (non-rotating) frame of reference as the stars move in their
orbits.  The instantaneous velocity ($\vec{v}_m$) and acceleration are
given by:
\begin{equation} \label{eq:part_vel}
\frac{d\vec{r}_m}{dt} = \vec{v}_m ,
\end{equation} 
and
\begin{equation} \label{eq:part_accel}
\frac{d\vec{v}_m}{dt} = \frac{\vec{F}_m}{m} 
	= - \frac{ G M_1 m\,(\vec{r}_m-\vec{r}_1) }{ |\vec{r}_m-\vec{r}_1|^3}
	  - \frac{ G M_2 m\,(\vec{r}_m-\vec{r}_2) }{ |\vec{r}_m-\vec{r}_2|^3} ,
\end{equation}
where the subscripts 1, 2, and $m$ refer to the two stars and the
particle respectively, $\vec{F}_m$ is the gravitational force on the
particle, $M$ is the mass of a star, and $\vec{r}$ is a position
vector. The $x$ and $y$ components of equations~\ref{eq:part_vel} and
\ref{eq:part_accel} are integrated by a Runge-Kutta algorithm~\cite{numrec} 
to obtain the trajectory of the particle. The initial coordinate of
the particle is taken to be a point on the surface of the donor (whose
radius is defined to be the size of its critical lobe at periastron
and held constant around the orbit) on the line between the centers of
the two stars.  The initial velocity of the particle is the vector sum
of the instantaneous orbital velocity of the donor, the velocity of a
particle on the surface due to rotation of the donor, and possibly an
additional radial velocity away from the surface (see below).

\subsection{Final Outcome of Particle Trajectories} 

The possible final outcomes of the particle trajectory are as follows:
\begin{enumerate}
\item
It is assumed that the accretion disk almost fills the instantaneous
critical lobe of the neutron star at periastron (i.e., 80\% of
the distance between the neutron star and the L1 point at periastron),
and that the disk maintains that size at all other orbital phases. The
particle is assumed to have {\bf entered the disk} if it passes closer
than this distance from the neutron star at any point in the orbit.
\item
The donor is likewise assumed to fill its critical lobe at periastron
and maintain that radius throughout the orbit. The particle is
considered to be {\bf recaptured} if it re-enters that radius relative
to the donor at any point in the orbit.
\item
The particle is assumed to be {\bf ejected} from the binary (or at
least no longer a likely contributor to mass transfer) if its distance
from the CM exceeds twice the semimajor axis of the orbit.
\item
After two full binary orbits, if none of the above three cases have
occurred, the outcome is {\bf undetermined}.
\end{enumerate}

In all cases, the orbital period was taken to be that of Cir~X-1,
16.55~days, and the mass of the neutron star was taken to be
1.4~\msun.  For simplicity, the donor star is assumed to rotate
uniformly at the average Keplerian frequency ($\omega_{rot} = 2 \pi /
16.55~{\rm days}$). Also, a donor mass of 1~\msun\ has been adopted
for the discussion below, where results are shown for two
eccentricities: 0.5 and 0.8.  Such values might be close to actual
values for Cir~X-1, but are mainly chosen for illustrative purposes.

\begin{figure}
\begin{centering}
\PSbox{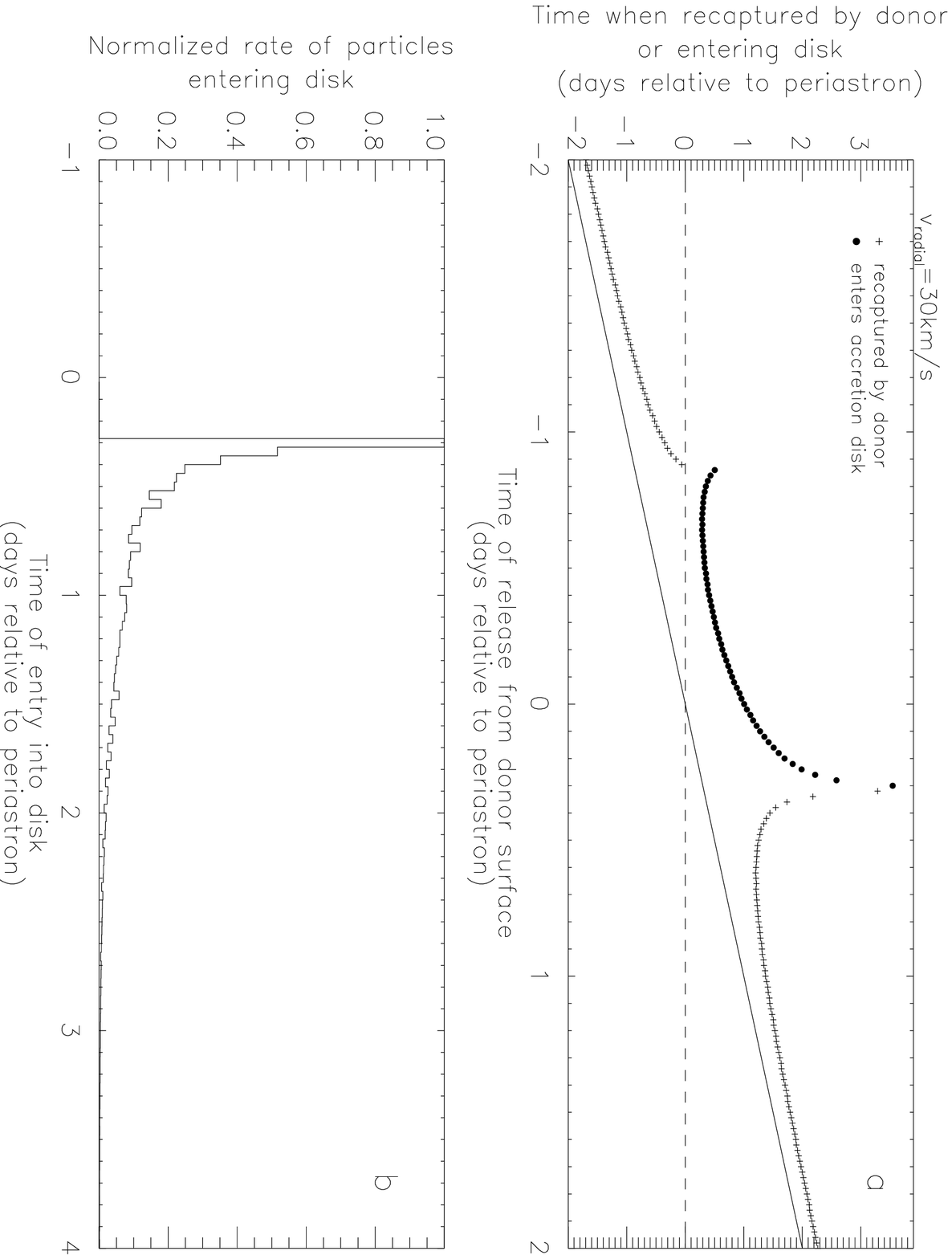 hscale=52 vscale=52 angle=90 voffset=7 hoffset=385}{4.8in}{3.45in}
\vspace{0.2in}
\PSbox{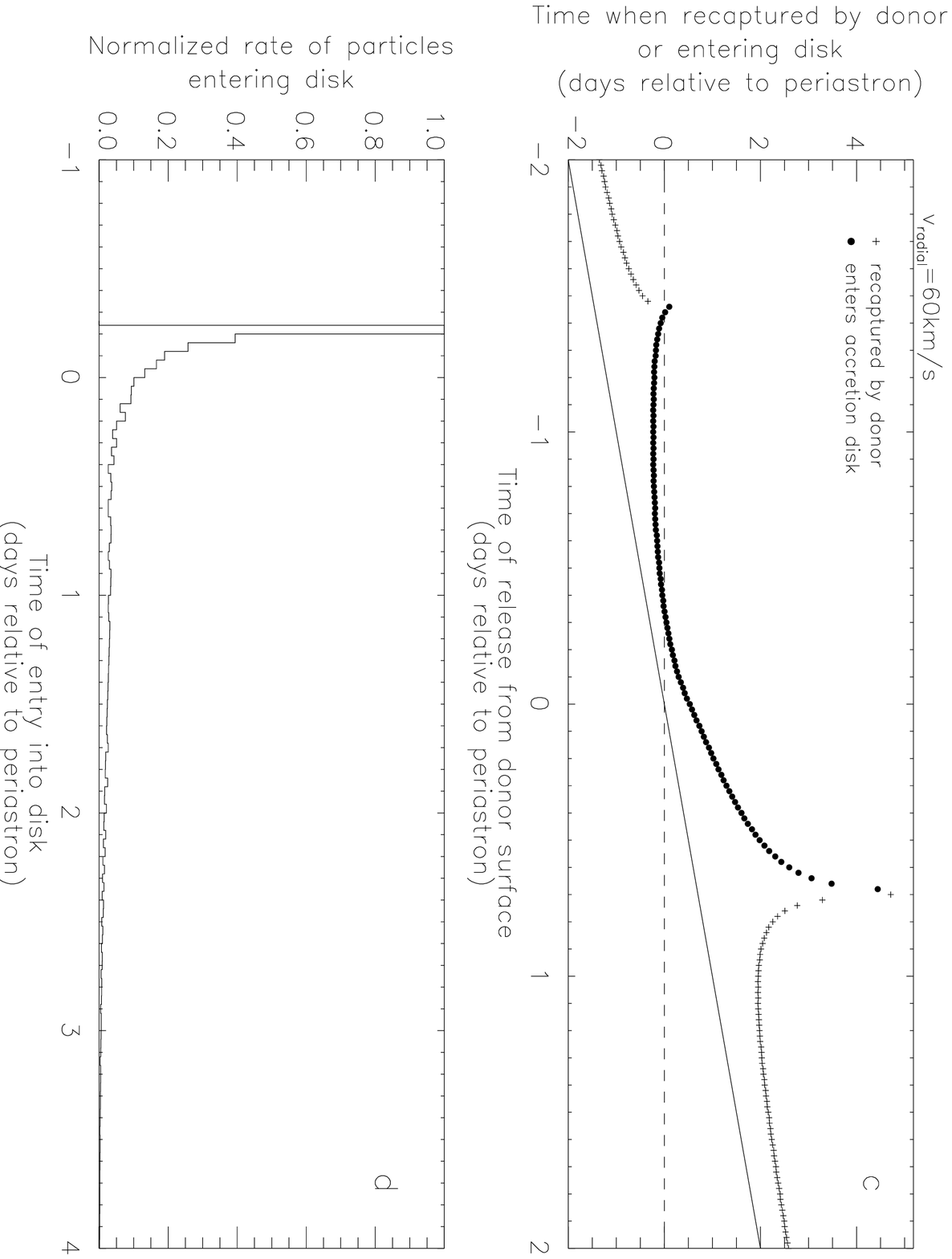 hscale=52 vscale=52 angle=90 voffset=-20 hoffset=385}{4.8in}{3.45in}
\caption{
Panels (a) and (c): stop time of particles in a system with e=0.5 (and
P=16.55~d) versus time of release from donor (1 \msun), with the final
outcome (recapture or entry into disk) indicated.  Panels (b) and (d):
normalized rate of particles entering the accretion disk versus their
arrival time relative to periastron.  The top panels, (a) and (b), are
for an initial radial velocity from the surface of 30~km/s, and the
bottom panels, (c) and (d), are for 60~km/s.  In (a) and (c), the
solid diagonal line has a slope of unity and indicates zero elapsed
time between release and the final outcome; the horizontal dashed line
indicates a final outcome occurring at periastron. }
\label{fig:particles_e0.5}
\end{centering}
\end{figure}

Figure~\ref{fig:particles_e0.5} shows the final outcome of particles
which were released from the surface of the donor at different orbital
phases (evenly spaced in time) of a system with $e=0.5$.  The top
panels (Figures~\ref{fig:particles_e0.5}a, b) are for an initial
radial velocity off the surface of 30~km/s, or about 15\% of the
escape velocity from the star (if it were isolated).  In
Figure~\ref{fig:particles_e0.5}a, one can see the arrival time as a
function of start time, where the different symbols indicate where the
particle entered the accretion disk and where it was recaptured by the
donor. Only particles that begin transfer during the interval $-0.9$
to $+0.3$ days relative to periastron eventually enter the disk, and
most of the particles arrive at the disk
(Figure~\ref{fig:particles_e0.5}b) closely spaced in time near day~0.3
or shortly after.  Figures~\ref{fig:particles_e0.5}c
and~\ref{fig:particles_e0.5}d show that similar behavior occurs for
particles with a larger radial velocity from the surface of 60~km/s
($\sim$30\% of the escape velocity from the donor). The larger initial
velocity allows particles to reach the disk over a somewhat longer
portion of the orbit, and also results in most of the particles
reaching the disk {\em before} periastron.

The fact that mass transfer is most effective for particles released
before periastron is related to the relative velocity of the stars in
their eccentric orbit.  Particles released as the stars approach
periastron naturally move toward the neutron star, and are more likely
to be captured by the disk. Particles released after periastron have
to try to ``catch up'' to a receding neutron star, and after a certain
point will simply be recaptured by the donor. The asymmetric profile
of particles entering the disk is also related to the relative motion
of the stars. Particles released shortly before periastron are carried
along the orbit of the donor and their arrival times are thus focussed
near the time of periastron.

\begin{figure}
\begin{centering}
\PSbox{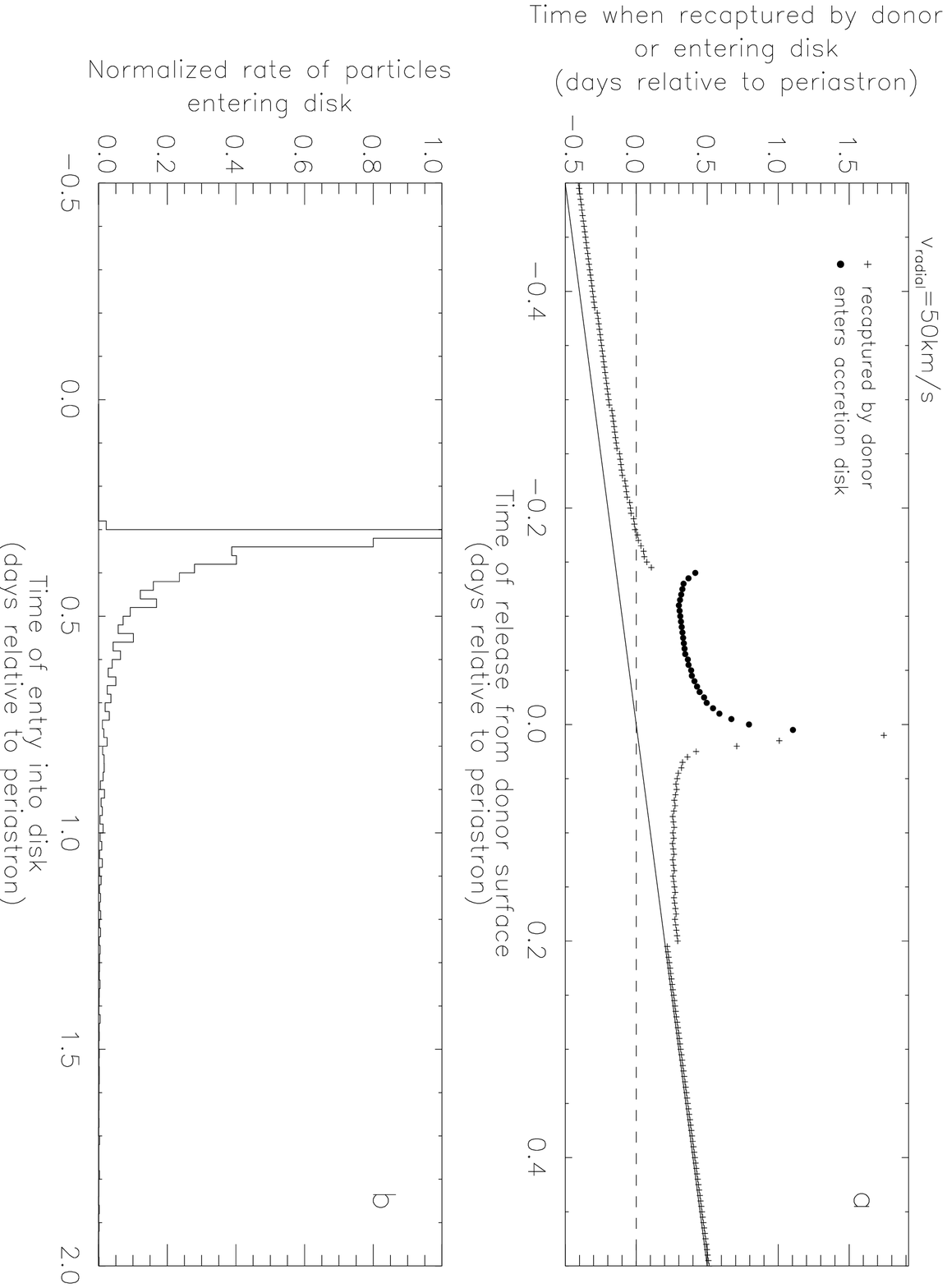 hscale=52 vscale=52 angle=90 voffset=7 hoffset=385}{4.8in}{3.45in}
\vspace{0.2in}
\PSbox{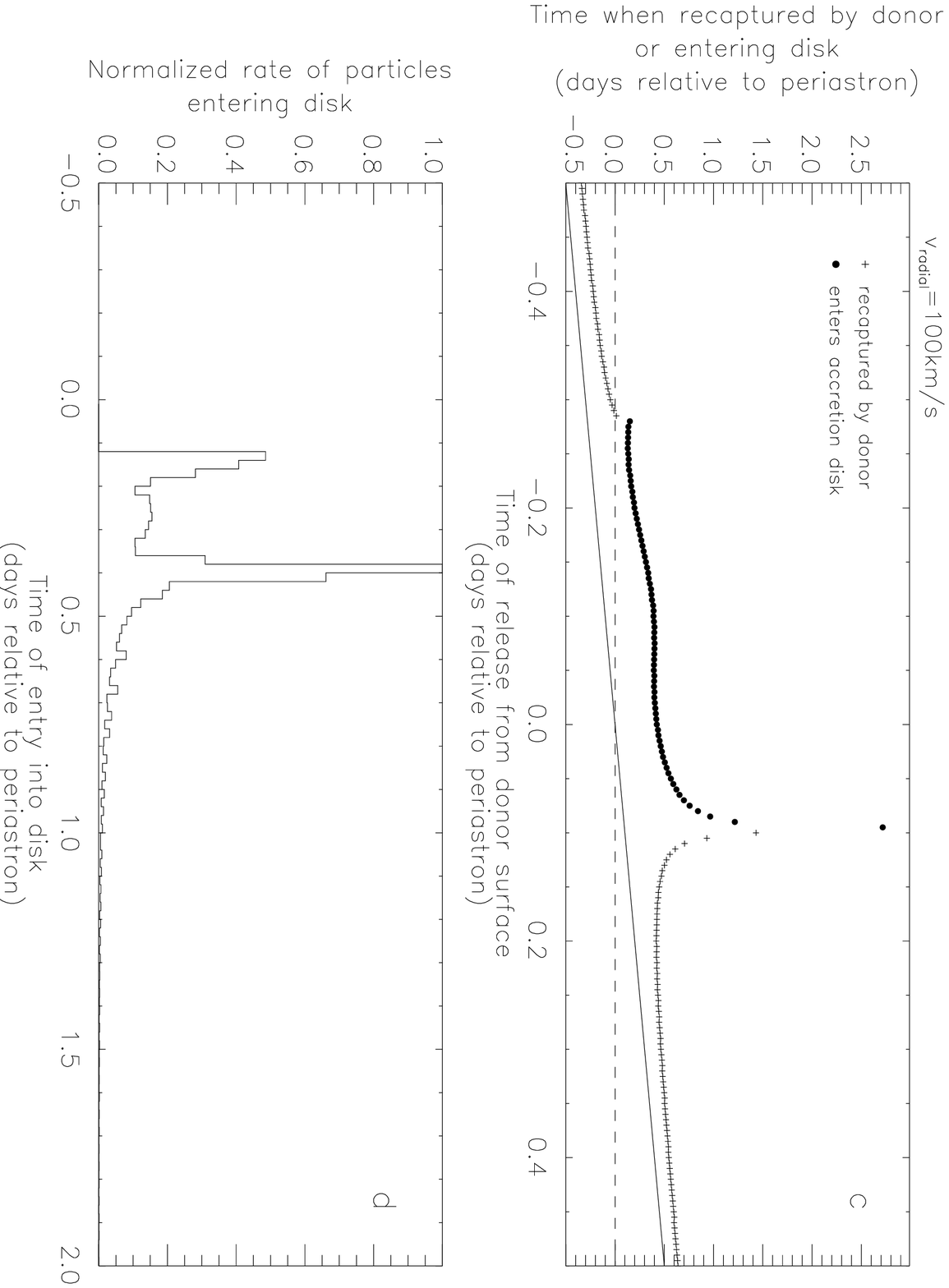 hscale=52 vscale=52 angle=90 voffset=-20 hoffset=385}{4.8in}{3.45in}
\caption{
Panels (a) and (c): stop time of particles in a system with e=0.8 (and
P=16.55~d) versus time of release from donor (1 \msun), with the final
outcome (recapture or entry into disk) indicated.  Panels (b) and (d):
normalized rate of particles entering the accretion disk versus their
arrival time relative to periastron.  The top panels, (a) and (b), are
for an initial radial velocity from the surface of 50~km/s, and the
bottom panels, (c) and (d), are for 100~km/s.  In (a) and (c), the
solid diagonal line has a slope of unity and indicates zero elapsed
time between release and the final outcome; the horizontal dashed line
indicates a final outcome occurring at periastron. }
\label{fig:particles_e0.8}
\end{centering}
\end{figure}

Figures~\ref{fig:particles_e0.8}a and~\ref{fig:particles_e0.8}b show
the outcome of a similar transfer scenario as
Figure~\ref{fig:particles_e0.5}, but with a larger eccentricity
($e=0.8$) and with each particle receiving an initial release velocity
of 50~km/s radially away from the surface of the donor (again about
15\% of the escape velocity). The higher eccentricity confines mass
transfer to a narrower range of orbital phase, mainly a release within
0.15~d prior to periastron passage.

Figures~\ref{fig:particles_e0.8}c and~\ref{fig:particles_e0.8}d also
show the outcome of transfer in a system with $e=0.8$, but with a
larger radial velocity of 100~km/s ($\sim$30\% of the escape velocity
from the donor). In this case the accretion rate shows two sharp peaks
after periastron passage, separated by only 0.25~d
(Figure~\ref{fig:particles_e0.8}d). The combined high orbital velocity
approaching periastron in the eccentric orbit and the high initial
velocity from the surface cause some particles to initially
``overshoot'' the neutron star, eventually to be captured in the disk,
but only after a delay that concentrates the arrival time of many
particles near day~0.4

\begin{figure}
\begin{centering}
\PSbox{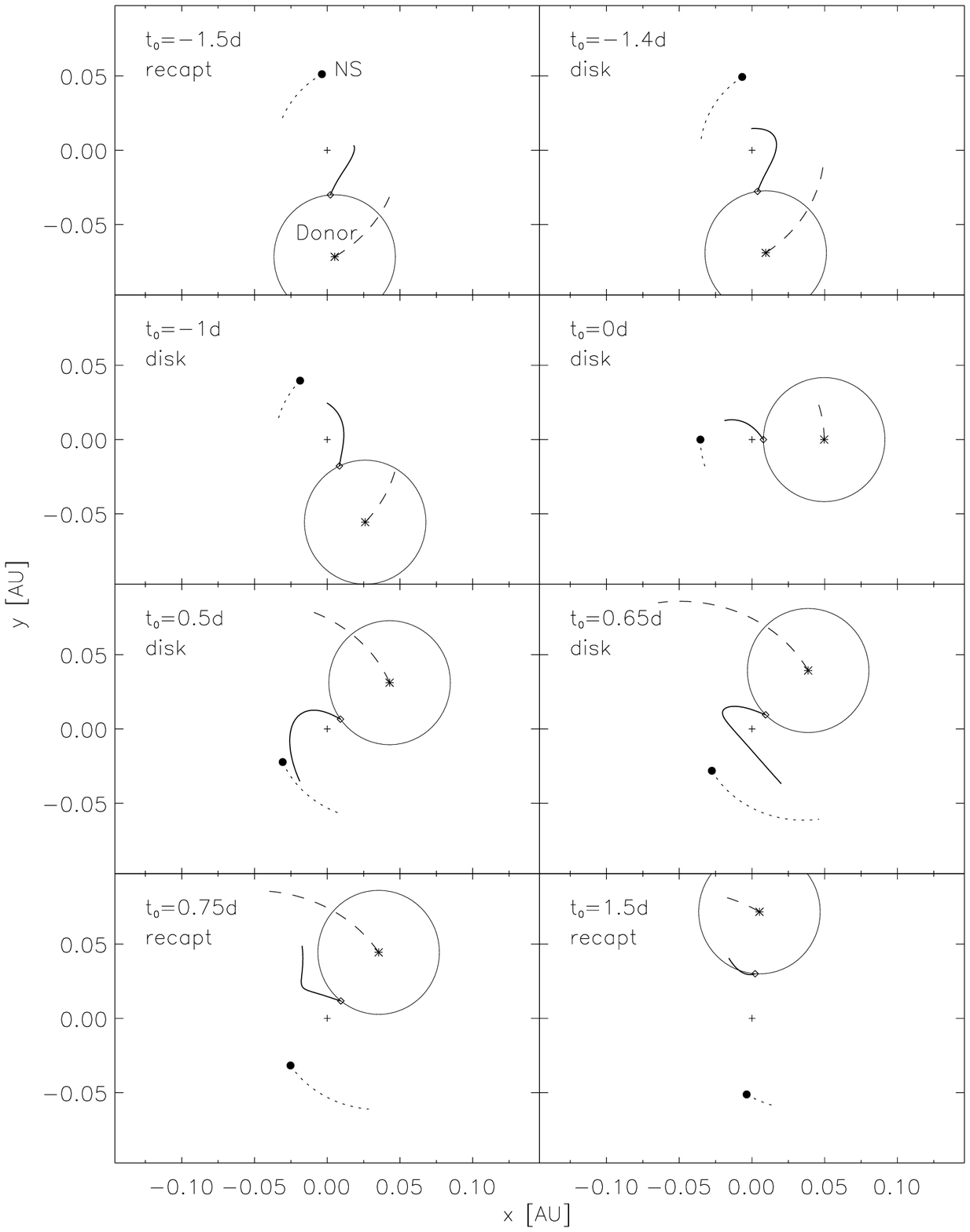 hscale=90 vscale=90 voffset=-40 hoffset=-10}{5.9in}{7.15in}
\caption{
Trajectory of a particle moving in an eccentric binary system
($M_{don}$=1 \msun, e=0.5), where the particle starts with an initial
radial velocity of 60~km/s off the surface of the donor at a time
$t_0$ relative to periastron. The plot is in a non-rotating frame with
the CM at the origin ($+$).  The location of the particle, donor star,
and neutron star at $t_0$ are indicated by the diamond, asterisk, and
solid dot respectively. The subsequent motion of the particle until
enters the accretion disk or is recaptured is shown as a solid curve,
and the motion of the donor and neutron star in their orbits during
that time is indicated by a dashed or dotted ellipse segment
respectively.  }
\label{fig:trajectories}
\end{centering}
\end{figure}

Eight illustrative trajectories for some of the test particles shown
in Figure~\ref{fig:particles_e0.5}c ($e=0.5$), released at various
orbital phases with $v_{radial} = 60$~km/s, are shown in
Figure~\ref{fig:trajectories}. These plots show how the motion of the
two stars in their eccentric orbits can produce a wide variety of
trajectories for a particle transferred from the surface of the donor.

\subsection{Comparison to the X-ray Intensity Profile}

The sharp rise time of the number of particles entering the accretion
disk in Figures~\ref{fig:particles_e0.5} and~\ref{fig:particles_e0.8}
resembles the sharp rise in X-ray intensity from Cir~X-1 after phase
zero (not necessarily periastron) in the ASM light curve (see
Chapter~\ref{ch:observs}). These peaks are quite narrow (generally
only a fraction of a day) compared to the $\sim$3-5 days typically
observed in the X-ray light curve.  However, the actual intensity
profile depends on additional factors, such as the duration of time
that matter spends in the disk before falling onto the neutron
star. These results serve primarily as an example of what can be done
once more is known about the parameters of the Cir~X-1 system, such as
the actual eccentricity and the mass and stellar type of the donor.

%%%%%%%%%%%%%%%%%%%%%%%%%%%%%%%%%%%%%%%%%%%%%%%%%%%%%%%%%%%%%%%%%%%%%%%%%%%%%%%
\section{Mass Transfer via a Stellar Wind}
\label{sec:wind}

When the donor star in an X-ray binary does not fill its Roche lobe, a
stellar wind or other outflow must be responsible for mass
transfer. This is the case in some high-mass X-ray binaries, where the
donor star experiences significant mass loss via a wind from the
surface. Since a high-mass companion for Cir~X-1 cannot yet be ruled
out, I also studied wind transfer in an eccentric binary. I developed
a computer code to study the mass transfer rate as a function of
orbital phase. The resulting theoretical mass accretion profiles can
be compared with the observed X-ray intensity profiles, which at least
in some circumstances may be a measure of the accretion rate.
 
\subsection{Bondi-Hoyle Accretion}

Material in the stellar wind passing within a critical radius of the
neutron star is sufficiently deflected by the gravitational field of
the neutron star to eventually be captured and accrete onto the
surface (first presented by Bondi \& Hoyle in
1944~\cite{bondihoyle}). Since the wind material flows around both
sides of the neutron star (in the orbital plane; the problem is
actually three-dimensional) the net angular momentum of the captured
matter is small and accretion occurs without forming an accretion disk.
The critical radius, i.e., the maximum
accretion radius ($r_{acc}$), is the radius inside which the
gravitational potential energy due to the neutron star exceeds the
kinetic energy of the particles in the wind due to the relative
velocity of the wind and neutron star
($\vec{v}_{rel}=\vec{v}_{wind}-\vec{v}_{orb}$) and the thermal particle
motion (related to the sound speed $v_{s}$):
\begin{equation} \label{eq:wind_potential}
\frac{G M_{ns} m}{r_{acc}} \simeq \frac{1}{2} m v_{rel}^2 + \frac{1}{2} m v_{s}^2,
\end{equation}
where $M_{ns}$ is the mass of the neutron star and $m$ is the mass of
a wind particle (mainly protons). Solving
equation~\ref{eq:wind_potential} for the critical accretion radius
gives
\begin{equation} \label{eq:r_acc}
r_{acc} \simeq \frac{2 G M_{ns}}{v_{rel}^2+v_{s}^2}.
\end{equation}
The mass accretion rate is then 
\begin{equation} \label{eq:mdot_acc}
\dot{M}_{acc} = \rho(r) v_{rel} \pi r_{acc}^2,
\end{equation}
where $\rho(r)$ is the density of the wind at a distance $r$ from the
center of the donor star, i.e., the separation between the stars.
Conservation of mass provides an expression that relates the mass loss
rate of the donor due to wind ($\dot{M}_w$), $v_{wind}$,
and $\rho(r)$:
\begin{equation} \label{eq:mdot_wind}
\dot{M}_w = \rho(r) v_{wind} 4 \pi r^2.
\end{equation}
Combining equations~\ref{eq:r_acc}, \ref{eq:mdot_acc}, and
\ref{eq:mdot_wind} gives the fraction of the wind captured by the
neutron star:
\begin{equation} \label{eq:mdotfract}
\frac{ \dot{M}_{acc} }{ \dot{M}_w } = 
   \frac{G^2 M_{ns}^2}{r^2} \frac{v_{rel}}{v_{wind}} \frac{1}{(v_{rel}^2+v_s^2)^2}. 
\end{equation}

\subsection{Wind Accretion Profile in an Eccentric Binary} 

For a given set of binary parameters (masses, orbital period, and
eccentricity), the positions and velocities of the two stars around
their orbits can be calculated according to the expressions given in
section~\ref{sec:keplerorb}. The mass of the neutron star is taken to
be 1.4~\msun\ and a period of 16.55~days is used.

High-mass stars such as OB supergiants typically have high wind speeds
of 1000-2000~km/s. A wind speed of 1000~km/s is used in the case
discussed below. The relative velocity between the wind and the
neutron star is calculated from the wind speed and the relative
orbital velocity of the stars.

For a wind temperature $T$, the sound speed is given by $v_s =
\sqrt{\gamma k T / m}$, where $k$ is Boltzmann's constant and
$\gamma=5/3$ for a monatomic gas.  A typical wind temperature of
10000~K thus corresponds to a sound speed of about 12~km/s. This is
much less than the wind speeds in this calculation. So, the wind
temperature does not contribute significantly to the final results,
but a value of 10000~K has been adopted for concreteness.

In this calculation, wind propagation delays are ignored, i.e.\ the
wind velocity vector is calculated using the {\em current} position
and velocity of the donor star.  This simplification is valid when the
wind speed is significantly larger than the orbital velocity of the
donor.  In the limit of very high wind speed, $v_{rel} \approx v_w \gg
v_s$, and so equation~\ref{eq:mdotfract} reduces to
\begin{equation} \label{eq:mdotfract_highvw}
\frac{ \dot{M}_{acc} }{ \dot{M}_w } = 
	\frac{G^2 M_{ns}^2}{r^2} \frac{1}{v_w^4}.
\end{equation}
Thus, for a given wind speed, it is primarily the distance between the
stars that determines the fraction of the wind captured. However,
equation~\ref{eq:mdotfract} is still used in the code, since the
orbital motion of the neutron star will produce an asymmetry of the
profile relative to phase zero.

In an eccentric orbit the separation between the stars changes with
orbital phase ($\theta$) according to
\begin{equation} \label{r_theta}
r(\theta) = a \frac{1 - e^2}{1 + e \cos \theta} 
	= \left(\frac{P}{2 \pi}\right)^{2/3} G^{1/3} (M_{ns} + M_{don})^{1/3} 
	\frac{1 - e^2}{1 + e \cos \theta},
\end{equation}
where the Kepler's third law was used in the second equality.  Thus
for a fixed orbital period and high wind speed, the wind capture
fraction is weakly dependent on the mass of the donor star, strongly
dependent on eccentricity, and varies with orbital phase as
$(1+e\cos\theta)^2$.  Thus, only the normalization, and not the shape,
of the wind accretion profiles change significantly with different
donor masses or (high) wind speeds.

\begin{figure}
\begin{centering}
\PSbox{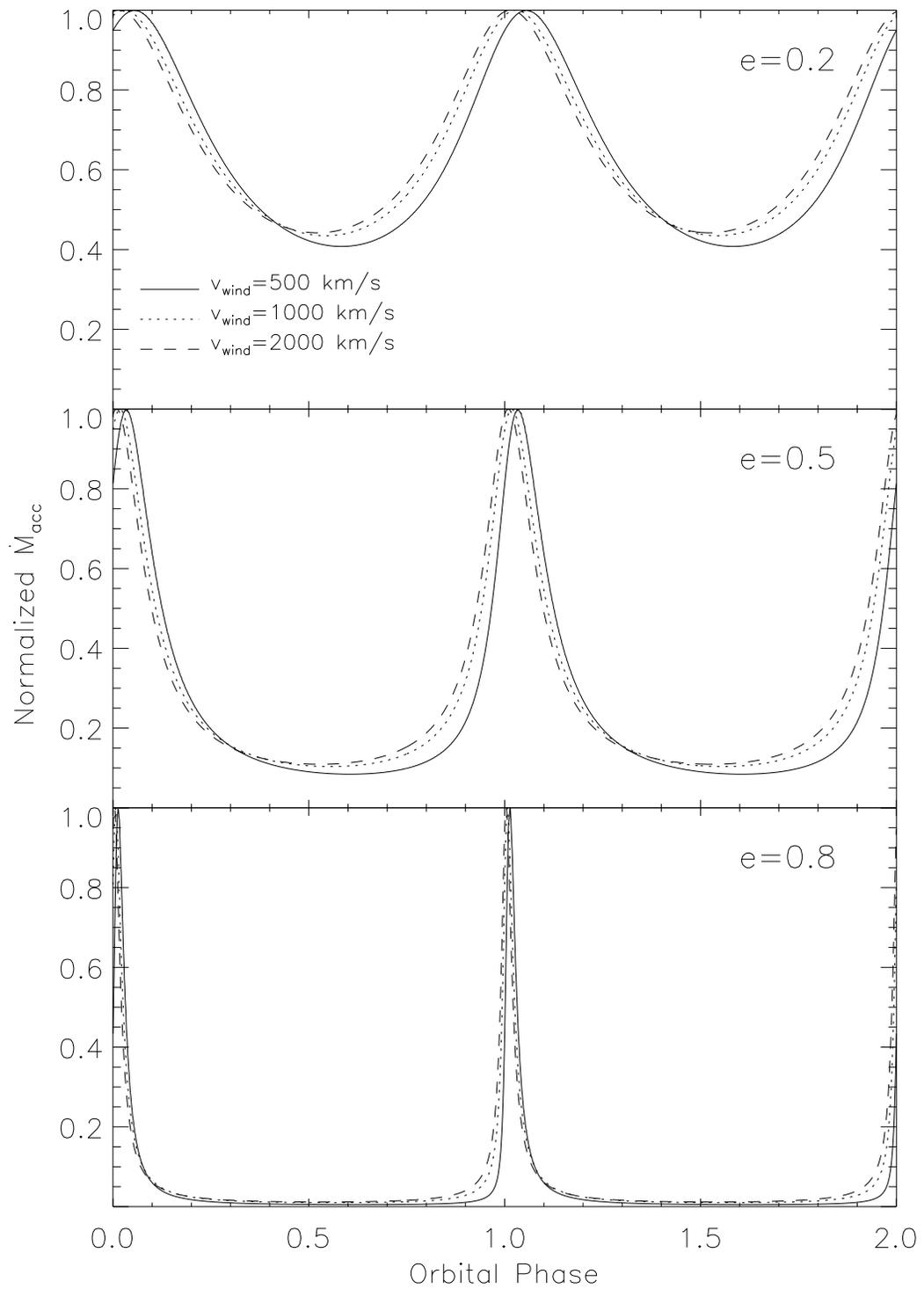 hscale=90 vscale=90 voffset=-10 hoffset=-15}{5.9in}{7.5in}
\caption{
Wind accretion rate versus orbital phase for three eccentricities and
wind speeds for a 5~\msun\ donor.  Each curve is normalized by its
peak value; the actual wind capture fraction (relative to the
mass-loss rate of the donor) decreases significantly with higher
velocity.}
\label{fig:windmdot}
\end{centering}
\end{figure}

The profile of mass accretion rate versus orbital phase for a 5~\msun\
donor is shown in Figure~\ref{fig:windmdot} for three values of
eccentricity (0.2, 0.5, 0.8) and three wind speeds (500, 1000,
2000~km/s). As expected, a low eccentricity produces a broad profile,
while a high eccentricity shows a much narrower peak. The peak value
occurs slightly after phase zero due to the relative velocity of the
wind and neutron star near periastron, and this shift is largest when
the wind speed is slowest.

\subsection{Comparison to the X-ray Intensity Profile}

The X-ray light curve of Cir~X-1 as observed by the ASM (see
Chapter~\ref{ch:observs}) shows that flaring is usually limited to a
quarter to a third of each 16.55-d cycle, suggesting at least a
moderate eccentricity. The flaring is complex and varies significantly
from cycle to cycle. In contrast to the wind profiles in
Figure~\ref{fig:windmdot}, many X-ray cycles show an asymmetric
profile, with the intensity declining more slowly than the rise. Some
cycles even show a secondary peak well separated from the first. Such
behavior cannot be accommodated in the simple wind model presented
here. Moreover, the X-ray light curves show a substantial baseline
intensity level that does not occur in simulations with a large
eccentricity.

\chapter{Eccentric Binary Evolution}
\label{ch:bin_evol}

\section{Overview}

In order to explore the possible evolutionary history of Circinus~X-1,
I have developed a binary evolution code to model the evolution of a
neutron-star and low-mass companion ($M<3$~\msun) in an eccentric
orbit. Although a high-mass companion cannot be ruled out based on the
observed optical and infrared magnitudes (see
section~\ref{sec:counterpart}), the X-ray behavior is most consistent
with a low-mass X-ray binary, and thus I only consider systems with a
low-mass companion in this model.  This code was originally developed
by C.~M.~Becker to model accreting white dwarfs in circular binary
orbits~\cite{becker97}.  In order to keep the model simple, yet
include eccentricity, I assumed that mass transfer (via Roche-lobe
overflow) occurs only at periastron and that the transfer is
conservative (no mass lost from the system). I have also added the
effects of tidal dissipation on the orbit and allow a system to evolve
even when not transferring mass.  I then used this code as the basis
for a Monte Carlo population synthesis study of low-mass X-ray
binaries that might resemble Cir~X-1. I show that the number of
systems in the Galaxy with parameters similar to those of Cir~X-1
should be of order unity, consistent with the unique status of the
source as an eccentric LMXB with a high accretion rate.

%%%%%%%%%%%%%%%%%%%%%%%%%%%%%%%%%%%%%%%%%%%%%%%%%%%%%%%%%%%%%%%%%%%%%%
\section{Binary Evolution Theory}

\subsection{Evolution of Binary Parameters}

% Keplerian elements
The Keplerian elements of the binary system can be specified by the
mass of the neutron star ($M_{ns}$), the mass of the donor star
($M_{don}$), the orbital angular momentum ($J_{orb}$), and the
eccentricity ($e$). The orbital period ($P_{orb}$), angular frequency
($\Omega_{orb}$), and semimajor axis ($a$) are related to these four
quantities by straightforward application of Kepler's third law
\begin{equation} \label{eq:keplers3rd}
\left(\frac{P_{orb}}{2\pi}\right)^2 = \Omega_{orb}^{-2} = \frac{a^3}{G M_T}
\end{equation}
and the expression for the orbital angular momentum
\begin{equation} \label{eq:Jorb}
J_{orb} = G^{1/2}M_{ns}M_{don}M_{T}^{-1/2}a^{1/2}(1-e^2)^{1/2},
\end{equation}
where $M_T = M_{ns} + M_{don}$.

Mass transfer will occur when the donor star fills its critical
potential lobe.  For simplicity, I assume that the mass transfer is
conservative, i.e., all matter lost by the donor is accreted by the
neutron star: $ \delta M_{ns} = -\delta M_{don} $ and $ \delta M_{T} =
0 $.  Furthermore, since the critical potential lobe will be smallest
at periastron, it is assumed that mass transfer occurs {\em only} at
periastron. This is likely to be a good approximation for large
eccentricities but clearly cannot hold true in the limit of zero
eccentricity. For the purpose of studying systems that resemble
Cir~X-1, systems with very low or zero eccentricity (i.e.,
$e\simlt$0.01) are not of interest.

% Components of radius evolution
Because mass transfer depends on the size of the donor relative to its
critical potential lobe, it is necessary to follow the evolution of
the donor star radius.  There are several means by which the donor
radius may change, each of which will be discussed in more detail
below. At this point I will only mention what those mechanisms are:
(1) The radius of the donor changes over time due to the nuclear
evolution of its core.  (2) The star responds adiabatically to the
removal of mass. (3) The radius, if not at equilibrium, adjusts toward
the equilibrium radius on a thermal time scale.  The net effect of
these contributions on radius evolution can be expressed as:
\begin{equation} \label{eq:delRs}
\frac{\delta R_{don}}{R_{don}} = 
	\left( \frac{\delta R_{don}}{R_{don}} \right)_{nuc} 
	+ \left( \frac{\delta R_{don}}{R_{don}} \right)_{th}
	+ \left( \frac{\delta R_{don}}{R_{don}} \right)_{ad}.
\end{equation}

% Components of angular momentum loss
Orbital angular momentum is lost through tidal dissipation, magnetic
braking, and gravitational radiation.
\begin{equation} \label{eq:delJs}
\frac{\delta J}{J} = \left( \frac{\delta J}{J} \right)_{tid} 
	+ \left( \frac{\delta J}{J} \right)_{mb} 
	+ \left( \frac{\delta J}{J} \right)_{gr}.
\end{equation}
Likewise, these mechanisms produce a change in eccentricity. Unlike
angular momentum (which is assumed to be conserved during mass
transfer), the eccentricity changes due to mass transfer. Thus,
the total change in eccentricity is given by:
\begin{equation} \label{eq:deles}
\frac{\delta e}{e} = \left( \frac{\delta e}{e} \right)_{\dot{m}} 
	+ \left( \frac{\delta e}{e} \right)_{tid} 
	+ \left( \frac{\delta e}{e} \right)_{mb} 
	+ \left( \frac{\delta e}{e} \right)_{gr}.
\end{equation}
I will discuss each of these mechanisms in detail below. The final two
terms in equation~\ref{eq:deles} are expected to occur on much longer time
scales than the first two terms and are thus ignored.

% % % % % % % % % % % % % % % % % % % % % % % % % % % % % % % % % % % 

\subsection{Evolution of the Donor Star Radius}

\subsubsection{Nuclear Evolution}

The equilibrium radius of a star depends on its total mass, as well as
the mass of its core produced by nuclear evolution. An expression for
this radius is given by Rappaport et~al.~\cite{rappaport95}:
\begin{equation}
\label{eqn:radeq}
\frac{R_{eq}}{R_\odot}  \,\simeq\,
0.85 \left(\frac{M_{don}}{M_\odot}\right)^{0.85} +
\frac{4950 M_c^{4.5}}{(1+4M_c^4)},
\end{equation}
where $M_c$ is the mass of the core expressed in solar units.

As the core mass grows through nuclear evolution, the radius of the
donor will also increase. Taking the derivative of
equation~\ref{eqn:radeq} and dividing by the donor radius gives the
fractional change in the radius due to nuclear evolution:
\begin{equation} \label{eqn:drr_nuc}
\left(\lgder{R_{don}}\right)_{nuc} = 
	\frac{\delta R_{eq}}{R_{don}} =
	\frac{R_{eq} - 0.85 (M_{don}/M_\odot)^{0.85}}{R_{don}}
        \left[ 4.5 - \frac{16 M_c^4}{(1+4M_c^4)} \right] 
	\lgder{M_c}
\end{equation}

The nuclear evolution of the core is straightforward to compute since
in a red giant the luminosity is generated from nuclear burning in a
thin shell surrounding a helium core. The change in core mass is
determined by the rate at which hydrogen is burned in a shell around
the helium core. In the low-mass giants being considered here,
hydrogen is fused into helium via the proton-proton chain, which
converts 0.7\% of the hydrogen mass into energy. Thus, the core gains
mass at a rate proportional to the luminosity of the donor
($L_{don}$). Becker gives an expression for this rate~\cite{becker97}:
\begin{equation}
\label{eqn:dcore_dt}
\dot{M}_c = 1.47{\times}10^{-11} f_c \left(\frac{L_{don}}{L_\odot}\right)
\left(\frac{X}{0.7}\right)^{-1}\;\;\mbox{\mspyr.}
\end{equation}
where X is the hydrogen abundance (mass fraction), taken to be
$\sim$0.75 and $f_c$ is an ad~hoc constant multiplicative factor of~2
to make this simple evolutionary model more consistent with the
results of more sophisticated codes.

An approximate analytic formula for the luminosity of a star on the
red giant branch is given by the following expression, adapted from
the result of Eggleton \cite{eggleton:book}:
\begin{equation}
\label{eqn:lumin}
\frac{L_{don}}{L_\odot}  \simeq
2 \left(\frac{M_{don}}{M_\odot}\right)^{4} + 
\frac{2{\times}10^5 (M_c/M_\odot)^6}{1+2.5 (M_c/M_\odot)^4+ 3 (M_c/M\odot)^5}. 
\end{equation}

% % % % % % % % % % % % % % % % % % % % % % % % 
\subsubsection{Thermal Readjustment Toward Equilibrium}

If the radius of the donor star differs from its thermal equilibrium
radius, it will adjust toward the equilibrium radius on a
thermal timescale. For example, if the critical potential lobe of the
star restricts its radius to be smaller than it would be if it were in
thermal equilibrium, the star will expand on a thermal time scale,
driving mass transfer.  The following relation is used to describe the
readjustment of the stellar radius of the donor on a thermal
(Kelvin-Helmholtz) timescale:
\begin{equation}
\label{eqn:drr_eq}
\left(\lgder{R_{don}}\right)_{th} \simeq 
	\frac{R_{eq} - R_{don}}{R_{don} \tau_{KH}}\;{\delta}t\;,
\end{equation}
where
\begin{equation}
\label{eqn:tau_kh}
\tau_{KH} \simeq \frac{3 G M^2}{7 R L} \simeq 1.5{\times}10^7
\left(\frac{M_{don}}{M_\odot}\right)^2
\left(\frac{R_{don}}{R_\odot}\right)^{-1}
\left(\frac{L_{don}}{L_\odot}\right)^{-1}
{\rm yrs}.
\end{equation}

% % % % % % % % % % % % % % % % % % % % % % % % 
\subsubsection{Adiabatic Response to Mass Loss}

A star will respond adiabatically to mass loss. Rappaport,
Joss, and Webbink~\cite{rappaport82} relate the resulting change of
the radius to the change in mass through a factor $\xi_{ad}$:
\begin{equation} \label{eq:delR_ad}
\left( \frac{\delta R_{don}}{R_{don}} \right)_{ad} = 
	\xi_{ad} \frac{\delta M_{don}}{M_{don}}.
\end{equation}
Di\,Stefano et~al.~\cite{distefano96} performed detailed evolution
studies with a Henyey-type code including mass loss. They found that
for stars with a low core mass ($M_{cor}<0.2 M_{\odot}$), the radius
shrinks when mass is removed, but for higher core mass, the mass loss
causes the star to expand (resulting in unstable transfer). They
parameterized their results for $\xi_{ad}$ with respect to core mass:
\begin{equation}
\label{eqn:def_xiad}
\xi_{ad} =\tilde{\xi}_{ad} \left[ 1 - \left(\frac{M_c}{\tilde{M}_c}\right)^2\right] , 
\end{equation}
where $\tilde{\xi}_{ad} = 4.0$ and $\tilde{M}_c = 0.2$ \msun.

% % % % % % % % % % % % % % % % % % % % % % % % % % % % % % % % % % % % % % % % 
\subsection{Evolution of Orbital Parameters}

\subsubsection{Tidal Evolution}
\label{sec:tides}

In a binary system, tidal friction will dissipate energy while
conserving the total angular momentum. In particular, in a system
where one star is compact (e.g. a neutron star), the sum of the total
orbital energy plus the rotational energy of the ``donor'' star
(actually not necessarily donating matter in this discussion) will
decrease while angular momentum is exchanged between the orbit and the
rotation of the star. In the weak friction model, it is assumed that
the star is tidally deformed into the equipotential surface it would
have formed a constant time $\tau$ ago in the absence of the tidal
friction, i.e., a tidal bulge that is offset from the line between the
centers of the stars. The rate of change of the orbital and rotational
parameters due to this tidal dissipation is given by Hut~\cite{hut81}
(using variables as defined above):
% Original version as in Hut:
%\begin{equation} \label{eq:dadt_tides}
%\frac{da}{dt} = -6 \frac{k}{T} q(1+q) \left(\frac{R_{don}}{a}\right)^8
%		\frac{a}{(1-e^2)^{15/2}} 
%		\left[ f_1(e^2) - (1-e^2)^{3/2} f_2(e^2)\frac{\Omega}{n}\right],
%\end{equation}
%\begin{equation} \label{eq:dedt_tides}
%\frac{de}{dt} = -27 \frac{k}{T} q(1+q) \left(\frac{R_{don}}{a}\right)^8
%		\frac{e}{(1-e^2)^{13/2}} 
%		\left[ f_3(e^2) - \frac{11}{8}(1-e^2)^{3/2} f_4(e^2)\frac{\Omega}{n}\right],
%\end{equation}
%\begin{equation} \label{eq:dspindt_tides}
%\frac{d\Omega}{dt} = 3 \frac{k}{T} \frac{q^2}{r_g^2} \left(\frac{R_{don}}{a}\right)^6
%		\frac{n}{(1-e^2)^{15/2}} 
%		\left[ f_2(e^2) - (1-e^2)^{3/2} f_5(e^2)\frac{\Omega}{n}\right],
%\end{equation}
% WARNING: HUT'S q IS DEFINED OPPOSITE THAT OF THIS CHAPTER, SO HAD TO INVERT
\begin{equation} \label{eq:dadt_tides}
\frac{da}{dt} = -6 \frac{G M_{don} k \tau}{R_{don}^3} \frac{1+q}{q^2} 
		\left(\frac{R_{don}}{a}\right)^8
		\frac{a}{(1-e^2)^{15/2}} 
		\left[ f_1(e^2) - (1-e^2)^{3/2} f_2(e^2)\frac{\Omega_{rot}}{\Omega_{orb}}\right],
\end{equation}
\begin{equation} \label{eq:dedt_tides}
\frac{de}{dt} = -27 \frac{G M_{don} k \tau}{R_{don}^3} \frac{1+q}{q^2}
		\left(\frac{R_{don}}{a}\right)^8
		\frac{e}{(1-e^2)^{13/2}} 
		\left[ f_3(e^2) - \frac{11}{8}(1-e^2)^{3/2} f_4(e^2)\frac{\Omega_{rot}}{\Omega_{orb}}\right],
\end{equation}
and
\begin{equation} \label{eq:dspindt_tides}
\frac{d\Omega_{rot}}{dt} = 3 \frac{G M_{don} k \tau}{R_{don}^3} 
		\frac{1}{q^2 r_g^2} \left(\frac{R_{don}}{a}\right)^6
		\frac{\Omega_{orb}}{(1-e^2)^{15/2}}
		\left[ f_2(e^2) - (1-e^2)^{3/2} f_5(e^2)\frac{\Omega_{rot}}{\Omega_{orb}}\right],
\end{equation}
where
\begin{displaymath} 
\begin{array}{l}
f_1(e^2) = 1 + \frac{31}{2}e^2 + \frac{255}{8}e^4+\frac{185}{16}e^6+\frac{25}{64}e^8 \\
f_2(e^2) = 1 + \frac{15}{2}e^2 + \frac{45}{8}e^4+\frac{15}{16}e^6 \\
f_3(e^2) = 1 + \frac{15}{4}e^2 + \frac{15}{8}e^4+\frac{5}{64}e^6 \\
f_4(e^2) = 1 + \frac{3}{2}e^2 + \frac{1}{8}e^4 \\
f_5(e^2) = 1 + 3 e^2 + \frac{3}{8}e^4 .
\end{array}
\end{displaymath}
Also, $q = M_{don}/M_{ns}$ is the mass ratio and $\Omega_{rot}$ and
$\Omega_{orb}$ are the rotational and mean orbital angular
velocities. The dimensionless radius of gyration, $r_g$, is related to
the moment of inertia of the donor star by $I=r_g^2 M_{don}
R_{don}^2$, and the apsidal motion constant k is also related to the
structure of the donor. A typical value for the radius of gyration is
$r_g^2 = 0.1$ for the low-mass relatively unevolved stars considered
here.

For stars with an outer convective zone (i.e.\ low-mass stars with
$M~\simlt$~1--2~\msun), Lecar gives an approximate expression for the product of
the apsidal motion constant and the tidal time lag~\cite{lecar76}:
\begin{equation} \label{eq:ktau}
k\tau \approx 
	25  \frac{\eta \lambda v_{conv} \rm{(km/s)}}{g/g_\odot}~\rm{s},
\end{equation}
where $g=G M / R^2$ is the gravitational acceleration. The convective
velocity, $v_{conv}$, is typically $\sim$1~km/s for a wide range of
convective stars.  Furthermore, $\eta$ and $\lambda$ are the
fractional mass and fractional depth respectively of the outer
convection zone of the donor star. Lecar gives a table of these two
quantities for main-sequence stars with effective temperatures of in
the range of $T$=5200--7200~K~\cite{lecar76}. To facilitate
interpolation and extrapolation of values for $\eta\lambda$, an
analytic function was fit to the table entries, resulting in the
following relation:
\begin{equation} \label{eq:etalambda}
\log_{10} (\eta \lambda) = \left\{ \begin{array}{ll}
  10.66 - 2.7907\e{-6}\,T^2 + 7.5778\e{-10}\,T^3 - 5.899\e{-14}\,T^4 & 
	T\geq5000 \rm{K}\\
  -1.25375 & T < 5000 K .
\end{array} \right. 
\end{equation}
For stars with effective temperatures below 5000~K, the tidal
circularization time scale is very short and systems containing such
stars will circularize before other mechanisms affect the orbital
evolution. In contrast, for stars with effective temperatures above
7200~K, the time scale is very long and tidal dissipation will not play
a significant role in the evolution of the orbital parameters.

% % % % % % % % % % % % % % % % % % % % % % % 
\subsubsection{Effect of Mass Transfer on the Orbit}
\label{sec:del_e_mdot}

The effect of mass transfer in a circular orbit is relatively simple
to calculate: if angular momentum is conserved, then the change in the
semimajor axis of the orbit depends uniquely on the change in the
masses of the stars. Mass transfer onto a star less massive than the
donor star will decrease their separation, and transfer onto a more
massive star will increase the orbital separation.  In an eccentric
orbit, the angular momentum depends on both the eccentricity as well
as the semimajor axis, and thus conservation of angular momentum is
insufficient to determine how both orbital parameters change. So,
Eggleton has derived a prescription for the effect of mass transfer on
the orbital eccentricity~\cite{eggleton:book}.  Mass transfer that
occurs at orbital angle $\theta$ (position angle of one star relative
to the other, where $\theta=0$ corresponds to periastron) results in a
change in eccentricity given by:
\begin{equation} \label{eq:del_e(theta)}
\delta e = - 2\,(1-q)(\cos e + e)\frac{\delta M_{don}}{M_{don}}.
\end{equation}
When mass is transferred at a constant rate around the entire orbit,
the average change in eccentricity is zero, since over a complete
orbit $\ave{\cos\theta} = -e$.  In the other extreme, when mass
transfer occurs only at periastron (as in a highly eccentric orbit), 
\begin{equation} \label{eq:del_e_peri}
\delta e = - 2\,(1-q)(1+e)\frac{\delta M_{don}}{M_{don}}.
\end{equation}

Since $\delta M_{don} < 0$ during mass transfer and $q = M_{don}/M_{ns}$,
equation~\ref{eq:del_e_peri} shows that mass transfer will decrease
the eccentricity when $M_{don}>M_{ns}$ and will increase the
eccentricity when the mass ratio is reversed.

% % % % % % % % % % % % % % % % %
\subsubsection{Magnetic Braking}

{\em This section on magnetic braking is reproduced from
Becker~\cite{becker97}.}

Most stars have some magnetic field, and the interaction of the field
lines with the ejected matter increases the effective ``lever arm''
and, therefore, the amount of specific angular momentum carried away by
the stellar wind. Thus a small amount of mass loss can, in principle,
extract a significant amount of angular momentum from the rotation of
the star.  If tidal interactions between the donor and companion keep
the donor star co-rotating with the orbit, then magnetic braking can
result in angular momentum losses from the orbit.

Skumanich \cite{skumanich72} and Smith \cite{smith79} independently
determined a rotation rate/age relation for isolated stars, mostly
with masses about 1 \msun, in clusters of various ages.  From this
relation an empirical magnetic braking torque was inferred. Verbunt
\& Zwaan~\cite{verbunt81} applied this to binary systems as a
way to drive angular momentum loss.  I use a slightly modified version
of this braking law as given by Rappaport et~al.~\cite{rappaport83}:
\begin{equation}
\label{eqn:djj_mb}
\left(\lgder{J}\right)_{mb} = -1.73{\times}10^{-27} f_{mb} 
\left[\frac{R_\odot^4 (R_{don}/R_\odot)^\gamma M_T^{1/3}}
{G^{2/3} M_{wd} P^{10/3}}\right]\;{\delta}t\;,
\end{equation}
where $\gamma$ is the parameterized magnetic braking index (typically
$\gamma=4$), $f_{mb}$ is a multiplicative constant of order unity, and
``cgs units'' are used.  In the process of applying magnetic braking to
binaries, a number of liberties were taken in applying the results of
Skumanich and Smith.  The use of the magnetic braking law in binaries
extends to shorter periods and different masses for stars than in the
original data. The parameter $\gamma$ is used in place of any real
knowledge of how the braking depends on stellar radius since most of
the stars used to develop the original relation were G stars with more
or less the same radius.

\subsubsection{Gravitational Radiation}

{\em This section on gravitational radiation is reproduced from
Becker~\cite{becker97}.}

Gravitational radiation arises from a mass distribution with a time
varying quadrupole moment, such as a pair of stars orbiting each
other.  For such a simple geometry (a circular orbit), the rate of
angular momentum loss is given by the Einstein quadrupole radiation
formula:
\begin{equation}
\label{eqn:djj_gr}
\left(\lgder{J}\right)_{gr} =
-\frac{32}{5}(4\pi^2)^{4/3}\frac{G^{5/3}}{c^5}\frac{M_{wd}M_{don}}{M_T^{1/3}
P^{8/3}}\;{\delta}t\;, 
\end{equation}
where $P$ is the orbital period \cite{landau_lifshitz}.  For
the systems we will consider, with periods typically larger than 10
hours, gravitational radiation alone drives mass transfer too slowly
to have any significant effect.

% % % % % % % % % % % % % % % % % % % % % % % % % % % % % % % % % % % % % % % % 
\subsection{Stable Mass Transfer}

Mass transfer can act as a sort of regulator, preventing the donor
radius from exceeding the critical potential lobe radius (stable
transfer). If mass transfer is unable to prevent the donor from
overfilling the critical potential lobe, transfer will be unstable.
If mass transfer (or another mechanism) eventually results in the
donor underfilling its critical potential lobe, mass transfer will
stop.  Thus, during stable mass transfer the donor radius can
continuously adjust so that it always equals the Roche lobe radius.
Based on the assumption of stable transfer, I will now derive the
expression that governs the rate of mass transfer.

For the case of circular orbits, the Roche-lobe radius is related to
the semimajor axis of the orbit through a function $f(q)$ of the mass
ratio $q = M_{don}/M_{ns}$:
\begin{equation} \label{eq:roche_a}
R_{L} = f(q)\cdot a.
\end{equation}
An analytic approximation of this
function has been derived by P.~Eggleton~\cite{eggleton:roche}:
\begin{equation} \label{eq:f(q)}
f(q) = \frac{0.49 q^{2/3}}{0.6 q^{2/3} + \ln(1+q^{1/3})}.
\end{equation}
For non-zero eccentricity, a plausible approximation to the effective
critical potential lobe size is obtained by replacing the semimajor
axis in equation~\ref{eq:roche_a} with the instantaneous separation
$d$.  A more precise measure for size of the critical potential lobe
can be obtained by calculating the location of points on the critical
surface of an effective potential that accounts for the eccentricity
of the orbit, e.g., as given by Avni~\cite{avni76} (see
section~\ref{sec:eccRochePoten}). However, for simplicity and speed in
the evolution code, I will use equation~\ref{eq:roche_a} with the
instantaneous separation.

In a Keplerian orbit, the separation as a function of angle
$(\theta)$ from periastron is given by
\begin{equation}
d = \frac{a\,(1-e^{2})}{1 + e\,\cos\theta}.
\end{equation}
Thus at periastron, $d = a\,(1-e)$.  As described above, mass transfer
via Roche-lobe overflow requires that the radius of the donor star
equal the radius of the critical potential lobe at periastron radius:
\begin{equation} \label{eq:RdonRL}
R_{don} \equiv R_{L} = f(q) \cdot a \cdot (1-e).
\end{equation}
Taking the logarithmic derivative of equation~\ref{eq:RdonRL} shows
how changes in these parameters are coupled:
\begin{equation} \label{eq:delRdon}
\frac{\delta R_{don}}{R_{don}} = 
	(1+q) \frac{d\,\ln f}{d\,\ln q} \frac{\delta M_{don}}{M_{don}} 
	+ \frac{\delta a}{a} - \frac{\delta e}{1-e}.
\end{equation}
The logarithmic derivative of $f(q)$ is calculated from
equation~\ref{eq:f(q)} to be:
\begin{equation}\label{eq:dlnf_dlnq}
\frac{d \ln f}{d \ln q} = \frac{2}{3} \left\{ 
\frac{ \ln(1+q^{1/3}) - 0.5\frac{q^{1/3}}{1+q^{1/3}}}
{0.6q^{2/3} + \ln(1+q^{1/3})}
\right\}.
\end{equation}

Since I have chosen to work with orbital angular momentum rather than
the semimajor axis, the logarithmic derivative of the orbital angular
momentum (equation~\ref{eq:Jorb}) is needed:
\begin{equation} \label{eq:delJ}
\frac{\delta J}{J} = \frac{\delta M_{ns}}{M_{ns}} 
		+ \frac{\delta M_{don}}{M_{don}}
		- \frac{1}{2}\frac{\delta M_{T}}{M_{T}} 
		+ \frac{1}{2}\frac{\delta a}{a}
		- \frac{e\,\delta e}{1-e^2}.
\end{equation}
The masses in this relation can be re-expressed in terms of only the
donor mass and the mass ratio:
\begin{equation} \label{eq:delJ2}
\frac{\delta J}{J} = (1 - q)\,\frac{\delta M_{don}}{M_{don}} 
		+ \frac{1}{2}\frac{\delta a}{a}
		- \frac{e\,\delta e}{1-e^2}.
\end{equation}

Combining equations~\ref{eq:delRdon} and~\ref{eq:delJ2} to eliminate
the semimajor axis gives:
\begin{equation} \label{eq:delJ3}
\frac{\delta J}{J} = 
	\left[ 1 - q - \frac{1+q}{2}\frac{d\,\ln f}{d\,\ln q} \right]\,
	\frac{\delta M_{don}}{M_{don}} 
	+ \frac{1}{2}\frac{\delta R_{don}}{R_{don}}
		+ \frac{1}{2}\frac{e}{(1+e)}\frac{\delta e}{e}.
\end{equation}

Using equations \ref{eq:delRs}, \ref{eq:delJs}, and \ref{eq:deles} to
explicitly include each source of evolution into~\ref{eq:delJ3}:
\begin{eqnarray} \label{eq:del_mdon_allcomps}
	  2 \left( \frac{\delta J}{J} \right)_{mb} 
	+ 2 \left( \frac{\delta J}{J} \right)_{gr}
	+  2 \left( \frac{\delta J}{J} \right)_{tid} 
	- \frac{e}{(1+e)} %\left\{
	   \left( \frac{\delta e}{e} \right)_{tid} 
	- \left( \frac{\delta R_{don}}{R_{don}} \right)_{nuc} 
	- \left( \frac{\delta R_{don}}{R_{don}} \right)_{th}
= \nonumber \\ 
	2\left[1 - q - \frac{1+q}{2}\,\frac{d\, \ln f}{d\, \ln q}\right] \, 
		\frac{\delta M_{don}}{M_{don}} 
	+ \left( \frac{\delta R_{don}}{R_{don}} \right)_{ad}
	+ \frac{e}{(1+e)} \left( \frac{\delta e}{e} \right)_{\dot{m}} 
 , \mbox{\ }
\end{eqnarray}
where all the explicitly time-dependent terms have been moved to the
left hand side, and terms directly proportional to mass loss appear on
the right hand side.

Now using equations \ref{eq:delR_ad} and \ref{eq:del_e(theta)} to
replace the last two terms in equation~\ref{eq:del_mdon_allcomps} with
their dependencies on mass transfer, we find
\begin{eqnarray} \label{eq:del_mdon_final}
	  2 \left( \frac{\delta J}{J} \right)_{mb} 
	+ 2 \left( \frac{\delta J}{J} \right)_{gr}
	+  2 \left( \frac{\delta J}{J} \right)_{tid} 
	- \frac{e}{(1+e)} %\left\{
	   \left( \frac{\delta e}{e} \right)_{tid} 
	- \left( \frac{\delta R_{don}}{R_{don}} \right)_{nuc} 
	- \left( \frac{\delta R_{don}}{R_{don}} \right)_{th}
= \nonumber \\ 
	\left[\, \xi_{ad} + 2(1 - q) - 2\frac{ \ave{\cos\theta}+e}{1+e} (1 - q)
	       - (1+q)\frac{d\, \ln f}{d\, \ln q}\right] \, 
	    \frac{\delta M_{don}}{M_{don}} .
\end{eqnarray}

The time-dependent left-hand side of this equation, referred to as the
``numerator'', governs the time scale for mass transfer. The
right-hand side depends on the mass transferred, rather than time.
The term in square brackets on the right-hand side (the
``denominator''), determines if mass transfer is stable.  As mentioned
in section~\ref{sec:del_e_mdot}, mass transfer limited to periastron 
results in $\ave{\cos\theta} = 1$, so that the second and third terms
in the denominator cancel. For transfer around the entire orbit, when
$\ave{\cos\theta} =-e$, the third term of the denominator vanishes.

All of the terms in the numerator are time dependent and can be
calculated from the expressions given in the discussion above. The
expressions given for gravitational radiation and magnetic braking are
strictly correct only for circular orbits. However, these mechanisms
of angular momentum loss are small in a system with a relatively long
orbital period such as Cir~X-1 (days), and thus the circular-orbit
approximation is sufficient to estimate the magnitude of these
quantities.

% % % % % % % % % % % % % % % % % % % % % % % % % % % % % % % % % % % % % % % % 
% % % % % % % % % % % % % % % % % % % % % % % % % % % % % % % % % % % % % % % % 
\section{Binary Evolution Code}

\subsection{Code Algorithm}

The parameters that need to be specified at the start of the evolution
are the inital donor mass and the eccentricity and semimajor axis of
the orbit. The initial core mass can be varied, but was generally
taken to be zero and evolved as part of the code. The initial neutron
star mass was taken to be 1.4~\msun\ in all cases. The initial spin of
the donor was assumed to be slow relative to the orbital frequency
($\Omega_{rot}/\Omega_{orb}=\mscinot{1}{-3}$ for concreteness) and
allowed to spin-up via tidal effects.  The donor star was assumed to
start in equilibrium ($R_{don}=R_{eq}$).  It generally underfilled its
initial periastron critical potential lobe, but in some cases just
filled the critical lobe.  Mass transfer started immediately in the
latter case.

A minimum of 10 time steps are used per decade of
time. Smaller time steps are chosen as necessary so that no parameter
changes by more than 0.25\% during each step, and $10^7$~yr is the
largest time step allowed. For each time step:
\begin{enumerate}
\item The new core mass is determined from the simplified stellar 
evolution code.
\item The equilibrium radius of the donor is determined from its 
	total and core mass.
\item The change in donor radius due to nuclear evolution, 
thermal adjustment, and adiabatic response to mass loss is calculated.
\item The effects of tidal dissipation, magnetic braking, and 
	gravitational radiation on the orbit are calculated.
\item If the donor star fills its critical potential lobe, determine 
the amount of mass transfer necessary to keep the donor radius no
larger than the radius of critical potential lobe (calculate numerator
and denominator from equation~\ref{eq:del_mdon_final}).
\item Update all the system parameters based on the above changes.
\end{enumerate}
The evolution is stopped if any of the following conditions occur:
\begin{enumerate}
\item The donor star is stripped ($M_{don}-M_{cor} < 0.05$~\msun).
\item The mass of the donor star drops below 0.1~\msun.
\item The denominator of equation~\ref{eq:del_mdon_final} is negative (mass 
	transfer is unstable).
\item The binary circularizes ($e<0.01$), and thus is no longer a candidate 
	to produce a system that resembles Cir~X-1.
\item The binary comes close to being unbound. When 
$a$ becomes extremely large and $e$ approaches unity, the system has
become nearly unbound or has unphysical parameters.  Thus I stop the
code when $e>0.99$ and refer to such systems as being ``unbound''.
\item More than 10 billion years have elapsed (longer than the age 
of most stars in the Galaxy).
\end{enumerate}

% % % % % % % % % % % % % % % % % % % % % % % % % % % % % % % % % % % % % % % % 
\subsection{Results}

As mentioned in section~\ref{sec:del_e_mdot}, mass transfer tends to
decrease the eccentricity while the donor star is more massive than
the neutron star. In some cases, this can lead to circularization of
the orbit. However, in many cases mass transfer eventually inverts the
mass ratio (or $M_{don}<M_{ns}$ to begin with) so that the
eccentricity and the semimajor axis are increased during continued
mass transfer, often nearly unbinding the system.  Systems that start
with the donor star in contact with its critical potential lobe
transfer mass at a high rate and usually unbind (see
equation~\ref{eq:del_e_peri}) in $\sim$\tento{8} years.

\begin{figure}
\begin{centering}
\PSbox{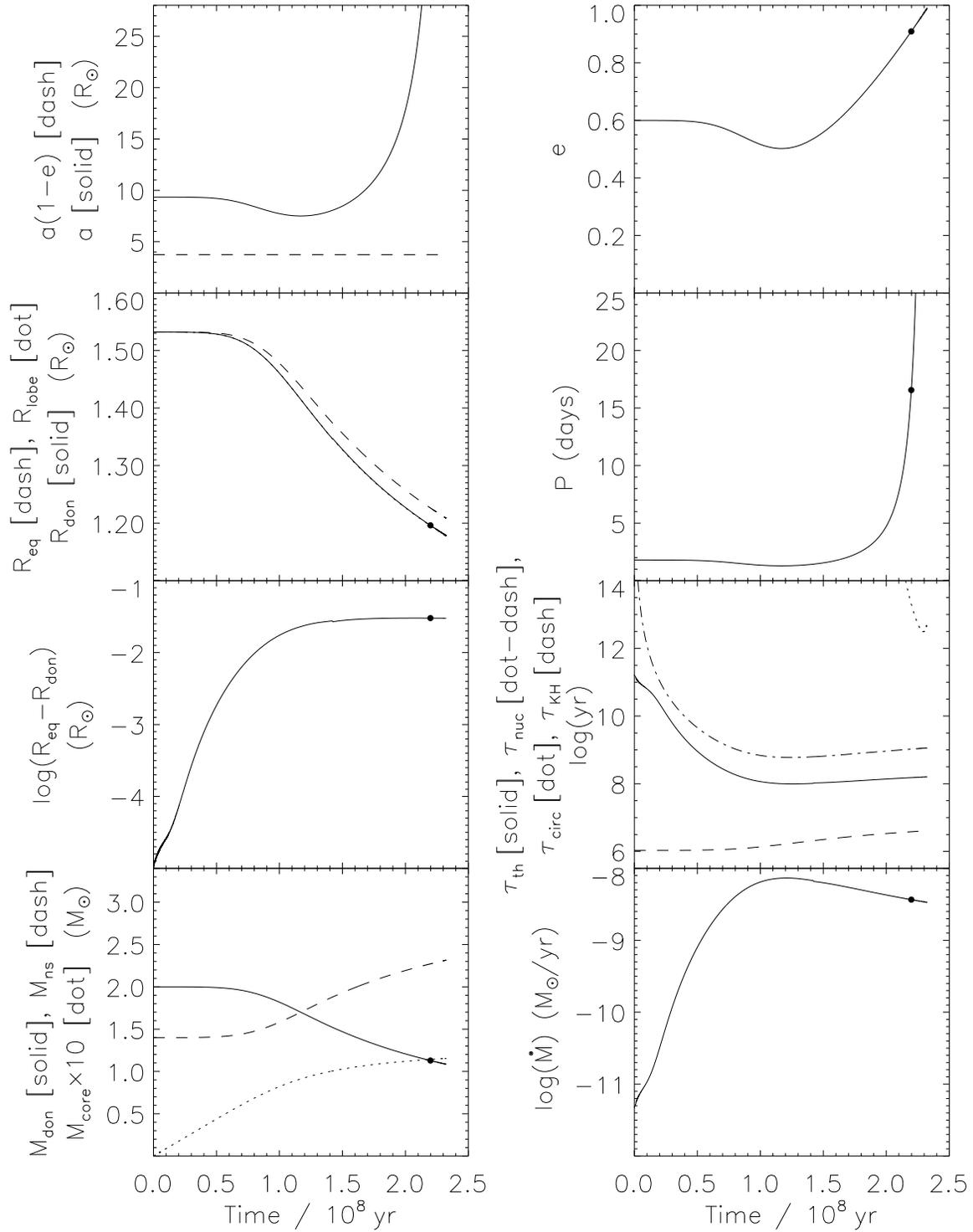 
  hoffset=-20 voffset=-14 hscale=90 vscale=90}{5.85in}{7.5in}
\caption{
Results of a binary evolution calculation for a system started with
the donor star just filling its critical potential lobe, with initial
values: $M_{don}=2\,{\rm M}_\odot$, $a=9.34\,{\rm R}_\odot$, and
$e=0.6$. The donor remains in contact with its critical potential lobe
throughout the evolution; thus, the curve for $R_{lobe}$ coincides
with the curve for $R_{don}$.  A solid dot on some of the curves marks
the time corresponding to an orbital period of 16.55~days.}
\label{fig:evolve_md2_mc0_e0.6_a9}
\end{centering}
\end{figure}

Figure~\ref{fig:evolve_md2_mc0_e0.6_a9} shows the binary evolution
calculation results for a 2~\msun\ donor that starts with donor star
in contact with its critical potential lobe. In this situation, the
thermal expansion of the donor towards the equilibrium radius
dominates over other mechanisms, as is evident from the panel in
Figure~\ref{fig:evolve_md2_mc0_e0.6_a9} showing the various time
scales ($\tau_{th} \equiv R_{don}/(\dot{R}_{don})_{th}$,
$\tau_{nuc} \equiv R_{don}/(\dot{R}_{don})_{nuc}$, $\tau_{circ}
\equiv 1/\dot{e}_{tid}$, and $\tau_{KH}$ is given by
equation~\ref{eqn:tau_kh}). (The thermal expansion time scale is
related to the Kelvin-Helmholtz time scale, but is longer by a factor
shown in equation~\ref{eqn:drr_eq}.)
%$\tau_{mb} \equiv J/\dot{J}_{mb}$ 

Mass transfer initially decreases the eccentricity and semimajor axis
(such that periastron separation remains unchanged since transfer
occurs only at periastron). After \scinot{1.2}{8}~yr, 0.3~\msun\ has
been transferred, making the neutron star more massive than the
donor. Continued mass transfer then gradually increases the
eccentricity and semimajor axis (again such that periastron distance,
$a(1-e)$, remains approximately constant) until the system is nearly
unbound (after about \scinot{3}{8}~yr). 

This system maintains a moderately high mass transfer rate during most
of its entire lifetime, and its period increases to the level of
Cir~X-1 (16.55~d) and beyond (the time corresponding to a 16.55-d
period is indicated with a dot on several of the curves in
Figure~\ref{fig:evolve_md2_mc0_e0.6_a9}).  Thus this system might go
through a phase that resembles Cir~X-1. At the point where the period
is near 16~days, its eccentricity is quite high, about 0.9, and the
mass transfer rate is \scinot{3.5}{9}~\msun/yr.

\begin{figure}
\begin{centering}
\PSbox{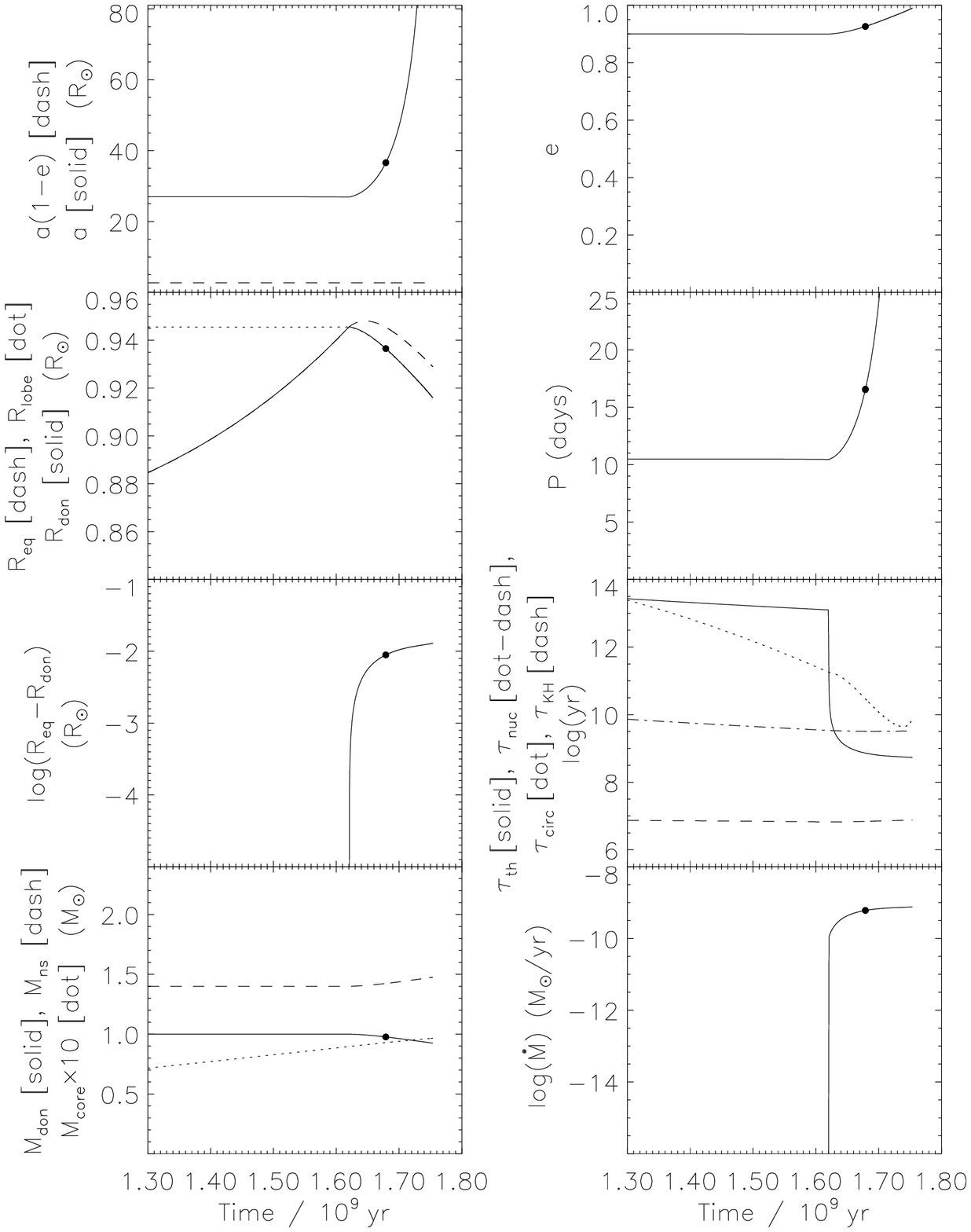 
  hoffset=-20 hscale=90 vscale=90}{5.85in}{7.65in}
\caption{
Results of binary evolution calculation for a system started with the
donor underfilling its critical potential lobe, with initial values:
$M_{don}=1\,{\rm M}_\odot$, $a=27\,{\rm R}_\odot$, and $e=0.9$.  
}
\label{fig:evolve_md1_mc0_e0.9_a27}
\end{centering}
\end{figure}

The results of a calculation starting with a 1~\msun\ donor initially
underfilling its critical potential lobe are shown in
Figure~\ref{fig:evolve_md1_mc0_e0.9_a27}. This system evolves
uneventfully for over a billion years before the nuclear evolution of
the core eventually brings the donor into contact with its critical
potential lobe.  (The time scales for magnetic braking and tides to
affect the orbit are much longer than the nuclear time scale for most
of the system's lifetime.) Once in contact, the donor begins mass
transfer and continues for about \scinot{1.3}{8}~yr until the system
is nearly unbound.  This system would also go through a phase of
moderate mass transfer ($\sim$\tento{-9}~\msun/yr) and high
eccentricity with a period near 16~days.

%%% Results as a function of separation

\begin{figure}
\begin{centering}
\PSbox{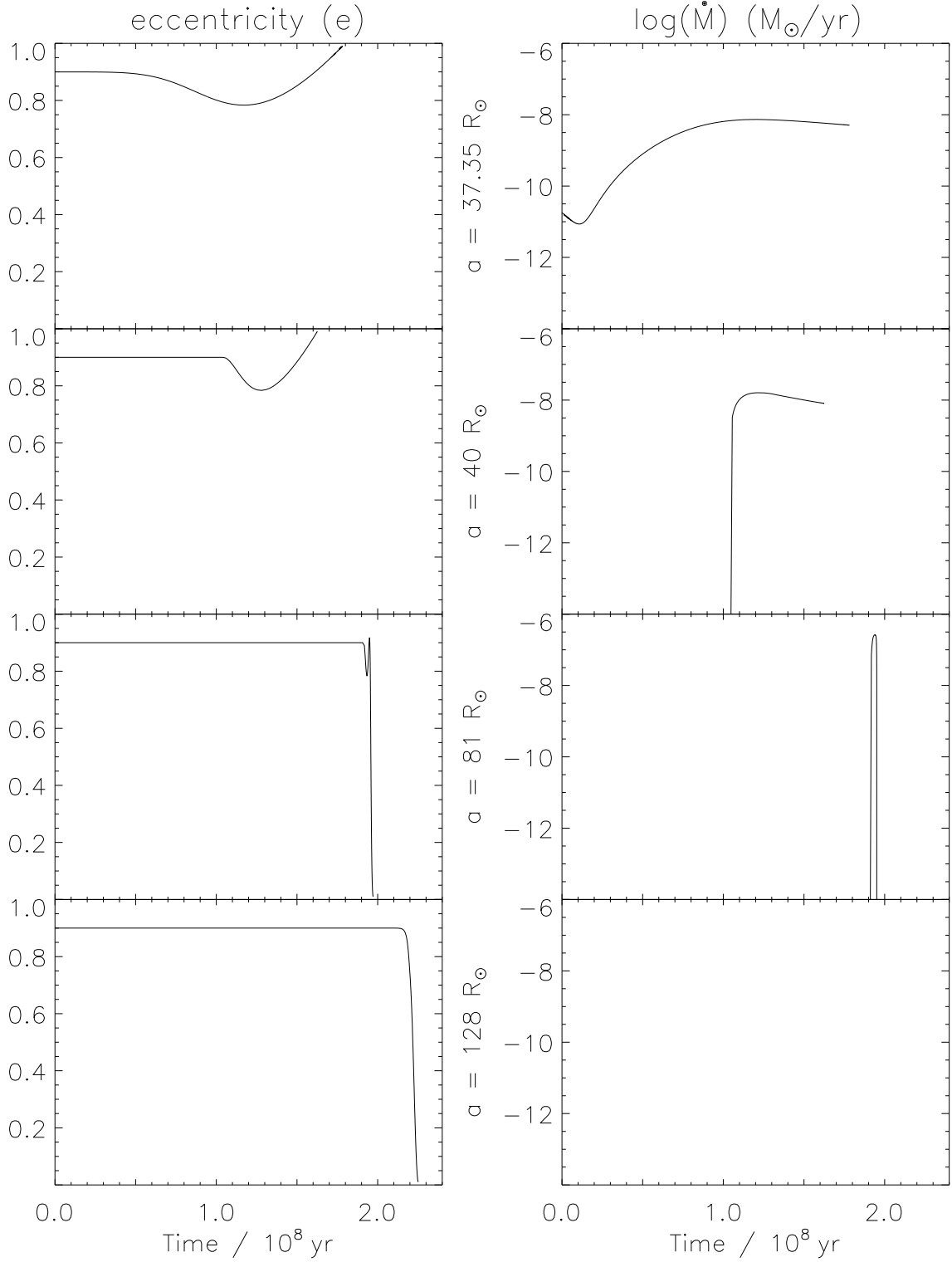 
  hoffset=-20 voffset=-7 hscale=90 vscale=90}{5.85in}{7.65in}
\caption{
Eccentricity and mass transfer rate versus time for binaries with
initial values $M_{don}=2$~\rsun, $e=0.9$, and semimajor axes ($a$)
between 37 and 128~\rsun.
}
\label{fig:evolve5emdot}
\end{centering}
\end{figure}

The results for mass transfer depend strongly on the size of the
critical potential lobe, and thus depend on the separation of the
stars at periastron ($a(1-e)$).  Here I illustrate this effect by
holding $e$ constant and varying $a$. Figure~\ref{fig:evolve5emdot}
shows the evolution of eccentricity and mass transfer rate for four
systems which differ only in their initial semimajor axes (initially,
$M_{don}=2~$\msun\ and $e=0.9$).
For the system with the smallest orbit ($a = 37$~\rsun), the donor
star initially just fills its critical potential lobe, similar to the
system in Figure~\ref{fig:evolve_md2_mc0_e0.6_a9}, and thus transfers
mass continuously until it unbinds after about \scinot{1.8}{8}~yr.

The second system in Figure~\ref{fig:evolve5emdot} ($a = 40$~\rsun)
initially underfills its critical potential lobe, similar to that in
Figure~\ref{fig:evolve_md1_mc0_e0.9_a27}. Nuclear evolution eventually
brings the donor into contact, at which point mass transfer
begins. Unlike the system in Figure~\ref{fig:evolve_md1_mc0_e0.9_a27},
the donor in this binary is initially more massive than the neutron
star, so the eccentricity dips before ultimately increasing towards
unity.

Skipping the third binary for the moment, the final system in
Figure~\ref{fig:evolve5emdot} has a large semimajor axis ($a =
128$~\rsun) and the donor is far from filling its critical potential
lobe. The radius of the donor increases due to nuclear evolution of
the core, resulting in a cooler effective temperature. The tidal time
scale strongly depends on the temperature through the convection-shell
parameters $\eta$ and $\lambda$ (see section~\ref{sec:tides}).  Thus,
the tidal time scale drops rather sharply after
$\sim$\scinot{2}{8}~yrs, resulting in rapid circularization of the
orbit. The donor star never fills its critical potential lobe; thus,
no mass transfer occurs.

The third binary in Figure~\ref{fig:evolve5emdot} ($a = 81$~\rsun) is
intermediate between the second and fourth. It initially underfills
its critical potential lobe, but comes into contact though nuclear
evolution. The eccentricity initially dips and then rises, as in the
second system. However, the mass transfer eventually results in
reduced tidal time scale (shorter than in the fourth system). Tidal
effects then rapidly reduce the eccentricity and the orbit
circularizes, as in the fourth system. Tidal circularization also
approximately doubles the periastron separation, pulling the donor out
of contact with its critical potential lobe and turning off mass
transfer about \scinot{4}{6}~yrs after it began.

% % % % % % % % % % % % % % % % % % % % % % % % % % % % % % % % % % % % % % % % 
 % % % % % % % % % % % % % % % % % % % % % % % % % % % % % % % % % % % % % % % 

\section{Monte-Carlo Population Synthesis}

\subsection{Systems Resembling Cir~X-1}

The results of individual binary evolution calculations show that
systems resembling Cir~X-1 can be produced. However, an entire
population of LMXBs must be studied in in order to determine the
likelihood of actually observing such a system.

Terman, Taam, \& Savage~\cite{terman96} have conducted a population
synthesis study of LMXBs, following an initial distribution of
zero-age binaries (with the more massive star $>8$~\msun) from the
main sequence, through a common-envelope phase, to just after the
subsequent supernova explosion of the more evolved star. Systems that
remain bound after the supernova explosion and do not immediately
merge produce a population of low-mass main sequence stars
($\simlt3$~\msun) with a neutron-star companion (pre-LMXBs) that are
expected to eventually transfer mass through evolution of the orbit
and/or the companion star.

Using the binary evolution code described in the previous section, I
have conducted a Monte-Carlo simulation of the evolution of pre-LMXBs
throughout their entire lifetime as potential eccentric LMXBs.  The
initial population was drawn from the pre-LMXB population derived by
Terman, Taam, \& Savage~\cite{terman96}, which yielded a relatively
uniform distribution of systems in a specific range of $e$ and $\log
a$.  Thus, I chose the eccentricity from a linear distribution in the
range $e=0.01$ to $0.99$, and the semimajor axis from a logarithmic
distribution in the range $a=1.25$~\rsun\ to $400$~\rsun. Combinations
outside the pre-LMXB distribution of Terman, Taam, \& Savage were
discarded. The initial neutron-star mass was fixed to be $1.4
M_{\odot}$ in all cases. Donor star masses were randomly chosen in the
range $M_{don}=0.8$ to $3 M_{\odot}$, and all donors were assumed to
initially have zero core mass. Donor masses that corresponded to an
equilibrium radius larger than the periastron critical potential lobe
size were discarded as being systems that merge immediately or
otherwise fail to contribute to the LMXB population

For this study, \tento{5} systems were evolved as described in the
previous section. The following criteria were established to classify
a system as one that resembled Cir~X-1: an orbital period within
$\sim$20\% of 16.55~days ($13~{\rm d} < P < 20~{\rm d}$), a high mass
transfer rate ($10^{-9}~{\rm M}_\odot$/yr $< \dot{M} < 10^{-7}~{\rm
M}_\odot$/yr), and at least a moderate eccentricity ($e > 0.1$).

Cir~X-1 is 6--10~kpc away, yet still quite bright in X-rays, so any
similar such system should also be visible from almost anywhere in the
Galaxy. Since there is only one system detected with parameters
meeting the above criteria, there is probably only about one such
system in the Galaxy at this time.

Of the \tento{5} systems, 42\% were discarded immediately due to the
donor being too large for the orbit. Of the remaining systems,
(pre-LMXBs) 15\% resembled Cir~X-1 at some point in their lifetime. On
average, each of the pre-LMXBs eventually resembled Cir~X-1 for
\scinot{4.0}{5}~yrs. Terman, Taam, \& Savage~\cite{terman96} predict a
pre-LMXB birth rate of $\sim$\scinot{3}{-6}--\tento{-5} yr$^{-1}$ in
the Galaxy.  Multiplying the predicted birth rate by the average
lifetime in a state resembling Cir~X-1 results in an expectation of
1--4 systems similar to Cir~X-1 in the Galaxy at any given time.
Thus, the fact that Cir~X-1 exists but is unique is consistent with
binary evolution calculations.

\subsection{Predicted System Parameters for Cir~X-1}

In addition to determining the lifetime of systems that resemble Cir~X-1,
the Monte Carlo simulation was also used to study the distribution of
system parameters corresponding to 16.55 day orbital periods and
$10^{-9}~{\rm M}_\odot$/yr $< \dot{M} < 10^{-7}~{\rm M}_\odot$/yr.

\begin{figure}
\begin{centering}
\PSbox{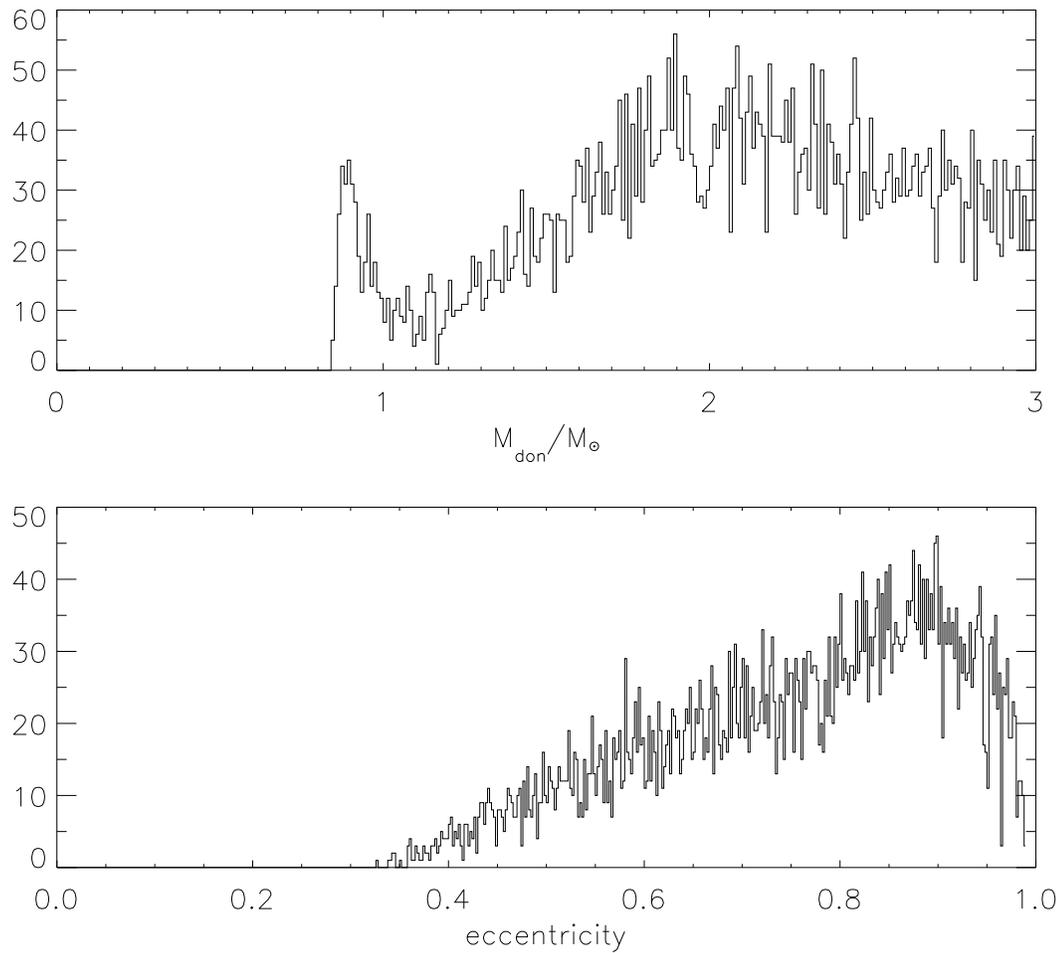
  hoffset=-20 hscale=90 vscale=90 voffset=-200}{5.85in}{5.35in}
\caption{
Histograms of number of binaries versus {\em initial} $M_{don}$ and $e$ which
eventually reach $P=16.55$~d and 
$\mscinot{1}{-9}~{\rm M}_\odot$/yr $ < \dot{M} <
\mscinot{1}{-7}~{\rm M}_\odot$/yr.
The y-axis normalization is arbitrary.}
\label{fig:evolve_cirx1_init} 
\end{centering}
\end{figure}

Histograms of the number of binaries versus the {\em initial} donor
mass and eccentricity for binaries that eventually reach $P=16.55$~d
and $10^{-9}~{\rm M}_\odot$/yr $<
\dot{M} < 10^{-7}~{\rm M}_\odot$/yr are shown in 
Figure~\ref{fig:evolve_cirx1_init}. The initial donor mass
distribution shows that most progenitors of systems resembling Cir~X-1
have donor stars more massive than the neutron star. As I will show
below, these systems come into contact and transfer mass due to the
nuclear evolution of the donor. There is also smaller population of
low-mass systems (the peak below 1~\msun) which evolve into contact
through angular momentum losses (tides and magnetic braking).  The
initial eccentricity distribution includes all values above $e=0.33$
and is skewed toward higher eccentricity, peaking near $e=0.9$.

\begin{figure}
\begin{centering}
\PSbox{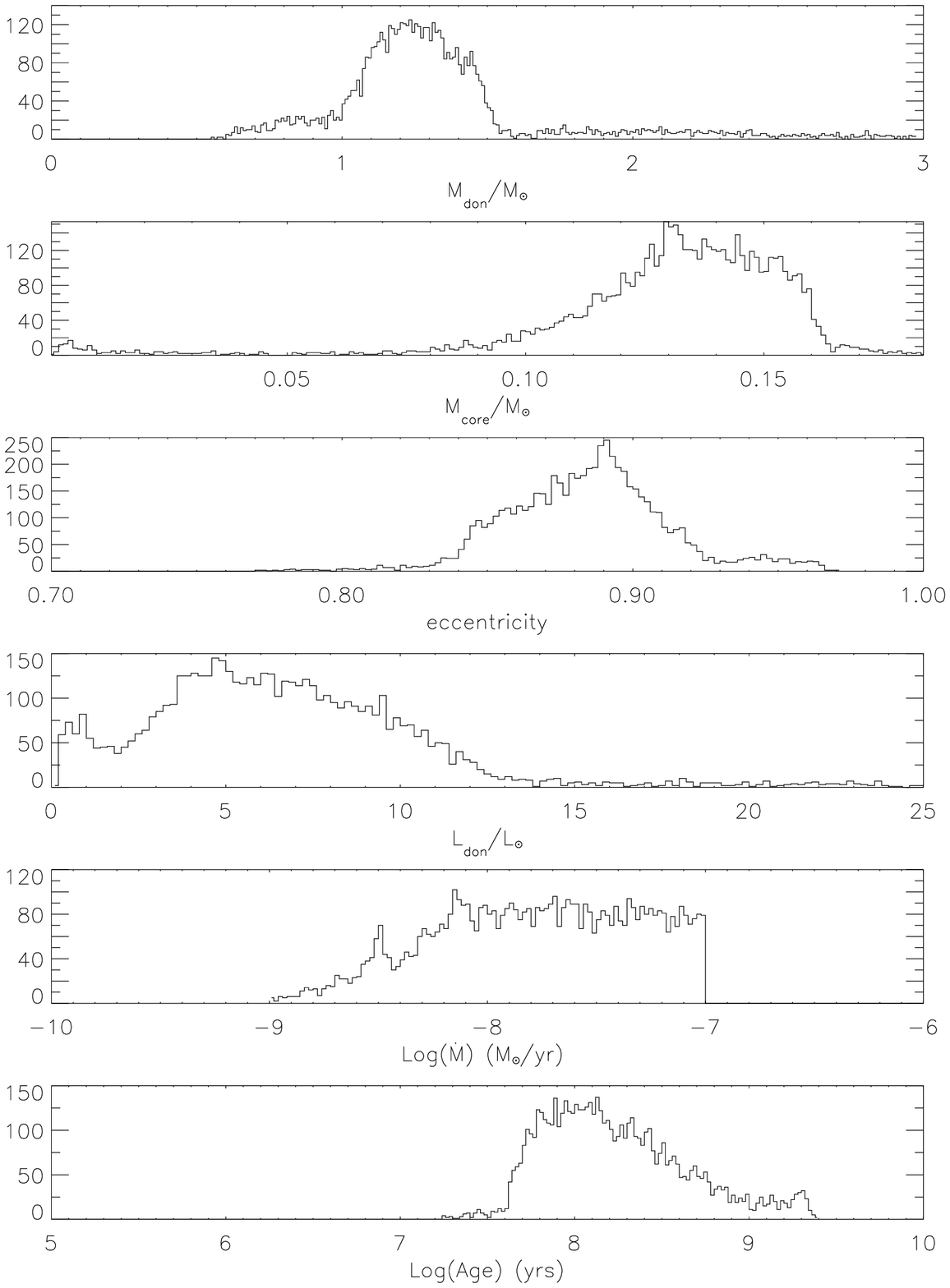
  hoffset=-20 hscale=90 vscale=90}{5.85in}{7.65in}
\caption{
Histograms of the number of binaries versus various system parameters
at the time when systems resemble Cir~X-1, i.e., $P=16.55$ d and
$\mscinot{1}{-9}~{\rm M}_\odot$/yr $ <
\dot{M} < \mscinot{1}{-7}~{\rm M}_\odot$/yr. 
The y-axis normalization is arbitrary.}
\label{fig:evolve_cirx1_final} 
\end{centering}
\end{figure}

Figure~\ref{fig:evolve_cirx1_final} shows histograms of the number of
systems versus various system parameters for binaries {\em at the time
they reach a state resembling Cir~X-1}: $P=16.55$~d and $10^{-9}~{\rm
M}_\odot$/yr $<\dot{M} < 10^{-7}~{\rm M}_\odot$/yr.  
The donor mass distribution peaks at about 1.25~\msun\ and is almost
entirely confined to $M_{don} < 1.5$~\msun.  This is a reflection of
the fact that most systems that reach 16.55~d periods during mass
transfer are in the stage of increasing orbital period, which can
occur only after the donor has become less massive than the neutron
star.  The core mass distribution shows that most systems have
developed a moderate core of 0.10--0.16~\msun, indicating that nuclear
evolution is driving transfer in most of the systems. A small fraction
of donor stars with a low core mass correspond to the lower-mass
systems which evolve mainly through angular momentum losses.

The eccentricity distribution is constrained to high values, showing a
peak between $e=0.83--0.93$. A tail extending above $e=0.93$
corresponds to the angular-momentum driven systems.  The fact that no
systems with low eccentricity have a 16.55-day orbit during mass
transfer is due to the large periastron separation for a system with
$a \approx 30$--40~\rsun\ (as required for two stars with nearly equal
masses, $\sim$ 1~\msun, and $P=16.55$~d) and small eccentricity. For
such systems, the critical lobe exceeds the donor size and no mass
transfer will occur.

The luminosity of the donor star in most systems is less than about
13~${\rm L}_\odot$, as expected for a low-mass star that is not highly
evolved. The optical and infrared magnitudes of the donor are
discussed below.

The mass transfer rate is roughly uniform between
\tento{-8}--\tento{-7}~\msun/yr (\scinot{3}{-8}~M$_\odot$/yr is
approximately the Eddington limit for a 1.4~\msun\ neutron star) and
falls off gradually below \tento{-8}~\msun/yr.  There is also a
narrow peak in the distribution at about
\scinot{3}{-9}~M$_\odot$/yr, corresponding to the angular-momentum
driven systems.

The ages of systems when reaching a stage that resembles Cir~X-1 are typically 
\scinot{4}{7} to \scinot{1}{9}. Systems older than about 
\scinot{1}{9} are mainly the angular-momentum
driven systems. All of the systems are significantly older than the
young age ($\sim$30000--100000~yr) that would be implied for Cir~X-1
if it were associated with the nearby supernova remnant. Thus this
model implies that association with the supernova remnant is
inconsistent with the assumption that Cir~X-1 has a low-mass
companion.

\begin{figure}
\begin{centering}
\PSbox{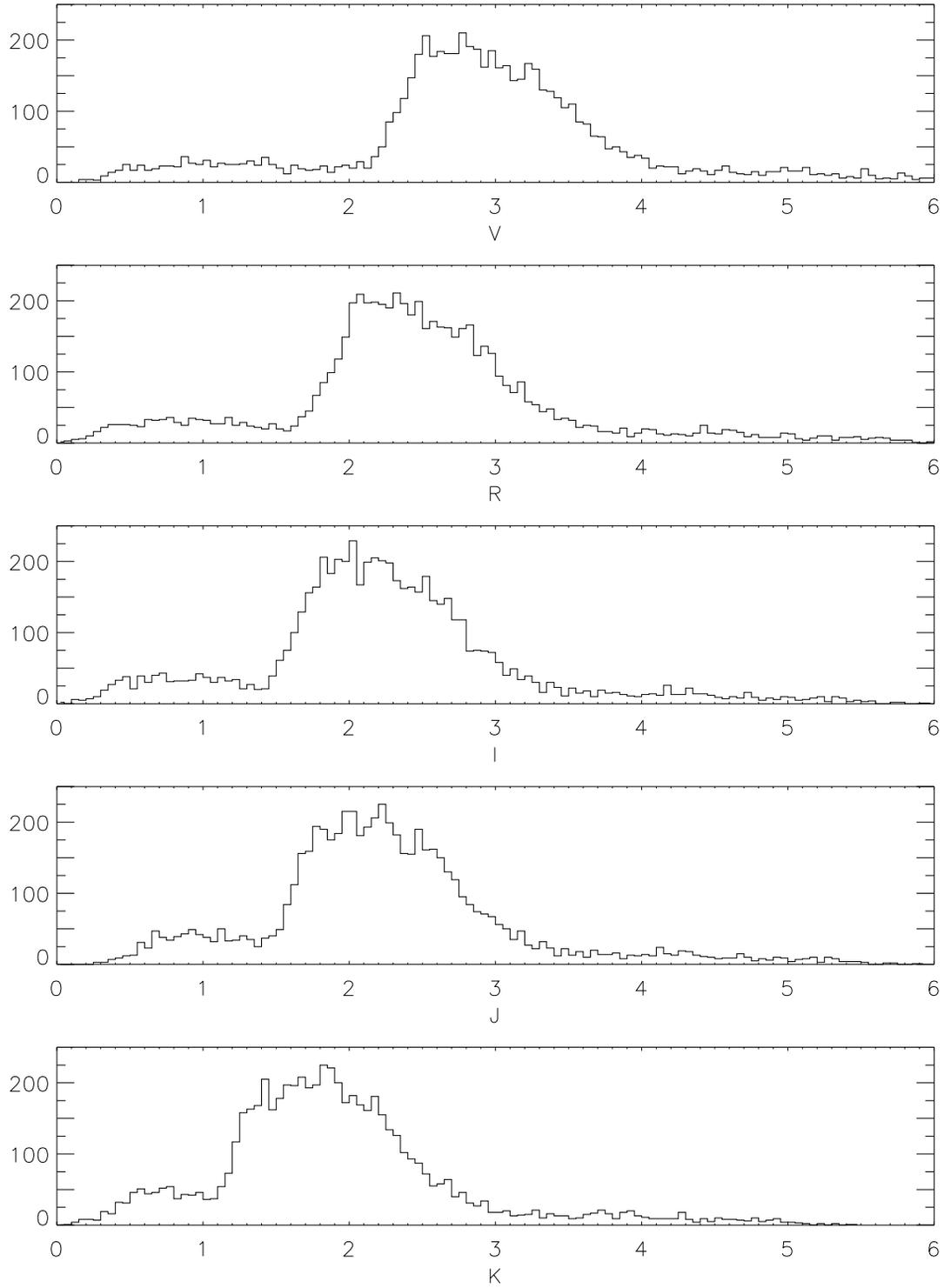
  hoffset=-20 hscale=90 vscale=90}{5.85in}{7.65in}
\caption{
Optical and infrared absolute magnitudes (V,R,I,J,K) of the donor star
in systems that resemble Cir~X-1, i.e., those from
Figure~\protect{\ref{fig:evolve_cirx1_final}}.  The y-axis
normalization of the histograms is arbitrary.}
\label{fig:evolve_cirx1_mags} 
\end{centering}
\end{figure}

The optical and infrared absolute magnitudes (V, R, I, J, \& K) of the
donor star the systems that resemble Cir~X-1 are shown in
Figure~\ref{fig:evolve_cirx1_mags}. Most donors have a V~magnitude of
2--4. Assuming a distance of 8~kpc for Cir~X-1 (see
section~\ref{sec:distance}), the distance modulus is -14.5 magnitudes.
Using this and a visual extinction of $A_{V}\sim4$ (see
section~\ref{sec:counterpart}), the mean absolute visual magnitude of
$V\sim3$ in Figure~\ref{fig:evolve_cirx1_mags} corresponds to a
apparent magnitude of $m_V \sim 21.5$, which is very similar to the
non-flaring component of the observed visual
magnitude~\cite{moneti92}.  A higher visual extinction would further
increase the apparent visual magnitude, making the donor star fainter
than the observed magnitude and requiring a contribution from the
accretion disk. 

The K-band absolute magnitude in Figure~\ref{fig:evolve_cirx1_mags} is
$\sim$1--3 for most donors.  Applying the same distance modulus
correction as above (and ignoring the much smaller extinction in at
this longer wavelength), gives an estimated apparent K~magnitude of
about 16.5, which is more than 4 magnitudes fainter than the observed
non-flaring value~\cite{glass94}. Thus, the K-band flux of a low-mass
companion is much too faint to account for the observed flux, implying
a significant contribution from the accretion disk (X-ray heating of
the companion would be a variable contribution due to the different
viewing angles around the orbit). Thus, the optical and infrared
magnitudes of donor stars in the modeled systems are not inconsistent
with (brighter than) the observed fluxes.

\section{Conclusions}

The eccentric binary evolution code developed for this chapter has
shown that a system such as Cir~X-1 can be naturally produced with a
low-mass companion.  A population synthesis study based on this code
has shown that the number of systems expected to resemble Cir~X-1 is
consistent with its unique status in the Galaxy. These results also
strongly favor a high eccentricity and donor less massive than the
neutron star as current system parameters. Currently the orbital
period is the only well-established orbital parameter of the system,
making detailed comparison of the evolution results to actual values
difficult. For example, in most cases, systems reached 16.55-d periods
during mass transfer while unbinding (P increasing). However, if the
current ephemeris equation (equation~\ref{eq:ephem}) is to be taken
literally, the period may be decreasing rapidly. If such a change
becomes well established, then this would be a significant constraint
on models for the system.  Furthermore, if Cir~X-1 truly is a young
LMXB (\tento{5} yr) associated with a nearby supernova remnant (see
Chapter~\ref{ch:intro}), then it must have evolved through a different
scenario (i.e., $M_{don} > 3$~\msun) than the LMXB model developed
here, which generally requires ages of at least \scinot{4}{7}~yr.

This evolution code (or a slightly modified version that allows
circular orbits) can also be applied to other systems and can be used
as the basis for a population synthesis study of LMXB properties in
general. Some of this spin-off work is already in progress.

\chapter{Conclusions}
\label{ch:conclusions}

The All-Sky Monitor light curve of Cir~X-1 has shown consistent
flaring above a steady, bright baseline, as well as significant
variety among intensity profiles of each cycle.  I have developed
computer codes to calculate theoretical accretion-rate profiles for an
eccentric binary in which mass transfer occurs via Roche-lobe overflow
or via a stellar wind. These profiles generally show much less
structure than the X-ray intensity profile. The theoretical profiles
show that mass transfer is strongly peaked near periastron (in the
case of a stellar wind and moderate eccentricity) or entirely confined
to phases near periastron (for Roche-lobe overflow). The fact that the
intensity of Cir~X-1 maintains a high baseline level at all orbital
phases might be related to the gradual emptying of the accretion disk
(not included in the models). More difficult to understand is the
strong mid-phase flares that sometimes occur in the X-ray light curve,
when the neutron star and donor star are believed to be relatively far
apart (near apastron).

I developed a binary evolution code for an eccentric low-mass X-ray
binary. About 15\% of systems drawn from a post-supernova
population of pre-LMXBs~\cite{terman96} were found to possess
parameters that might be similar to those of Cir~X-1 during some stage
of their evolution. On average, each LMXB system resembled Cir~X-1 for
\scinot{4}{5}~yrs. This result implies that the number of systems in
the Galaxy that should currently resemble Cir~X-1 should be of order a
unity, consistent with its unique status as an eccentric LMXB with a
high accretion rate.
 
Our multi-frequency campaigns showed that the current X-ray, radio,
and infrared flaring all occur within a day of each other, shortly
after phase zero. Unfortunately recent radio flares have been very
weak, allowing only marginal detection of the flares. Clearly, a
return of bright radio flares would be a welcomed opportunity to study
correlated behavior in more detail. Infrared photometry on timescale
shorter that hours might also prove useful in future campaigns.
Optical and IR spectroscopy carried out during our second
multi-frequency campaign showed evidence for emission lines with
high-velocity Doppler shifts ($\sim$300--400~km/s). Compared with
similar measurements made with the Hubble Space
Telescope~\cite{mignani97}, these observations strongly suggest the
need for a spectroscopy study around an entire 16.55-d cycle in order
to search for evidence of the binary orbit. Another promising future
project is the attempt to detect the angular velocity (proper motion)
of the system away from the supernova remnant with which it might be
associated~\cite{stewart93}, thus establishing that the system is
young. This project will require precise radio astrometry using VLBI
(very long baseline interferometry). Preliminary work for the proper
motion study was recently carried out at ATCA by R.~Fender.

In this project I have used extensive \RXTE\ observations of
Circinus~X-1 to characterize much of the complex behavior exhibited by
the source. Based on several timing and spectral studies, I
classify the variability we have observed from Cir~X-1 into three
categories: (1) absorption dips, (2) motion {\em along} a Z-source
hardness-intensity track, and (3) shifts {\em of} the Z-source
hardness-intensity track. I will now summarize the main
characteristics associated with each type of variability.

Absorption dips generally occur within the 0.5~d before to 1.5~d after
phase zero and last seconds to hours in duration.  During dips, the
intensity drops below the baseline intensity level while hardness
ratios initially increase, but then return to a low value during strong
dips. A faint component ($\sim$10\% of the total flux) remains
unaffected by the heavy absorption, and is probably due to scattering
by material distributed over a larger region than the X-ray source
itself. This conclusion is similar to that presented by Brandt
et~al.~\cite{brandt96} based on \ASCA\ observations of a single
low-to-high transition. Here I have shown that multiple dips occur
near phase zero in a complex and cycle-dependent manner, as might be
expected for obscuration due to inhomogeneities in the mass transfer
stream or intermittent ``bumps'' on the outer edge of the accretion
disk. I have shown that the evolution of the energy spectrum
throughout transitions into dips is consistent with a variably
absorbed bright component and faint unobscured component, and that an
iron emission line present both inside and outside dips appears to be
associated with the faint (scattered) component.  I have also
identified the curved tracks that absorption dips produce in
color-color and hardness intensity diagrams. The tracks are similar to
those due to dips in the black-hole candidates GRO~J1655-40 and
4U~1630-47~\cite{kuulkers98,tomsick98}. Having recognized spectral
tracks due to absorption, I was able to focus other studies on other
behavior that may be more intrinsic to the source.

I have identified much of the current behavior of Cir~X-1 as that of a
Z-source low-mass X-ray binary.  Hardness-intensity diagrams show that
the track for Cir~X-1 (e.g., Figure~\ref{fig:june97_hid20reg}) differs
somewhat from the canonical ``Z'' track (see
Figure~\ref{fig:z_atoll_cc_pds}), but the timing characteristics on
the branches strongly support the Z-source interpretation. In
hardness-intensity diagrams, the ``horizontal branch'' (HB) in Cir~X-1
becomes almost vertical on the left end.  Power density spectra of
Cir~X-1 from times corresponding to the vertical portion of the HB
show horizontal-branch QPOs that evolve in frequency between 1.3~Hz
and 30~Hz (increasing in frequency moving down the branch). On the
horizontal portion of the HB, this QPO occurs at $\sim$30--35~Hz and
fades in strength, becoming only a ``knee'' in the power spectrum. A
broad peak begins to arise near 4~Hz in power spectra from the
horizontal portion of the HB. This peak becomes a prominent
normal-branch QPO on the NB, and remains centered near 4~Hz. The
flaring branch for Cir~X-1 generally turns above rather than below the
normal branch (although it's orientation depends on the energy bands
used in the hardness ratio). No QPOs are present in power spectra from
the flaring branch. 

The energy spectrum on the horizontal branch of Cir~X-1 is fit well
with a two-component model consisting of a soft disk blackbody and a
higher-temperature blackbody (presumably from closer to the
surface). In this model, motion along the HB is mainly associated with
an increasing inner radius of the disk, implying that as the luminosity
increases across the HB, the inner edge of the disk moves further away
from the surface (contrary to expectations that the higher luminosity,
and thus probably higher mass accretion rate, should result in a {\em
decrease} in the radius of the inner edge of the disk).  Energy
spectra on the normal branch indicate that the hard blackbody fades
away, leaving only the disk blackbody. On the lower NB, a feature in
the spectrum develops above 10~keV. This feature becomes more
prominent on the flaring branch.

When the intensity of Cir~X-1 is at the ``quiescent'' (1.0~Crab)
level, it is generally on the vertical portion of the HB. Much of the
``flaring'' activity above the ``quiescent'' level corresponds to
motion across the HB. From the HB/NB apex, the intensity can decrease
either down the normal branch (with hardness correlated with
intensity) or back across the HB. The lower end of the normal branch
is often at a similar intensity level to that of the vertical HB;
however, from the low-intensity end of the normal branch, small bursts
or ``mini-flares'' occur, resulting in motion up the flaring branch
and producing very low frequency noise (VLFN) in the power spectrum.
Bildsten~\cite{bildsten93,bildsten95} has proposed that VLFN in low
mass X-ray binaries is produced by small patches on the neutron-star
surface burning intermittently due to slow non-convective combustion,
rather than the rapid ignition of the entire surface which occurs in
type~I bursts. The ``flares'' or ``bursts'' seen on the flaring branch
in Z sources, and Cir~X-1 in particular, may be part of a continuous
range from weak VLFN bursts, to small flares, to type I bursts.

I have shown that some of the variability in Cir~X-1 is due to shifts
of the ``Z'' track in hardness-intensity diagrams. These shifts are at
least in part related to the cycling hardness ratios we observe with
the All-Sky Monitor, i.e., gradual hardening of the spectrum during
much of the 16.55-d cycle and more rapid hardening of the spectrum at
the onset of the flaring state.  There is a tendency for
horizontal-branch behavior to occur at the low-intensity, hard portion
of the cycle, and normal or flaring-branch behavior to occur during
high-intensity, soft portions of the cycle.  Thus, evolution of the
timing properties versus orbital phase is related to both the shifting
Z-track and motion around the Z track.  Clearly this can produce
behavior more complicated than we observed in our first set of
observations, presented in Chapter~\ref{ch:mar96paper}.

Although there is still much to be understood about the physical
mechanisms responsible for the correlated spectral and timing behavior
in Cir~X-1, this project has demonstrated that these unanswered
questions for Cir~X-1 are related to similar issues in an entire
class of systems (Z~sources). In fact, some of the unique features of
the behavior of Cir~X-1 may prove useful in future studies of the Z
and atoll classes of low-mass X-ray binaries. 

If the baseline intensity level of Cir~X-1 decreases from its current
bright level, it will provide an opportunity to test the hypothesis
that Cir~X-1 shows atoll-source behavior at lower mass-accretion
rates. Establishing the existence of a source showing the behavior of both
classes under different conditions would be helpful in determining the
physical distinctions between the classes (e.g., different magnetic
fields would be ruled out).

Aspects of the behavior of Cir~X-1 that differ from that of typical
Z~sources may also provide important constraints on models for
Z-source behavior. For example, models of horizontal-branch
oscillations (typically observed at 13--60~Hz) will have to address
what physical property (e.g., mass, spin, or magnetic field of the
compact object) of Cir~X-1 results in significantly lower-frequency
HBOs (1.3--35~Hz). Also, the large shifts of the Z-track in
hardness-intensity diagrams for Cir~X-1, and the fact that such shifts
are associated with the 16.55-d orbital cycle of Cir~X-1, are not
explained by the hypothesis that mass accretion rate increases along
the Z~track. Kuulkers
et~al.~\cite{kuulkers96:gx340+0,kuulkers96:cygx-2,kuulkers94:gx5-1}
have suggested that secular shifts in several Z sources are due to
inclination effects. However, the shifts in those sources were not
observed to occur periodically as in Cir~X-1, nor were the shifts as
large (see section~\ref{sec:feb97_other_sources}).  Another
possibility is, if accretion occurs both through a disk {\em and}
through a spherical inflow (or other means), then the accretion rate
of each component may be responsible for one aspect of the observed
behavior.

In the past, the unique status of Cir~X-1 among X-ray binaries has
often resulted in its being relegated to the discussion of ``other
sources''.  The results presented here show that Cir~X-1 has much in
common with other low-mass X-ray binaries and that its unusual
features may provide the opportunity to study a system
with somewhat atypical physical parameters.

\appendix
\chapter{Fourier Timing Techniques}
\label{ch:pds}

\section{Introduction}

Temporal analysis in {X}-ray astronomy is generally accomplished through
examination of the Fourier power density spectrum (PDS) and complex
cross spectrum. The following is a brief review of some of the
standard techniques that are directly applicable to RXTE data (see
van~der~Klis~\cite{klis89:fourier}, e.g., for a more complete
discussion of standard techniques). In addition, specific formulae are
derived to show how the analysis software handles such complications
as data gaps and variable number of detectors.

The analysis software we developed at M.I.T. operates on time series
data stored in ``DS (DataStream)'' format, which dynamically changes
storage format to reduce file sizes, and allows data to be piped
between programs to chain together several tasks. These features
result in smaller files and faster processing than the standard FTOOLS
package provided by NASA. Useful information can be stored as keywords
in header blocks of DS files. Since the timing programs in particular
rely heavily on using keyword parameters, keyword names will be
referenced in the throughout this appendix, which is intended to
provide documentation to our software.
\vspace*{1.5in}

\section{Power Density Spectra}

\subsection{Leahy Normalization}

The discrete Fourier transform of a time series of N bins (keyword
NUMPOINT) containing counts $x_{k}$ $(k=0,..,N-1)$ is
\begin{equation}
a_{j} = \sum_{k=0}^{N-1}x_{k}e^{2\pi i j k / N}, {\hskip 0.5 in} j=-N/2,...,N/2-1.
\end{equation}
Frequency bin $j$ corresponds to a frequency of $\nu=j/T=j/Nt_{b}$,
where $T$ is the total duration of the time series and $t_{b}$ is the
duration of a single time bin. The normalization used by Leahy
et~al.~\cite{leahy83} defines the power in frequency bin $j$ to be
\begin{equation}
P_{j} = \frac{2}{N_{ph}}|a_{j}|^{2}, {\hskip 0.5 in} j=0,...,N/2
\end{equation}
where $N_{ph}$ is the total number of photons in the transform. In
this normalization, the distribution of powers for Poisson counting
noise has a mean value of 2 and obeys the $\chi^{2}$ distribution with
2 degrees of freedom. This has the advantage of straightforward
estimation of the significance of high points in the PDS.

\subsection{Estimate of the Poisson Noise Level}
Real time series data always contains counting noise ($x = x_{source}
+ x_{noise}$).  This adds power $P = P_{source} + P_{noise}$.
%$a = a_{source} + a_{noise}$, 
%$P = P_{source} + P_{noise}$

In the absence of detector effects, Poisson counting
noise produces a constant power equal to 2 in the Leahy
normalization. However, detector dead-time effects result in a
Poisson-level that varies with Fourier frequency.  Zhang
et~al.~\cite{zhang95,zhang96} (see also~\cite{morgan97}) have
calculated this effect 
%for paralyzable dead time 
to be
\begin{equation} \label{eq:poissonlevel}
<P_{noise}(\nu)> = 2[1-2r_{0}\tau_{d}(1-\frac{\tau_{d}}{2t_{b}})]
-2\frac{N-1}{N}r_{0}\tau_{d}(\frac{\tau_{d}}{t_{b}})\cos(2\pi t_{b}\nu)
+2r_{vle}r_{0}\tau_{vle}^{2}[\frac{\sin(\pi\tau_{vle}\nu)}{\pi\tau_{vle}\nu})]^{2},
\end{equation}
where $r_{0}$ is the count rate per PCU detector, $\tau_{d}$ is the
dead time per event (10 $\mu$s for the PCA), $t_{b}$ is the time bin
size, and N is the number of time bins in the transformed segment. The
last term accounts for dead time due to very large events (VLEs),
occurring at a rate $r_{vle}$ per PCU with a dead-time window
($\tau_{vle}$) of 55 $\mu$s or 155 $\mu$s for each VLE~\cite{zhang96}.

\subsection{Fractional RMS Variability}
Dividing a Leahy-normalized PDS by the mean count rate from the source
results in a normalization giving the square of the fractional
root-mean-squared variability per Hz. In this normalization, the
square root of the integrated power under a PDS feature (such as a QPO
peak) gives the fractional rms variability of the intensity due to
that feature. 

The average {\it total} count rate per second (keyword RATE) is
\begin{equation}
R = \frac{1}{t_{b}} \frac{\sum_{k=1}^{N}x_{k}}{N} 
  = \frac{\bar{x}}{t_{b}}.
\end{equation}
However, to obtain the mean rate of the {\it source} $R^{'}$, the
background $b$ must be subtracted:
\begin{equation}
R^{'} = \frac{1}{t_{b}} \frac{\sum_{k=1}^{N}(x_{k} - b_{k})}{N}
      = \frac{1}{t_{b}} (\bar{x}-\bar{b}) = R - B,
\end{equation}
where $B$ is the average background rate per second (keyword BKGD).
Generally, the average background rate is estimated from slew data or
other means modeled with software provided with the FTOOLS package.

\subsection{Estimates of Power Variance}
The variance of powers in the Leahy normalization is 4 (so
$\sigma=2=\mu$).  The variance of the mean power from $M$ consecutive
frequency bins or from $W$ PDSs is:
\begin{equation}
\sigma_{\mu}=\frac{2}{\sqrt{MW}}.
\end{equation}
However, intrinsic noise from the source often causes the actual
variance of the averaged powers to exceed the theoretical
variance. Thus, a useful error estimate is derived from the {\em
sample} variance resulting from averaging $MW$ powers $P_{i},
i=1,..,MW$ (dropping the frequency index for clarity):
\begin{equation}
s^{2}=\frac{1}{MW-1}\sum_{i=1}^{MW}(P_{i}^{2}-<P>^{2})
\end{equation}
The sample variance of the mean power is 
\begin{equation}
s_{\mu}^{2}=\frac{s^{2}}{MW}=\frac{1}{MW(MW-1)}\sum_{i=1}^{MW}(P_{i}^{2}-<P>^{2})
\end{equation}

\subsection{Data Gaps}
Ideally, each PDS should be derived from a continuous time series with
no gaps. However, gaps are common in real data, making it necessary to
work with partially filled (but not sparse) time series'. The simplest
way to handle data gaps is to fill the empty bins with the mean number
of counts per bin. (In practice the mean is subtracted from the data
to suppress the DC term in the PDS, so unoccupied bins are filled with
zeros.) To compensate for an incomplete time series, averages should
be weighted by the fraction exposure $f$, the fraction of occupied
bins in the time series (keyword FILL). I.e., in averaging $M$
frequency bins from each of $W$ power spectra, each with fractional
exposure $f_{w}$,
\begin{equation}
<P>=\frac{\sum_{m=1}^{M}\sum_{w=1}^{W}f_{w}P_{mw}}{\sum_{m=1}^{M}\sum_{w=1}^{W}f_{w}}
\end{equation}
Likewise, the sample variance is 
\begin{equation}
s^{2}=\frac{1}{{\sum_{m=1}^{M}\sum_{w=1}^{W}f_{w}}-1}\sum_{m=1}^{M}\sum_{w=1}^{W}(P_{mw}^{2}-<P>^{2})
\end{equation}
which can be re-expressed as 
\begin{equation}
s^{2}=\frac{ {\sum_{m=1}^{M}\sum_{w=1}^{W}f_{w}} }{ {\sum_{m=1}^{M}\sum_{w=1}^{W}f_{w}} - 1 }(<P^{2}>-<P>^{2})
\end{equation}
where
$<P^{2}>=\frac{\sum_{m=1}^{M}\sum_{w=1}^{W}f_{w}P_{mw}^{2}}{\sum_{m=1}^{M}\sum_{w=1}^{W}f_{w}}$.
Finally, the sample variance of the mean power is
\begin{equation}
s_{\mu}^{2}=\frac{ s^{2} }{ {\sum_{m=1}^{M}\sum_{w=1}^{W}f_{w}} } =
\frac{ <P^{2}>-<P>^{2} }{ {\sum_{m=1}^{M}\sum_{w=1}^{W}f_{w}} - 1 }
\end{equation}
Both $<P>$ and $<P^{2}>$ are saved in the PDS files and used in error
estimates (COL keywords with PDS and PDS{\char94}2 labels).

\subsection{Variable Number of Detectors}

For instrument safety, one or more of the five proportional counter
units (PCUs) of the PCA may be shut off during an
observation. Transforms are not performed for time segments which
include a discontinuous detector turn-on/turn-off. However, separate
PDSs with different numbers of PCUs turned on, $\bar{p}_{w}$ (keyword
NUMPCU), can still be averaged if care is taken to handle the various
parameters properly. When averaging $W$ PDSs the combined average of
the number of PCUs on, $\bar{p}_{w}$ (keyword NUMPCU), is weighted by
the total number of occupied bins $n_{w}=W_{w}N_{w}f_{w}$ (from
keywords NUMFFT $\times$ NUMPOINT $\times$ FILL):
\begin{equation}
\bar{p} = \frac{ \sum_{w=1}^{W}{n_{w}\bar{p}_{w}} }{ \sum_{w=1}^{W}n_{w} } = 
\frac{ \sum_{w=1}^{W}{n_{w}\bar{p}_{w}} }{ n_{tot} }
\end{equation}
where $n_{tot}$ is the total number of occupied bins.
Alternately, the PCU average can be
expressed as $\bar{p}=\sum_{p=1}^{5}{n_{p}p}/n_{tot}$, where $n_{p}$ is the
number of bins with $p$ PCUs on. 

For the rms normalization, the relevant rates are those actually
observed, regardless of the number of detectors involved.  Thus, the
combined average rate and background rate from W PDSs, each with
average rate $\bar{x}_{w}$ and background rate $\bar{b}_{w}$, are
simply weighted by the number of occupied bins:
\begin{equation}
\bar{x} = \frac{ \sum_{w=1}^{W}{n_{w}\bar{x}_{w}} }{ n_{tot} }, {\hskip 0.5 in} 
\bar{b} = \frac{ \sum_{w=1}^{W}{n_{w}\bar{b}_{w}} }{ n_{tot} }.
\end{equation}

Deadtime effects in each PCU are independent of the other PCUs, so the
average rate per PCU $r_{0}$ used in the Poisson-level estimate
(equation~\ref{eq:poissonlevel}) should be weighted by the number of
PCUs on:
\begin{equation}
r_{0} = \frac{1}{t_{b}} \frac{ \sum_{p=1}^{5}n_{p}p\frac{\bar{x}_{p}}{p} }
		    { \sum_{p=1}^{5}n_{p}p }
	= \frac{1}{t_{b}} \frac{\sum_{p=1}^{5}n_{p}\bar{x}_{p}}{n_{tot}\bar{p}}
\end{equation}
where $\bar{x}_{p}$ is the average number of counts in all time bins
having $p$ PCUs on.  The numerator of the last term above is simply
the total number of counts, which is equal to $n_{tot}{\bar{x}}$. Thus, the 
average rate per PCU reduces to 
\begin{equation}
r_{0} = \frac{1}{t_{b}} \frac{ \bar{x} }{ \bar{p} } = \frac{R}{\bar{p}}.
\end{equation}
The Poisson-level estimate also requires the average product per PCU
of the VLE rate and the good count rate. 
For a constant number of PCUs $p$:
\begin{equation}
<r_{vle}r_{0}>_{p}
  =\frac{1}{t_{b}^2}
   \frac{ \sum_{k=1}^{n_{tot}} p \frac{v_{k}}{p} \frac{x_{k}}{p} } 
        { \sum_{k=1}^{n_{tot}}p } 
  =\frac{1}{t_{b}^2} \frac{\sum_{k=1}^{n_{tot}}v_{k}x_{k}}{n_{tot}p^2}, 
\end{equation}
where $v_{k}$ is the VLE rate per bin. The average over different
numbers of PCUs must be based on the number of occupied bins for each $p$:
\begin{equation}
<r_{vle}r_{0}> = 
  \frac{ \sum_{p=1}^{5}n_{p}p<r_{vle}r_{0}>_{p} }
       { \sum_{p=1}^{5}n_{p}p }
	= \frac{ \sum_{p=1}^{5}n_{p}p<r_{vle}r_{0}>_{p} }
               { n_{tot}\bar{p} }.
\end{equation}
(The keyword VLEP saves this VLE product as $\bar{p}^2<r_{vle}r_{0}>$
to be consistent with the RATE and BKGD keywords, which are not per
PCU.)

\section{Complex Cross Spectra}

The Fourier transforms of a time series in two energy channels can be
used to produce a complex cross spectrum in addition to the two power
density spectra. Using the same notation as for the PDS (see above),
the cross spectrum derived from two Fourier transforms $a_{j}$ and
$b_{j}$ is defined as
\begin{equation}
C_j = \comprod{a_j}{b_j}, {\hskip 0.5 in} j=0,...,N/2.
\end{equation}

\subsection{Time Lags}
The argument of the complex-valued cross spectrum is the phase delay
(in radians) between intensity fluctuations in the two channels (once
again dropping the frequency subscript):
\begin{equation}
\delta\phi = \arctan\left(\frac{Im(C)}{Re(C)}\right)
	   = \arctan\left(\frac{Re(a) Im(b) - Im(a) Re(b)}
			       {Re(a) Re(b) + Im(a) Im(b)}\right).
\end{equation}
This is easily converted to a time delay via $\delta t = \delta\phi /
2 \pi \nu$.  A positive value of $\delta t$ indicates that intensity
fluctuations in the second channel lag those in the first.

\subsection{Coherence Function}

The magnitude of the cross spectrum can be used to derive the
coherence function between two channels. The coherence function is a
measure of the degree of linear correlation between the two time
series at each Fourier frequency~\cite{vaughan97}.
For noiseless signals, the coherence function is defined as
\begin{equation} \label{eq:coher}
\gamma^2 
	= \frac{|<\comprod{a}{b}>|^2}{<|\comprod{a}{a}|><|\comprod{b}{b}|>}, 
	= \frac{|<C>|^2}{<P_1><P_2>} 
\end{equation}
where $P_1$ and $P_2$ are the power spectra derived from $a$ and $b$
respectively and the angle brackets denote an average over $m$
independent measurements (ideally $m\rightarrow\infty$). 

For signals with Poison counting noise, the quantities used in
equation~\ref{eq:coher} should be for the source only, thus 
\begin{equation} \label{eq:coher_noise1}
\gamma^2 = \frac{| < \comprod{(a-a_{noise})}{(b-b_{noise})} > |^2}
	{<P_1-P_{1,noise}><P_2-P_{2,noise}>} 
\end{equation}
\begin{equation} \label{eq:coher_noise2}
\gamma^2 = \frac{|<\comprod{a}{b}> - <\comprod{a}{b_{noise}}> - 
	  	<\comprod{a_{noise}}{b}> + <\comprod{a_{noise}}{b_{noise}}>|^2}
	     {<P_1-P_{1,noise}><P_2-P_{2,noise}>} 
\end{equation}
\begin{equation} \label{eq:coher_noise3}
\gamma^2 = \frac{|<C> - <\comprod{a}{b_{noise}}> - 
	  	<\comprod{a_{noise}}{b}> + <C_{noise}>|^2}
	     {<P_1-P_{1,noise}><P_2-P_{2,noise}>}.
\end{equation}

According to Vaughan and Nowak~\cite{vaughan97}, in the Gaussian limit
the numerator in equation~\ref{eq:coher_noise3} can be expressed as 
\begin{equation} \label{eq:gausslim}
|<C_{source}>|^2 = |<C>|^2 - <n^2>,
\end{equation}
where $n^2 \equiv 
( P_{1,source}P_{2,noise} + P_{1,noise}P_{2,source} + P_{1,noise}P_{2,noise})/m $.  Thus, the intrinsic source coherence
function can be expressed in terms of the observed powers and cross
spectra and the estimated Poisson noise powers:
\begin{equation} \label{eq:coher_noise4}
\gamma^2 = \frac{|<C>|^2 - ( P_{1,source}P_{2,noise} + P_{1,noise}P_{2,source} +
		P_{1,noise}P_{2,noise} ) / m}
	     {<P_1-P_{1,noise}><P_2-P_{2,noise}>}.
\end{equation}

\chapter{Standard PCA Light Curves}
\label{ch:std_pcalc}

This appendix contains standard 20-kilosecond light curves of all our
1996--1997 PCA observations of Cir~X-1. By displaying all data with a
uniform scale, the wide range of variability and intensity levels is
apparent.

Each 20-ks panel is labelled along the right side with the letter of
the study with which it was associated (A--H, see
Table~\ref{tab:pcaobs}) and several time labels referring to time zero
of that panel: the truncated Julian date (JD-2450000.5), the calendar
date (YR-MO-DAY), the mission elapsed time (MET = seconds since 1994
Jan.\ 1.0), and the orbital phase (see equation~\ref{eq:ephem}).

The counting rates are derived from 16~s time bins over the full PCA
energy band and are the average rates from PCUs 0, 1, and 2, which
were operating during all observations (PCUs 3 and 4 were often
switched off). A counting rate of 2.6~kcts/s corresponds to an
intensity of $\sim$1.0~Crab (1060~$\mu$Jy at 5.2~keV).  Background is
not subtracted but is negligible. The data have not been corrected for
detector deadtime; such corrections would increase the baseline 1~Crab
level by about 5\% and the peak 2.7~Crab level by about 13\%.

\begin{centering}
\PSbox{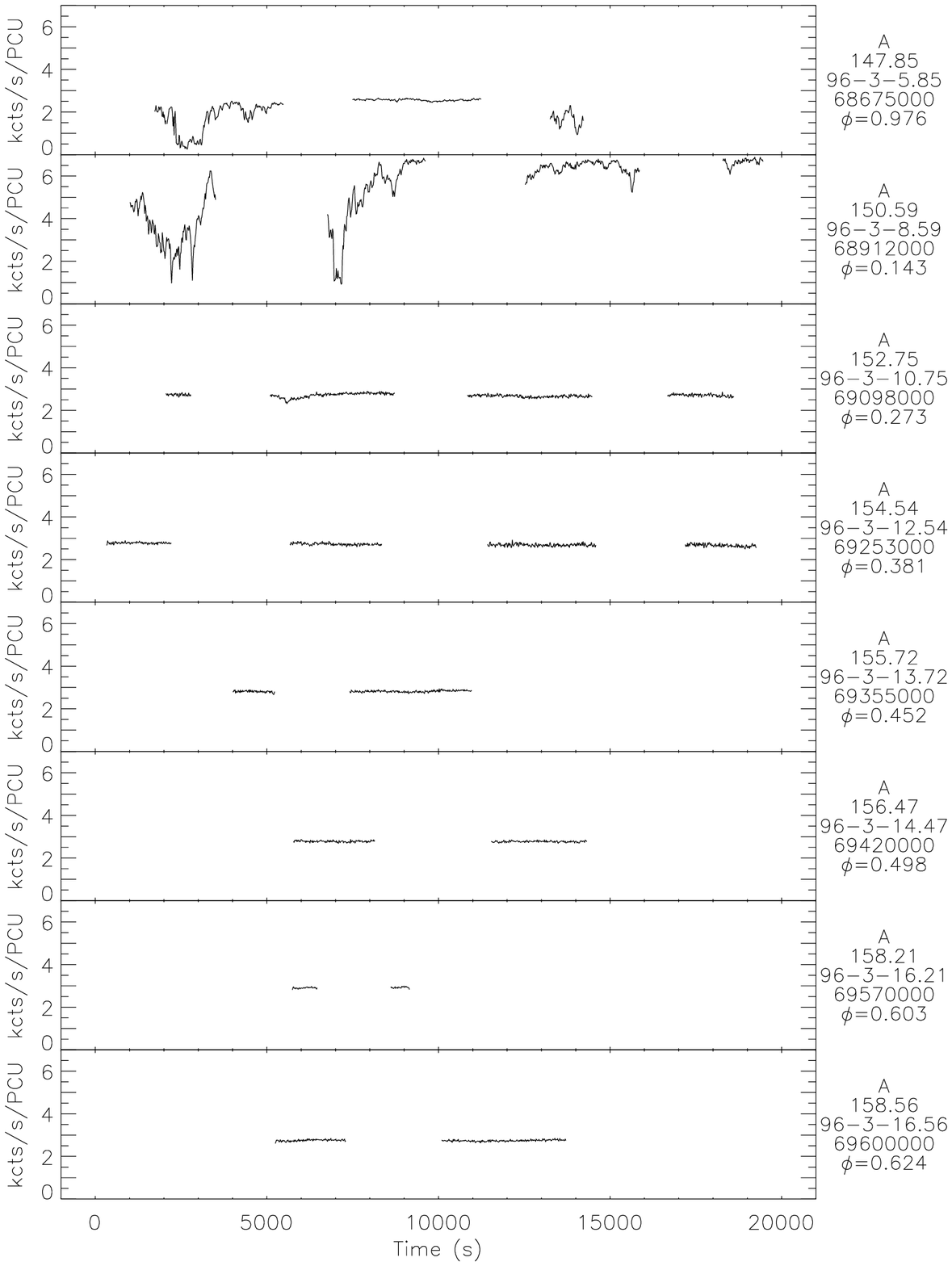 voffset=-78 hoffset=-92}{6.5in}{8.55in}
\PSbox{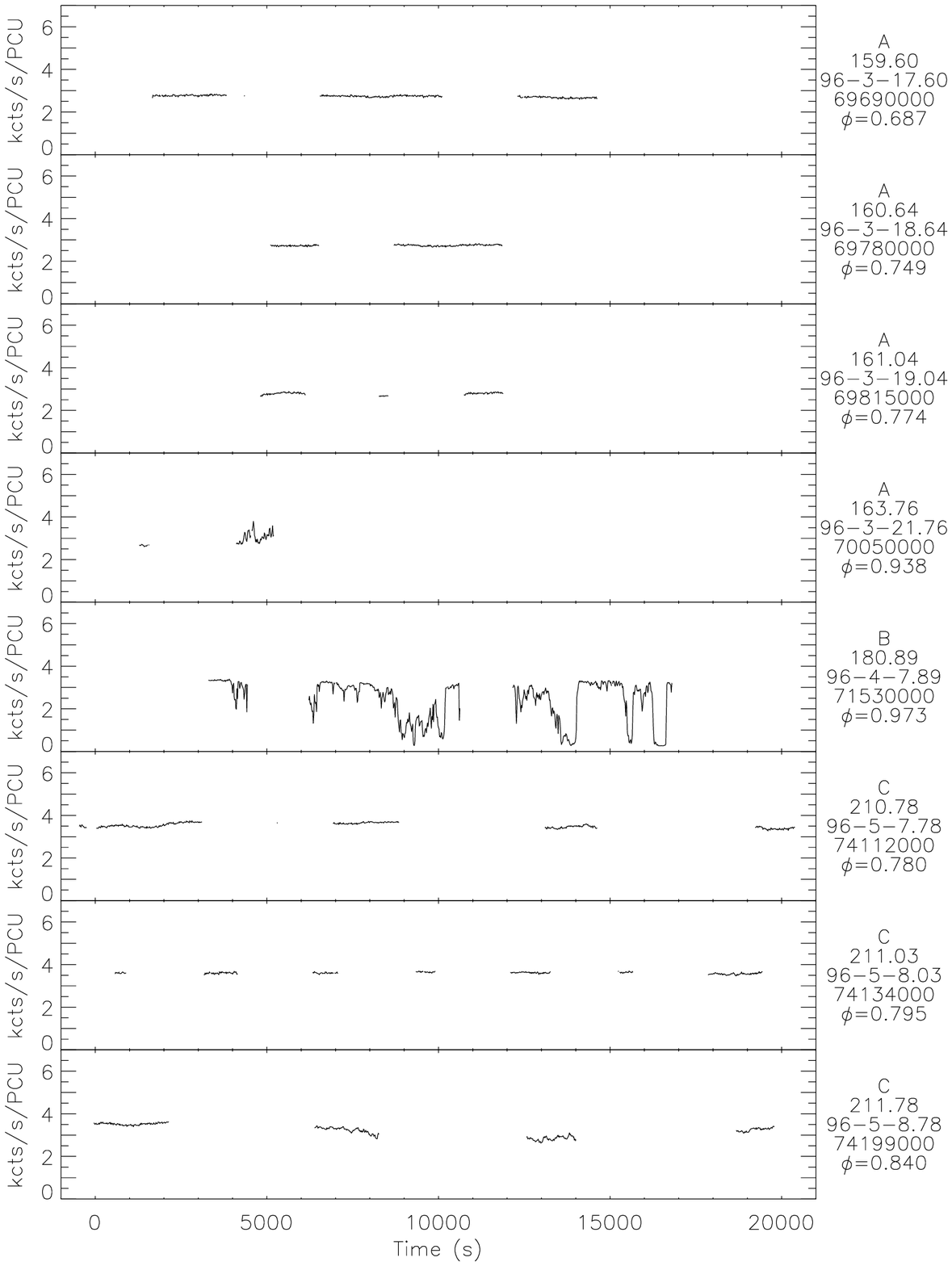 voffset=-78 hoffset=-92}{6.5in}{8.55in}
\PSbox{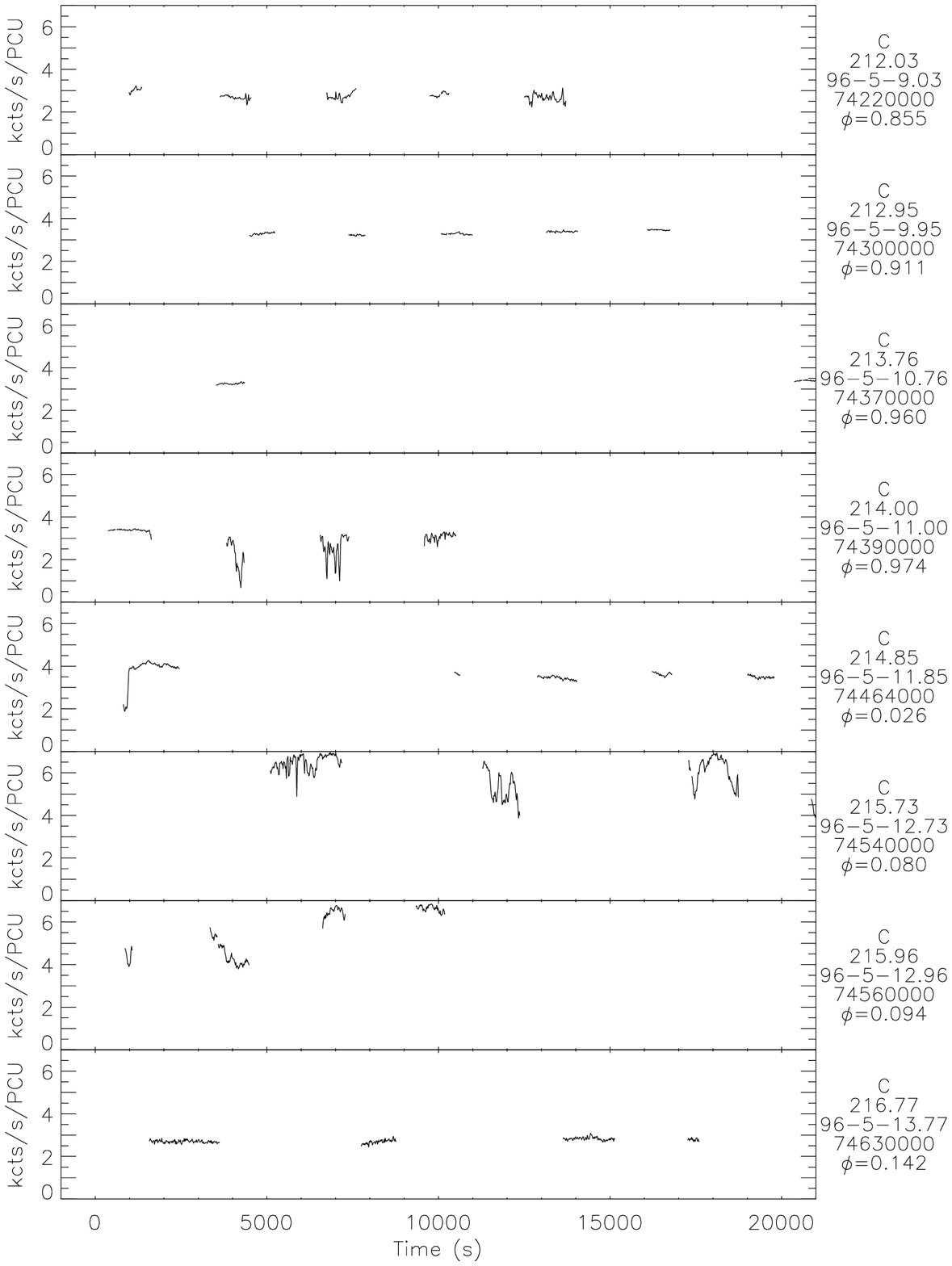 voffset=-78 hoffset=-92}{6.5in}{8.55in}
\PSbox{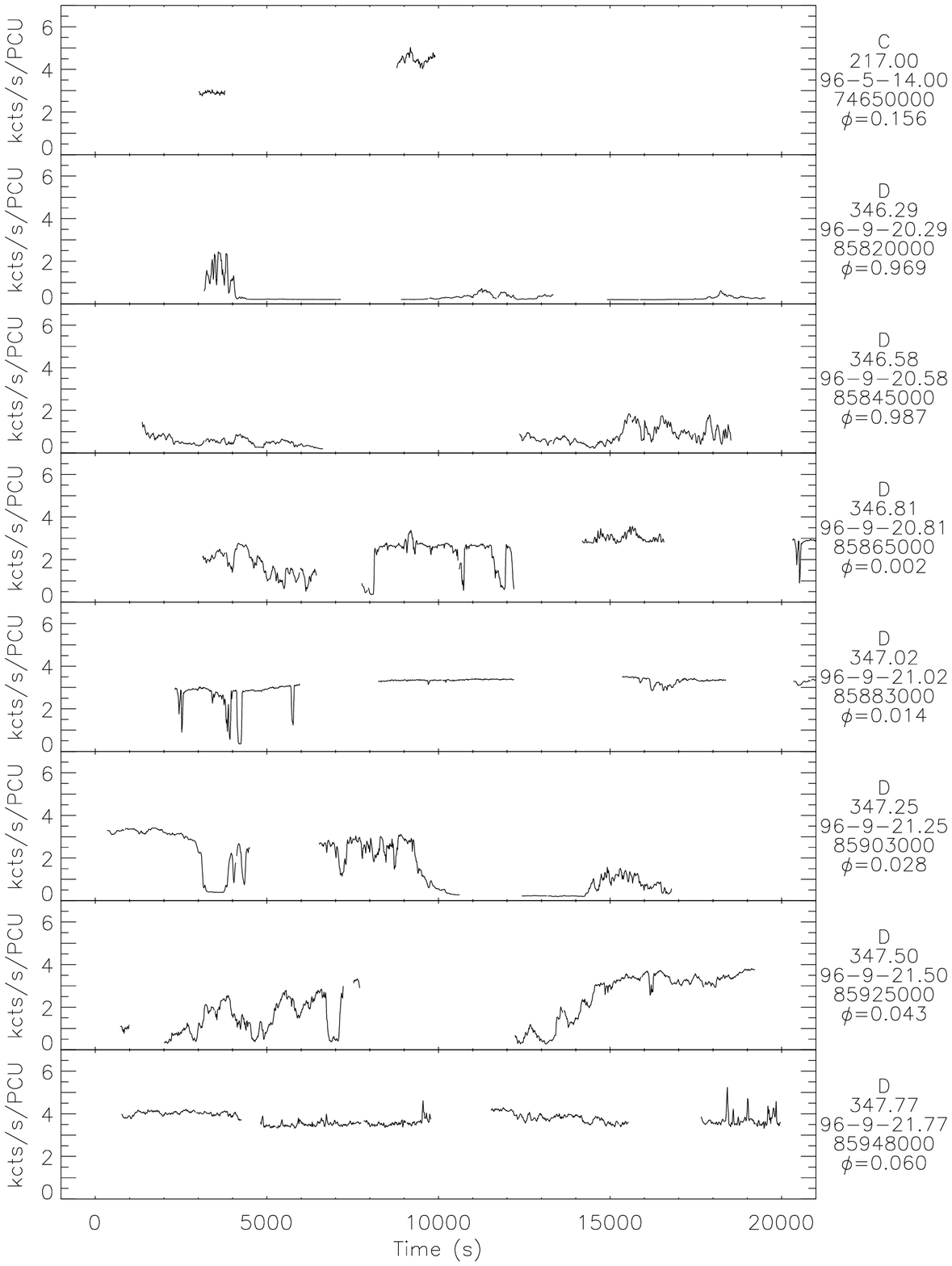 voffset=-78 hoffset=-92}{6.5in}{8.55in}
\PSbox{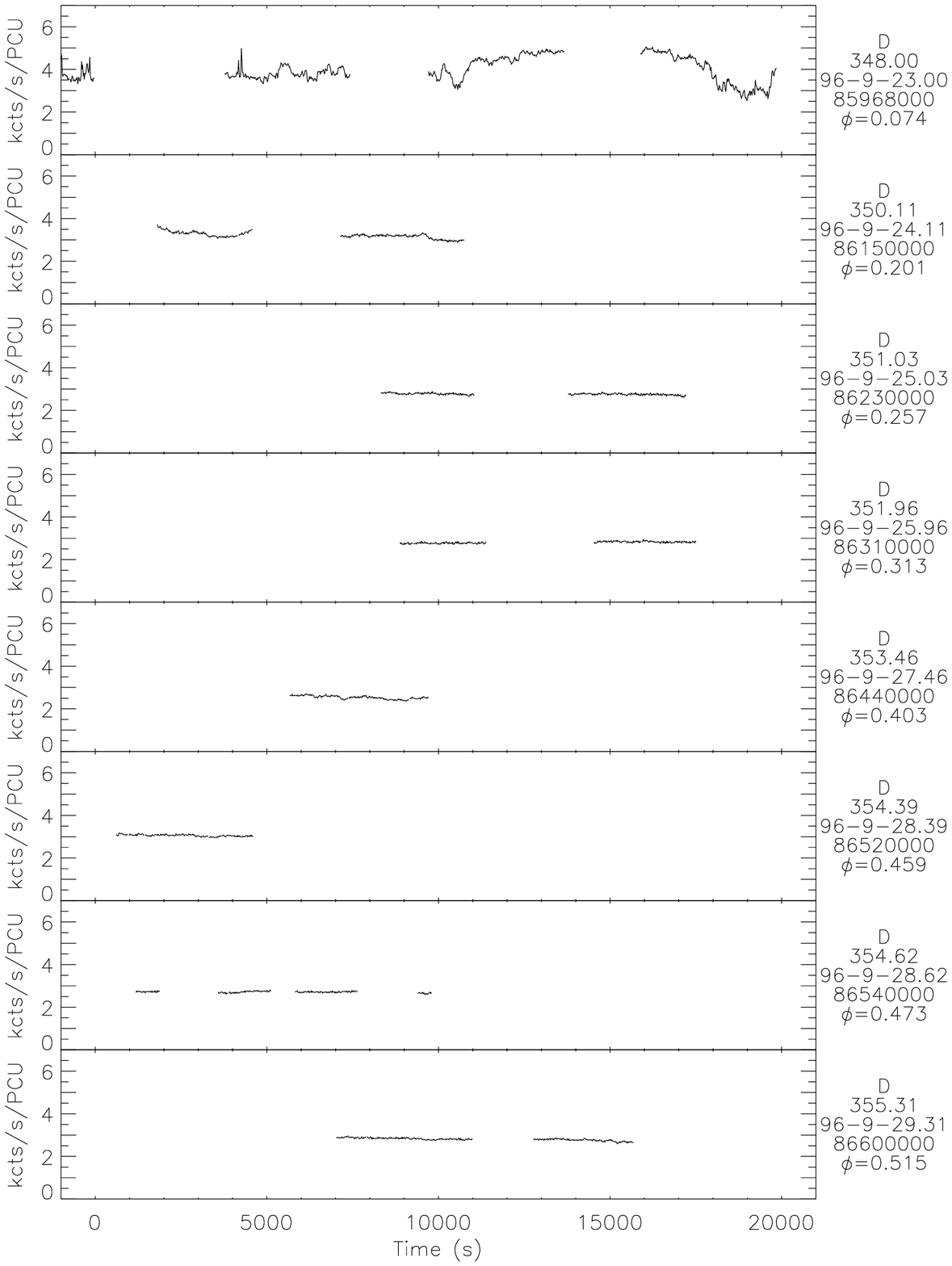 voffset=-78 hoffset=-92}{6.5in}{8.55in}
\PSbox{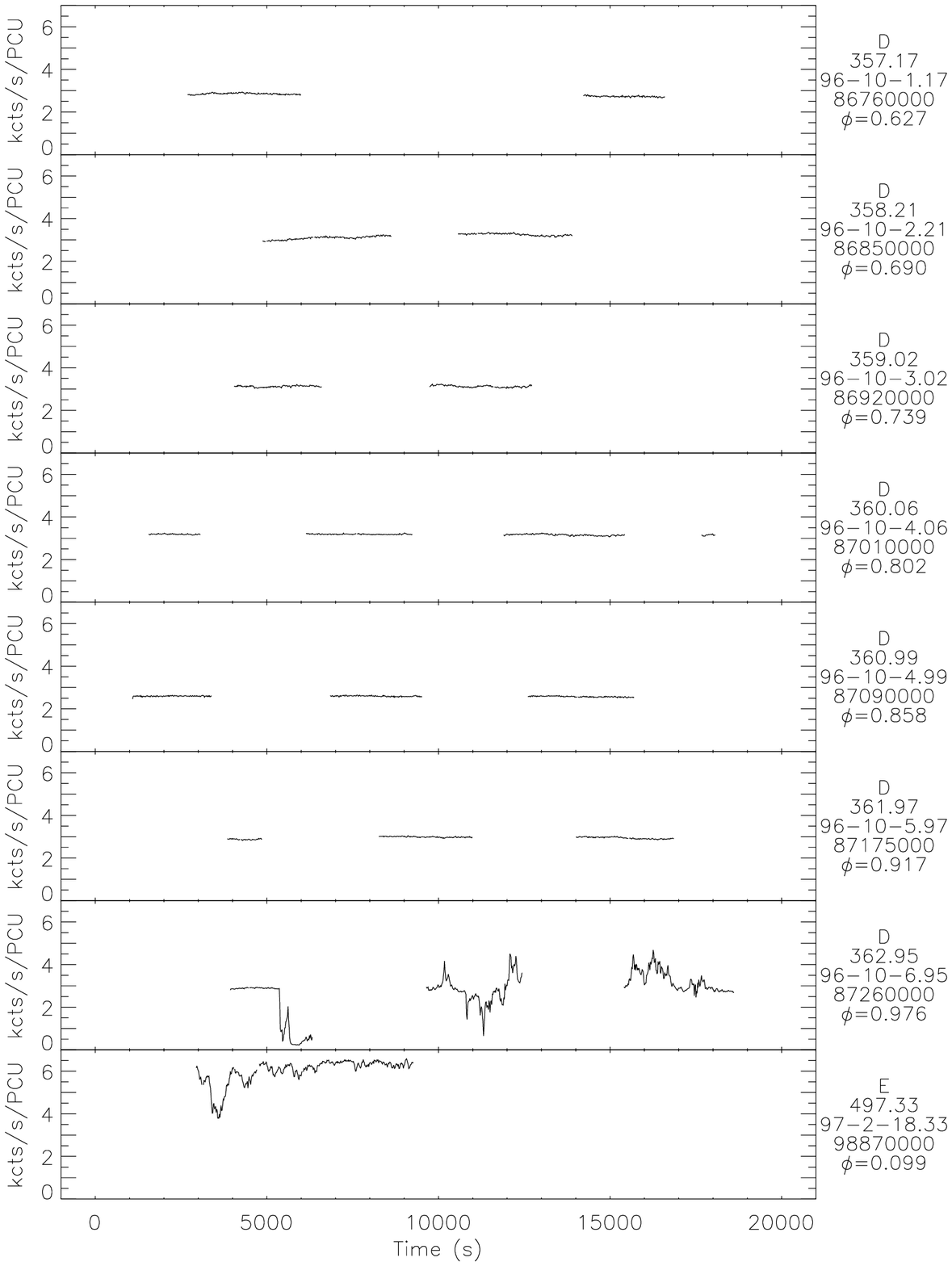 voffset=-78 hoffset=-92}{6.5in}{8.55in}
\PSbox{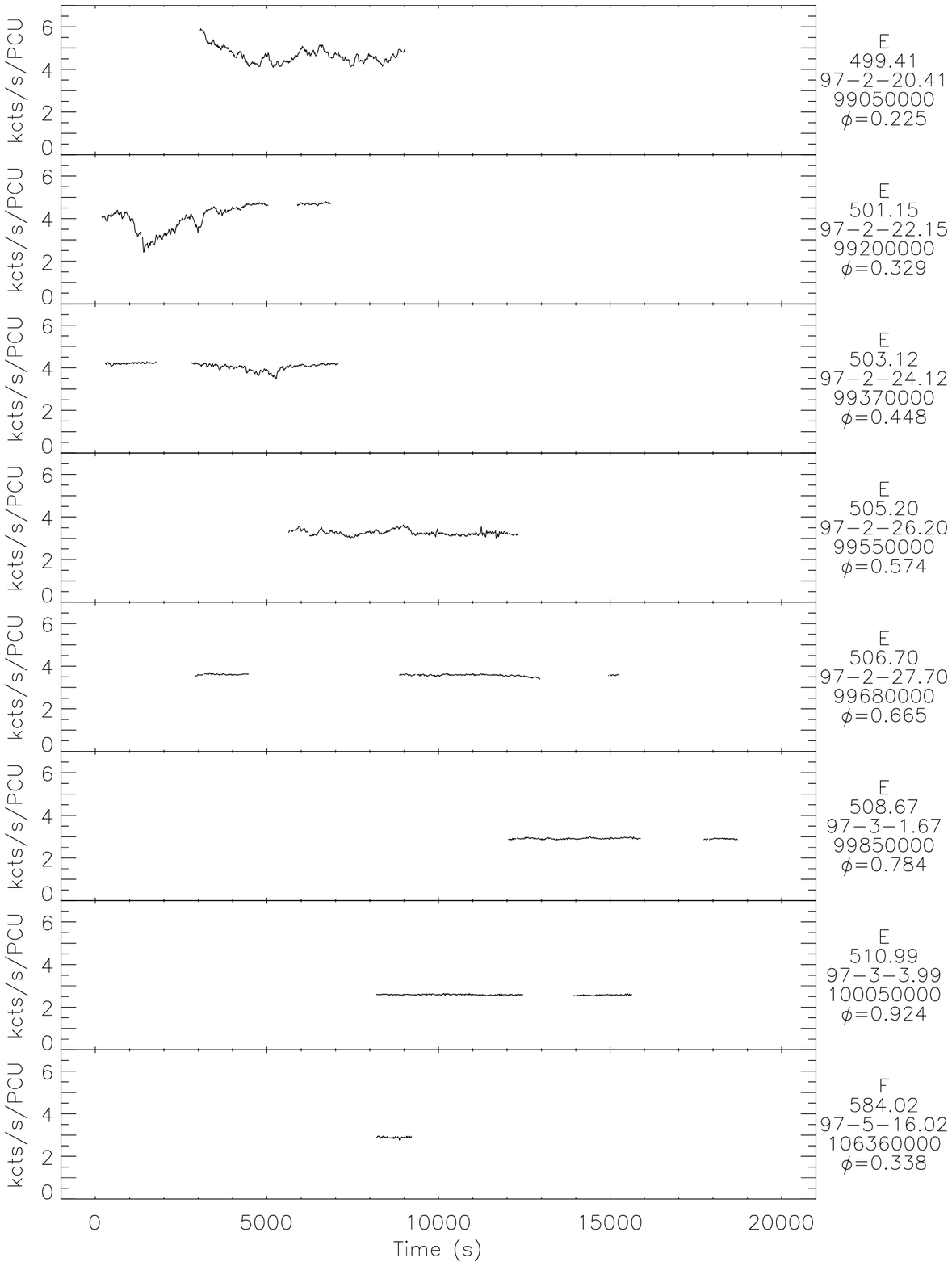 voffset=-78 hoffset=-92}{6.5in}{8.55in}
\PSbox{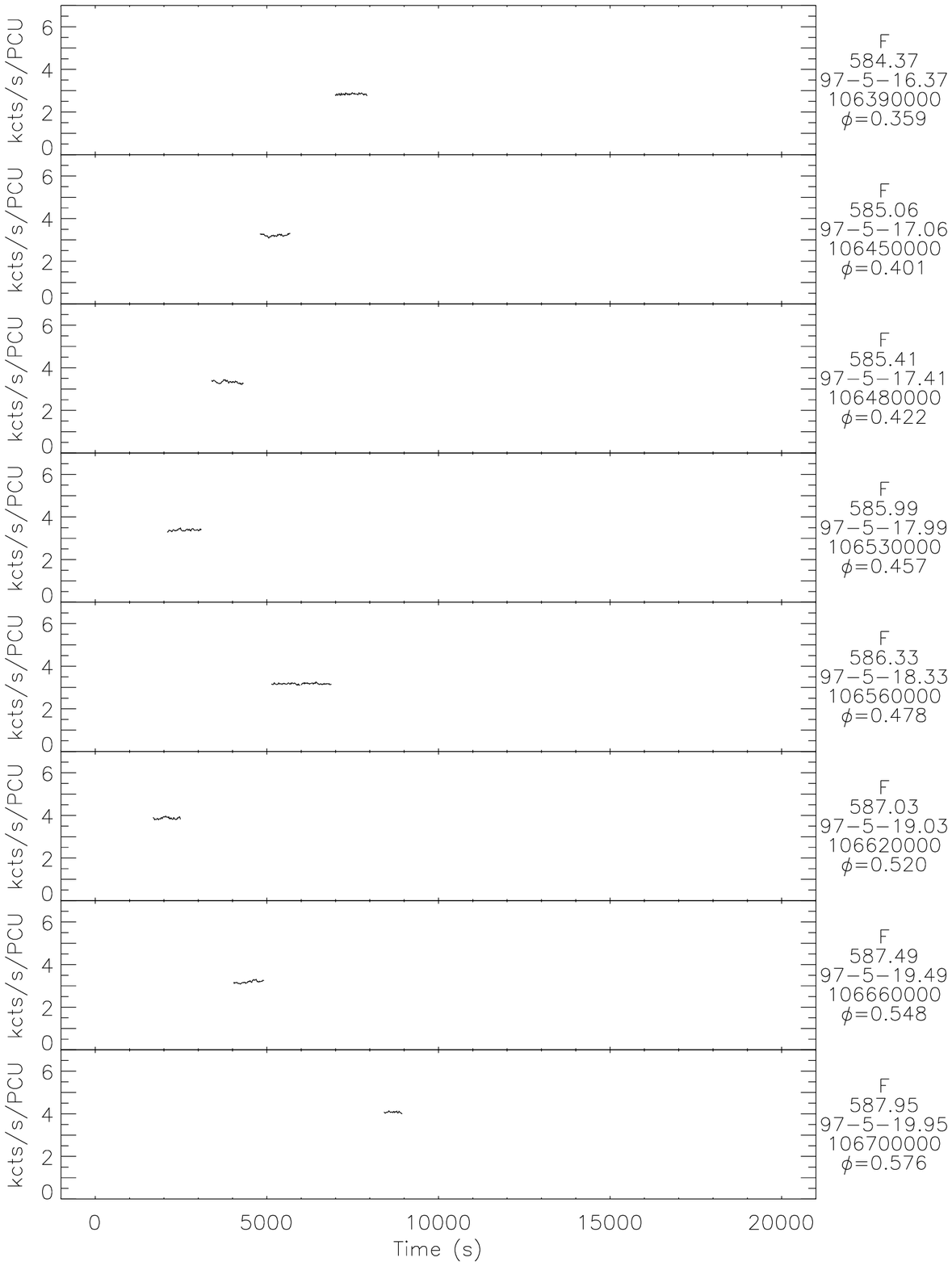 voffset=-78 hoffset=-92}{6.5in}{8.55in}
\PSbox{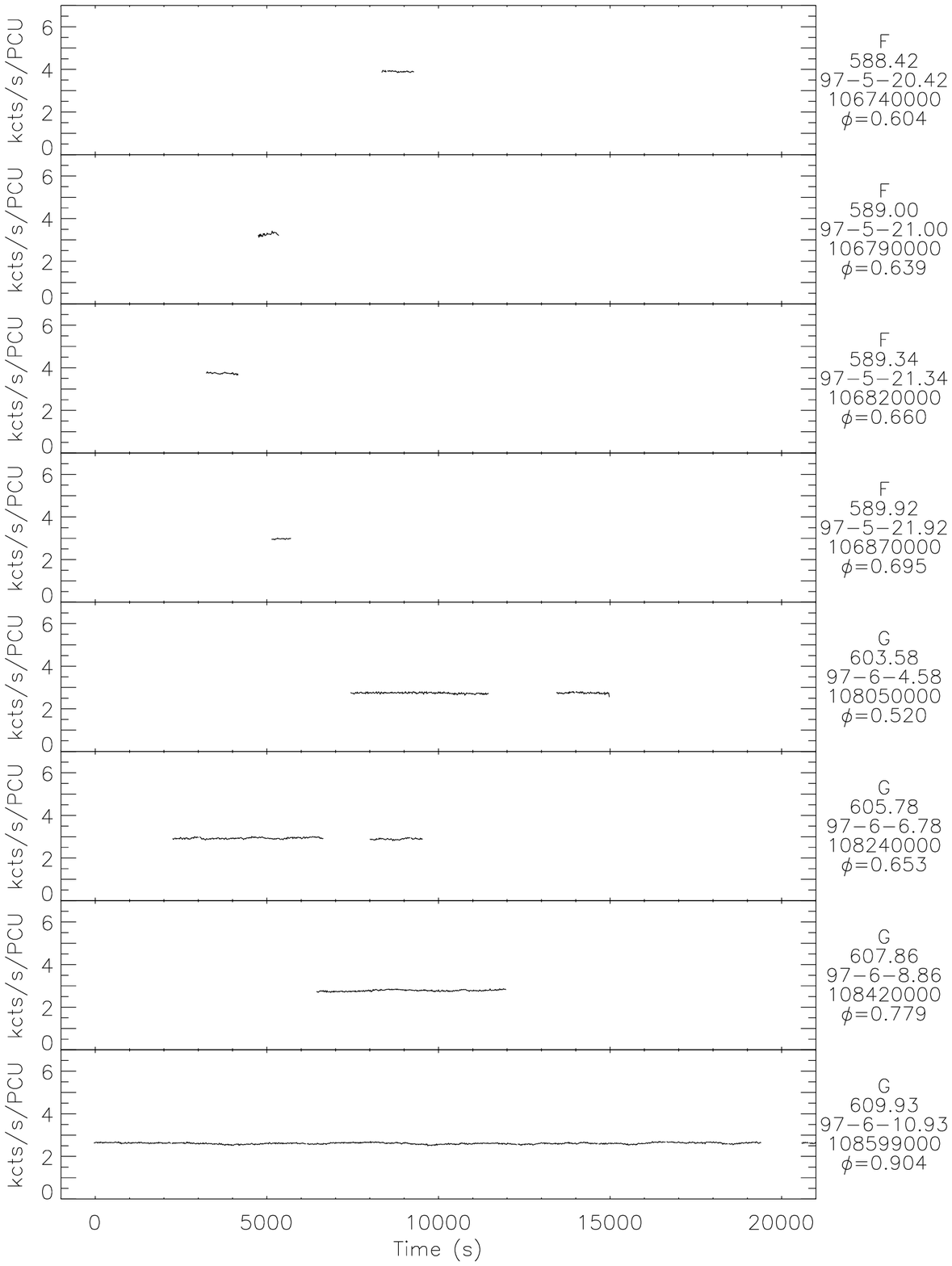 voffset=-78 hoffset=-92}{6.5in}{8.55in}
\PSbox{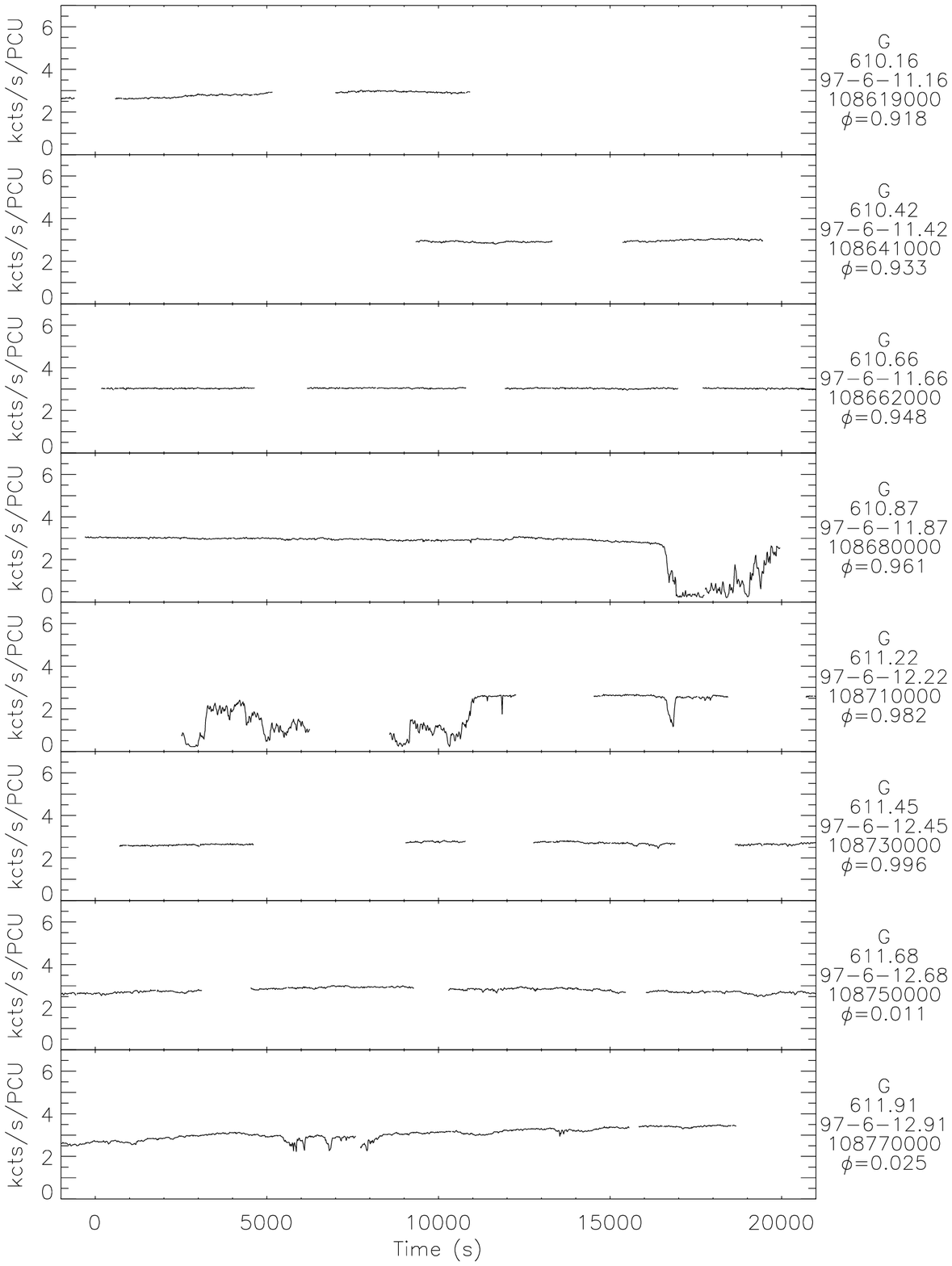 voffset=-78 hoffset=-92}{6.5in}{8.55in}
\PSbox{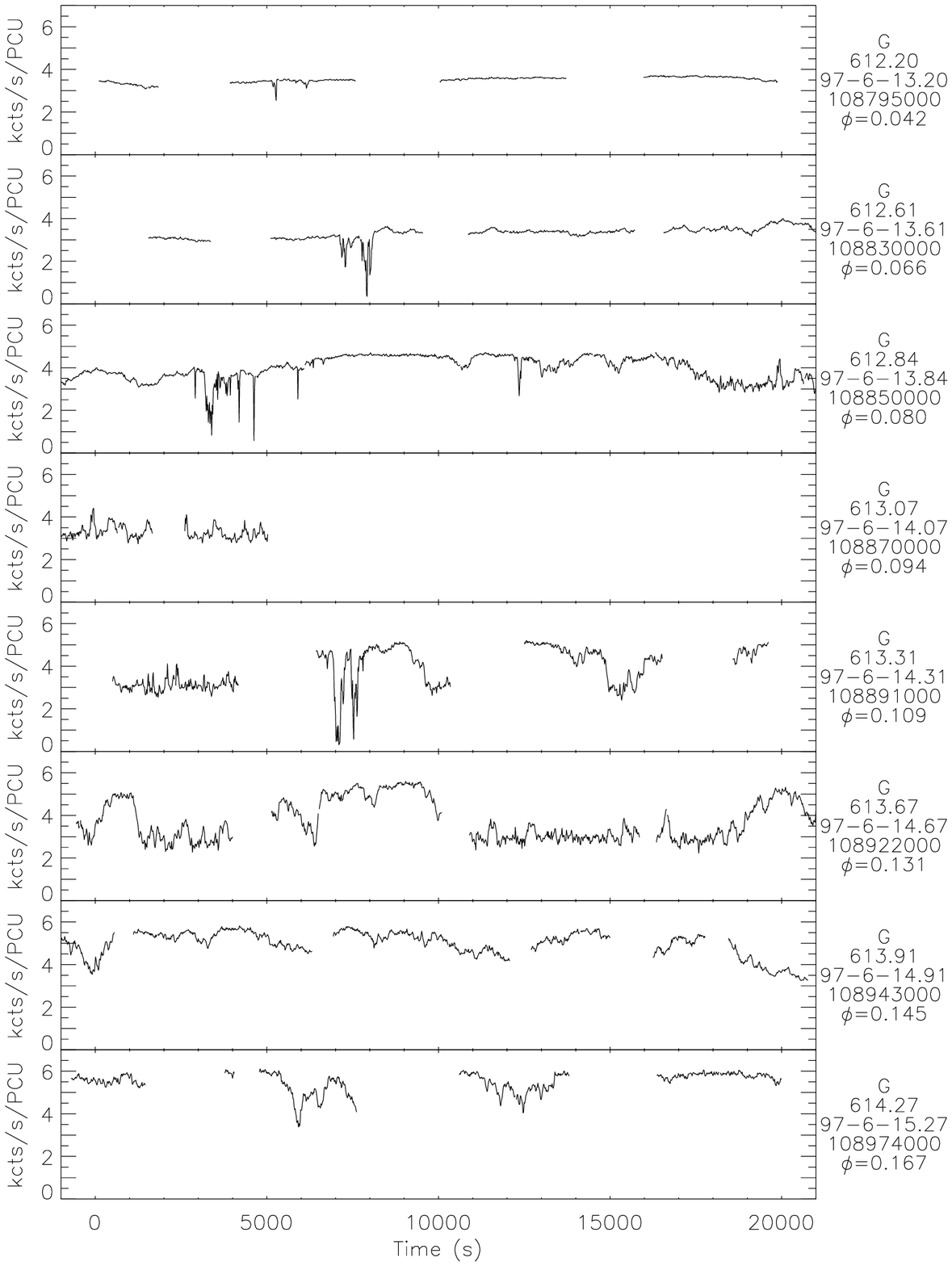 voffset=-78 hoffset=-92}{6.5in}{8.55in}
\PSbox{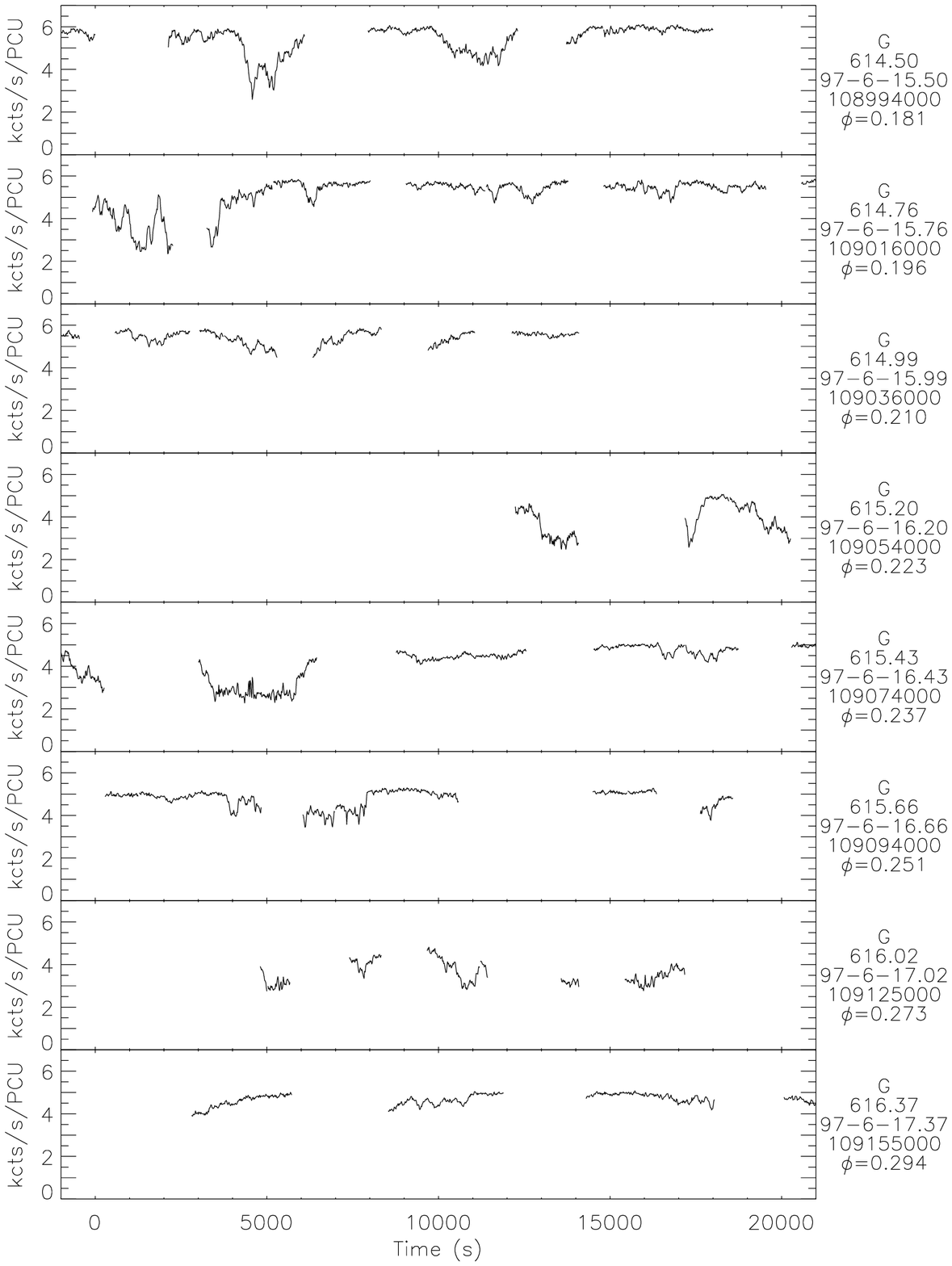 voffset=-78 hoffset=-92}{6.5in}{8.55in}
\PSbox{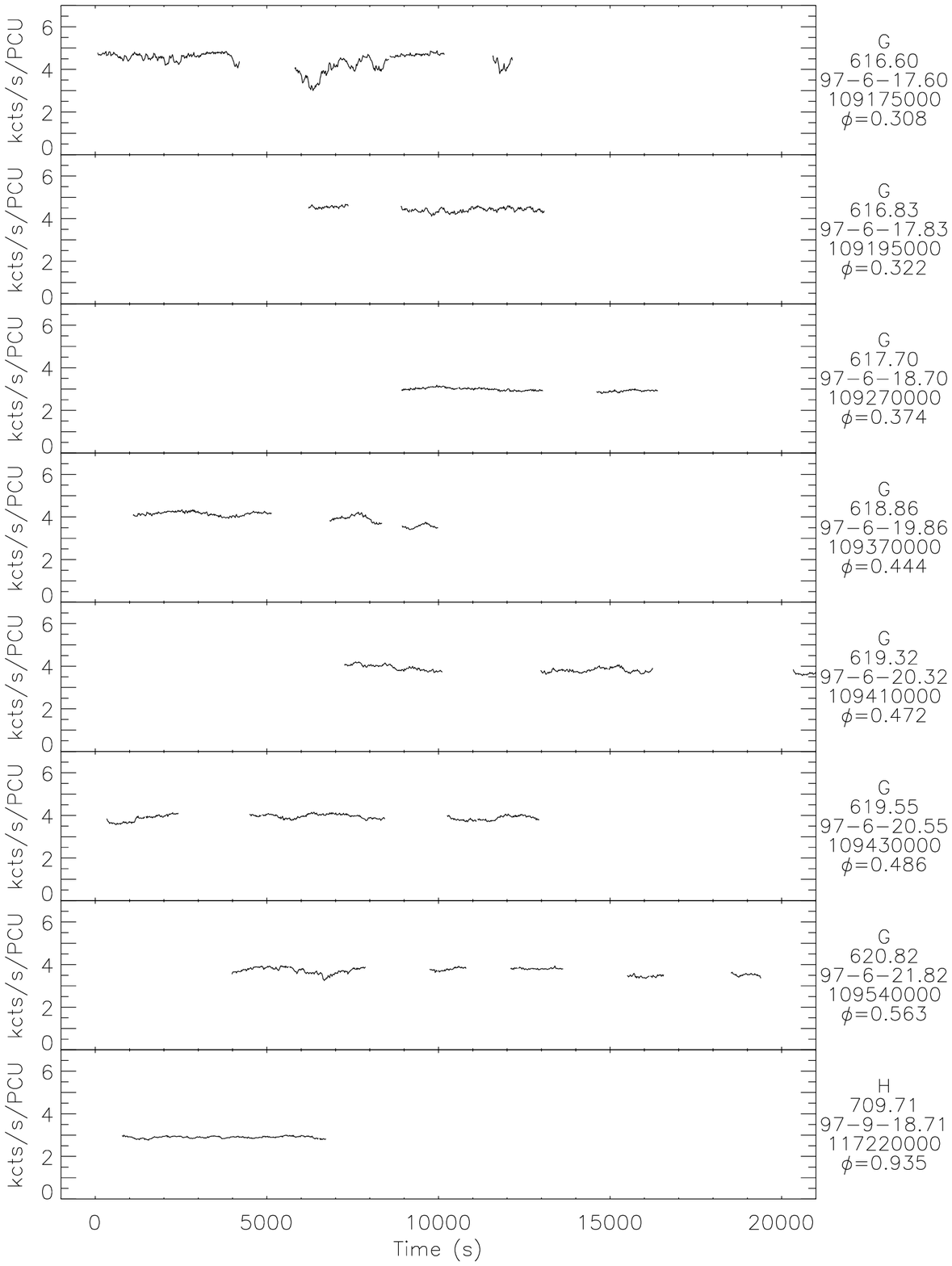 voffset=-78 hoffset=-92}{6.5in}{8.55in}
\PSbox{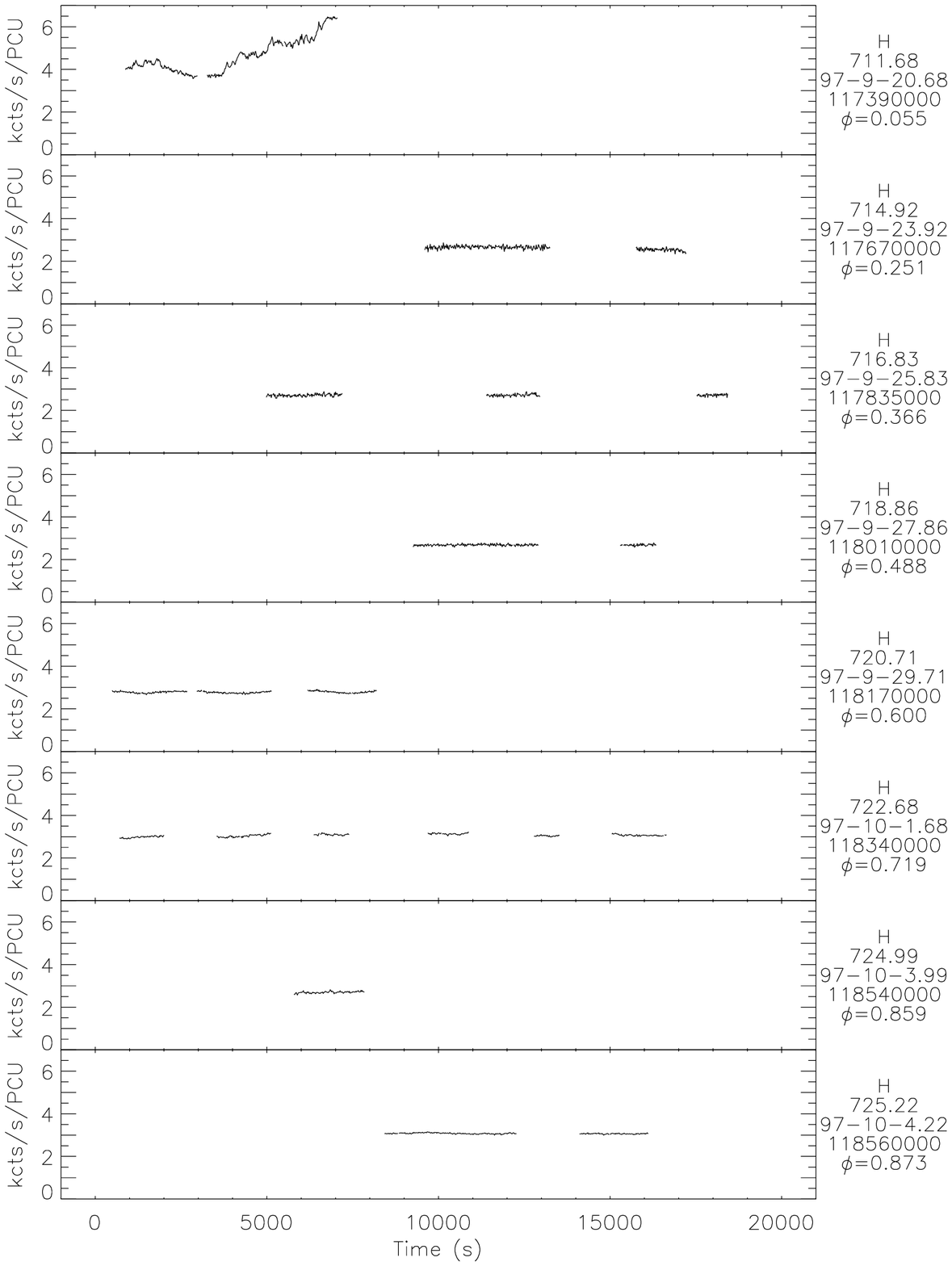 voffset=-78 hoffset=-92}{6.5in}{8.55in}
\end{centering}

%% This defines the bibliography file (main.bib) and the bibliography style.
%% If you want to create a bibliography file by hand, change the contents of
%% this file to a `thebibliography' environment.  For more information 
%% see section 4.3 of the LaTeX manual.
\bibliography{main}
\bibliographystyle{plain}

% causes all entries to be included in source list, even if not cited.
\nocite{*} 

\end{document}